\documentclass[12pt, a4paper]{article}

\RequirePackage[l2tabu, orthodox]{nag}

\usepackage{silence}
\ActivateWarningFilters[pdftoc]

\usepackage[top=18mm,bottom=18mm,left=25mm,right=25mm]{geometry}

\usepackage{amsmath,amssymb,amsfonts}

\usepackage[table]{xcolor}

\usepackage{centernot}

\newcommand{\aline}{\hspace{.5cm}\hfil\rule{0.93\textwidth}{.4pt}}

\usepackage{makeidx}
\makeindex
\usepackage[nottoc]{tocbibind}

\newcommand{\myindex}[1]{\textbf{\boldmath #1}}

\usepackage{pgf}
\usepackage{tikz}
\usetikzlibrary {shadings} 
\usetikzlibrary{shapes}
\usetikzlibrary{shapes.misc}
\tikzset{cross/.style={cross out, draw=black, thick, fill=none, minimum size=2*(#1-\pgflinewidth), inner sep=0pt, outer sep=0pt}, cross/.default={3pt}}
\newcommand{\uvertex}{
\coordinate (base) at (0, 1);
\draw (0, 0) node[fill, circle, minimum size = 1.3mm, inner sep = 0]{};
  \draw (3, 0) node [cross]{};
  \draw (0, 3) node[fill, circle, minimum size = 1.3mm, inner sep = 0]{};
  \draw (3, 3) node[fill, circle, minimum size = 1.3mm, inner sep = 0]{};
}
\newcommand{\qvertex}{
\coordinate (base) at (0, 1);
\draw (0, 0) node [cross]{};
  \draw (3, 0) node [cross]{};
  \draw (0, 3) node[cross]{};
  \draw (3, 3) node[cross]{};
}
\newcommand{\vertices}{
\coordinate (base) at (0, 1);
\draw (0, 0) node[fill, circle, minimum size = 1.3mm, inner sep = 0]{};
  \draw (3, 0) node[fill, circle, minimum size = 1.3mm, inner sep = 0]{};
  \draw (0, 3) node[fill, circle, minimum size = 1.3mm, inner sep = 0]{};
  \draw (3, 3) node[fill, circle, minimum size = 1.3mm, inner sep = 0]{};
}

\usepackage{multirow}

\usepackage{xcolor}

\newcommand*{\colorboxed}{}
\def\colorboxed#1#{%
  \colorboxedAux{#1}%
}
\newcommand*{\colorboxedAux}[3]{%
  \begingroup
    \colorlet{cb@saved}{.}%
    \color#1{#2}%
    \boxed{%
      \color{cb@saved}%
      #3%
    }%
  \endgroup
}

\usepackage[colorlinks=true,linktoc=all,linkcolor=black,citecolor=red,urlcolor=blue,backref=page]{hyperref}

\usepackage{etoolbox}
\makeatletter
\patchcmd{\BR@backref}{\newblock}{\newblock(}{}{}
\patchcmd{\BR@backref}{\par}{)\par}{}{}
\makeatother

\newcommand{\kac}[2]{
 \draw[latex-latex] (#1-.3, .5) -- (.5, .5) -- (.5, #2-.3);
 \node [below] at (#1-.3, .4) {$r$};
 \node [left] at (.4, #2-.3) {$s$};
 \foreach[parse=true] \r in {1,...,#1-1}{
 \draw (\r, .6) -- (\r, .4) node[below] {$\r$};
 }
 \foreach[parse=true] \s in {1,...,#2-1}{
 \draw (.6, \s) -- (.4, \s) node[left] {$\s$};
 }
 \def\p{#1};
 \def\q{#2};
 }
 \newcommand{\thiskac}[3]{
 \node at (#1, #2) {$#3$};
 \node at (\p-#1, \q-#2) {$#3$};
 }

\renewcommand{\arraystretch}{1.5}

\title{\bfseries Exactly solvable conformal field theories}

\author{Sylvain Ribault \vspace{2mm}
\\
{\normalsize CEA Saclay, Institut de Physique Th\'eorique}
 \\
 {\footnotesize \ttfamily sylvain.ribault@ipht.fr }
}

\begin{document}

\maketitle

\begin{abstract}
We review 2d CFT in the bootstrap approach, and sketch the known exactly solvable CFTs with no extended chiral symmetry: Liouville theory, (generalized) minimal models, limits thereof, and loop CFTs, including the $O(n)$, Potts and $PSU(n)$ CFTs. 

Exact solvability relies on local conformal symmetry, and on the existence of degenerate fields. We show how these assumptions constrain the spectrum and correlation functions. We discuss how crossing symmetry equations can be solved analytically and/or numerically, leading to analytic expressions for structure constants in terms of the double Gamma function.  

In the case of loop CFTs, we sketch the corresponding statistical models, and derive the relation between statistical and CFT variables. We review the resulting combinatorial description of correlation functions, and discuss what remains to be done for solving the CFTs. 
\end{abstract}

\clearpage

\tableofcontents

\hypersetup{linkcolor=blue}

\numberwithin{equation}{section}
\setcounter{section}{-1}

\pagebreak

\section{Introduction}

\subsubsection*{Exact solvability in CFT}

In 1997, Di Francesco, Mathieu and Sénéchal published a book called Conformal field theory \cite{fms97}, which dealt almost exclusively with conformal field theory \textit{in two dimensions}. Since then, there has been much progress on CFT in $d\geq 3$, thanks in particular to the development and systematic use of the conformal bootstrap approach \cite{prv18}. However, when it comes to CFTs that we can solve exactly, the nontrivial cases are still all in $d=2$, and their solvability relies on 2 ingredients:
\begin{itemize}
 \item An infinite-dimensional algebra of conformal transformations, called the Virasoro algebra. In general dimension, the algebra $\mathfrak{so}(d+1,1)$ of conformal transformations of the flat Euclidean space $\mathbb{R}^d$ is finite-dimensional. 
 \item Degenerate fields: fields whose products with any other fields may be written as sums of finitely many fields, modulo conformal symmetry. Degenerate fields can be constructed not only in $d=2$, but also in general dimension, where they are called weight-shifting operators \cite{kks17}. Under an OPE with a weight-shifting operator, the conformal dimension of a field is shifted by an integer. But except in special cases, the spectrum is not stable under such shifts. So weight-shifting operators cannot be consistenly added to the CFT, and do not constrain it, although they are still useful for studying conformal blocks. 
\end{itemize}
In $d=2$, these ingredients give us access to powerful bootstrap techniques, analytic and numerical, which underlie the solvability of physically well-motivated CFTs such as the Ising model or more generally the $O(n)$ model.
Even then, solvable CFTs are very special cases. Starting from a simple CFT and following a renormalization group flow, we can easily end up at a fixed point whose CFT is intractable \cite{ab24}.

\subsubsection*{Not much exact solvability in $d\geq 3$}

In $d\geq 3$, there are no known techniques that could lead to exact solutions for the Ising or $O(n)$ models. The main bootstrap techniques are numerical instead, and rely on the assumption of unitarity \cite{prv18}. This not only restricts the CFTs that can be addressed, but also leads to inequalities rather than equalities, making it harder to reach exact results. 
Exact solvability is restricted to toy models: CFTs that are constructed for the purpose of allowing exact calculations, rather than for answering physical questions: 
\begin{itemize}
 \item Mean field theory, also known as the theory of generalized free fields. In this CFT, fields come in infinite series, with conformal dimensions that differ by integers. Weight-shifting operators can be used for computing structure constants, whose exact expressions are written in terms of the Gamma function \cite{kks18}.
 \item The Parisi--Sourlas uplift leads to the construction of a $d=4$ CFT from any $d=2$ CFT. If the initial $d=2$ CFT is exactly solved, this leads to exact expressions for a class of correlation functions in the $d=4$ CFT \cite{tre24}. It remains to be seen whether the full $d=4$ CFT is exactly solvable.
\end{itemize}
There are other examples of exact results \cite{vtrz+25}, most notably in CFTs with supersymmetry, but they are restricted to special observables, and do not imply that all correlation functions of the CFT can be exactly computed. Nevertheless, some exact results might be hints of exact solvability:
\begin{itemize}
 \item In $d=4$, the planar conformal fishnet theory is a non-supersymmetric CFT where quite a few conformal dimensions and correlation functions can be exactly computed \cite{gkk18}. However the theory is admittedly a toy model, and is still far from being exactly solved.
 \item Harmonic analysis on groups or group quotients provides a generic way of constructing CFTs, and generalizations thereof. Of course, the resulting theories are typically not exactly solvable \cite{kmp21}. However, harmonic analysis on $H_3^+=SL_2(\mathbb{C})/SU_2$ can be interpreted as an exactly solvable $d=2$ CFT with only global conformal symmetry, also known as the light asymptotic limit of Liouville theory \cite{rib14}. This raises the possibility that other group quotients might give rise to exactly solvable $d\geq 3$ CFTs. 
\end{itemize}

\subsubsection*{Solvable CFTs in $d=2$}

We will first deal with CFTs that are considered solved, in that we exactly know the 3-point structure constants, and can compute any correlation function to arbitrary precision:
\begin{itemize}
 \item Liouville theory,
 \item (generalized) minimal models, and limits thereof. 
\end{itemize}
We will also propose a unified treatment of loop CFTs, which are not solved but are believed to be solvable, because we have sufficiently many exact results. By loop CFTs we mean a large class of CFTs, with various global symmetry groups, which describe critical limits of various statistical models in various phases.

The known solvable CFTs have many applications to statistical physics, in particular to percolation, to polymers, and more recently to active hydraulics \cite{jcpb23}. Liouville theory and minimal models have well-known applications to quantum gravity in 2d. Applications of non-diagonal solvable CFTs such as loop CFTs to quantum gravity in 3d are limited by the bound on the central charge $\Re c < 13$. (See Section \ref{sec:dotc}.) A CFT that would be holographically dual to quantum gravity in 3d Anti-de Sitter space would need to exist for $c\to+\infty$, while for 3d de Sitter space we would need $\Re c = 13$ \cite{god24, cemr24}.

We will not cover solved or solvable CFTs that have extended chiral symmetries: symmetry algebras strictly larger than the conformal algebra, while being left-right factorized like the conformal algebra. Some CFTs of that type can be solved using the methods that we will cover, for example certain Wess--Zumino--Witten models.

\subsubsection*{How to solve CFTs in $d=2$}

When a CFT is solvable, it can typically be solved in several different ways. In this text, we will only use the conformal bootstrap approach, and omit other historically relevant techniques, in particular:
\begin{itemize}
 \item The modular bootstrap consists in exploiting modular invariance of the torus partition function. This object has the advantage of being technically simple, since it involves only the spectrum and not the structure constants. The corresponding drawback is that modular invariance is neither necessary nor sufficient for consistency of the CFT.
 \item Coulomb gas integrals can be used for deriving exact results in some models, by writing them as perturbed free field theories. With this technique, the conservation of momentum in the free theories limits the correlation functions that can be computed. In some cases, this limitation can be overcome to some extent: for example, torus partition functions in loop CFTs can be computed as infinite combinations of partition functions of free theories with different radiuses of compactification \cite{fsz87}.
\end{itemize}
We believe that the conformal bootstrap approach is technically simpler, more general, and more powerful. In particular, our main objects are correlation functions on the sphere, and we do not have a rigid notion of which fields belong to a CFT. For example, in the Ising model, we can introduce disorder fields or compute cluster connectivities, without having to worry about modular invariance of the torus partition function.

\subsubsection*{Conformal bootstrap approach}

Solving a CFT means determining its spectrum and structure constants. We deduce spectra from assumptions such as
\begin{itemize}
 \item the existence of degenerate fields,
 \item single-valuedness of correlation functions,
 \item analyticity in the central charge or in conformal dimensions, which in particular allows us to find CFTs as limits of sequences of other CFTs. 
\end{itemize}
As first demonstrated by Teschner in the case of Liouville theory \cite{tes95},
these assumptions also lead to constraints on structure constants, which may or may not be enough for determining them analytically. If they are not enough, we can determine structure constants numerically by solving crossing symmetry equations. And from sufficiently precise numerical results, we can deduce analytic formulas.

While a CFT can be defined as a solution of bootstrap equations, the spectrum and structure constants of solvable CFTs are determined by solving only a small subset of equations. And it is very hard to prove that they solve all equations. In other words, the bootstrap approach is better at addressing uniqueness than existence of CFTs. In the case of Liouville theory with $c\geq 25$, a rigorous probabilistic construction is known. In the probabilistic approach, the theory is solved using bootstrap ideas, and it can be proved that all bootstrap equations are obeyed \cite{gkr24}. For most other solvable CFTs, existence is not proved, but it is supported by various types of evidence, such as lattice constructions, field theory constructions, or numerical checks of bootstrap equations.

\subsubsection*{Related resources}

We strive to give simple derivations of nontrivial results, without introducing too many auxiliary objects that are later discarded. The calculations are not all done in detail in this text, but they are doable with reasonable effort. Readers who want formal exercises are directed to earlier texts in the same spirit: the minimal review \cite{rib16}, and the longer review \cite{rib14}. Further exercises can be found in a Wikiversity course \cite{rib25}, which is mostly based on this text.

Technical terms that are defined in this text can be found in the index. We also use standard terms that we do not define, because adequate definitions are easily found in Wikipedia or other resources. For example, the definitions of an \textit{irreducible representation}, an \textit{indecomposable representation} or a \textit{projective representation} are found in Wikipedia as the first results of searches for these terms.

\subsubsection*{Acknowledgements}

\begin{itemize}
 \item This text is based on lecture notes for a course given at IPhT Saclay in March-April 2024. Video recordings are available on \href{https://www.youtube.com/playlist?list=PLrLctLPAdPNtD7yFOcfdNx2J-A_CE1svr}{IPhT-TV}. I am grateful to the course organizers Riccardo Guida, Pierfrancesco Urbani and Monica Guica, and to the participants. I am also grateful to the students who followed my M2 ICFP course in 2025 and 2026.
 \item I would like to thank Linnea Grans-Samuelsson, Jesper Jacobsen, Santiago Migliaccio, Nikita Nemkov, Rongvoram Nivesvivat, Marco Picco, Paul Roux, Hubert Saleur, and Raoul Santachiara for collaboration on loop CFTs and related subjects. 
 \item I am grateful to Antoine Bourget, Victor Godet, Jesper Jacobsen, Gregory Korchemsky, Paul Roux, Ingo Runkel, Hubert Saleur, Xi Yin and Bernardo Zan for helpful discussions and correspondence.
 \item I would like to thank Max Downing, Kay Wiese, Rongvoram Nivesvivat and Paul Roux for comments and suggestions on the manuscript. 
 \item I am grateful to the 2 anonymous SciPost reviewers for their valuable suggestions, which led to many corrections and clarifications, and to the editor Monica Guica.
 \item This work is partly a result of the project ReNewQuantum, which received funding from the European Research Council.
\end{itemize}

\section{Basics of 2d CFT}\label{sec:bo}

In this section we introduce conformal symmetry, fields, operator product expansions, and correlation functions. In particular, we emphasize the 2 ingredients of exact solvability: \textit{local} conformal symmetry and \textit{degenerate} fields. We begin with the Virasoro algebra, which describes local conformal transformations. 

\subsection{The Virasoro algebra and its representations}\label{sec:vir}

\subsubsection{The Virasoro algebra}

Let us consider the Riemann sphere $\overline{\mathbb{C}}=\mathbb{C}\cup \{\infty\}$, equipped with the metric $ds^2 = dzd\bar z$. By definition, a \myindex{conformal transformation}\index{conformal transformation} of a Riemannian manifold is a transformation that preserves angles. On any open subset of the Riemann sphere, any holomorphic map $z\to f(z)$ is conformal, because it transforms the metric into itself, up to a scalar factor: 
\begin{align}
 dzd\bar z\to dfd\bar f = |f'(z)|^2 dzd\bar z\ .
\end{align}
Conversely, any conformal map is holomorphic. 
In a quantum field theory on the Riemann sphere, states live on a constant time slice $|z|=1$, so we consider conformal transformations of that circle, equivalently of $\mathbb{C}^*= \mathbb{C}\backslash \{0\}$.
Let the \myindex{Witt algebra}\index{Witt algebra} be the algebra of infinitesimal conformal transformation of $\mathbb{C}^*$: this algebra is infinite-dimensional, with the basis 
\begin{align}
 \left(\ell_n\right)_{n\in\mathbb{Z}}  \quad \text{with} \quad \ell_n = -z^{n+1}\frac{\partial}{\partial z}\ ,
 \label{lpz}
\end{align}
and the commutation relations 
\begin{align}
 [\ell_n,\ell_m] = (n-m)\ell_{m+n}\ .
\end{align}
The generators of the Witt algebra include the translation generator $\ell_{-1} = -\frac{\partial}{\partial z}$, and the dilation generator $\ell_0 = -z\frac{\partial}{\partial z}$. In fact, $(\ell_{-1},\ell_0,\ell_1)$ generate the infinitesimal conformal transformations of the Riemann sphere.
The corresponding Lie group is the group of \myindex{global conformal transformations}\index{conformal transformation!---global} $PSL_2(\mathbb{C})$, whose elements act as 
\begin{align}
 z \mapsto \frac{az+b}{cz+d}\quad , \quad (a,b,c,d\in \mathbb{C},\ ad-bc\neq 0)\ .
 \label{abcd}
\end{align}
Now, in a quantum theory, symmetries act projectively on states. And projective representations of an algebra are equivalent to representations of that algebra's central extension. 
Therefore, the algebra that describes local conformal transformations in conformal field theory is the Witt algebra's central extension, called the Virasoro algebra. The \myindex{Virasoro algebra}\index{Virasoro!---algebra} $\mathfrak{V}$ has generators $(L_n)_{n\in\mathbb{Z}}$ that correspond to the Witt algebra generators, plus a central generator. We assume that the central generator has only one eigenvalue in a given CFT, called the \myindex{central charge}\index{central charge} $c\in\mathbb{C}$. We replace the central generator with $c$ in the commutation relations:
 \begin{align}
  \boxed{[L_n,L_m] = (n-m)L_{n+m} +\frac{c}{12}(n-1)n(n+1)\delta_{n+m,0}} \ .
  \label{vir}
 \end{align}
Then we consider $c$ as a parameter not only of a CFT, but also of the underlying Virasoro algebra.

\subsubsection{Highest-weight representations}\label{sec:hwr}

In a conformal field theory, the space of states is a representation of the Virasoro algebra, and can be decomposed into indecomposable representations. But which indecomposable representations are physically relevant? To answer this question, we will focus on the properties of the dilation generator $L_0$ of the Virasoro algebra. Conceptually, this is because $L_0$ can be interpreted as the energy operator. Technically, this is because $L_0$ controls the convergence of operator product expansions, as we will see in Section \ref{sec:csope}. More specifically, a necessary condition for convergence is that the eigenvalues of $L_0$ be bounded from below in any indecomposable representation, whether or not $L_0$ is diagonalizable in that representation. We will always assume that this condition holds.

The eigenvalues of $L_0$ are called \myindex{conformal dimensions}\index{conformal!---dimension}. Under the action of the Virasoro generator $L_n$, the conformal dimension decreases by $n$. For any vector $V$ in a representation of the Virasoro algebra, we indeed have
\begin{align}
 L_0V{} = \Delta V{} \quad \Rightarrow\quad  L_0 L_nV{} = L_nL_0V{} + [L_0, L_n] V{}  = (\Delta-n)L_nV{} \ .
\end{align}
In an indecomposable representation $\mathcal{R}$, the $L_0$-eigenstate with the lowest eigenvalue must therefore be annihilated by $L_{n>0}$. This eigenstate is a \myindex{primary state}\index{primary!---state}, where we define a primary state $V_\Delta$ of conformal dimension $\Delta$ by 
\begin{align}
  \boxed{L_0 V_\Delta = \Delta V_\Delta \quad , \quad L_{n>0} V_\Delta = 0}\ .
 \end{align}
Let us introduce a basis of \myindex{creation operators}\index{creation operator},
\begin{align}
 \mathcal{L} = \left\{\prod_{i=1}^k L_{-n_i} \right\}_{k\in\mathbb{N},\ 0<n_1\leq \dots \leq n_k}=\left\{1, L_{-1}, L_{-1}^2, L_{-2},\cdots \right\}\ ,
 \label{lcm}
\end{align}
where the unit operator $1\in \mathcal{L}$ is obtained in the case $k=0$.
Let  $N=|L|=\sum_{i=1}^k n_i \in\mathbb{N}$ be the  \myindex{level}\index{level} of $|L|$, then we write
\begin{align}
 \mathcal{L}_N = \Big\{ L\in\mathcal{L}\Big| |L|=N\Big\}\ . 
\label{ln}
 \end{align}
Let us sketch our basis of creation operators up to the level $N=4$, with arrows representing the action of the Virasoro generators $L_{-1},L_{-2},L_{-3},L_{-4}$:
 \begin{align}
 \begin{tikzpicture}[scale = .3, baseline=(current  bounding  box.center)]
  \draw[-latex, very thick] (20, 0) -- (20, -27) node [right] {$N$};
  \foreach \x in {0, ..., 4}
  {
  \draw [dotted] (-20, {-6*\x}) -- (20, {-6*\x}) node [right] {${\x}$};
  }
  \node[fill = white] at (0, 0) (0) {$1$};
  \node[fill = white] at (-3,-6) (1) {$L_{-1}$};
  \node[fill = white] at (-6, -12) (11) {$L_{-1}^2$};
  \node[fill = white] at (-9, -18) (111) {$L_{-1}^3$};
  \node[fill = white] at (-12, -24) (1111) {$L_{-1}^4$};
  \node[fill = white] at (0,-12) (2) {$L_{-2}$};
  \node[fill = white] at (-3,-18) (12) {$L_{-1}L_{-2}$};
  \node[fill = white] at (-6,-24) (112) {$L_{-1}^2L_{-2}$};
  \node[fill = white] at (0,-24) (22) {$L_{-2}^2$};
  \node[fill = white] at (6,-18) (3) {$L_{-3}$};
  \node[fill = white] at (6,-24) (13) {$L_{-1}L_{-3}$};
  \node[fill = white] at (12,-24) (4) {$L_{-4}$};
  \draw[-latex] (0) -- (1);
  \draw[-latex] (1) -- (11);
  \draw[-latex] (11) -- (111);
  \draw[-latex] (111) -- (1111);
  \draw[-latex] (0) -- (2);
  \draw[-latex] (0) -- (3);
  \draw[-latex] (0) -- (4);
  \draw[-latex] (2) -- (12);
  \draw[-latex] (12) -- (112);
  \draw[-latex] (2) -- (22);
  \draw[-latex] (3) -- (13);
 \end{tikzpicture}
 \label{verma}
\end{align}
Operators such as
$L_{-2}L_{-1} = L_{-1}L_{-2} -L_{-3}$ are linear combination of elements of our basis, and are therefore not displayed.

Let $V_\Delta$ be a primary state of dimension $\Delta$.
Any state of the type $L V_\Delta$ with $L\in\mathcal{L}$ is called a \myindex{descendant state}\index{descendant!---state} of $V$ of level $|L|$. Its conformal dimension is $\Delta+|L|$.
By extension, a linear combination of descendant states is also called a descendant state.

A primary state $V{} \in \mathcal{R}$ generates a subrepresentation $\mathcal{R}_{V{}}= \text{Span}(LV{})_{L\in\mathcal{L}}\subset\mathcal{R}$, which is the space of its descendant states. If $\mathcal{R}=\mathcal{R}_V$ for some $V\in\mathcal{R}$, then $\mathcal{R}$ is called a \myindex{highest-weight representation}\index{highest-weight representation}\index{representation!highest-weight---}. Then $L_0$ is diagonalizable in $\mathcal{R}$, and $\mathcal{R}$ is indecomposable. If moreover the states $(LV{})_{L\in\mathcal{L}}$ are linearly independent, then $\mathcal{R}_{V{}}$ is called the \myindex{Verma module}\index{Verma module} $\mathcal{V}_\Delta$, for $\Delta$ the conformal dimension of $V$.
If $(LV{})_{L\in\mathcal{L}}$ are not linearly independent, then $\mathcal{R}_{V{}}$ is the quotient representation of $\mathcal{V}_\Delta$ by a nontrivial subrepresentation, namely by the kernel of the canonical surjective map $\mathcal{V}_\Delta\to \mathcal{R}$.

It is sometimes useful to consider representations that are not highest-weight. This is in particular the case of \textbf{logarithmic representations}\index{logarithmic!---representation}\index{representation!logarithmic---}, which we define as indecomposable representations where $L_0$ is bounded from below but not diagonalizable. A logarithmic representation is always reducible, because the highest-weight representation generated by the $L_0$-eigenvector with the smallest eigenvalue is a nontrivial subrepresentation. For example, a simple logarithmic representation can be constructed from the formal $\Delta$-derivative of a primary state $V_\Delta$, which obeys $L_0 V'_\Delta = \Delta V'_\Delta + V_\Delta$.

\subsubsection{Singular vectors of Verma modules}\label{sec:nvvm}

In a representation of the Virasoro algebra, we define a \myindex{singular vector}\index{singular vector} as a primary state that is a descendant of another primary state. Singular vectors play an important structural role: a highest-weight representation is irreducible if and only if it has no singular vector.

Let us look for singular vectors in the Verma module $\mathcal{V}_\Delta$. This can be done level by level, starting from level $1$. That level is generated by a single state $L_{-1}V_\Delta$. To determine whether that state is primary, let us determine how it behaves under the action of $L_{n\geq 1}$:
\begin{align}
L_n L_{-1}V_\Delta = [L_n, L_{-1}] V_\Delta = (n+1) L_{n-1}V_\Delta = 
\left\{\begin{array}{ll} 0 &  \quad \text{if } n\geq 2\ , \\ 2\Delta V_\Delta & \quad \text{if } n = 1\ . \end{array}\right. 
\end{align}
Therefore, $\mathcal{V}_\Delta$ has a singular vector at level $1$ if and only if $\Delta=0$. Next, consider the case of level $2$, where we write states as
\begin{align}
 X{} = \left(\lambda L_{-1}^2 + \mu L_{-2}\right) V_\Delta\ ,
\end{align}
where $\lambda$ and $\mu$ are complex coefficients. It is easy to show that $L_{n\geq 3}X{}=0$, and we compute 
\begin{subequations}
\label{l1x}
\begin{align}
 L_1X{} &= \left((4\Delta+2)\lambda + 3\mu\right) L_{-1}V_\Delta\ ,
\\
L_2 X{} &= \left(6\Delta \lambda+(4\Delta+\tfrac12 c) \mu\right)V_\Delta\ .
\end{align}
\end{subequations}
Therefore, the conditions $L_1X{}=L_2 X{}=0$ for $X{}$ to be a singular vector boil down to a system of 2 linear equations for the coefficients $(\lambda,\mu)$. The system has a nonzero solution if and only if its determinant vanishes, which amounts to
\begin{align}
 \Delta = \frac{1}{16}\left( 5-c\pm\sqrt{(c-1)(c-25)} \right) \ .
 \label{dpm}
\end{align}
Solving CFTs will lead to fairly complicated formulas, which would become even more complicated if we tolerated a square root at this early stage.
To get rid of the square root, we introduce a new notation for the central charge: a parameter $\beta$ defined by
\begin{align}
 \boxed{c = 1- 6\left(\beta - \beta^{-1}\right)^2 } \ .
 \label{cb}
\end{align}
Then the conformal dimensions \eqref{dpm} become 
$
 \Delta = -\frac12 + \frac34\beta^{\pm 2}
$, and the corresponding singular vectors are $X\propto \left(L_{-1}^2 -\beta^{\pm 2}L_{-2}\right)V_\Delta$.
The price to pay for this simplification is that the 4 values $\beta,-\beta,\beta^{-1},-\beta^{-1}$ all correspond to the same central charge. 

More generally, there is an infinite family of conformal dimensions $(\Delta_{(r, s)})_{r,s\in\mathbb{N}^*}$ such that $\mathcal{V}_{\Delta_{(r, s)}}$ has a singular vector $L_{\langle r,s\rangle} V_{\Delta_{(r, s)}}$ at level $N=rs$. The first few cases are:
\begin{align}
\renewcommand{\arraystretch}{1.5}
\begin{array}{|c|c|c|l|}
\hline 
N & (r,s) & \Delta_{(r,s)} & L_{\langle r,s\rangle} 
\\
\hline\hline
1 & (1,1) & 0  & L_{-1}
\\
\hline
\multirow{2}{*}{2} & 
(2,1) & -\frac12 +\frac{3}{4} \beta^2  & L_{-1}^2 -\beta^2 L_{-2}
\\
\cline{2-4}
& (1,2) & -\frac12 + \frac{3}{4}\beta^{-2}  & L_{-1}^2 -\beta^{-2} L_{-2} 
\\
\hline
\multirow{2}{*}{3} &
(3,1 ) &  -1 +2 \beta^2  &L_{-1}^3 -4\beta^2 L_{-1}L_{-2}+2\beta^2(2\beta^2+1)L_{-3}
\\
\cline{2-4}
& (1,3 ) &  -1 +2 \beta^{-2}  &L_{-1}^3 -4\beta^{-2} L_{-1}L_{-2}+2\beta^{-2}(2\beta^{-2}+1)L_{-3}
\\
\hline
\multirow{5}{*}{4} &
(4,1 ) &  -\frac32 +\frac{15}{4} \beta^2  & 
\begin{array}{r}
 L_{-1}^4 -10\beta^2L_{-1}^2L_{-2} 
 +2\beta^2\left(12\beta^2+5\right) L_{-1}L_{-3}
  \\
  +9\beta^4 L_{-2}^2 
 -6\beta^2\left(6\beta^4+4\beta^2+1\right)L_{-4}
\end{array}
\\
\cline{2-4}
& (2,2) & \frac34\left(\beta-\beta^{-1}\right)^2 & 
\begin{array}{l}
L_{-1}^4-2\left(\beta^2+\beta^{-2}\right)L_{-1}^2L_{-2} 
+\left(\beta^2-\beta^{-2}\right)^2 L_{-2}^2
\\ 
+2\left(1+\left(\beta+\beta^{-1}\right)^2\right) L_{-1}L_{-3} 
 -2\left(\beta+\beta^{-1}\right)^2 L_{-4}
\end{array}
\\
\cline{2-4}
& (1,4 ) &  -\frac32 +\frac{15}{4} \beta^{-2} &
\begin{array}{r}
 L_{-1}^4 -10\beta^{-2}L_{-1}^2L_{-2} 
 +2\beta^{-2}\left(12\beta^{-2}+5\right) L_{-1}L_{-3}
  \\
  +9\beta^{-4} L_{-2}^2 
 -6\beta^{-2}\left(6\beta^{-4}+4\beta^{-2}+1\right)L_{-4}
\end{array}
\\
\hline
\end{array}
\label{ars}
\end{align}
In these cases as in the general case, the dimensions $\Delta_{(r,s)}$ and singular vectors $L_{\langle r,s\rangle}$ have simple behaviours under the transformations $\beta\to -\beta$ and $\beta\to \beta^{-1}$, which leave the central charge \eqref{cb} invariant:
\begin{align}
 \Delta_{(r,s)}(\beta)&=\Delta_{(r,s)}(-\beta)=\Delta_{(s,r)}(\beta^{-1})\ ,
 \\
 L_{\langle r,s\rangle}(\beta)&=L_{\langle r,s\rangle}(-\beta)=L_{\langle s,r\rangle}(\beta^{-1})\ .
\end{align}
Let us introduce a new notation for the conformal dimension: the \myindex{momentum}\index{momentum} $P$ defined by
\begin{align}
 \boxed{\Delta = \frac{c-1}{24} + P^2}\ .
 \label{dp}
\end{align}
In this notation, a Verma module is now called $\mathcal{V}_P$. The \myindex{reflection}\index{reflection} $P\to -P$ leaves
$\Delta$ invariant, therefore $\mathcal{V}_P=\mathcal{V}_{-P}$. For any $r,s\in\mathbb{N}^*$, the dimension $\Delta_{(r, s)}$ corresponds to the momenta
\begin{align}
 \boxed{P_{(r, s)} = \frac12\left(r\beta -s\beta^{-1}\right)}\ ,
 \label{prs}
\end{align}
and it follows that the singular vector $L_{\langle r,s\rangle} V_{\Delta_{(r, s)}}$ is a primary state of dimension
\begin{align}
 \boxed{\Delta_{(r, s)} + rs = \Delta_{(r, -s)}}\ , 
 \label{drms}
\end{align}
We will provide a derivation of $P_{(r, s)}$ using operator product expansions in Section \ref{sec:dope}. On the other hand, the expressions of $L_{\langle r,s\rangle} $ for $rs\geq 3$ are complicated and not particularly useful, see \cite{wat24}.

For $\beta^2\in\mathbb{C}\backslash \mathbb{Q}$, the Verma module $\mathcal{V}_\Delta$ has a singular vector if and only if $\Delta=\Delta_{(r, s)}$ for some $r,s\in\mathbb{N}^*$, in which case it has only 1 singular vector. For $\beta^2\in\mathbb{Q}$ however, a Verma module can have several singular vectors. If $\beta^2 = \frac{q}{p}$ with $p,q\in\mathbb{N}^*$, we indeed have the identity
\begin{align}
 \Delta_{(r+p,s+q)} = \Delta_{(r,s)}\ ,
 \label{rpsq}
\end{align}
which together with $\Delta_{(r,s)}=\Delta_{(-r,-s)}$ implies coincidences of the type $\Delta_{(r,s)}=\Delta_{(r',s')}$ with $r',s'\in\mathbb{Z}$. 
Given $r_1,s_1\in \mathbb{N}^*$, the reducible Verma module $\mathcal{V}_{\Delta_{(r_1,s_1)}}$ has a singular vector $L_{\langle r_1,s_1\rangle} V_{\Delta_{(r,s)}}$. If $pq>0$, or if $pq<0$ with $\left\lfloor \frac{r_1}{|p|}\right\rfloor \neq \left\lfloor \frac{s_1}{|q|}\right\rfloor$ and $\left\lceil \frac{r_1}{|p|}\right\rceil \neq \left\lceil \frac{s_1}{|q|}\right\rceil$, then the dimension $\Delta_{(r_1,-s_1)}$ \eqref{drms} of this singular vector is of the type $\Delta_{(r_2,s_2\rangle}$ with $r_2,s_2\in\mathbb{N}^*$, leading to another singular vector $L_{\langle r_2,s_2\rangle}L_{\langle r_1,s_1\rangle} V_{\Delta_{(r,s)}}$.
It turns out that all singular vectors of Verma modules are of the type $L_{\langle r_1,s_1\rangle} V_{\Delta_{(r,s)}}$ or $L_{\langle r_2,s_2\rangle}L_{\langle r_1,s_1\rangle} V_{\Delta_{(r,s)}}$. Assuming singular vectors are normalized as $L_{\langle r,s\rangle} = L_{-1}^{rs} +\cdots$,
given a singular vector of the type $L_{\langle r_3,s_3\rangle}L_{\langle r_2,s_2\rangle}L_{\langle r_1,s_1\rangle} V_{\Delta_{(r_1,s_1)}}$, we indeed have
\begin{align}
 \exists r_4,s_4\in \mathbb{N}^*\ , \ L_{\langle r_3,s_3\rangle}L_{\langle r_2,s_2\rangle}L_{\langle r_1,s_1\rangle} V_{\Delta_{(r_1,s_1)}} = L_{\langle r_4,s_4\rangle} V_{\Delta_{(r_4,s_4)}}\ .
 \label{lllel}
\end{align}
To see this, let $\epsilon,\eta\in\{+,-\}$ be the signs such that $P_{(r_1,-s_1)}=\epsilon P_{(r_2,s_2)}$ and $P_{(r_2,-s_2)}=\eta P_{(r_3,s_3)}$, and let $r_4=|\epsilon r_1-r_2+\eta r_3|$ and $s_4=|\epsilon s_1+s_2+\eta s_3|$, then $\sum_{i=1}^3 r_is_i=r_4s_4$, i.e. the 2 singular vectors in Eq. \eqref{lllel} have the same level.

To summarize, numbers of singular vectors depend a lot on $\beta^2$. If $\beta^2<0$, then $\{\Delta_{(r, s)}\}_{r,s\in\mathbb{N}^*}$ is bounded from above, and Eq. \eqref{drms} implies that a Verma module can only have finitely many singular vectors. If $\beta^2\in \mathbb{Q}_{>0}$, then by Eq. \eqref{rpsq} the existence of a singular vector implies the existence of infinitely many others:
\begin{align}
 \begin{array}{|c||c|c|c|}
  \hline 
  \text{Value of }\beta^2 & \mathbb{C}\backslash \mathbb{Q} & \mathbb{Q}_{>0} & \mathbb{Q}_{<0}
  \\
  \hline 
  \text{Value of }c & \text{generic} & c\leq 1 , c\in\mathbb{Q} &  c\geq 25, c\in\mathbb{Q}
  \\
  \hline 
  \#\text{singular vectors in } \mathcal{V}_{\Delta_{(r, s)}} & 1 & \infty & \text{finite}
  \\
  \hline 
 \end{array}
\end{align}
To illustrate the 2 cases $\beta^2\in \mathbb{Q}_{>0}$ and $\beta^2\in \mathbb{Q}_{<0}$, let us plot the points $(r,s)\in \mathbb{N}^*\times \mathbb{N}^*$, with a green shade that increases with the conformal dimension $\Delta_{(r,s)}$. We also draw a red arrow for the vector $(p, q)$, along which $\Delta_{(r,s)}$ is constant according to Eq. \eqref{rpsq}:
\begin{align}
 \begin{tikzpicture}[baseline=(base), scale = .5]
 \coordinate (base) at (0, 2);
\begin{scope}
 \clip (0, 0) rectangle (10, 6);
 \begin{scope}[transform canvas={rotate=31}]
 \fill[top color=green] (0,0) rectangle (11.6,5.2);
 \end{scope}
 \begin{scope}[transform canvas={rotate=31}]
 \fill[bottom color=green, top color = white] (0,-5.2) rectangle (11.6,0);
 \end{scope}
 \end{scope}
  \draw[latex-latex] (0, 6) node [left] {$s$} -- (0, 0) node[below left] {$0$} -- (10, 0) node [below] {$r$};
  \draw[thick, red, -latex] (0, 0) -- (5, 3);
  \foreach \r in {1,...,10}{
  \foreach \s in {1,...,6}{
  \filldraw (\r, \s) circle [radius = 1.3pt];
  }}
  \node at (5, -2) {$\beta^2 = \frac35>0$};
 \end{tikzpicture}
\qquad \qquad 
\begin{tikzpicture}[baseline=(base), scale = .5]
 \coordinate (base) at (0, 2);
 \begin{scope}
  \clip (0, 0) rectangle (10, 6);
  \begin{scope}[transform canvas={rotate=-31}]
  \fill[bottom color=green, top color = white] (-7,0) rectangle (13,10.3);
  \end{scope}
 \end{scope}
  \draw[latex-latex] (0, 6) node [left] {$s$} -- (0, 0) node[below left] {$0$} -- (10, 0) node [below] {$r$};
  \draw[thick, red, -latex] (5, 0) -- (0, 3);
  \foreach \r in {1,...,10}{
  \foreach \s in {1,...,6}{
  \filldraw (\r, \s) circle [radius = 1.3pt];
  }}
  \node at (5, -2) {$\beta^2 = -\frac35<0$};
 \end{tikzpicture}
\end{align}

\subsubsection{Vanishing and non-vanishing singular vectors}

When a representation $\mathcal{R}$ has a singular vector $X{}=LV{}$, we can define a quotient representation $\frac{\mathcal{R}}{\mathcal{R}_{X{}}}$ by setting the singular vector to zero.
By an abuse of terminology, we say that $\frac{\mathcal{R}}{\mathcal{R}_{X{}}}$ has a \myindex{vanishing singular vector}\index{singular vector!vanishing---}, which really means that its primary state $V{}$ obeys the \myindex{singular vector equation}\index{singular vector!---equation} $LV{}=0$. On the other hand, $\mathcal{R}$ itself has a non-vanishing singular vector $X{}$.

For example, let us set $L_{-1}V_0=0$ in $\mathcal{V}_0$. The resulting quotient representation has a basis made of states $LV_0$ such that the creation operator $L\in\mathcal{L}$ does not involve $L_{-1}$. Let us sketch this basis up to the level $N=4$, for comparison with a Verma module \eqref{verma}:
 \begin{align}
 \begin{tikzpicture}[scale = .15, baseline=(current  bounding  box.center)]
  \draw[-latex, very thick] (20, 0) -- (20, -28) node [right] {$N$};
  \foreach \x in {0, ..., 4}
  {
  \draw [dotted] (-6, {-6*\x}) -- (20, {-6*\x}) node [right] {${\x}$};
  }
  \node[fill = white] at (0, 0) (0) {$1$};
  \node[fill = white] at (0,-12) (2) {$L_{-2}$};
  \node[fill = white] at (0,-24) (22) {$L_{-2}^2$};
  \node[fill = white] at (6,-18) (3) {$L_{-3}$};
  \node[fill = white] at (12,-24) (4) {$L_{-4}$};
  \draw[-latex] (0) -- (2);
  \draw[-latex] (0) -- (3);
  \draw[-latex] (0) -- (4);
  \draw[-latex] (2) -- (22);
 \end{tikzpicture}
 \label{r11}
\end{align}
If $\mathcal{R}$ has several singular vectors, we can define various quotient representations by setting some singular vectors to zero. A \myindex{degenerate representation}\index{degenerate!---representation}\index{representation!degenerate---} is a representation that has at least 1 vanishing singular vector. A degenerate representation is partly degenerate (= reducible) if it contains some non-vanishing singular vectors, and \myindex{fully degenerate}\index{fully degenerate!---representation}\index{representation!fully degenerate---} (= irreducible) if all singular vectors vanish.

For $r,s\in\mathbb{N}^*$, the level-$rs$ singular vector $L_{\langle r,s\rangle}V_{\Delta_{(r,s)}}\in \mathcal{V}_{\Delta_{(r,s)}}$ is a primary state of dimension $\Delta_{(r,-s)}$, according to Eq. \eqref{drms}. Together with its descendants, this state generates a Verma module $\mathcal{V}_{\Delta_{(r,-s)}}\subset \mathcal{V}_{\Delta_{(r,s)}}$.
The Verma $\mathcal{V}_{\Delta_{(r,s)}}$ is therefore reducible, as is any highest-weight representation with a non-vanishing singular vector.
Setting the singular vector to zero, we obtain the degenerate representation
\begin{align}
 \mathcal{R}^d_{\langle r,s\rangle} 
 =\frac{\mathcal{V}_{\Delta_{(r,s)}}}{\mathcal{V}_{\Delta_{(r,-s)}}}\ . 
 \label{rvv}
\end{align}
If $\beta^2\in \mathbb{Q}$, the Verma module $\mathcal{V}_{\Delta_{(r,s)}}$ in general has several singular vectors, and $\mathcal{R}^d_{\langle r,s\rangle}$ may be partly degenerate. Let us call $\mathcal{R}^f_{\langle r,s\rangle}$ the fully degenerate quotient of  $\mathcal{V}_{\Delta_{(r,s)}}$: this quotient can generally not be obtained by setting 1 singular vector to zero, but it is always enough to set 2 singular vectors to zero, with all other singular vectors being descendants of these two. Schematically,
\begin{align}
 \mathcal{R}^f_{\langle r,s\rangle} = \frac{\mathcal{V}_{\Delta_{(r,s)}}}{\mathcal{V}_{\Delta_{(r_1,-s_1)}}+ \mathcal{V}_{\Delta_{(r_2,-s_2)}}}\ .
\end{align}
We define the \myindex{degenerate state}\index{degenerate!---state} $V^d_{\langle r,s\rangle}$ and \myindex{fully degenerate state}\index{fully degenerate!---state} $V^f_{\langle r,s\rangle}=V^f_{\langle r_1,s_1\rangle}=V^f_{\langle r_2,s_2\rangle}$ as the primary states of the representation $\mathcal{R}^d_{\langle r,s\rangle}$ and 
$\mathcal{R}^f_{\langle r,s\rangle}$. These states obey the singular vector equation
\begin{align}
L_{\langle r, s\rangle} V^d_{\langle r,s\rangle} = L_{\langle r, s\rangle} V^f_{\langle r,s\rangle} = 0\ .
\label{lvdz}
\end{align}
Let us summarize these notations:
\begin{align}
\renewcommand{\arraystretch}{1.3}
 \begin{array}{|c|r|l|}
  \hline
  \text{Primary state} & \text{Highest-weight representation} & \text{Irreducible?}
  \\
  \hline\hline
  V_\Delta   &  \text{Verma module } \mathcal{V}_\Delta & \text{if and only if } \Delta\notin\{\Delta_{(r,s)}\}_{r,s\in\mathbb{N}^*}
  \\
  \hline
  V^d_{\langle r,s\rangle} & \text{degenerate rep. } \mathcal{R}^d_{\langle r,s\rangle} & \text{if } \beta^2\notin \mathbb{Q}
  \\
  \hline
  V^f_{\langle r,s\rangle} & \text{fully degenerate rep. } \mathcal{R}^f_{\langle r,s\rangle} & \text{yes}
  \\
  \hline
 \end{array}
\end{align}
We only wrote a sufficient condition of irreducibility for $\mathcal{R}^d_{\langle r,s\rangle}$: a necessary condition would be more complicated.
It can happen that $V^d_{\langle r,s\rangle}=V^f_{\langle r,s\rangle}$, in which case we may use the former notation to emphasize the existence of a vanishing descendant, or the latter notation to emphasize the absence of non-vanishing singular vectors.

\subsubsection{The Shapovalov form}

We will now introduce the Shapovalov form, which will be useful for computing operator product expansions and conformal blocks. We will see that the Shapovalov from has zeros related to singular vectors, and this will lead to poles of the operator product expansions and conformal blocks.

Let us define an involution on the Virasoro algebra by 
\begin{align}
 L^*_n = L_{-n} \ . 
\end{align}
This involution obeys $[L_m,L_n]^* = -[L_m^*,L_n^*]$. It can be extended to the universal enveloping algebra by $\left(LL'\right)^* = L'^*L^*$. Using this involution, we define the \myindex{Shapovalov form}\index{Shapovalov form} on any highest-weight representation as the symmetric bilinear form such that
\begin{align}
 \big(LV_\Delta\big| L'V_\Delta\big) = \left\{\begin{array}{ll} 0 & \text{if } |L|\neq |L'|\ , 
                                                \\
                                              S_{L,L'}(\Delta) & \text{if } |L|=|L'|\ , \quad \text{where} \quad L^*L'V_\Delta = S_{L,L'}(\Delta) V_\Delta\ .  
                                             \end{array}\right.
\label{shap}
\end{align}
At each level $N\in\mathbb{N}$, the Shapovalov form is described by a matrix that depends on $\Delta$ and $c$. In the case of a Verma module at levels $N=0,1,2$, these matrices are
\begin{align}
 S_{1,1} = 1 \ \  , \ \   S_{L_{-1},L_{-1}} = 2\Delta \ \  , \ \  \begin{bmatrix} S_{L_{-1}^2,L_{-1}^2} & S_{L_{-1}^2,L_{-2}} \\ S_{L_{-1}^2,L_{-2}} & S_{L_{-2},L_{-2}}\end{bmatrix} = 
 \begin{bmatrix} 4\Delta(2\Delta+1) & 6\Delta \\ 6\Delta & 4\Delta+\frac{c}{2}\end{bmatrix}\ .
 \label{sh012}
\end{align}
For the Shapovalov form, any singular vector or descendant thereof is a \myindex{null vector}\index{null vector}, i.e. a vector orthogonal to the whole representation. In particular,
\begin{align}
 \forall L, L'\in \mathcal{L}\ , \quad \big(LL_{\langle r,s\rangle} V_{\Delta_{(r,s)}}\big|L' V_{\Delta_{(r,s)}}\big) =0 \ . 
\end{align}
A highest-weight representation is irreducible if and only if at each level $N\in\mathbb{N}$, its Shapovalov matrix is invertible.

For a Verma module at generic values of the central charge, $S_{LL_{\langle r,s\rangle},L'}(\Delta)$ has a simple zero at $\Delta=\Delta_{(r,s)}$. This implies that the inverse Shapovalov form at level $N$ has simple poles for $\Delta\in \left\{\Delta_{(r,s)}\right\}_{rs\leq N}$. In a basis of level-$rs$ creation operators that includes $L_{\langle r,s\rangle}$, the level-$rs$ inverse Shapovalov form behaves as
\begin{align}
 S^{-1}_{L,L'}(\Delta) \underset{\Delta\to \Delta_{(r,s)}} = \frac{1}{\Delta-\Delta_{(r,s)}} \frac{\delta_{L,L_{\langle r,s\rangle}}\delta_{L',L_{\langle r,s\rangle}}}{S'_{L_{\langle r,s\rangle},L_{\langle r,s\rangle}}(\Delta_{(r,s)})}  + O(1) \ . 
 \label{smo}
\end{align}
In fact, the level-$N$ inverse Shapovalov form can be computed explicitly in terms of $\left\{L_{\langle r,s\rangle}\right\}_{rs\leq N}$ \cite{fqs24}.

Using the same coefficients $\big(LV_\Delta\big| L'V_\Delta\big)$ \eqref{shap}, we can define a sesquilinear form instead of a bilinear form. This sesquilinear form is Hermitian if its coefficients are real, i.e. if $\Delta,c\in\mathbb{R}$. Our highest-weight representation is called \myindex{unitary}\index{representation!unitary---} if that Hermitian form is positive definite. Unitarity plays an important role in some applications of CFT, but it has no bearing on exact solvability, so we will not elaborate further on that subject.

\subsection{Fields and operator product expansions}\label{sec:fope}

In quantum field theory, there is a \myindex{state-field correspondence}\index{state-field correspondence} or operator-state correspondence. Taking Euclidean time to be the radial coordinate around a point $z_0$, states live on spheres of radius $R$ while fields live at $z_0$, so in the limit $R\to 0$ a state gives rise to a field.
In the presence of scale invariance, we do not lose information by sending $R\to 0$, so the correspondence is bijective. In particular, in a conformal field theory, we identify the space of fields with the \myindex{space of states}\index{space of states}  or \myindex{spectrum}\index{spectrum}  --- a vector space on which the Virasoro algebra acts. To a state $V$, the correspondence associates a field $V(z)$, which depends on a position $z\in\overline{\mathbb{C}}$ in the Riemann sphere. Primary states, descendant states and degenerate states respectively give rise to \myindex{primary fields}\index{primary!---field}\index{field!primary---}, \myindex{descendant fields}\index{descendant!---field}\index{field!descendant---} and \myindex{degenerate fields}\index{degenerate!---field}\index{field!degenerate---}.

In the bootstrap approach, which is axiomatic rather than constructive, fields need not be constructed from other objects. Rather, we use fields as a convenient notation for stating properties of correlation functions, which we will introduce in Section \ref{sec:cor}. In order to facilitate comparisons, let us indicate how fields are constructed in other approaches:
\begin{itemize}
 \item \textbf{Fields as operators:} Fields may be constructed as linear maps from the spectrum to itself. Since the Virasoro algebra also acts on the spectrum, there is a natural action of the Virasoro algebra on fields. Namely, the action that we will write as $L_nV(z)$ is defined as the operator commutator $[L_n,V(z)]$. The $N$-point correlation function that we will write as $\left<\prod_{i=1}^N V_i(z_i)\right>$ is then constructed as $\left<0\middle| \prod_{i=1}^N V_i(z_i)\middle|0\right>$, where $\left|0\right>$ is the vacuum state, on which the $N$ operators act in some order. The formalism of vertex operator algebras provides an algebraic construction of fields as operators.
 \item \textbf{Fields as functionals:} In a functional integral approach such as the Coulomb gas formalism, fields are constructed from a fundamental free field $\varphi(z)$, in terms of which the functional integration measure $D\varphi(z)$ is defined. A primary fields $V_\Delta(z)$ is constructed as an exponential $e^{\alpha\varphi(z)}$, where $\alpha$ depends on the dimension $\Delta$ and on the central charge. Descendant fields also involve derivatives of $\varphi(z)$.
\end{itemize}

\subsubsection{The energy-momentum tensor}

The dependence of fields on the coordinate $z$ is determined by the assumption that the Virasoro generator $L_{-1}$ generates translations, just like the Witt algebra generator $\ell_{-1}$ \eqref{lpz}:
\begin{align}
  \boxed{\frac{\partial}{\partial z} V(z) = L_{-1} V(z)}  \ .
  \label{pvlv}
 \end{align}
 Since fields depend on $z$, the Virasoro algebra that acts on fields also depends on $z$. This can be made explicit using the notation
\begin{align}
   L_n V(z) = L_n^{(z)} V(z) \ . 
\end{align}
Applying Eq. \eqref{pvlv} to $L_n^{(z)} V(z)$, we find how the Virasoro generator $L_n^{(z)}$ depends on $z$,
 \begin{align}
 \frac{\partial}{\partial z} L_n^{(z)} = [L_{-1}^{(z)},L_n^{(z)}]= -(n+1)L_{n-1}^{(z)}\ ,\qquad (\forall n\in\mathbb{Z})\ .
\end{align}
This can be rewritten as the $z$-independence of a formal Laurent series that combines all the generators $(L_n^{(z)})_{n\in\mathbb{Z}}$, 
\begin{align}
 \forall y\in\mathbb{C}\backslash \{z\}\ , \qquad \frac{\partial}{\partial z} \sum_{n\in\mathbb{Z}} \frac{L_n^{(z)}}{(y-z)^{n+2}} = 0\  .
\end{align}
The formal Laurent series is then called the \myindex{energy-momentum tensor}\index{energy-momentum tensor}, 
\begin{align}
  \boxed{T(y) = \sum_{n\in\mathbb{Z}} \frac{L_n^{(z)}}{(y-z)^{n+2}}} \ .
  \label{tl}
 \end{align}
 A priori, the energy-momentum tensor only makes sense when acting on a field $V(z)$,
 \begin{align}
 T(y)V(z) = \sum_{n\in\mathbb{Z}} \frac{L_n V(z)}{(y-z)^{n+2}}\quad , \quad L_n V(z) = \frac{1}{2\pi i} \oint_{z}dy\ (y-z)^{n+1} T(y)V(z)\ .
 \label{lvtv}
\end{align}
However, we will shortly interpret $T(y)$ as a field, and $T(y)V(z)$ as an operator product expansion of 2 fields. The statement that $V_\Delta(z)$ is a primary field of dimension $\Delta$ is equivalent to the OPE 
\begin{align}
 \boxed{T(y)V_\Delta(z) = \frac{\Delta}{(y-z)^2} V_\Delta(z) + \frac{1}{y-z} \frac{\partial}{\partial z} V_\Delta(z) + O(1)}\ ,
 \label{tvd}
\end{align}
where we used Eq. \eqref{pvlv} and wrote the regular terms as $O(1)$.

\subsubsection{Operator product expansions}\label{sec:ope}

The fundamental axiom of conformal field theory, which underlies the conformal bootstrap approach, is the existence of \myindex{operator product expansions (OPEs)}\index{OPE}. This axiom states that there exists a bilinear product on the space of fields, 
\begin{align}
  \boxed{V_1(z_1)V_2(z_2) = \sum_{k} C^k_{12}(z_1,z_2) V_k(z_2)}\ ,
  \label{ope}
 \end{align}
 for some family $(V_k)_{k}$ of linearly independent states, and
some functions $C^k_{12}(z_1,z_2)$ called \myindex{OPE coefficients}\index{OPE!---coefficient}. Operator product expansions generally exist in quantum field theory, but they are particularly useful in conformal field theory because they hold for any $z_1$ in a finite neighbourhood of $z_2$ (but $z_1\neq z_2$), and not just asymptotically as $z_1\to z_2$. The finite neighbourhood in question depends on the correlation function in which we perform the OPE, and also on the family $(V_k)_{k}$.

We further assume that the product of fields is commutative,
\begin{align}
 \boxed{V_1(z_1)V_2(z_2) = V_2(z_2)V_1(z_1)}\ .
 \label{comm}
\end{align}
This implies that OPEs are \myindex{commutative}\index{OPE!---commutativity} and \myindex{associative}\index{OPE!---associativity}. However, the resulting constraints on the coefficients $C^i_{12}(z_1,z_2)$ are not particularly simple, because of having to choose a position ($z_2$ in our conventions) for the fields in the right-hand side of the OPE \eqref{ope}. 

We also assume that there exists an \myindex{identity field}\index{identity field} $I(z)$ whose OPE with any other field is trivial, namely $I(z_1)V(z_2) = V(z_2)$. (The corresponding state is called the \myindex{vacuum state}\index{vacuum state}.) Let us consider the energy-momentum tensor $T$ as a field, then we have
$
 T(y)I(z) = T(y) = T(z)  + O(y-z)
$. 
Comparing this with Eq. \eqref{lvtv} applied to $I(z)$, we find 
\begin{align}
 L_{n\geq -1} I(z) = 0 \quad , \quad L_{-2}I(z)=T(z) \ .
\end{align}
In particular, the identity field is a degenerate primary field of conformal dimension zero, with a vanishing singular vector at level $1$, i.e. $I(z) \propto V^d_{\langle 1,1\rangle}(z)$. And from the expression of $T(z)$ as a descendant of that field, we deduce
\begin{align}
 T(y)T(z) = \frac{\frac{c}{2}}{(y-z)^4} + \frac{2T(z)}{(y-z)^2} + \frac{\partial T(z)}{y-z} + O(1)\ .
\label{tt}
\end{align}
This OPE is equivalent to the commutation relations of the Virasoro algebra \eqref{vir}.

\subsubsection{Conformal symmetry constraints on OPEs}\label{sec:csope}

For technical simplicity, we focus on an OPE of 2 primary fields. In that OPE, we single out the contributions of a primary field $V_\Delta$ and its descendants:
\begin{align}
 V_{\Delta_1}(z_1)V_{\Delta_2}(z_2) \supset \sum_{L\in \mathcal{L}} C_{\Delta_1,\Delta_2}^{\Delta, L}(z_1,z_2) LV_\Delta(z_2) \ ,
\end{align}
where $\mathcal{L}$ \eqref{lcm} is the set of creation operators. On both sides of the OPE, let us insert $\frac{1}{2\pi i}\oint_{z_1,z_2} dy (y-z_2)^{m+1}T(y)$ with $m\in\mathbb{Z}$, where the integration contour encloses both $z_1$ and $z_2$. Using Eqs. \eqref{lvtv} and \eqref{tvd}, we find
\begin{align}
 \left(L_m^{(z_2)} +z_{12}^{m+1}\partial_{z_1} + (m+1)z_{12}^m\Delta_1\right) V_{\Delta_1}(z_1)V_{\Delta_2}(z_2) \supset \sum_{L\in \mathcal{L}} C_{\Delta_1,\Delta_2}^{\Delta, L}(z_1,z_2)L_mLV_\Delta(z_2) \ .
 \label{lmope}
\end{align}
We consider 3 cases:
\begin{itemize}
 \item $\boxed{m=-1}$\ : We use Eq. \eqref{pvlv} for $V_{\Delta_2}(z_2)$ and $LV_{\Delta}(z_2)$, and perform the OPE again on the left-hand side:
 \begin{align}
 \left(\partial_{z_1}+\partial_{z_2}\right) \sum_{L\in \mathcal{L}} C_{\Delta_1,\Delta_2}^{\Delta, L}(z_1,z_2)LV_\Delta(z_2) = \sum_{L\in \mathcal{L}} C_{\Delta_1,\Delta_2}^{\Delta, L}(z_1,z_2)\partial_{z_2} LV_\Delta(z_2) \ .
 \end{align}
 This leads to $(\partial_{z_1}+\partial_{z_2})C^{\Delta,L}_{\Delta_1,\Delta_2}(z_1,z_2)=0$, i.e. translation invariance of the OPE coefficients.
 \item $\boxed{m=0}$\ : We use $L_0LV_{\Delta}(z_2)= (\Delta+|L|) V_{\Delta}(z_2)$, and perform the OPE again on the left-hand side. This leads to $(z_{12}\partial_{z_1}+\Delta_1+\Delta_2-\Delta-|L|)C^{\Delta,L}_{\Delta_1,\Delta_2}(z_1,z_2) = 0$, and we deduce the dependence of OPE coefficients on $z_1,z_2$:
 \begin{align}
  C^{\Delta,L}_{\Delta_1,\Delta_2}(z_1,z_2) = z_{12}^{\Delta+|L|-\Delta_1-\Delta_2} C^{\Delta,L}_{\Delta_1,\Delta_2}\ . 
 \end{align}
\item $\boxed{m\geq 1}$\ : We use $L_m^{(z_2)}V_{\Delta_2}(z_2)=0$, and perform the OPE again on the left-hand side. The action of the differential operator on the OPE coefficients is now known, and we find 
\begin{align}
 \sum_{L\in\mathcal{L}} C^{\Delta,L}_{\Delta_1,\Delta_2}z_{12}^{|L|+m}(\Delta+|L|+m\Delta_1-\Delta_2) LV_{\Delta}= \sum_{L\in\mathcal{L}} C^{\Delta,L}_{\Delta_1,\Delta_2}z_{12}^{|L|}L_m LV_{\Delta}\ .
\end{align}
Extracting the coefficient of $z_{12}^N$ for some $N\in\mathbb{N}$, we find 
\begin{align}
 L_m W_N = \left(\theta_{m}+N-m\right)W_{N-m} \quad \text{with} \quad \left\{\begin{array}{l} W_N = \sum_{|L|=N}C^{\Delta,L}_{\Delta_1,\Delta_2} LV_{\Delta}\ , \\  \theta_{m} = m\Delta_1-\Delta_2 + \Delta\ . \end{array}\right.
 \label{opew}
\end{align}
For example, $L_1W_1 = \theta_1 W_0$, which amounts to $2\Delta C^{\Delta,L_{-1}}_{\Delta_1,\Delta_2} = (\Delta+\Delta_1-\Delta_2) C^{\Delta,1}_{\Delta_1,\Delta_2}$.
\end{itemize}
These linear equations for the OPE coefficients $C^{\Delta,L}_{\Delta_1,\Delta_2}$ are called \myindex{OPE Ward identities}\index{Ward identity!OPE---}. 
Let us look for a solution $\left(f^{\Delta,L}_{\Delta_1,\Delta_2}\right)_{L\in\mathcal{L}}=\left(f^L\right)_{L\in\mathcal{L}}$, normalized by setting $f^{1}=1$, equivalently $W_0 = V_\Delta$. Iterating Eq. \eqref{opew}, we can compute $L^*W_N$ for any creation operator $L\in\mathcal{L}_N$. On the other hand, $L^*W_N = \sum_{L'\in\mathcal{L}_N} S_{L,L'} f^{L'}V_\Delta$ where $S_{L,L'}$ is the Shapovalov form \eqref{shap}.
This leads to the linear system 
\begin{align}
 \forall L\in \mathcal{L}_N, \ \sum_{L'\in\mathcal{L}_N} S_{L,L'} f^{L'} = g^L \quad \text{with} \quad 
 g^{L_{-n_1}L_{-n_2}\cdots L_{-n_k}}_{\Delta,\Delta_2,\Delta_1} = \prod_{i=1}^k \left(\theta_{n_i}+\textstyle{\sum}_{j=i+1}^k n_j\right)
 \ .
 \label{sgl}
\end{align}
Assuming the Shapovalov form is invertible at level $N$, the solution is
\begin{align}
 \boxed{f^{\Delta,L}_{\Delta_1,\Delta_2} = \sum_{L'\in\mathcal{L}_{N}} S_{L,L'}^{-1}(\Delta) g^{L'}_{\Delta,\Delta_2,\Delta_1}}\ .
 \label{fsg}
\end{align}
For example, we have $g^{L_{-1}} = \theta_{1}$, $g^{L_{-1}^2} = \theta_{1}(\theta_{1}+1)$ and $g^{L_{-2}}=\theta_{2}$.
Inverting the Shapovalov form \eqref{sh012}, we deduce
\begin{subequations}
 \label{fl12}
 \begin{align}
  f^{L_{-1}} & = \frac{\Delta+\Delta_1-\Delta_2}{2\Delta}\ ,
  \label{fl1}
  \\
  \left[\begin{smallmatrix}
 f^{L_{-1}^2} \\ f^{L_{-2}} 
 \end{smallmatrix}\right] 
 & =
   \frac{1}{16(\Delta-\Delta_{(2,1)})(\Delta-\Delta_{(1,2)})}
\left[\begin{smallmatrix} 2+\frac{c}{4\Delta} & -3 \\ -3 & 4\Delta+2 \end{smallmatrix}\right]
\left[\begin{smallmatrix} (\Delta+\Delta_1-\Delta_2)(\Delta+\Delta_1-\Delta_2+1) \\ \Delta+2\Delta_1-\Delta_2 \end{smallmatrix}\right] \ .
\label{fl2}
 \end{align}
\end{subequations}
If $\Delta\notin\left\{\Delta_{(r,s)}\right\}_{rs\in\mathbb{N}^*}$, the Shapovalov form is invertible over the whole Verma module $\mathcal{V}_\Delta$. The solution $\left(f^L\right)_{L\in\mathcal{L}}$ of the OPE Ward identities is unique, and the contributions of $V_\Delta$ and its descendants in an OPE of 2 primary fields read
\begin{align}
 \boxed{V_{\Delta_1}(z_1)V_{\Delta_2}(z_2) \supset  C_{\Delta_1,\Delta_2}^{\Delta} z_{12}^{\Delta-\Delta_1-\Delta_2} \Bigg(V_\Delta(z_2) +\sum_{L\in \mathcal{L}\backslash\{1\}}z_{12}^{|L|} f^{\Delta,L}_{\Delta_1,\Delta_2} LV_\Delta(z_2)\Bigg)}
 \ ,
 \label{prope}
\end{align}
where the \myindex{OPE structure constant}\index{OPE!---structure constant}\index{structure constant!OPE---} $C^\Delta_{\Delta_1,\Delta_2}$ is left undetermined by conformal symmetry, while the dependence on $z_1,z_2$ and the coefficients $f^{\Delta,L}_{\Delta_1,\Delta_2}$ are universal quantities that are in principle known. 

If $\Delta=\Delta_{(r,s)}$, there is a singular vector $L_{\langle r,s\rangle} V_{\Delta_{(r,s)}}$ at level $rs$ in the Verma module $\mathcal{V}_{\Delta_{(r,s)}}$. This singular vector manifests itself as a simple pole at $\Delta=\Delta_{(r,s)}$ of the inverse Shapovalov form \eqref{smo}, so that $(f^L)_{L\in\mathcal{L}_{N\geq rs}}$ also have a simple pole at $\Delta=\Delta_{(r,s)}$. On the other hand, if a fully degenerate field $V^f_{\langle r,s\rangle}$ appears in an OPE, then the basis of descendant fields that appear in the OPE is a subset of $(LV^f_{\langle r,s\rangle})_{L\in\mathcal{L}}$, and their coefficients are again universal quantities. But as we will see in Section \ref{sec:dope}, a degenerate field can appear in the OPE $V_{\Delta_1}V_{\Delta_2}$, only under certain conditions on $\Delta_1$ and $\Delta_2$.

In the case of the OPE coefficients $C^{\Delta,L}_{\Delta_1,\Delta_2}(z_1,z_2)$, we have just seen that the primary field determines the contributions of its descendant fields, provided there is no singular vector among these descendants, i.e. provided the primary field generates an irreducible representation. In Section \ref{sec:csc}, we will also see that correlation functions of primary fields determine the correlation functions of their descendant fields, with no caveats about singular vectors this time. This property of Virasoro symmetry does not in general hold for larger chiral symmetry algebras such as W-algebras.

The form of the dependence on $z_1,z_2$ provides an a posteriori justification for our assumption in Section \ref{sec:hwr} that $L_0$-eigenvalues be bounded from below. We have indeed seen that an $L_0$-eigenvector $LV_\Delta$ of dimension $\Delta+|L|$ comes with a factor $z_{12}^{\Delta+|L|}$. If $L_0$-eigenvalues were not bounded from below, the sum over states could not converge in a neighbourhood of $z_1=z_2$.

\subsubsection{Fusion rules and fusion products}\label{sec:dope}

Because of the poles in OPE coefficients, there are constraints on the fields that can appear in the OPE $V_{\Delta_1}V_{\Delta_2}$. Let us focus on the pole at $\Delta_{(1,1)}=0$. Assuming $\beta^2\notin \mathbb{Q}$ for simplicity, there are 2 types of primary fields of dimension $0$: the non-degenerate field $V_0$, which generates the Verma module $\mathcal{V}_0$, and the degenerate field $V^d_{\langle 1,1\rangle}$, which obeys the singular vector equation $L_{-1}V^d_{\langle 1,1\rangle}=\partial V^d_{\langle 1,1\rangle}=0$ and generates the degenerate representation $\mathcal{R}^d_{\langle 1,1\rangle}$ \eqref{r11}.
\begin{itemize}
 \item If $\Delta_1\neq \Delta_2$, then $\operatorname{Res}_{\Delta=0}f^{\Delta,L_{-1}}_{\Delta_1,\Delta_2}\neq 0$, and we cannot impose the normalization $C^{\Delta,1}_{\Delta_1,\Delta_2}=0$: rather, we may choose $C^{L_{-1}}\neq 0$ while $C^1=0$. Therefore, the non-vanishing singular vector $L_{-1}V_0\propto V_1$ may appear in the OPE, together with its descendants, while $V_0$ itself and its other descendants give vanishing contributions. In particular, neither $V^d_{\langle 1,1\rangle}$ nor its descendants can contribute.
 \item If $\Delta_1=\Delta_2$, then $\operatorname{Res}_{\Delta=0}f^{\Delta,L_{-1}}_{\Delta_1,\Delta_2}= 0$. The equation $L_1W_1=\theta_{1,1}W_0$ \eqref{opew} reduces to $0=0$, and does not determine $C^{L_{-1}}$ from $C^1$. In particular, our OPE can include $V^d_{\langle 1,1\rangle}$.
\end{itemize}
What about an OPE of $V_0$ or $V^d_{\langle 1,1\rangle}$ with another primary field? The OPE $V_0V_{\Delta_2}$ is just a special case of the OPE \eqref{prope}, where arbitrary primary fields may appear.
But the OPE $V^d_{\langle 1,1\rangle}V_{\Delta_2}$ must obey 
\begin{align}
 0 = \frac{\partial}{\partial z_1} V^d_{\langle 1,1\rangle}(z_1)V_{\Delta_2}(z_2) \supset C_{\langle 1,1\rangle,\Delta_2}^{\Delta} \frac{\partial}{\partial z_1} z_{12}^{\Delta-\Delta_2} \big(V_\Delta(z_2) +\cdots \big) \ ,
\end{align}
therefore $V_\Delta$ may appear only provided $\Delta=\Delta_2$. This agrees with the remark (in Section \ref{sec:ope}) that $V^d_{\langle 1,1\rangle}$ is proportional to the identity field. 

Let us introduce the \myindex{fusion rules}\index{fusion!---rule} as a convenient notation for writing which primary fields can appear in OPEs:
\begin{align}
 V^d_{\langle 1,1\rangle} V_{\Delta} = V_\Delta \quad , \quad V_{\Delta_1} V_{\Delta_2} \ni V^d_{\langle 1,1\rangle} \implies \Delta_1=\Delta_2\ . 
 \label{vov}
\end{align}
Just like OPEs themselves, fusion rules are commutative and associative.

Fusion rules of primary fields have an algebraic counterpart: the \myindex{fusion product}\index{fusion!---product} of the corresponding representations of the Virasoro algebra \cite{kr18}.
In terms of the fusion product, the fusion rule \eqref{vov} reads
\begin{align}
 \mathcal{R}^d_{\langle 1,1\rangle} \times \mathcal{V}_\Delta = \mathcal{V}_\Delta \quad , \quad \mathcal{R}^d_{\langle 1,1\rangle} \subset \mathcal{V}_{\Delta_1}\times \mathcal{V}_{\Delta_2} \implies \Delta_1=\Delta_2\ .
 \label{rtv}
\end{align}
Given 2 representations $\mathcal{R}_i,\mathcal{R}_j$ of the Virasoro algebra with a given central charge, the fusion product $\mathcal{R}_i \times \mathcal{R}_j$ is a representation of the same Virasoro algebra. This allows the fusion product to describe the behaviour of OPEs in a given CFT. The decomposition into indecomposables $\mathcal{R}_i \times \mathcal{R}_j = \sum_k m_{i,j}^k \mathcal{R}_k$ is the algebraic counterpart of the OPE. The coefficient $m_{i,j}^k\in \mathbb{N}\cup\{\infty\}$ is called a \myindex{fusion multiplicity}\index{fusion!---multiplicity}.
The price to pay is that the fusion product has a complicated definition that involves the fields' positions.

In contrast, given representations $\mathcal{R}_1,\mathcal{R}_2$ of two copies of the Virasoro algebra with central charges $c_1,c_2$, the tensor product $\mathcal{R}_1\otimes \mathcal{R}_2$ is a representation of the Virasoro algebra with central charge $c_1+c_2$. The tensor product is useful for defining the product of two CFTs, with a spectrum that is the tensor product of their two spectra.

The construction and properties of the fusion product are well-established in the case of irreducible Kac table representations at $\beta^2 \in \mathbb{Q}_{>0}$ \cite{gab99}. Then the case of degenerate representations and Verma modules at generic $\beta^2\in \mathbb{C}\backslash\mathbb{Q}$ can be deduced by taking limits and doing analytic continuation in $\beta^2$. To actually compute fusion products, it will be technically easier to begin with the second case in Section \ref{sec:fpdr}, and deduce the first case in Section \ref{sec:amm}. We will accept that in both cases the fusion product is commutative and associative, with multiplicities $m_{i,j}^k\in \{0,1\}$ that are  invariant under permutations of the 3 indices. In fact, $m_{i,j}^k\in \{0,1\}$ is an algebraic consequence of the result of Section \ref{sec:csope} that in OPEs, primary fields determine descendant fields, provided the fields transform in irreducible representations of the Virasoro algebra.

\subsubsection{Fusion products of degenerate representations}\label{sec:fpdr}

In this section we assume $\beta^2\notin\mathbb{Q}$, so that the degenerate fields $V^d_{\langle r,s\rangle}$ are fully degenerate, with only 1 vanishing singular vector.
Let us derive the fusion rules that involve the degenerate field $V^d_{\langle 2,1\rangle}$. These fusion rules are encoded in the 3 equivalent conditions 
\begin{align}
V^d_{\langle 2,1\rangle} V_{\Delta_1}\ni V_{\Delta_2} \iff V_{\Delta_1}V_{\Delta_2} \ni V^d_{\langle 2,1\rangle} \iff \underset{\Delta=\Delta_{(2,1)}}{\operatorname{Res}} f^{\Delta,L_{-1}^2}_{\Delta_1,\Delta_2}=0\ . 
\end{align}
The latter condition can be evaluated explicitly using Eq. \eqref{fl2}, leading to a constraint on $\Delta_1,\Delta_2$. Equivalently, the OPE $V_{\Delta_1}(z_1)V^d_{\langle 2,1\rangle}(z_2)$ is constrained by $L_{\langle 2,1\rangle}V^d_{\langle 2,1\rangle}=0$ with $L_{\langle 2,1\rangle} = L_{-1}^2 -\beta^2L_{-2}$ \eqref{ars}. Assuming $V_{\Delta_1}(z_1)V^d_{\langle 2,1\rangle}(z_2)\ni V_{\Delta_2}$, 
using Eq. \eqref{pvlv} and Eq. \eqref{lmope}, we deduce 
\begin{multline}
 V_{\Delta_1}(z_1)L_{\langle 2,1\rangle}V^d_{\langle 2,1\rangle} \supset \left(\frac{\partial^2}{\partial z_2^2} + \frac{\beta^2}{z_{12}}\frac{\partial}{\partial z_1} -\frac{\beta^2\Delta_1}{z_{12}^2}+L_{-2}\right)
 \\
 C_{\Delta_1,\langle 2,1\rangle}^{\Delta_2} z_{12}^{\Delta_2-\Delta_1-\Delta_{(2,1)}}\Big(V_{\Delta_2}(z_2)+O(z_{12})\Big)\ .
\end{multline}
This OPE must vanish, including the contribution of the primary field $V_{\Delta_2}(z_2)$, therefore 
$\left(\frac{\partial^2}{\partial z_2^2} + \frac{\beta^2}{z_{12}}\frac{\partial}{\partial z_1} -\frac{\beta^2\Delta_1}{z_{12}^2}\right)
 z_{12}^{\Delta_2-\Delta_1-\Delta_{(2,1)}} =0$. This amounts to 
\begin{align}
 \beta^2 \left(\Delta_{(2,1)}+¨2\Delta_1-\Delta_2\right) = \left(\Delta_{(2,1)}+\Delta_1-\Delta_2\right)\left(\Delta_{(2,1)}+1+\Delta_1-\Delta_2\right)\ .
\end{align}
Using the expression \eqref{ars} for $\Delta_{(2,1)}$, and replacing conformal dimensions with momenta \eqref{dp}, this equation reduces to
\begin{align}
 \prod_{\pm,\pm}\left(\tfrac{\beta}{2}\pm P_1\pm P_2\right) = 0\ . 
\end{align}
This leads to the fusion rules of $V^d_{\langle 2,1\rangle}$, from which the fusion rules of $V^d_{\langle 1,2\rangle}$ are deduced by $\beta\to \beta^{-1}$. We write these fusion rules in terms of fusion products, where we label Verma modules by their momenta:
\begin{align}
 \boxed{\mathcal{R}^d_{\langle 2,1\rangle}\times \mathcal{V}_P = \sum_\pm \mathcal{V}_{P\pm \frac{\beta}{2}}} \quad, \quad \boxed{\mathcal{R}^d_{\langle 1,2\rangle}\times \mathcal{V}_P = \sum_\pm \mathcal{V}_{P\pm \frac{1}{2\beta}}}\ . 
 \label{rvvp}
\end{align}
From these simple formulas, all the discrete spectra of Section \ref{sec:sesc} will follow. For the moment, let us deduce the fusion products of higher degenerate representations $\mathcal{R}^d_{\langle r,s\rangle}$, using the associativity of the fusion product. We start with the remark that a highest-weight representation $\mathcal{R}$ is degenerate if and only if its fusion product with any Verma module $\mathcal{R}\times \mathcal{V}_P$ has finitely many terms:
\begin{itemize}
 \item If $\mathcal{R}=\mathcal{R}^d_{\langle r,s\rangle}$ is degenerate, the corresponding field $V^d_{\langle r,s\rangle}$ obeys
 \begin{align}
V^d_{\langle r,s\rangle} V_{\Delta_1}\ni V_{\Delta_2} \iff V_{\Delta_1}V_{\Delta_2} \ni V^d_{\langle r,s\rangle} \iff\forall L\in \mathcal{L},\  \underset{\Delta=\Delta_{(r,s)}}{\operatorname{Res}} f^{\Delta,L}_{\Delta_1,\Delta_2}=0\ ,
\end{align}
 as we saw already for $(r,s)=(1,1)$ and $(r,s)=(2,1)$. (If the residues vanish for $|L|=rs$, then they vanish for all $L\in\mathcal{L}$.) This condition is a polynomial equation of order $rs$ on $\Delta_1,\Delta_2$.

 \item If $\mathcal{R}$ is not degenerate, OPEs of the corresponding primary field
 only obey constraints from conformal symmetry, which do not restrict the primary fields that can appear.
\end{itemize}
It follows that a fusion product of degenerate representations is itself a finite sum of degenerate representations. 
For example, according to Eq. \eqref{rvvp}, the product $\mathcal{R}^d_{\langle 2,1\rangle}\times\mathcal{R}^d_{\langle 2,1\rangle}$ is the sum of 2 representations of momenta $\{P_{(2,1)} \pm \frac{\beta}{2}\} = \{P_{(1,1)},P_{(3,1)}\}$. These representations must be degenerate, therefore $\mathcal{R}^d_{\langle 2,1\rangle}\times\mathcal{R}^d_{\langle 2,1\rangle}=\mathcal{R}^d_{\langle 1,1\rangle}+\mathcal{R}^d_{\langle 3,1\rangle}$. By associativity of $\mathcal{R}^d_{\langle 2,1\rangle}\times\mathcal{R}^d_{\langle 2,1\rangle}\times \mathcal{V}_P$, we then deduce
$
 \mathcal{R}^d_{\langle 3,1\rangle} \times \mathcal{V}_P = \mathcal{V}_{P-\beta}+\mathcal{V}_P+\mathcal{V}_{P+\beta}
$. Iterating this reasoning over $r$ or $s$, we obtain 
\begin{align}
 \mathcal{R}^d_{\langle 2,1\rangle}\times \mathcal{R}^d_{\langle r,s\rangle} 
 \underset{r\geq 2}{=} \sum_\pm \mathcal{R}^d_{\langle r\pm 1,s\rangle} 
 \quad , \quad 
 \mathcal{R}^d_{\langle 1,2\rangle}\times \mathcal{R}^d_{\langle r,s\rangle} 
 \underset{s\geq 2}{=}\sum_\pm \mathcal{R}^d_{\langle r,s\pm 1\rangle} \ .
 \label{rrrs}
\end{align}
This allows us to find the fusion product of any degenerate representation with a Verma module, by iteration over $r$ and $s$:
\begin{align}
 \boxed{\mathcal{R}^d_{\langle r,s\rangle}\times \mathcal{V}_P = \sum_{i=-\frac{r-1}{2}}^{\frac{r-1}{2}} \sum_{j=-\frac{s-1}{2}}^{\frac{s-1}{2}} \mathcal{V}_{P+i\beta +j\beta^{-1}}}\ ,
 \label{rrsvp}
\end{align}
where the sums run by increments of $1$. Since this sum has $rs$ terms, the Verma module $\mathcal{V}_{\Delta_{(r,s)}}$ must have a singular vector at level $rs$, a result that we stated in Section \ref{sec:nvvm}, and which we now derived using fusion products. Fusion can even be used to compute the singular vectors themselves \cite{fms97}.

We can then find the fusion product $\mathcal{R}^d_{\langle r_1,s_1\rangle}\times \mathcal{R}^d_{\langle r_2,s_2\rangle}$, either by iterating over $r_1$ and $s_1$, or by summing the degenerate representations whose momenta appear in both  $\mathcal{R}^d_{\langle r_1,s_1\rangle}\times \mathcal{V}_{P_{(r_2,s_2)}}$ and $\mathcal{R}^d_{\langle r_2,s_2\rangle}\times \mathcal{V}_{P_{(r_1,s_1)}}$:
\begin{align}
  \boxed{\mathcal{R}^d_{\langle r_1,s_1 \rangle} \times \mathcal{R}^d_{\langle r_2,s_2 \rangle} = \sum_{r\overset{2}{=}|r_1-r_2|+1}^{r_1+r_2-1}\ \sum_{s\overset{2}{=}|s_1-s_2|+1}^{s_1+s_2-1} \mathcal{R}^d_{\langle r,s \rangle}}\ ,
  \label{rrsr}
\end{align}
where the sums now run by increments of $2$. The number of terms is $\min(r_1,r_2)\min(s_1,s_2)$. For example,
\begin{align}
 \mathcal{R}^d_{\langle 5,2\rangle}\times \mathcal{R}^d_{\langle 3,3\rangle} = \mathcal{R}^d_{\langle 3,2\rangle} + \mathcal{R}^d_{\langle 5, 2\rangle} + \mathcal{R}^d_{\langle 7,2\rangle} +\mathcal{R}^d_{\langle 3,4\rangle}+\mathcal{R}^d_{\langle 5,4\rangle} + \mathcal{R}^d_{\langle 7,4\rangle} \ .
\end{align}

\subsection{Correlation functions}\label{sec:cor}

Correlation functions are observables of quantum field theory: quantities that we can in principle compute, and compare with experimental data. We will focus on $N$-point \myindex{correlation functions}\index{correlation function} (or \myindex{$N$-point functions}\index{N-point function@$N$-point function} for short) on the Riemann sphere: functions of $N$ distinct positions $z_1,z_2,\dots, z_N\in \overline{\mathbb{C}}$, which are written in terms of $N$ fields as
\begin{align}
 \Big< V_1(z_1) V_2(z_2) \cdots V_N(z_N)\Big>\ . 
\end{align}
Correlation functions are assumed to depend linearly on fields: in particular, this implies $\partial_{z_1}\left<  V_1(z_1) \cdots \right> = \left< \partial_{z_1} V_1(z_1) \cdots \right>$.
All the assumptions and results of Section \ref{sec:fope} may be understood as constraints on correlation functions. 
In 2d CFT, there are other observables that we will not consider, such as: 
\begin{itemize}
 \item Correlation functions on higher-genus Riemann surfaces, starting from the torus.
 \item Correlation functions on Riemann surfaces with boundaries, starting from the disc.
 \item Correlation functions in the presence of defect lines. 
 \item Entanglement entropy. 
\end{itemize}

\subsubsection{Conformal symmetry constraints}\label{sec:csc}

In order to derive the conformal symmetry constraints for an $N$-point function of primary fields $Z$, it is convenient to consider the $N+1$-point function $Z(y)$ that also includes the energy-momentum tensor $T(y)$:
\begin{align}
 Z = \left< \prod_{i=1}^N V_{\Delta_i}(z_i)\right> \quad , \quad Z(y) = \left< T(y)\prod_{i=1}^N V_{\Delta_i}(z_i)\right>\ .
\end{align}
The function $Z(y)$ is meromorphic, with $N$ poles at $y=z_i$, where its behaviour is constrained by the OPE $T(y)V_{\Delta_i}(z_i)$ \eqref{tvd}. To fully determine $Z(y)$, it remains to determine its behaviour at $y=\infty$. Because our fields live on the Riemann sphere, they must be smooth at $y=\infty$: in the case of $T(y)$, we assume that this means 
\begin{align}
 T(y) \underset{y\to\infty} = O\left(\frac{1}{y^4}\right)\ .
 \label{tinf}
\end{align}
This implies first of all that $\lim_{y\to \infty} Z(y)=0$, therefore $Z(y)$ is completely determined by its residues at $y=z_i$:
\begin{align}
 Z(y) = \sum_{i=1}^N \left(\frac{\Delta_i}{(y-z_i)^2} +\frac{1}{y-z_i}\frac{\partial}{\partial z_i}\right) Z\ .
 \label{zy}
\end{align}
Furthermore, Eq. \eqref{tinf} implies the vanishing of the coefficients of $y^{-1},y^{-2},y^{-3}$ in the Laurent expansion of $Z(y)$ near $y=\infty$, leading to the 3 \myindex{global Ward identities}\index{Ward identity!global---} 
\begin{align}
 \sum_{i=1}^N \partial_{z_i} Z = \sum_{i=1}^N \left(z_i \partial_{z_i} + \Delta_i\right) Z = \sum_{i=1}^N \left(z_i^2 \partial_{z_i} + 2\Delta_iz_i\right) Z = 0\ .
 \label{gward}
\end{align}
Equivalently, the behaviour of $Z$ under global conformal transformations \eqref{abcd} must be
\begin{align}
 \boxed{\left< \prod_{i=1}^N V_{\Delta_i}(z_i) \right>
 = \prod_{i=1}^N (cz_i +d)^{-2\Delta_i}\left< \prod_{i=1}^N  V_{\Delta_i}\left(\frac{az_i+b}{cz_i+d}\right) \right> }\ .
 \label{zgc}
\end{align}
The global Ward identities are valid not only for primary fields, but also for \myindex{quasi-primary fields}\index{quasi-primary field}\index{field!quasi-primary---}: $L_0$-eigenvectors that are annihilated by $L_1$ but not necessarily by $L_{n\geq 2}$.
In particular, applying Eq. \eqref{zgc} to the transformation $z\mapsto \frac{1}{z}$, we find $V_\Delta(z)\underset{z\to\infty}{=} O(z^{-2\Delta})$, which allows us to define 
\begin{align}
 V_\Delta(\infty) = \lim_{z\to\infty} z^{2\Delta}V_\Delta(z)\ . 
 \label{vdi}
\end{align}
Furthermore, 
according to the OPE $T(y)T(z)$ \eqref{tt}, the energy-momentum tensor $T$ is a quasi-primary field of dimension $\Delta=2$, whose assumed behaviour at $\infty$ \eqref{tinf} is consistent with Eq. \eqref{vdi}. 
Using Eq. \eqref{zgc} for a correlation function that involves $T$, we deduce a definition of descendant fields at $\infty$, 
\begin{align}
 L_n V_\Delta(\infty) = \frac{1}{2\pi i} \oint_\infty dy\ y^{1-n} T(y)V_\Delta(\infty)\ .
 \label{lnia}
\end{align}
(Compare with $L_nV_\Delta(z)$ \eqref{lvtv}.) 
For any $a\in\mathbb{C}$, we could alternatiely define $L_n V_\Delta(\infty)$ as $\frac{1}{2\pi i} \oint_\infty dy\ (y-a)^{1-n} T(y)V_\Delta(\infty)$. This would amount to using the tranformation $z\mapsto \frac{1}{z-a}$ instead of $z\mapsto \frac{1}{z}$.
We interpret $\frac{1}{z-a}$ as a \myindex{local coordinate}\index{local coordinate} near $z=\infty$. On general Riemann surfaces, correlation functions that involve descendant fields depend on choices of local coordinates, whereas correlation functions of primary fields do not, see \cite[Section 3]{br24}. On the Riemann sphere, there is
a canonical local coordinate near any point $z\in\mathbb{C}$, but not near $z=\infty$, where we chose the local coordinate $\frac{1}{z}$.

Let us solve the global Ward identities in the cases $N=1,2,3,4$. We have $3$ linear differential equations in the $N$ variables $z_i$, and we write their solutions up to $z_i$-independent prefactors:
\begin{itemize}
 \item $\boxed{N=1}$\ : The identities amount to $\partial_{z_1} Z = \Delta_1 Z=0$, and their solution is 
 \begin{subequations}
 \begin{align}
  \Big<V_{\Delta_1}(z_1)\Big>\propto \delta_{\Delta_1,0}\ . 
 \end{align}
 \item $\boxed{N=2}$\ : Again, there are more Ward identities than variables, leading to $Z\neq 0\implies \Delta_1=\Delta_2$, and to the solution
 \begin{align}
 \Big< V_{\Delta_1}(z_1)V_{\Delta_2}(z_2) \Big> \propto \delta_{\Delta_1,\Delta_2} z_{12}^{-2\Delta_1} \ .
 \label{2pt}
\end{align}
\item $\boxed{N=3}$\ : With as many Ward identities as variables, we have a unique solution with no constraints on the conformal dimensions:
\begin{align}
 \left< \prod_{i=1}^3 V_{\Delta_i}(z_i) \right> \propto z_{12}^{\Delta_3-\Delta_1-\Delta_2} z_{13}^{\Delta_2-\Delta_1-\Delta_3} z_{23}^{\Delta_1-\Delta_2-\Delta_3}\ .
 \label{3pt}
\end{align}
\item $\boxed{N=4}$\ : The solution involves an arbitrary function $G(z)$ of the \myindex{cross-ratio}\index{cross-ratio} $z=\frac{z_{12}z_{34}}{z_{13}z_{24}}$:
\begin{align}
 \left< \prod_{i=1}^4 V_{\Delta_i}(z_i) \right> 
 \propto z_{13}^{-2\Delta_1} z_{23}^{\Delta_1-\Delta_2-\Delta_3+\Delta_4} z_{24}^{-\Delta_1-\Delta_2+\Delta_3-\Delta_4} z_{34}^{\Delta_1+\Delta_2-\Delta_3-\Delta_4} G\left(\frac{z_{12}z_{34}}{z_{13}z_{24}}\right)\ .
 \label{4pt}
\end{align}
\end{subequations}
A 4-point function is therefore completely determined by its dependence on $1$ of its $4$ variables. In particular, it can be convenient to set $3$ positions to the fixed values $(z_2,z_3,z_4)=(0,\infty,1)$, in which case
\begin{align}
 \Big< V_{\Delta_1}(z) V_{\Delta_2}(0)V_{\Delta_3}(\infty)V_{\Delta_4}(1) \Big> \propto G(z)\ . 
\end{align}
\end{itemize}
In addition to global Ward identities, there are \myindex{local Ward identities}\index{Ward identity!local---} , which follow from $\oint_\infty dy\ \epsilon(y)Z(y)=0$, where $\epsilon(y)$ is a meromorphic function such that $\epsilon(y)\underset{y\to\infty}{=}O(y^2)$, with poles at $y=z_i$. Local Ward identities allow us to determine correlation functions of descendant fields from the corresponding correlation functions of primary fields. In the case $\epsilon(y) = \frac{1}{(y-z_1)^{n-1}}$ with $n\geq -1$, we compute $\oint_\infty = \sum_{i=1}^N\oint_{z_i}$ using Eq. \eqref{lvtv} for $L_{-n}V_{\Delta_1}(z_1)$ as well as Eq. \eqref{tvd} for $T(y)V_{\Delta_j}(z_j)$ if $j\geq 2$, we obtain the Ward identity
\begin{align}
 \left<L_{-n}V_{\Delta_1}(z_1)\prod_{j=2}^N V_{\Delta_j}(z_j)\right> = \sum_{j=2}^N \left(-\frac{1}{z_{j1}^{n-1}}\frac{\partial}{\partial z_j} +\frac{n-1}{z_{j1}^n} \Delta_j\right) \left<\prod_{j=1}^N V_{\Delta_j}(z_j)\right>\ . 
 \label{lwi}
\end{align}
If $n\geq 2$ this is a local Ward identity, while for $n=-1,0,1$ we recover global Ward identities. 
In particular, for any creation operator $L\in\mathcal{L}$ we can compute the ratio
\begin{align}
 g^L_{\Delta_1,\Delta_2,\Delta_3} = \frac{\left< LV_{\Delta_1}(0) V_{\Delta_2}(\infty)V_{\Delta_3}(1)\right>}{\left< V_{\Delta_1}(0) V_{\Delta_2}(\infty)V_{\Delta_3}(1)\right>}\ .
 \label{gl}
\end{align}
Let us show that this coincides with the quantity that appeared in OPE Ward identities \eqref{sgl}. 
In the 3-point function $\left< L'V_\Delta(0)V_{\Delta_2}(\infty)V_{\Delta_1}(1)\right>\propto g^{L'}_{\Delta,\Delta_2,\Delta_1}$, we write the OPE $V_{\Delta_1}(1)V_{\Delta_2}(\infty)$ as a linear combination of descendant fields $LV_{\Delta}(\infty)$. Since our local coordinate near $\infty$ is $y=\frac{1}{z}$, we obtain a series in powers of $y_1-y_2=1$, which may seem large until we remember that the field $L'V_{\Delta}(0)$ is infinitely far away at $y=\infty$. 
Assuming for simplicity $\left<V_\Delta(0) V_\Delta(\infty) \right>=1$, 
we obtain
\begin{align}
 g^{L'}_{\Delta,\Delta_2,\Delta_1}= \sum_{L\in\mathcal{L}} f^{\Delta,L}_{\Delta_1,\Delta_2} \Big<L'V_\Delta(0) LV_\Delta(\infty) \Big> \ .
 \label{gfr}
\end{align}
By the definition \eqref{lnia} of descendant fields at $\infty$, the 2-point function coincides with the Shapovalov form, 
\begin{align}
 \boxed{S_{L,L'}(\Delta)  = \Big<L'V_\Delta(0) LV_\Delta(\infty) \Big> =  \Big<L^*L'V_\Delta(0) V_\Delta(\infty) \Big>} \ , 
 \label{sll}
 \end{align}
which vanishes if $|L|>|L'|$ because $V_\Delta$ is primary, and also if $|L|<|L'|$ because $LV_\Delta$ is at $\infty$. So we recover the expression \eqref{sgl} of $g^L$ in terms of OPE coefficients and the Shapovalov form. 

\subsubsection{Single-valuedness and chiral factorization}

We assume that correlation functions are \myindex{single-valued}\index{single-valuedness} on the Riemann sphere. Equivalently, we assume that any 2 fields are \myindex{mutually local}\index{mutual locality}, i.e. that they have trivial monodromies around each other: $V_1(ze^{2\pi i})V_2(0)=V_1(z)V_2(0)$. 

The correlation functions of Section \ref{sec:csc} are of the form $z^\Delta$, which is multivalued unless $\Delta\in\mathbb{Z}$. Single-valued functions like $|z|^{2\Delta}=z^\Delta\bar{z}^\Delta$ are not written in terms of powers of $z$, but also involve $\bar z$. In the literature, the notation $f(z,\bar z)$ is sometimes used for arbitrary functions of $z$, with $f(z)$ reserved to holomorphic functions, i.e. functions such that $\frac{\partial}{\partial \bar z} f(z) =0$. We will not use the notation $f(z,\bar z)$, which is redundant since $\bar z$ is itself a function of $z$.

The generators $\ell_n$ \eqref{lpz} of local conformal transformations are only valid when acting on holomorphic functions: for more general functions, the generators are
\begin{align}
 \ell_n +\bar{\ell}_n \quad , \quad i(\ell_n -\bar \ell_n)\  .
 \label{epbe}
\end{align}
In fact we want the algebra of conformal transformations to act on a complex vector space: the space of complex-valued functions on $\overline{\mathbb{C}}$ for the Witt algebra, the space of states of the CFT for the Virasoro algebra. We should therefore complexify our symmetry algebra:
\begin{align}
 \text{Span}_\mathbb{R}\left(\ell_n +\bar{\ell}_n , i(\ell_n -\bar \ell_n)\right)_{n\in\mathbb{Z}}\ \underset{\text{complexify}}{\longrightarrow}\ \text{Span}_\mathbb{C}\left(\ell_n , \bar \ell_n\right)_{n\in\mathbb{Z}}=\text{Witt}_\mathbb{C}\times \overline{\text{Witt}}_\mathbb{C}\ .
\end{align}
After taking central extensions, we obtain 2 copies of the Virasoro algebra over $\mathbb{C}$: a \myindex{left-moving}  Virasoro algebra\index{left-moving} $\mathfrak{V}$ with generators $L_n$, and a \myindex{right-moving} Virasoro algebra\index{right-moving} $\bar{\mathfrak{V}}$ with generators $\bar{L}_n$, such that $[L_m,\bar L_n]=0$. For simplicity we assume that the left and right central charges coincide.
In the case of the algebra of global conformal transformations, we start with the 6-dimensional real Lie algebra $\mathfrak{sl}_2(\mathbb{C})$, with generators \eqref{epbe} with $n=-1,0,1$, and the complexification yields $\mathfrak{sl}_2(\mathbb{C})^\mathbb{C} = \mathfrak{sl}_2(\mathbb{C})\times \mathfrak{sl}_2(\mathbb{C})$.

In Sections \ref{sec:vir} and \ref{sec:fope}, we have been doing \myindex{chiral CFT}\index{CFT!chiral---}\index{chiral CFT}, which means considering $\mathfrak{V}$ only. In a full CFT, the symmetry algebra is the \myindex{conformal algebra}\index{conformal!---algebra} $\mathfrak{V}\times \bar{\mathfrak{V}}$. When it comes to representations, Verma modules of $\mathfrak{V}\times \bar{\mathfrak{V}}$ are simply $\mathcal{V}_\Delta\otimes \bar{\mathcal{V}}_{\bar \Delta}$, where we write a bar over representations of $\bar{\mathfrak{V}}$. There also exist representations of $\mathfrak{V}\times \bar{\mathfrak{V}}$ that are not factorized, in particular logarithmic representations \cite{nr20}.

Fields now belong to representations of $\mathfrak{V}\times \bar{\mathfrak{V}}$, and they obey the right-moving equation $\frac{\partial}{\partial \bar z} V(z) = \bar L_{-1} V(z)$, in addition to the left-moving equation \eqref{pvlv}. This allows us to define a right-moving energy-momentum tensor $\bar T(y) = \sum_{n\in\mathbb{Z}} \frac{\bar L_n^{(z)}}{(\bar y-\bar z)^{n+2}}$. While $T$ is holomorphic, $\bar T$ is antiholomorphic:
\begin{align}
 \partial_{\bar z} T(z) = 0 \quad , \quad \partial_z \bar T(z) = 0 \ . 
\end{align}
Let $V_{\Delta,\bar\Delta}(z)$ be a field that is primary with respect to the conformal algebra, with the left dimension $\Delta$ and the right dimension $\bar\Delta$. The difference of these dimensions is called the \myindex{conformal spin}\index{conformal!---spin}\index{spin!conformal---}
\begin{align}
 S = \Delta -\bar\Delta\ . 
 \label{sdd} 
\end{align}
According to Eq. \eqref{zgc}, 
under a rotation $z\mapsto e^{i\theta}z$, corresponding to a global conformal transformation $\left(\begin{smallmatrix} a & b \\ c & d\end{smallmatrix}\right)= \left(\begin{smallmatrix} e^{\frac{i}{2}\theta} & 0 \\ 0 &  e^{-\frac{i}{2}\theta}\end{smallmatrix}\right)$, a primary field transforms as 
\begin{align}
 V_{\Delta,\bar\Delta}(z) \to  e^{iS\theta} V_{\Delta,\bar\Delta}\left(e^{i\theta} z\right) \ .
 \label{veis}
\end{align}
In order to write correlation functions, let us define the \myindex{modulus squared notation}\index{modulus squared notation} for functions of the central charge, conformal dimensions, and positions,
\begin{align}
 \left| f(c,\Delta, z)\right|^2 = f(c,\Delta, z)f(c,\bar\Delta,\bar z)\ ,
 \label{fdz2}
\end{align}
where $z,\bar z$ are complex conjugates, while $\Delta,\bar\Delta$ are generally not. 
For simplicity we write $V_i$ for a primary field of dimensions $\Delta_i,\bar\Delta_i$.
The solution of the left- and right-moving global Ward identities for 2-point functions is the modulus squared of Eq. \eqref{2pt},
\begin{align}
\boxed{ \Big< V_1(z_1)V_2(z_2) \Big> = B_{12} \left|\delta_{\Delta_1,\Delta_2} z_{12}^{-2\Delta_1}\right|^2} \ ,
\label{2ptb}
\end{align}
where the constant, $z_i$-independent coefficient $B_{12}$ is now called the \myindex{2-point structure constant}\index{structure constant!2-point---}. Similarly, 3-point functions are given by the modulus squared of Eq. \eqref{3pt},
\begin{align}
 \boxed{  \Big< V_1(z_1)V_2(z_2)V_3(z_3) \Big> = C_{123} \left|z_{12}^{\Delta_3-\Delta_1-\Delta_2} z_{13}^{\Delta_2-\Delta_1-\Delta_3} z_{23}^{\Delta_1-\Delta_2-\Delta_3}\right|^2 }\ ,
 \label{3ptc}
\end{align}
where $C_{123}$ is the \myindex{3-point structure constant}\index{structure constant!3-point---}. Single-valuedness of the 2-point and 3-point functions implies that spins obey
\begin{align}
 S_i\in\frac12 \mathbb{Z} \quad , \quad S_1+S_2+S_3 \in\mathbb{Z}\ . 
 \label{sihz}
\end{align}
Now we have assumed that fields commute \eqref{comm}, but the $z_i$-dependent factors of the 2-point and 3-point functions are not invariant under permutations. This implies that structure constants behave as
\begin{align}
 B_{21} = (-)^{2S_1} B_{12} \quad ,\quad C_{\sigma(1)\sigma(2)\sigma(3)} = \text{sign}(\sigma)^{S_1+S_2+S_3} C_{123}\ , 
 \label{b21}
\end{align}
where $\sigma$ is a permutation of $\{1,2,3\}$. In particular, for the 2-point function of a field with itself, we have $B_{12}=B_{21}$, which must vanish if the spin is half-integer $S_1\in \frac12+\mathbb{Z}$. Fields with half-integer spins should rather anti-commute with themselves, i.e. they should be fermionic. 

Finally, notice that $\left<V_1V_2\right> = \left<IV_1V_2\right>$, where $I$ is the identity field. Assuming for simplicity that $\left<V_1V_2\right>\neq 0$ implies
not only $(\Delta_1,\bar \Delta_1)=(\Delta_2,\bar\Delta_2)$ but in fact
$V_1=V_2$, we have
\begin{align}
 B_{jk}=\delta_{jk}B_j\ ,
 \label{djkbj}
\end{align}
which implies
\begin{align}
 \boxed{B_k = C_{Ikk}}\ .
 \label{bicii}
\end{align}

\subsubsection{Crossing symmetry and conformal blocks}\label{sec:cscb}

Let us systematically use OPEs for simplifying correlation functions. Knowing the contribution \eqref{prope} of left-moving descendants of primary fields, the complete OPE of 2 primary fields reads 
\begin{align}
 V_1(z_1)V_2(z_2) = \sum_{k\in\mathcal{S}^{12}} C_{12}^k \left|z_{12}^{\Delta_k-\Delta_1-\Delta_2}\sum_{L\in\mathcal{L}} z_{12}^{|L|}f_{\Delta_1,\Delta_2}^{\Delta_k,L}\right|^2 L\bar L V_k(z_2)\ ,
 \label{tope}
\end{align}
where the basis $\mathcal{L}$ of creation operators was defined in Eq. \eqref{lcm},
we assume that all fields are primaries or descendants thereof, and the \myindex{OPE spectrum}\index{spectrum!OPE---}\index{OPE!---spectrum} $\mathcal{S}^{12}$ is the set of primary fields that appear in our OPE.
The first few terms read
\begin{multline}
 V_1(z_1)V_2(z_2) = \sum_{k\in\mathcal{S}^{12}} C_{12}^k \left|z_{12}^{\Delta_k-\Delta_1-\Delta_2}\right|^2
 \times \Big\{V_k(z_2) + f_{\Delta_1,\Delta_2}^{\Delta_k,L_{-1}} z_{12} L_{-1}V_k(z_2)
 \\
 + f_{\bar\Delta_1,\bar\Delta_2}^{\bar\Delta_k,L_{-1}} \bar z_{12} \bar L_{-1}V_k(z_2) + f_{\Delta_1,\Delta_2}^{\Delta_k,L_{-1}}f_{\bar\Delta_1,\bar\Delta_2}^{\bar\Delta_k,L_{-1}} |z_{12}|^2 L_{-1}\bar L_{-1}V_k(z_2)
 +\cdots \Big\} \ .
\end{multline}
Let us insert this OPE in a 3-point function, and focus on the leading terms as $z_1\to z_2$. Using the 2-point function \eqref{2ptb}, we obtain
\begin{align}
  \Big< V_1(z_1)V_2(z_2)V_3(z_3) \Big> \underset{z_1\to z_2}{\sim} 
  \left|z_{12}^{\Delta_3-\Delta_1-\Delta_2}z_{23}^{-2\Delta_3}\right|^2 
  \sum_{k\in\mathcal{S}}
  C_{12}^k B_{k3}\ .
\end{align}
Comparing with the 3-point function \eqref{3ptc}, and using Eq. \eqref{djkbj}, this implies a relation between the 2-point, 3-point and OPE structure constants,
\begin{align}
 \boxed{C_{123} = C_{12}^3B_{3}} \ .
 \label{ccb}
\end{align}
If we now insert the OPE in the 4-point function $\left<V_1(z)V_2(0)V_3(\infty)V_4(1)\right>$, we obtain a linear combination of 3-point functions of the type 
\begin{align}
 \Big<L\bar{L}V_k(0) V_3(\infty)V_4(1)\Big> = \left|g^L_{\Delta_k,\Delta_3,\Delta_4}\right|^2 \Big<V_k(0) V_3(\infty)V_4(1)\Big>=\left|g^L_{\Delta_k,\Delta_3,\Delta_4}\right|^2 C_{k34} \ ,
\end{align}
where the action of $L,\bar L\in\mathcal{L}$ gives rise to the factors $g^L$ \eqref{gl}, which are determined by conformal symmetry. 
This allows us to write our 4-point function as 
\begin{align}
 \boxed{\Big<V_1(z)V_2(0)V_3(\infty)V_4(1)\Big> = \sum_{k\in\mathcal{S}^{(s)}} C_{12}^k C_{k34} \left|\mathcal{F}^{(s)}_{\Delta_k}\left(c\middle|\Delta_1,\Delta_2,\Delta_3,\Delta_4\middle|z\right)\right|^2}  \ ,
 \label{sdec}
\end{align}
where we introduce the $s$-\myindex{channel spectrum}\index{spectrum!channel---}\index{channel!---spectrum} $\mathcal{S}^{(s)}= \mathcal{S}^{12}\cap \mathcal{S}^{34}$, and the $s$-channel \myindex{Virasoro blocks}\index{block!Virasoro---}\index{Virasoro!---block}, which are known functions of $z$ as well as the central charge and conformal dimensions, of the type
\begin{align}
 \mathcal{F}_{\Delta}^{(s)}\left(c\middle|\Delta_1,\Delta_2,\Delta_3,\Delta_4\middle|z\right) =  z^{\Delta-\Delta_1-\Delta_2}\sum_{N=0}^\infty f_N(\Delta)z^N \ .
 \label{fsl}
\end{align}
The OPE that we have taken leads to coefficients $f_N(\Delta)$ that are combinations of $f^L$ and $g^L$. Using the relation \eqref{fsg}, these coefficients can alternatively be rewritten in terms of $g^L$ alone, at the price of introducing the inverse Shapovalov form:
\begin{align}
 f_N(\Delta) = \sum_{\substack{L\in\mathcal{L} \\ |L|=N}} f_{\Delta_1,\Delta_2}^{\Delta,L}g^L_{\Delta,\Delta_3,\Delta_4} = \sum_{\substack{L,L'\in\mathcal{L} \\ |L|=|L'|=N}} g^{L'}_{\Delta,\Delta_2,\Delta_1} S_{L,L'}^{-1}(\Delta) g^L_{\Delta,\Delta_3,\Delta_4} \ ,
 \label{fnd}
\end{align}
The first few coefficients may be assembled from Eqs. \eqref{fl12}, \eqref{sgl}:
\begin{subequations}
\begin{align}
& f_0(\Delta) = 1 \quad , \quad f_1(\Delta) = \frac{(\Delta+\Delta_1-\Delta_2)(\Delta+\Delta_4-\Delta_3)}{2\Delta} \ ,
\\
& f_2(\Delta) =  \frac{\left[\begin{smallmatrix} (\Delta+\Delta_4-\Delta_3)(\Delta+\Delta_4-\Delta_3+1) \\ \Delta+2\Delta_4-\Delta_3 \end{smallmatrix}\right]^T
\left[\begin{smallmatrix} 2+\frac{c}{4\Delta} & -3 \\ -3 & 4\Delta+2 \end{smallmatrix}\right]
\left[\begin{smallmatrix} (\Delta+\Delta_1-\Delta_2)(\Delta+\Delta_1-\Delta_2+1) \\ \Delta+2\Delta_1-\Delta_2 \end{smallmatrix}\right]}{16(\Delta-\Delta_{(2,1)})(\Delta-\Delta_{(1,2)})}
 \ .
 \label{f2d}
\end{align}
\end{subequations}
The $s$-channel decomposition \eqref{sdec} reduces the 4-point function to a combination of model-dependent structure constants, and universal Virasoro blocks. Repeatedly using OPEs, any $N$-point function can likewise be decomposed into $N$-point Virasoro blocks. However, these decompositions are not unique. In the case $N=4$, there are 3 possible decompositions, called the \myindex{$s$-channel}\index{channel!$s$-, $t$-, $u$-}, \myindex{$t$-channel} and \myindex{$u$-channel} decompositions, depending on which OPE we perform. For each channel, let us write the relevant OPE, the \myindex{4-point structure constants}\index{structure constant!4-point---} that appear as coefficients, the asymptotics of the Virasoro blocks, and a diagram that represents the decomposition:
\begin{align}
 \begin{array}{|r||c|c|c|}
  \hline 
  \text{Channel} & s & t & u
  \\ \hline
  \text{OPE} & V_1(z)V_2(0) & V_1(z)V_4(1) & V_1(z)V_3(\infty)
  \\ \hline 
  \text{4-point} & D^{(s)}_k = C_{12}^k C_{k34} & D^{(t)}_k = C_{41}^k C_{k23} & D^{(u)}_k = C_{31}^k C_{k42}
\\ \hline 
  \text{Asymptotics} & {\scriptstyle \mathcal{F}^{(s)}_{\Delta}(z)\ \underset{z\to 0}{\sim}\ z^{\Delta-\Delta_1-\Delta_2}} & {\scriptstyle\mathcal{F}^{(t)}_{\Delta}(z)\ \underset{z\to 1}{\sim}\ (1-z)^{\Delta-\Delta_1-\Delta_4}}
  & {\scriptstyle\mathcal{F}^{(u)}_\Delta(z)\  \underset{z\to \infty}{\sim}\ \left(\tfrac{1}{z}\right)^{\Delta+\Delta_1-\Delta_3}}
  \\
  \hline 
\text{Diagram} & 
\begin{tikzpicture}[baseline=(current  bounding  box.center), very thick, scale = .3]
\draw (-1,2) node [left] {$2$} -- (0,0) -- node [above] {$k$} (4,0) -- (5,2) node [right] {$3$};
\draw (-1,-2) node [left] {$1$} -- (0,0);
\draw (4,0) -- (5,-2) node [right] {$4$};
\end{tikzpicture}
& 
\begin{tikzpicture}[baseline=(current  bounding  box.center), very thick, scale = .3]
 \draw (-2,3) node [left] {$2$} -- (0,2) -- node [left] {$k$} (0,-2) -- (-2, -3) node [left] {$1$};
\draw (0,2) -- (2,3) node [right] {$3$};
\draw (0,-2) -- (2, -3) node [right] {$4$};
\end{tikzpicture}
& 
\begin{tikzpicture}[baseline=(current  bounding  box.center), very thick, scale = .3]
\draw (-1,2) node [left] {$2$} -- (0,0) -- node [above] {$k$} (4,0) -- (5,2) node [right] {$3$};
\draw (-1,-2) node [left] {$1$} -- (4,0);
\draw (0,0) -- (5,-2) node [right] {$4$};
\end{tikzpicture}
  \\
  \hline 
 \end{array}
 \label{stu}
\end{align}
By the commutativity axiom \eqref{comm}, the 4-point function \eqref{4pt} is invariant under permutations of the 4 fields, and this determines how Virasoro blocks behave under permutations. In particular, $t$-channel and $u$-channel blocks can be deduced from $s$-channel blocks via the relations
\begin{align}
 \mathcal{F}^{(t)}_\Delta(z) = \mathcal{F}^{(s)}_\Delta(\Delta_1,\Delta_4,\Delta_3,\Delta_2|1-z) \quad , \quad 
 \mathcal{F}^{(u)}_\Delta(z) = z^{-2\Delta_1}\mathcal{F}^{(s)}_\Delta(\Delta_1,\Delta_3,\Delta_2,\Delta_4|\tfrac{1}{z})\ .
 \label{stotu}
\end{align}
The decomposition \eqref{sdec} of a 4-point function involves $s$-channel \myindex{conformal blocks}\index{block!conformal---}\index{conformal!---block} of the form
\begin{align}
 \mathcal{G}^{(x)}_{\Delta,\bar\Delta}(z) = \left|\mathcal{F}_{\Delta}^{(x)}(z)\right|^2\ , \qquad (x=s,t,u)\ . 
 \label{gf2}
\end{align}
These conformal blocks are chirally factorized into Virasoro blocks because of our assumption that all fields are primaries or descendants. However, representations of the conformal algebra that are not factorized give rise to conformal blocks that are also not factorized: this is the case with the logarithmic blocks of Section \ref{sec:log}.

Let us now write the \myindex{crossing symmetry equations}\index{crossing symmetry equation} as the equality between the 3 decompositions of a 4-point function into structure constants and conformal blocks:
\begin{align}
 \boxed{\forall z\in \overline{\mathbb{C}}\backslash\{0,1,\infty\} \ , \quad 
 \sum_{k\in\mathcal{S}^{(s)}} D_k^{(s)} \mathcal{G}^{(s)}_{\Delta_k,\bar\Delta_k}(z) 
 = \sum_{k\in\mathcal{S}^{(t)}} D_k^{(t)} \mathcal{G}^{(t)}_{\Delta_k,\bar\Delta_k}(z) 
 = \sum_{k\in\mathcal{S}^{(u)}} D_k^{(u)} \mathcal{G}^{(u)}_{\Delta_k,\bar\Delta_k}(z)} \ . 
 \label{seteu}
\end{align}
Given 4 fields $V_1,V_2,V_3,V_4$, this is a system of infinitely many equations (parametrized by the cross-ratio $z$) for the channel spectra $\mathcal{S}^{(x)}$ and the 4-point structure constants $D_k^{(x)}$.
We may also consider the system of crossing symmetry for all 4-point functions, whose unknowns are the \myindex{CFT data}\index{CFT!---data}: the spectrum of the theory (which includes the channel spectra of all 4-point functions) and the 2-point and 3-point structure constants. CFT data determine OPEs, and crossing symmetry of all 4-point functions is equivalent to the associativity of OPEs, and therefore to the consistency of the CFT on the sphere.

\section{Sketching exactly solvable CFTs}\label{sec:sesc}

In this section we derive the spectra and fusion rules of exactly solvable CFTs. We begin with a general discussion of CFTs, their spectra, and the consequences of having degenerate fields. Historically, the CFTs we will discuss have been discovered using a variety of techniques:
\begin{itemize}
 \item the modular bootstrap for minimal models \cite{fms97},
 \item the Lagrangian approach for Liouville theory with $c\geq 25$ \cite{zz95},
 \item lattice constructions for the $O(n)$ and Potts models \cite{fsz87},
 \item the numerical bootstrap for Liouville theory with $c\leq 1$ \cite{rs15}.
\end{itemize}
Here we recover their spectra from assumptions about Virasoro representations, their fusion rules, and single-valuedness of correlation functions. Although our assumptions are simple, they are not unique or inevitable, and it can be interesting to relax or modify them. This approach can also be used in CFTs with extended symmetry algebras, see for example \cite{bp24}.

\subsection{Spectrum and degenerate fields}

\subsubsection{What is a CFT? What is a spectrum?}

We define a \myindex{conformal field theory}\index{conformal!---field theory} as a set of correlation functions that obey the assumptions of Section \ref{sec:bo}, most importantly Virasoro symmetry, the commutativity of fields, the existence of OPEs, and single-valuedness. We already know that on the sphere, a conformal field theory is characterized by a spectrum and 2-point and 3-point structure constants such that crossing symmetry is obeyed.

This definition works well in \myindex{rational CFTs}\index{CFT!rational---}\index{rational CFT}, i.e. CFTs whose spectra are generated by finitely many primary fields and their infinitely many descendants.
In more general CFTs, we have to distinguish various spectra associated to various objects. By an abuse of terminology, we now call spectrum a basis of primary fields:
\begin{align}
 \begin{array}{|c|c|l|}
  \hline 
  \text{Object} & \text{Spectrum}  & \text{Definition or constraint}
  \\
  \hline \hline 
  V_1V_2 & \text{OPE spectrum } \mathcal{S}^{12} & V_1V_2 \sim \sum_{k\in \mathcal{S}^{12}} V_k 
  \\
  \hline 
  \left<V_1V_2V_3V_4\right> & \text{Channel spectrum } \mathcal{S}^{(s)} &  \mathcal{S}^{(s)}= \mathcal{S}^{12}\cap \mathcal{S}^{34}
  \\
  \hline 
  \text{CFT} & \text{Physical spectrum } \mathcal{S} &  i,j\in\mathcal{S}\implies \mathcal{S}^{ij}\subset \mathcal{S}  
  \\
  \hline 
  \widehat{\text{CFT}} & \text{Extended spectrum } \widehat{\mathcal{S}} & \mathcal{S}\subset \widehat{\mathcal{S}}\ , \quad i,j\in\widehat{\mathcal{S}}\implies \mathcal{S}^{ij}\subset \widehat{\mathcal{S}}
  \\
  \hline 
 \end{array}
 \label{specs}
\end{align}
The \myindex{physical spectrum}\index{spectrum!physical---} is made of the fields whose correlation functions we are interested in. Our constraint on the physical spectrum is called \myindex{closure under OPE}\index{closure under OPE}. This constraint means that if a field appears in an OPE (or a fortiori in a channel spectrum), then its correlation functions exist.
We also introduce \myindex{extended spectra}\index{spectrum!extended---} as larger spectra that are also closed under OPE. There are various reasons for considering extended spectra:
\begin{itemize}
 \item In Section \ref{sec:csc}, we have used the energy-momentum tensor $T$ for deriving Ward identities. However, in Liouville theory, $T$ does not belong to the physical spectrum. Nevertheless, it always makes sense to insert $T$ in correlation functions, because it is a descendant of the identity field. 
 \item In the $O(n)$ model and in Liouville theory, we will use the degenerate field $V^d_{\langle 1,2\rangle}$ for constraining correlation functions, even though it does not belong to the physical spectrum. Correlation functions of the $O(n)$ model that involve $V^d_{\langle 1,2\rangle}$ are not single-valued: they belong to an extended CFT that obeys weaker assumptions. 
 \item The spectrum of Liouville theory is continuous, so that the OPE gives rise to an integral over the real momenta of primary fields. Since the integrand is meromorphic, the integration contour can be deformed in the complex plane. This leads to an extended spectrum of complex momenta.
\end{itemize}
We will say that \myindex{a field exists}\index{existence (of a field)} if it belongs to some extended spectrum, i.e. if it can be consistently added to our CFT.

\subsubsection{Diagonal and non-diagonal fields}

A simple way to satisfy the half-integer spin condition \eqref{sihz} is to have fields of spin zero. However, whenever degenerate fields exist, we will use the slightly different notion of a \myindex{diagonal field}\index{field!diagonal---}\index{diagonal!---field}: a field of spin zero whose OPEs with degenerate fields only produces fields of spin zero. Conversely, a \myindex{non-diagonal field}\index{field!non-diagonal---} is a field that either has nonzero spin, or is related to fields with nonzero spins by taking OPEs with degenerate fields. A \myindex{diagonal CFT}\index{CFT!diagonal---}\index{diagonal!---CFT} is a CFT whose spectrum only involves diagonal fields.

Let us summarize the primary fields that we will consider, together with their momenta, and the representations of the conformal algebra that they generate.
\begin{align}
 \begin{array}{|r|l|l|l|l|}
  \hline 
  \text{Name} & \text{Notation} & \text{Conditions} & (P, \bar P) & \text{Representation}
  \\
  \hline \hline 
  \text{Degenerate} &  V^d_{\langle r,s\rangle} &  r,s\in\mathbb{N}^* & \left(P_{(r,s)},P_{(r,s)}\right)  & \mathcal{R}^d_{\langle r,s\rangle}\otimes \bar{\mathcal{R}}^d_{\langle r,s\rangle} 
  \\
  \hline 
  \text{Fully degenerate} & V^f_{\langle r,s\rangle} & \begin{array}{l} r,s\in\mathbb{N}^* \\ \beta^2\in\mathbb{Q} \end{array} & \left(P_{(r,s)},P_{(r,s)}\right)  & \mathcal{R}^f_{\langle r,s\rangle}\otimes \bar{\mathcal{R}}^f_{\langle r,s\rangle} 
  \\
  \hline 
  \text{Diagonal} & V_P & P\in\mathbb{C} & \left(P,P\right) & \mathcal{V}_P\otimes \bar{\mathcal{V}}_P 
  \\
  \hline 
  \text{Non-diagonal} & V_{(r,s)} & 
  \begin{array}{l} r,s\in\mathbb{C} \\ 
  rs\in \tfrac12\mathbb{Z} \end{array} & \left(P_{(r,s)},P_{(r,-s)}\right) & 
  \begin{array}{l} \mathcal{V}_{P_{(r,s)}} \otimes \bar{\mathcal{V}}_{P_{(r,-s)}} \\ \text{(in general)} \end{array}
  \\
  \hline 
 \end{array}
 \label{fields}
\end{align}
We parametrize non-diagonal primary fields $V_{(r,s)}$ using the Kac indices defined by Eq. \eqref{prs}, instead of the left and right dimensions. 
The indices $r,s$ are found by solving the linear system $\renewcommand{\arraystretch}{0.9}\Big\{\begin{array}{l} \beta r-\beta^{-1}s = 2P\ , \\ \beta r+\beta^{-1}s = 2\bar P\ . \end{array}$
According to Eq. \eqref{drms}, the conformal spin of $V_{(r,s)}$ is 
\begin{align}
 \boxed{S_{(r,s)}= -rs} \ . 
 \label{srs}
\end{align}
Then $V_{(r,s)}$ generates the Verma module $\mathcal{V}_{P_{(r,s)}} \otimes \bar{\mathcal{V}}_{P_{(r,-s)}}$, unless that module has singular vectors, which occurs if $r,s\in\mathbb{Z}^*$, and also for other values of $r,s$ if $\beta^2\in\mathbb{Q}$. If there are singular vectors, $V_{(r,s)}$ may generate a degenerate representation, or belong to a logarithmic representation.

Our labelling of fields is redundant, because the reflection $P\to -P$ leaved the conformal dimension \eqref{dp} invariant. We therefore assume the \myindex{reflection relations}\index{reflection!---relation}
\begin{align}
 V_{P} = V_{-P} \quad , \quad V_{(r,s)} = V_{(-r,-s)} \ . 
 \label{refl}
\end{align}
In some cases we will allow a nontrivial \myindex{reflection coefficient}\index{reflection!---coefficient} $R_P$ such that $V_P=R_P V_{-P}$.

\subsubsection{Constraints from degenerate fields}\label{sec:dotc}

We will now study how the existence of diagonal degenerate fields constrains the spectrum. We will focus on the degenerate fields $\{V^d_{\langle 2,1\rangle}$ and $V^d_{\langle 1,2\rangle}\} $, whose repeated OPEs generate the full set of degenerate fields $\{V^d_{\langle r,s\rangle}\}_{r,s\in\mathbb{N}^*}$.  We will focus on 2 types of CFTs:
\begin{itemize}
 \item \myindex{CFTs with 2 degenerate fields}\index{CFT!---with 2 degenerate fields}, in the sense that both $V^d_{\langle 2,1\rangle}$ and $V^d_{\langle 1,2\rangle}$ exist. 
 \item \myindex{CFTs with 1 degenerate field}\index{CFT!---with 1 degenerate field} $V^d_{\langle 1,2\rangle}$. Because the invariance under $\beta\to \beta^{-1}$ is broken by choosing $V^d_{\langle 1,2\rangle}$ rather than $V^d_{\langle 2,1\rangle}$, such CFTs depend on $\beta^2$ \eqref{cb} rather than on the central charge $c$. 
\end{itemize}
There might be CFTs such as $V^d_{\langle 1,3\rangle}$ exists whereas $V^d_{\langle 1,2\rangle}$ does not, but no examples are known.

Let us determine the OPEs of degenerate fields with other primary fields.
The OPE of $V^d_{\langle 2,1\rangle}$ or $V^d_{\langle 1,2\rangle}$ with a given primary field is constrained by the fusion rules \eqref{rvvp}, which allow 2 left-moving momenta and 2 right-moving momenta, leading to 4 allowed primary fields. In the case of a diagonal field $V_P$, the requirement that only primary fields of spin zero appear singles out the 2 diagonal fields among the 4 allowed fields,
\begin{align}
 \boxed{V^d_{\langle 2,1\rangle} V_P \sim \sum_\pm V_{P\pm\frac{\beta}{2}}} \quad , \quad \boxed{V^d_{\langle 1,2\rangle} V_P \sim \sum_\pm V_{P\pm\frac{1}{2\beta}}} \ . 
 \label{vpope}
\end{align}
In the case of a non-diagonal field $V_{(r,s)}$, the 4 fields that can a priori appear are
\begin{subequations}
\begin{align}
 V^d_{\langle 2,1\rangle}V_{(r,s)} &\subset \sum_\pm V_{(r\pm 1,s)} + \sum_\pm V_{(r,s\pm \beta^2)}\ ,
 \label{vtovrs}
 \\
 V^d_{\langle 1,2\rangle}V_{(r,s)} &\subset \sum_\pm V_{(r\pm \beta^{-2},s)} + \sum_\pm V_{(r,s\pm 1)}\ .
 \label{votvrs}
\end{align}
\end{subequations}
By Eq. \eqref{sihz} all fields must have half-integer spins, so the following spin differences must also be half-integer whenever the corresponding fields exist:
\begin{subequations}
\begin{align}
 S_{(r\pm 1,s)} - S_{(r,s)} = \mp s &\quad , \quad S_{(r,s\pm \beta^2)} - S_{(r,s)} = \mp r\beta^2\ ,
 \label{sdiff1}
 \\
 S_{(r\pm \beta^{-2},s)} - S_{(r,s)} = \mp s\beta^{-2} &\quad , \quad S_{(r,s\pm 1)} - S_{(r,s)} = \mp r\ . 
 \label{sdiff2}
\end{align}
\end{subequations}
We now assume that $\beta^2\notin \mathbb{Q}$, and remember $rs\in\frac12\mathbb{Z}$. It follows that $V_{(r\pm 1,s)}$ and $V_{(r,s\pm 1)}$ can coexist neither with $V_{(r,s\pm \beta^2)}$ nor with $V_{(r\pm \beta^{-2},s)}$. Therefore, the OPEs $V^d_{\langle 2,1\rangle}V_{(r,s)}$ and $V^d_{\langle 1,2\rangle}V_{(r,s)}$ each contain at most 2 primary fields, out of the 4 that are allowed by fusion rules. For notational simplicity, we choose these 2 primary fields as follows: 
\begin{align}
 \boxed{V^d_{\langle 2,1\rangle}V_{(r,s)} \sim \sum_\pm V_{(r\pm 1,s)}} \quad ,\quad
 \boxed{V^d_{\langle 1,2\rangle}V_{(r,s)} \sim \sum_\pm V_{(r,s\pm 1)}}\ . 
 \label{vdvrs}
\end{align}
While they were derived under the assumption $\beta^2\notin \mathbb{Q}$, we will assume that these OPEs are valid for any value of the central charge. 

The OPEs \eqref{vdvrs} and \eqref{vpope} indicate which primary fields can appear: we now assume that in each one of these OPEs, the 2 fields that can appear do appear, i.e.  $V^d_{\langle 1,2\rangle},V_{(r,s)}\in\widehat{\mathcal{S}}\implies V_{(r,s\pm 1)}\in\widehat{\mathcal{S}}$. 
Then the conformal spin \eqref{srs} must remain half-integer under $s\to s\pm 1$, therefore
$r\in \frac12 \mathbb{Z}$. 
Moreover, using the full OPE Eq. \eqref{tope}, we can compute the monodromy
\begin{align}
 V^d_{\langle 1,2\rangle}\left(e^{2\pi i}z\right) V_{(r,s)} (0) = e^{2\pi i r} V^d_{\langle 1,2\rangle}\left(z\right) V_{(r,s)} (0) \ ,
 \label{dnmono}
\end{align}
where the factor is $e^{2\pi i(S_{(r,s\pm 1)}-S_{(r,s)})} =e^{2\pi ir}\in \{1,-1\}$. In a CFT where $r$ can be half-integer, the existence of $V^d_{\langle 1,2\rangle}$ therefore requires that we slightly relax the axiom of single-valuedness, which would imply that the factor is $1$.
In any case, 
if $\left<V^d_{\langle 1,2\rangle}\prod_{i=1}^NV_{(r_i,s_i)}\right>\neq 0$,
then the product of the monodromies of $V^d_{\langle 1,2\rangle}$ must be $1$, therefore $\sum_{i=1}^N r_i  \in \mathbb{Z}$. This conclusion also holds for $N$-point functions of the type $\left<\prod_{i=1}^NV_{(r_i,s_i)}\right>$, which may be deduced from our $N+1$-point function using an OPE. Therefore, if $V^d_{\langle 1,2\rangle}$ exists, then the first Kac index of non-diagonal fields $V_{(r,s)}$ is half-integer and conserved modulo integers:
\begin{subequations}
\begin{align}
 \boxed{V^d_{\langle 1,2\rangle}\in\widehat{\mathcal{S}} \quad \implies\quad  r\in \frac12\mathbb{Z} \quad , \quad \sum_{i=1}^N r_i  \in \mathbb{Z}}\ . 
 \label{sriz}
\end{align}
Similarly, if $V^d_{\langle 2,1\rangle}$ exists, then the same is true for the second Kac index,
\begin{align}
 \boxed{V^d_{\langle 2,1\rangle}\in\widehat{\mathcal{S}} \quad \implies\quad  s\in \frac12\mathbb{Z} \quad , \quad \sum_{i=1}^N s_i  \in \mathbb{Z}}\ . 
 \label{ssiz}
\end{align}
\end{subequations}
These constraints also apply in the presence of diagonal fields, if we set $r=s=0$ for any diagonal field. 

Finally, a necessary condition for an OPE $V_1V_2\sim \sum_{k\in\mathcal{S}^{12}}V_k$ \eqref{tope} to converge is that the total conformal dimension be bounded from below over the OPE spectrum $\mathcal{S}^{12}$,
\begin{align}
 \boxed{\inf_{k\in \mathcal{S}^{12}}\Re\left(\Delta_k+\bar\Delta_k\right) > -\infty} \ . 
 \label{iddb}
\end{align}
The total conformal dimension of a non-diagonal field $V_{(r,s)}$ is 
\begin{align}
 \Delta_{(r,s)}+\Delta_{(r,-s)} = \frac12\left[ r^2\beta^{2} + s^2\beta^{-2}  -\left(\beta-\beta^{-1}\right)^2\right]\ .
\end{align}
If $V^d_{\langle 1,2\rangle}$ exists, then the OPE \eqref{vdvrs} increments $s$ by 1, therefore $s$ is unbounded in real part i.e. $\sup_{V_{(r,s)}\in \widehat{\mathcal{S}}} \left(\Re s\right) = \infty$. We assume that this holds not only for the extended spectrum $\widehat{\mathcal{S}}$, but also for some OPE spectrum $\mathcal{S}^{12}$. For the OPE not to diverge, it is necessary that $\Re\beta^{-2}\geq 0$. We assume the slightly stronger constraint $\Re\beta^{-2}> 0$, or equivalently
\begin{align}
 \boxed{\Re\beta^2 > 0 \quad \text{i.e.} \quad \Re c < 13}\ .
 \label{rbp}
\end{align}
This constraint holds for all known non-diagonal CFTs with at least 1 degenerate field.

\subsection{Diagonal CFTs with 2 degenerate fields}\label{sec:diag}

\subsubsection{Generalized minimal models}\label{sec:gmm}

For $\beta^2\in\mathbb{C}\backslash\mathbb{Q}$, we define the \myindex{generalized minimal model (GMM)}\index{minimal model!generalized---}\index{GMM} as the CFT whose spectrum of primary fields is made of all diagonal degenerate fields:
\begin{align}
 \boxed{\mathcal{S}^\text{GMM} = \left\{ V^d_{\langle r,s\rangle}\right\}_{r,s\in\mathbb{N}^*}} \ . 
\end{align}
Thanks to the degenerate fusion rules \eqref{rrsr}, any OPE spectrum is finite, and therefore respects the convergence condition \eqref{iddb}.

For $\beta^2\in \mathbb{Q}$, the degenerate dimensions $\Delta_{(r,s)}$ are not all different, and there is no GMM. However, there exist other CFTs whose primary fields are diagonal and (partly or fully) degenerate, starting from A-series minimal models. One way to explore these CFTs is to take limits of GMM correlation functions as $\beta^2\to \beta^2_0\in \mathbb{Q}$. This can give rise not only to minimal model correlation functions, but also to logarithmic correlation functions \cite{rib18}. According to Section \ref{sec:hwr}, a logarithmic representation can be generated by a $\Delta$-derivative field $V'_\Delta$. Correlation functions of $V'_\Delta$ have logarithmic dependence on its position, because $\frac{\partial}{\partial \Delta} z^\Delta = z^\Delta \log(z)$: this is why logarithmic representations are called logarithmic. Therefore, the logarithmic correlation functions that appear as $\beta^2\to \beta^2_0\in \mathbb{Q}$ may belong to \myindex{logarithmic CFTs}\index{CFT!logarithmic---}\index{logarithmic!---CFT}, i.e. CFTs whose spectra include logarithmic representations.

Generalized minimal models exist only on the sphere. In particular, their torus partition functions are infinite, because $\lim_{r,s\to\infty} \Re\Delta_{(r,s)}\neq +\infty$. Nevertheless, GMMs are interesting because they are very simple and natural CFTs, which give rise to various other CFTs in certain limits --- starting from Liouville theory.

\subsubsection{Liouville theory}\label{sec:liou}

Let us consider the spectrum of a generalized minimal model with $\beta^2\in\mathbb{R}_{>0}\backslash\mathbb{Q}$. The momenta $P_{(r,s)}$ \eqref{prs} are dense in the real line, i.e. $\mathbb{R}=\text{closure}\big(\left\{ P_{(r,s)}\right\}_{r,s\in\mathbb{N}^*}\big)$. Any diagonal field with a real momentum can therefore be obtained as a limit of degenerate fields,
\begin{align}
 \forall P\in\mathbb{R}\ , \quad V_P = \lim_{\substack{r,s\to\infty \\ P_{(r,s)}\to P}} V^d_{\langle r,s\rangle}\ ,
 \label{vplim}
\end{align}
where the vanishing singular vector of $V^d_{\langle r,s\rangle}$ disappears as its level $rs$ tends to infinity.
It turns out that $V_P$ does not depend on how exactly we take the limit, and also that $V_P=V_{-P}$. This can be proved at the level of correlation functions, by decomposing them into structure constants and conformal blocks:
\begin{itemize}
 \item By definition, conformal blocks only depend on conformal dimensions.
 \item GMM structure constants only depend on $r,s$ via $P_{(r,s)}$, see Section \ref{sec:dtdf}.
\end{itemize}
If we apply the limit to both fields in the OPE $V^d_{\langle r_1,s_1\rangle}V^d_{\langle r_2,s_2\rangle}$, the resulting OPE is formally given by the degenerate fusion rules \eqref{rrsr} with $r_1,s_1,r_2,s_2\to\infty$, and we obtain
\begin{align}
 \boxed{V_{P_1}V_{P_2} \sim \frac12 \int_\mathbb{R} dP\ V_P} \ ,
 \label{idpv}
\end{align}
where the factor $\frac12$ is because $V_P=V_{-P}$. Now, remember from Section \ref{sec:csope} that OPE coefficients have poles when the momentum $P$ becomes degenerate, and these poles are dense in the real line. The integral in the OPE $V_{P_1}V_{P_2}$ is therefore formally divergent. But it can be regularized by shifting the integration line into the complex plane:
\begin{align}
 V_{P_1}V_{P_2} \sim \frac12 \int_{\mathbb{R}+i\epsilon} dP\ V_P\ .
 \label{ireg}
\end{align}
Let us call \myindex{Liouville theory}\index{Liouville theory} the resulting CFT. The spectrum of primary fields is formally
\begin{align}
 \boxed{\mathcal{S}^\text{Liouville} = \frac12 \left\{ V_P\right\}_{P\in \mathbb{R}}}\ .
 \label{sliou}
\end{align}
The regularized OPE \eqref{ireg} requires that we extend the spectrum to complex momenta. This will turn out to be possible, because correlation functions are holomorphic in $P$.

For the moment, we have defined Liouville theory for $\beta^2\in\mathbb{R}_{>0}\backslash\mathbb{Q}$. By taking limits, we can extend it to the half-line $c\leq 1$ i.e. $\beta^2\in\mathbb{R}_{>0}$. But we cannot do an analytic continuation to the rest of the complex $\beta^2$-plane, because of the poles of OPE coefficients. These poles lie on the real $P$-line if $\beta^2\in\mathbb{R}_{>0}$, but as soon as $c$ strays from this half-line, the poles spread out in the complex $P$-plane, and cross the integration line $\mathbb{R}+i\epsilon$ for any $\epsilon>0$. In the following figures, we draw the regions where poles are found in blue, and the poles $\pm P_{(1,1)}$ as blue circles; the integration line is in red:
\begin{align}
 \newcommand{\polewedge}[3]{
\begin{scope}[#1]
\node[blue, draw,circle,inner sep=1pt,fill] at (0, 0) {};
\filldraw[opacity = .1, blue] (0,0) -- (3, -5) -- (3, 5) -- cycle;
\end{scope}
}
\begin{array}{ccc}
\begin{tikzpicture}[scale = .4, baseline=(current  bounding  box.center)]
  \draw[ultra thick, blue, opacity = .3] (-4.5, 0) -- (4.5, 0);
  \draw (-4.5, 0) -- (4.5, 0);
  \draw[line width = .6mm, red] (-4.5, .5) -- (4.5, .5);
  \draw[-latex] (-4.5,0)  -- (4.5,0) node [below] {$P$};
  \draw (0, -4) -- (0, 4);
\node[below left] at (.2, .1) {$0$};
\node[blue, draw,circle,inner sep=1pt,fill] at (2, 0) {};
\node[blue, draw,circle,inner sep=1pt,fill] at (-2, 0) {};
 \end{tikzpicture}
 & 
 \begin{tikzpicture}[scale = .4, baseline=(current  bounding  box.center)]
  \draw (-4.5, 0) -- (4.5, 0);
  \draw[line width = .6mm, red] (-4.5, 0) -- (4.5, 0);
  \draw[-latex] (-4.5,0)  -- (4.5,0) node [below] {$P$};
  \draw (0, -4) -- (0, 4);
\node[below left] at (.2, .1) {$0$};
\polewedge{rotate = 90, shift = {(1,.7)}}{$-\frac{Q}{2}$}{below};
  \polewedge{rotate = -90, shift = {(1, .7)}}{$\frac{Q}{2}$}{above};
 \end{tikzpicture}
 &
 \begin{tikzpicture}[scale = .4, baseline=(current  bounding  box.center)]
 \draw[ultra thick, blue, opacity = .3] (0,1.5) -- (0,4);
 \draw[ultra thick, blue, opacity = .3] (0,-1.5) -- (0,-4);
  \draw (-4.5, 0) -- (4.5, 0);
  \draw[line width = .6mm, red] (-4.5, 0) -- (4.5, 0);
  \draw[-latex] (-4.5,0)  -- (4.5,0) node [below] {$P$};
  \draw (0, -4) -- (0, 4);
\node[below left] at (.2, .1) {$0$};
\node[blue, draw,circle,inner sep=1pt,fill] at (0, 1.5) {};
\node[blue, draw,circle,inner sep=1pt,fill] at (0, -1.5) {};
 \end{tikzpicture}
 \vspace{3mm}
 \\
\beta^2>0 & \beta^2\in\mathbb{C}\backslash \mathbb{R} & \beta^2<0
\end{array}
\end{align}
As a result, as functions of $\beta^2 \in \mathbb{R}_{>0}$, correlation functions of Liouville theory are real-analytic but not holomorphic.

Since we cannot do an analytic continuation from $\beta^2\in \mathbb{R}_{>0}$,
we simply define Liouville theory for $\beta^2\in \mathbb{C}\backslash\mathbb{R}_{\geq 0}$
by the spectrum \eqref{sliou} and OPE \eqref{idpv}, where we no longer need to shift the integration line. We expect correlation functions to be holomorphic in $\beta^2\in \mathbb{C}\backslash\mathbb{R}_{\geq 0}$, because poles of OPE coefficients never cross the integration line. We also expect that correlation functions are holomorphic in $P$. (In fact they turn out to be holomorphic over the complex $P$-plane, minus a discrete set of poles.) And most importantly for the analytic bootstrap approach, we assume that degenerate fields exist, so that we have the extended spectrum
\begin{align}
 \boxed{ \widehat{\mathcal{S}}^\text{Liouville} = \frac12 \left\{ V_P\right\}_{P\in \mathbb{C}} \cup \left\{ V^d_{\langle r,s\rangle}\right\}_{r,s\in\mathbb{N}^*} }\ .
 \label{stliou}
\end{align}
The resulting extended CFT therefore includes a generalized minimal model. As we will see in Section \ref{sec:dtdf}, there is at most one CFT that obeys our definition of Liouville theory. Proving that it exists amounts to proving that 4-point functions are crossing-symmetric. This is very hard to do analytically, see Section \ref{sec:alaa} for references.

\subsubsection{A-series minimal models}\label{sec:amm}

Let us look for rational, diagonal CFTs. 
To any degenerate representation of the Virasoro algebra, we can associate a diagonal primary field. We conjecture that this leads to a bijection 
\begin{align}
\begin{array}{c}
 \text{Finite sets of degenerate representations that are closed under fusion}
 \\
 \iff 
 \\
 \text{Rational, diagonal CFTs on the sphere}
\end{array}
\end{align}
To the best of our knowledge, there is no known counterexample to this conjecture (and no proof either). 
From the fusion rules \eqref{rrsr}, no nontrivial finite set of simply degenerate representations can be closed under fusion. We therefore need multiply degenerate representations, which implies $\beta^2\in\mathbb{Q}$.  

In the case $\beta^2=-\frac{q}{p}<0$, with $p,q\in\mathbb{N}^*$ coprime, there are no finite sets of degenerate fields that are closed under OPEs. To look for such sets, we could focus on the simplest doubly degenerate field, 
\begin{align}
 V^f_{\langle p+1,1\rangle}=V^f_{\langle 1,q+1\rangle}\ .
\end{align}
Repeated OPEs of this field with itself generate infinitely many fields of the type $V^f_{\langle kp+1,1\rangle}$ for any $k\in\mathbb{N}$, starting from
\begin{align}
 V^f_{\langle p+1,1\rangle}V^f_{\langle p+1,1\rangle}\sim V^f_{\langle 1,1\rangle} + V^f_{\langle 2p+1,1\rangle}\ .
 \label{rfppo}
\end{align}
To show that we obtain fully degenerate fields, we can use OPE associativity, as well as OPEs with generic diagonal fields \eqref{rrsvp}, starting from
\begin{align}
 V^f_{\langle p+1,1\rangle}  V_P \sim \sum_{\pm} V_{P\pm \frac12\sqrt{-pq}}\ ,
\end{align}
where $\{P\pm \frac12\sqrt{-pq}\}_\pm = \{P+i\beta\}_{i=-\frac{p}{2},\cdots, \frac{p}{2}} \cap \{P+j\beta^{-1}\}_{j=-\frac{q}{2},\cdots, \frac{q}{2}}$. 

We therefore focus on the case $\beta^2 = \frac{q}{p}>0$ with $p,q\in\mathbb{N}^*$ coprime. For simplicity, we will start with $\mathcal{R}^f_{\langle 2,1\rangle}$ and $\mathcal{R}^f_{\langle 1,2\rangle}$, and generate a spectrum by repeatedly taking fusion products of these representations. This yields the known A-series minimal model: we neglect other diagonal, rational CFTs that exist on the sphere.
Let us show that $\mathcal{R}^f_{\langle 2,1\rangle}$ and $\mathcal{R}^f_{\langle 1,2\rangle}$ generate the representations $\mathcal{R}^f_{\langle r,s\rangle}$  whose indices belong to the \myindex{Kac table}\index{Kac table}
\begin{align}
 \boxed{ K_{p, q}=  \big((0,p)\times (0,q)\big)\bigcap \big(\mathbb{N}\times \mathbb{N}\big) }\ .
 \label{kac}
\end{align}
We will in fact show that $\left\{\mathcal{R}^f_{\langle r,s\rangle}\right\}_{(r,s)\in K_{p, q}}$ is stable under fusion. Representations in this set are doubly degenerate, 
\begin{align}
 \mathcal{R}^f_{\langle r,s\rangle} = \mathcal{R}^f_{\langle p-r,q-s\rangle}\ , 
 \label{prqs}
\end{align}
and this leads to the following constraints on fusion rules \eqref{rrsr}
\begin{subequations}
\begin{align}
 \mathcal{R}^f_{\langle r_1,s_1\rangle} \times \mathcal{R}^f_{\langle r_2,s_2\rangle} &\subset \sum_{r\overset{2}{=} |r_1-r_2|+1}^{r_1+r_2-1}\sum_{s\overset{2}{=} |s_1-s_2|+1}^{s_1+s_2-1} \mathcal{R}^d_{\langle r,s\rangle} \ , 
 \\
 \mathcal{R}^f_{\langle r_1,s_1\rangle} \times \mathcal{R}^f_{\langle r_2,s_2\rangle} &\subset \sum_{r'\overset{2}{=} |r_1-r_2|+1}^{2p-r_1-r_2-1} \sum_{s'\overset{2}{=} |s_1-s_2|+1}^{2q-s_1-s_2-1}\mathcal{R}^d_{\langle r',s'\rangle} \ , 
\end{align}
\end{subequations}
where $\mathcal{R}^d_{\langle r,s\rangle}$ is a representation that may or may not be fully degenerate. Now, $\mathcal{R}^d_{\langle r,s\rangle} = \mathcal{R}^d_{\langle r',s'\rangle}$ implies $(r,s)=(r',s')$ or $(r,s)=(p-r',q-s')$, but the second equality is impossible due to parity: since $p$ and $q$ are coprime, one of them must be odd, say $p$, but $r\equiv r'\equiv |r_1-r_2|+1\bmod 2$. We therefore obtain the fusion rules 
\begin{align}
 \boxed{\mathcal{R}^f_{\langle r_1,s_1\rangle} \times \mathcal{R}^f_{\langle r_2,s_2\rangle} = \sum_{r\overset{2}{=} |r_1-r_2|+1}^{\min(r_1+r_2,2p-r_1-r_2)-1}\sum_{s\overset{2}{=} |s_1-s_2|+1}^{\min(s_1+s_2,2q-s_1-s_2)-1} \mathcal{R}^f_{\langle r,s\rangle}} \ , 
 \label{rfrf}
\end{align}
where the indices on the right-hand side belong to the Kac table $(r,s)\in K_{p,q}$. The number of terms is $\min(r_1,r_2,p-r_1,p-r_2)\min(s_1,s_2,q-s_1,q-s_2)$. To show that $\mathcal{R}^f_{\langle r,s\rangle}$ is fully degenerate, let us study the fusion of fully degenerate representations with Verma modules.
From Eq. \eqref{rrsvp},
\begin{align}
 \mathcal{R}^f_{\langle r,s\rangle}\times \mathcal{V}_P \subset \sum_{i=-\frac{r-1}{2}}^{\frac{r-1}{2}} \sum_{j=-\frac{s-1}{2}}^{\frac{s-1}{2}} \mathcal{V}_{P+i\beta +j\beta^{-1}} \cap \sum_{i'=-\frac{p-r-1}{2}}^{\frac{p-r-1}{2}} \sum_{j'=-\frac{q-s-1}{2}}^{\frac{q-s-1}{2}} \mathcal{V}_{P+i'\beta +j'\beta^{-1}}\ .
\end{align}
The modules $\mathcal{V}_{P+i\beta +j\beta^{-1}}$ and $\mathcal{V}_{P+i'\beta +j'\beta^{-1}}$ coincide if and only if their momenta are opposite or equal. If they are opposite, $P=-P_{(i+i',j+j')}$ must belong to the (reflected) Kac table, so it is not generic. If they are equal, we have $i\beta +j\beta^{-1} = i'\beta +j'\beta^{-1}$ i.e. $2(i-i')q = 2(j'-j)p$.
But parity forbids $(i,j)=(i',j')$, since $p$ or $q$ is odd. And we have $2|i-i'|<p$ and $2|j-j'|<q$, so our equality cannot be satisfied, and 
\begin{align}
 \mathcal{R}^f_{\langle r,s\rangle}\times \mathcal{V}_P = 0 \ . 
\end{align}
Therefore, a Kac table representation is fully degenerate if and only if its fusion with a generic Verma module $\mathcal{V}_P$ is zero.
By associativity of the fusion product, the product \eqref{rfrf} of 2 fully degenerate representations is again a combination of fully degenerate representations.

We therefore define the \myindex{A-series minimal model}\index{minimal model!A-series---} AMM$_{p,q}$ by its spectrum, made of fully degenerate fields in the Kac table:
\begin{align}
 \boxed{\mathcal{S}^{\text{AMM}_{p,q}} = \frac12\left\{ V^f_{\langle r,s\rangle} \right\}_{(r,s)\in K_{p,q}} \quad \text{with} \quad 2\leq q<p \quad \text{and} \quad p, q \text{ coprime}}\ .
 \label{samm}
\end{align}
The factor $\frac12$ accounts for the $\mathbb{Z}_2$ symmetry \eqref{prqs}. We assume $2\leq q$ for the Kac table to be non-empty, and $q<p$ because AMM$_{p, q}=$AMM$_{q,p}$. The field $V^f_{\langle 1,1\rangle}$ is the identity field.

For example, AMM$_{4,3}$ has the central charge $c=\frac12$ and describes some of the observables of the critical Ising model. Let us describe its 3 primary fields, and display their dimensions in the Kac table: 
\begin{align}
 \begin{array}{|c||c|c|c|}
 \hline 
  \text{Field} &  V^f_{\langle 1,1\rangle } = V^f_{\langle 3,2\rangle} & V^f_{\langle 1,2\rangle} = V^f_{\langle 3,1\rangle} & V^f_{\langle 2,1\rangle} = V^f_{\langle 2,2\rangle} 
  \\
  \hline 
  \text{Notation} &  I &  \epsilon & \sigma
  \\
  \hline 
  \text{Name} & \text{Identity} & \text{Energy} & \text{Spin} 
  \\
  \hline 
  \text{Dimension} & 0 & \frac12 & \frac{1}{16}
  \\
  \hline 
 \end{array}
\qquad 
  \begin{tikzpicture}[baseline=(current  bounding  box.center)]
  \kac{4}{3};
  \thiskac{1}{1}{0};
  \thiskac{2}{1}{\frac{1}{16}};
  \thiskac{3}{1}{\frac12};
 \end{tikzpicture}
\end{align}
Having only 3 primary fields, we can afford to write the complete fusion rules:
\begin{align}
\begin{array}{l}
 I\times I = I \ ,
\\ I\times \epsilon = \epsilon\ ,
\\ I\times \sigma = \sigma\ ,
\end{array}
\hspace{2cm}
\begin{array}{l}
 \epsilon\times \epsilon = I\ ,
\\ \epsilon\times \sigma = \sigma\ ,
\\ \sigma \times \sigma = I + \epsilon\ .
\end{array}
\end{align}
For example, AMM$_{7,4}$ has the central charges $c=-\frac{13}{14}$. Its $9$ primary fields have the following dimensions:
\begin{align}
 \begin{tikzpicture}[baseline=(current  bounding  box.center)]
  \kac{7}{4};
  \thiskac{1}{1}{0};
  \thiskac{2}{1}{-\frac{1}{14}};
  \thiskac{3}{1}{\frac17};
  \thiskac{4}{1}{\frac{9}{14}};
  \thiskac{5}{1}{\frac{10}{7}};
  \thiskac{6}{1}{\frac52};
  \thiskac{1}{2}{\frac{13}{16}};
  \thiskac{2}{2}{\frac{27}{112}};
  \thiskac{3}{2}{-\frac{5}{112}};
 \end{tikzpicture}
\end{align}

\subsubsection{Runkel--Watts-type CFTs}\label{sec:rwt}

Let us consider a limit of A-series minimal models where $p,q\to\infty$ such that $\frac{q}{p}\to \beta_0^2>0$. We first take the Kac table indices to be fixed: for any $r,s\in\mathbb{N}^*$, the minimal model AMM$_{p,q}$ has a
field $V^f_{\langle r,s\rangle}$
provided $p>r$ and $q>s$. In AMM$_{p,q}$, this field has vanishing singular vectors at levels $rs$ and $(p-r)(q-s)$ by Eq. \eqref{prqs}. The second singular vector disappears as its level tends to infinity, and the field tends to the simply degenerate field $V^d_{\langle r,s\rangle}$. Therefore, in this limit, we obtain all the primary fields of a GMM, and we write
\begin{align}
 \lim_{\substack{\frac{q}{p}\to \beta_0^2\\ r,s\text{ fixed}}} \text{AMM}_{p,q} = \text{GMM}\qquad \text{with} \qquad \lim_{\substack{\frac{q}{p}\to \beta_0^2\\ r,s\text{ fixed}}}
V^f_{\langle r,s\rangle} = V^d_{\langle r,s\rangle}\ .
 \label{mm-gmm}
\end{align}
Since we obtain arbitrary degenerate fields $V^d_{\langle r,s\rangle}$ in the limit, it is also possible to send the indices $r,s$ to $\infty$, and therefore to send the momenta $P_{(r,s)}$ to arbitrary values $P\in\mathbb{R}$, just as in Eq. \eqref{vplim}. Therefore, Liouville theory with $c\leq 1$ is not only a limit of generalized minimal models, but also a limit of minimal models:
\begin{align}
 \lim_{\substack{\frac{q}{p}\to \beta_0^2\\ P_{(r,s)}\to P}} \text{AMM}_{p,q} = \text{Liouville}\ . 
 \label{mm-liou}
\end{align}
However, if $\beta_0^2\in\mathbb{Q}$, it is also possible to fine-tune the behaviour of $p,q,r,s$ such that constraints from the minimal model's fusion rules survive in the limit. As a result, we can obtain not only Liouville theory with its trivial OPE \eqref{idpv}, but also another diagonal CFT with the same continuous spectrum but a nontrivial OPE, called a \myindex{Runkel--Watts-type theory}\index{Runkel--Watts-type theory} after the original example at $c=1$, discovered by Runkel and Watts \cite{rw01}.

To begin with, we write $\beta_0^2 = \frac{q_0}{p_0}$ with $p_0$ and $q_0$ coprime integers, so that $\mathbb{Z}\beta_0+\mathbb{Z}\beta_0^{-1}=\frac{1}{\sqrt{p_0q_0}}\mathbb{Z}$. There exist infinitely many pairs $(p,q)\in \mathbb{N}^2$ such that
\begin{align}
 qp_0-pq_0 = 1 \ .
\end{align}
Choosing such pairs amounts to fine-tuning the values of $p, q$, because there are many other ways to satisfy $\lim_{p,q\to\infty}\frac{q}{p}= \frac{q_0}{p_0}$, which is equivalent to $qp_0-pq_0 \ll p$. For fixed $r,s$, we compute the momentum
\begin{align}
 P_{(r,s)} = \frac{1}{2\sqrt{p_0q_0}}\left[rq_0-sp_0+ \frac12\left(\frac{r}{p}+\frac{s}{q}\right) + O\left(\frac{1}{p^2}\right)\right]\ . 
 \label{prsrwt}
\end{align}
We now fine-tune $r,s$ such that 
\begin{align}
 \lim(rq_0-sp_0), \lim\frac{r}{p} , \lim\frac{s}{q} <\infty \quad , \quad \lim\frac{r}{p} = \lim\frac{s}{q} \ . 
\end{align}
This fine-tuning can be achieved using an infinite sequence of integers $k\in\mathbb{N}$ such that
\begin{align}
\lim \frac{kp_0}{p} = \lim \frac{kq_0}{q}\in (0, 1) \qquad \text{then} \qquad
 \left\{\begin{array}{l} r = r_0 + k p_0\ , \\ s = s_0 + k q_0\ . \end{array}\right.
\end{align}
This leads to 
\begin{align}
 \lim_{p,q\to\infty} P_{(r,s)} = \frac{1}{2\sqrt{p_0q_0}}\left(n+x\right)\quad \text{with} \quad \left\{\begin{array}{l} n =r_0q_0 - s_0p_0 \in \mathbb{Z} \ , \\ x =  \lim\frac{r}{p} = \lim\frac{s}{q} \in (0, 1) \ . \end{array}\right.
\end{align}
Now what happens to the minimal model fusion rules $\mathcal{R}^f_{\langle r_1,s_1\rangle} \times \mathcal{R}^f_{\langle r_2,s_2\rangle}$ \eqref{rfrf} when we apply our limit to both representations? Let us examine the momenta \eqref{prsrwt} of the resulting representations $\mathcal{R}^f_{\langle r,s\rangle}$. The combination $n=rq_0-sp_0$ takes integer values such that $n\equiv n_1+n_2+p_0+q_0\bmod 2$, while
both $\frac{r}{p}$ and $\frac{s}{q}$, and therefore also $x=\frac12\left(\frac{r}{p}+\frac{s}{q}\right)$, are evenly distributed in the interval $(|x_1-x_2|, \min(x_1+x_2,2-x_1-x_2))$. Writing the primary fields as $V_{n,x} = \lim V_{(r,s)}^f$, we thus find the OPE
\begin{align}
 \boxed{V_{n_1,x_1}V_{n_2,x_2} \sim \sum_{n\in n_1+n_2+p_0+q_0+2\mathbb{Z}}
\int_{|x_1-x_2|}^{\min (x_1+x_2,2-x_1-x_2)} dx\ V_{n, x}}\ .
\end{align}
Of course, $V_{n,x}$ is just another notation for the diagonal field $V_P$, with $n,x$ the integer and fractional parts of $2\sqrt{p_0q_0}P$:
\begin{align}
 P=\frac{n+x}{2\sqrt{p_0q_0}} \quad , \quad n=\Big\lfloor 2\sqrt{p_0q_0}P \Big\rfloor \quad , \quad x = \text{frac}\Big(2\sqrt{p_0q_0}P \Big)\ .
\end{align}
The reflection relation $V_P=V_{-P}$ is now $V_{n,x}=V_{-n-1,1-x}$. 
The resulting Runkel--Watts-type CFT has the same spectrum as Liouville theory \eqref{sliou} $
 \mathcal{S}^{\text{RWT}_{p_0,q_0}} = \frac12 \left\{ V_P\right\}_{P\in \mathbb{R}}$. In contrast to Liouville theory, the momenta cannot be continued to complex values, where the OPE would no longer make sense. Notice however that if $x_1\neq x_2$, the OPE does not produce fields with degenerate momenta, i.e. fields of the type $V_{n,0}$. So there is no need to regularize the integral over $x$. Alternatively, the OPE may be written in a manifestly reflection-invariant way using the momenta $P_i$ \cite{rs15},
 \begin{align}
  \left<\prod_{i=1}^3 V_{P_i}\right>\neq 0 \implies \prod_{\pm,\pm} \sin\pi\left(\tfrac{p_0+q_0}{2}+\sqrt{p_0q_0}(P_1\pm P_2\pm P_3)\right) < 0 \ .
  \label{rwtz}
 \end{align}
It turns out that Runkel--Watts-type theories can also be obtained as limits of Liouville theory with $\beta^2\in \mathbb{C}\backslash \mathbb{R}$ \cite{sch03, mce07}. The limit of Liouville theory for $\beta^2\to \beta_0^2>0$ generally does not make sense, but it exists if $\beta_0^2$ is rational and the momenta are real. In this limit, the condition \eqref{rwtz} emerges from the 3-point structure constants of Liouville theory.

\subsection{Non-diagonal CFTs with 2 degenerate fields}

\subsubsection{D-series minimal models}

Let us look for rational, non-diagonal CFTs, built from the same representations as A-series minimal models: fully degenerate representations with indices in the Kac table \eqref{kac}, with $\beta^2=\frac{q}{p}\in\mathbb{Q}_{>0}$. For any non-diagonal field $V_{(r,s)}$, we require that the left and right dimensions $\Delta_{(r,\pm s)}$ both coincide with dimensions from the Kac table. To achieve this, a simple ansatz is 
\begin{align}
 (r,s) \in \Big[\left(-\tfrac{p}{2},\tfrac{p}{2}\right) \cap \left(\mathbb{Z}+\tfrac{p}{2}\right)\Big] \times \Big[\left(-\tfrac{q}{2},\tfrac{q}{2}\right) \cap \left(\mathbb{Z}+\tfrac{q}{2}\right)\Big] \ . 
\end{align}
This amounts to taking advantage of the identity $\Delta_{(r,s)}=\Delta_{(r+\frac{p}{2},s+\frac{q}{2})}$ for centering the Kac table at $(r,s)=(0,0)$, which makes the table invariant under $(r,s)\to (r,-s)$. It remains to impose the integer spin condition $rs\in\mathbb{Z}$. If $p,q$ are both odd, no field has integer spin. Without loss of generality, we therefore assume $p\in 2\mathbb{N}^*$ and $q\in 2\mathbb{N}+1$, 
and the integer spin condition reduces to $r\in 2\mathbb{Z}$. This completes the determination of the non-diagonal sector. 

Since $q$ is odd, all non-diagonal fields have $s\in \mathbb{Z}+\frac12$, and the conservation of $s\bmod \mathbb{Z}$ \eqref{ssiz} (with $s=0$ for diagonal fields) reduces to the \myindex{conservation of diagonality}\index{conservation of diagonality}:
\begin{align}
D\times D = D \quad , \quad D\times N = N \quad , \quad N\times N = D \ ,
\label{ddd}
\end{align}
where $D$ stands for diagonal and $N$ for non-diagonal fields. 
Moreover, we assume that the diagonal sector is generated by taking OPEs of non-diagonal fields. 
By the degenerate fusion rules \eqref{rrsr}, the diagonal sector involves representations $\mathcal{R}^f_{\langle r,s\rangle}$ with $r\in 2\mathbb{N}+1$. We therefore define a \myindex{D-series minimal model}\index{minimal model!D-series---} $\text{DMM}_{p,q}$ by the spectrum
\begin{align}
 \boxed{\mathcal{S}^{\text{DMM}_{p,q}} = \frac12\left\{ V^f_{\langle r,s\rangle} \right\}_{\substack{(r,s)\in K_{p, q}\\ r\equiv 1\bmod 2}} 
 \bigcup 
 \frac12\left\{V_{(r,s)} \right\}_{\substack{(r,s)\in (-\frac{p}{2},\frac{p}{2})\times (-\frac{q}{2},\frac{q}{2})\\ (r,s)\in 2\mathbb{Z}\times (\mathbb{Z}+\frac12)}}
}\ ,
\label{sdmm}
\end{align}
where not only $V^f_{\langle r,s\rangle}$ but also the non-diagonal fields $V_{(r,s)}$ are fully degenerate. This is valid for
\begin{align}
 \boxed{p\in 2\mathbb{N}+6 \quad , \quad q \in 2\mathbb{N}+3 \quad ,\quad p,q\text{ coprime}} \ ,
\end{align}
where the lower bounds $p\geq 6$ and $q\geq 3$ ensure that the non-diagonal sector contains fields with nonzero spins. If we allow spins to be half-integer, we obtain the extended spectrum
\begin{align}
 \widehat{\mathcal{S}}^{\text{DMM}_{p,q}} = \frac12\left\{ V^f_{\langle r,s\rangle} \right\}_{\substack{(r,s)\in (0,p) \times (0,q)\\ (r,s)\in \mathbb{N}\times \mathbb{N}}} 
 \bigcup 
 \frac12\left\{V_{(r,s)} \right\}_{\substack{(r,s)\in (-\frac{p}{2},\frac{p}{2})\times (-\frac{q}{2},\frac{q}{2})\\ (r,s)\in \mathbb{Z}\times (\mathbb{Z}+\frac12)}}
\ .
\end{align}
The corresponding CFT may be called a \myindex{fermionic minimal model}\index{minimal model!fermionic---} $\widehat{\text{DMM}}_{p,q}$ \cite{rw20}. The numbers of primary fields in these minimal models are:
\begin{align}
 \begin{array}{|c||c|c|c|c|}
 \hline 
  \multirow{2}{*}{Model} & \multirow{2}{*}{AMM$_{p,q}$} & \multicolumn{2}{|c|}{\text{DMM}_{p,q}} 
   & \multirow{2}{*}{$\widehat{\text{DMM}}_{p,q}$}
  \\
  \cline{3-4}
  & & p\equiv 0\bmod 4 & p\equiv 2\bmod 4 & 
  \\
  \hline \hline 
  \text{Diagonal} & \frac12(p-1)(q-1) & \multicolumn{2}{|c|}{\frac14 p(q-1)} & \frac12 (p-1)(q-1) 
  \\
  \hline 
  \text{Non-diagonal} & 0  & \frac14(p-2)(q-1) &
  \frac14p(q-1) 
  & \frac12 (p-1)(q-1) 
  \\
  \hline 
  \text{Total} & \frac12(p-1)(q-1) & 
   \frac12(p-1)(q-1) &
  \frac12 p(q-1) & (p-1)(q-1)
  \\
  \hline 
 \end{array}
\end{align}
The fusion rules of D-series and fermionic minimal models are obtained by combining the fully degenerate fusion rules \eqref{rfrf} with the conservation of diagonality:
\begin{subequations}
\begin{align}
 V^f_{\langle r_1,s_1\rangle}V_{(r_2,s_2)} &\sim \sum_{r\overset{2}{=}\max(r_2-r_1,r_1-r_2-p)+1}^{\min(r_1+r_2,p-r_1-r_2)-1} \sum_{s\overset{2}{=}\max(s_2-s_1,s_1-s_2-q)+1}^{\min(s_1+s_2,q-s_1-s_2)-1} V_{(r,s)} \ ,
 \label{vfvn}
 \\
 V_{(r_1,s_1)} V_{(r_2,s_2)} & \sim \sum_{r\overset{2}{=}|r_1-r_2|+1}^{p-|r_1+r_2|-1} \sum_{s\overset{2}{=}|s_1-s_2|+1}^{q-|s_1+s_2|-1} V^f_{\langle r,s\rangle}\ . 
 \label{vnvn}
\end{align}
\end{subequations}
Curiously, the diagonal field $V^f_{\langle \frac{p}{2}, s\rangle}$ has the same conformal dimensions as the non-diagonal field $V_{(0,s-\frac{q}{2})}$: this an instance of a nontrivial \myindex{field multiplicity}\index{field!---multiplicity}. However, these 2 fields have different OPEs. 

For example, let us compare the diagonal and non-diagonal minimal models AMM$_{6,5}$ and DMM$_{6,5}$, which both have $c= \frac45$, and respectively describe observables of the tetracritical Ising model and of the critical 3-state Potts model. The Virasoro representations that appear in the spectra of both models belong to the Kac table $K_{6,5}$. Each entry in the table correspond to a primary field of AMM$_{6,5}$. Since $p=6\equiv 2\bmod 4$, both sectors of DMM$_{6,5}$ are built from representations $\mathcal{R}^f_{\langle r,s\rangle}$ with $r\in 2\mathbb{N}+1$. We indicate this in the Kac table by having the $r\in 2\mathbb{N}+1$ columns both boxed (for the diagonal sector) and highlighted (for the non-diagonal sector):
\begin{align}
\text{AMM}_{6,5}\ \&\ \text{DMM}_{6,5} \ : \qquad 
 \begin{tikzpicture}[baseline=(current  bounding  box.center)]
  \kac{6}{5};
   \foreach\r in {1,3,5}{
  \filldraw[red, ultra thick, fill = red!10!white] (\r-.4, .6) rectangle (\r+.4, 4.4);
  }
  \thiskac{1}{1}{0};
  \thiskac{2}{1}{\frac18};
  \thiskac{3}{1}{\frac23};
  \thiskac{4}{1}{\frac{13}{8}};
  \thiskac{5}{1}{3};
  \thiskac{1}{2}{\frac25};
  \thiskac{2}{2}{\frac{1}{40}};
  \thiskac{3}{2}{\frac{1}{15}};
  \thiskac{4}{2}{\frac{21}{40}};
  \thiskac{5}{2}{\frac75};
 \end{tikzpicture}
\end{align}
The 10 primary fields of AMM$_{6,5}$ are
\begin{align}
 V^f_{\langle 1,1\rangle},V^f_{\langle 2,1\rangle},V^f_{\langle 3,1\rangle},V^f_{\langle 4,1\rangle},V^f_{\langle 5,1\rangle},V^f_{\langle 1,2\rangle},V^f_{\langle 2,2\rangle},V^f_{\langle 3,2\rangle},V^f_{\langle 4,2\rangle},V^f_{\langle 5,2\rangle}\ .
\end{align}
In DMM$_{6,5}$, only 6 of these fields are present, and they are complemented by 6 non-diagonal fields. For each one of these 12 fields, let us indicate the left and right Kac indices of the corresponding degenerate representations:
\begin{align}
& 
 \begin{array}{|c||c|c|c|c|c|c|}
  \hline 
  \text{Field} & V^f_{\langle 1,1\rangle} & V^f_{\langle 3,1\rangle} & V^f_{\langle 5,1\rangle}
  & V^f_{\langle 1,2\rangle} & V^f_{\langle 3,2\rangle} & V^f_{\langle 5,2\rangle} 
  \\
  \hline
  \text{Indices} & (1,1)(1,1) & (3,1)(3,1) & (5,1)(5,1) & (1,2)(1,2) & (3,2)(3,2) &
  (5,2)(5,2) 
  \\
  \hline 
 \end{array}
 \nonumber
 \\ & 
  \begin{array}{|c||c|c|c|c|c|c|}
  \hline 
  \text{Field} & V_{(0,\frac12)} & V_{(0,\frac32)} & V_{(2,-\frac32)} & V_{(2,-\frac12)} & V_{(2,\frac12)} & V_{(2,\frac32)} 
  \\
  \hline
  \text{Indices} & (3,2)(3,2) & (3,1)(3,1) &  (5,1)(1,1) & (5,2)(1,2)  & (1,2)(5,2)  &  (1,1)(5,1)
  \\
  \hline 
 \end{array}
 \label{dmm65}
\end{align}
We therefore have coincidences of conformal dimensions for $V_{(0,\frac12)}\simeq V^f_{\langle 3,2\rangle}$ and $V_{(0,\frac32)}\simeq V^f_{\langle 3,1\rangle}$. The OPEs however do not coincide, for example
\begin{align}
 V^f_{\langle 3,1\rangle} V^f_{\langle 3,1\rangle} \sim V^f_{\langle 1,1\rangle} + V^f_{\langle 3, 1\rangle} + V^f_{\langle 5,1\rangle} \quad , \quad V^f_{\langle 3,1\rangle}V_{(0,\frac32)} \sim V_{(2,-\frac32)} + V_{(0,\frac32)} + V_{(2,\frac32)}\ . 
\end{align}
As another example, let us compare the diagonal and non-diagonal minimal models AMM$_{8,5}$ and DMM$_{8,5}$, which both have $c=-\frac{7}{20}$. The Virasoro representations that appear in the spectra of both models belong to the Kac table $K_{8,5}$. Each entry in the table correspond to a primary field of AMM$_{8,5}$. Since $p=8\equiv 0\bmod 4$, the diagonal sector of DMM$_{8,5}$ is built from representations $\mathcal{R}^f_{\langle r,s\rangle}$ with $r\in 2\mathbb{N}+1$ (boxed), while the non-diagonal sector is built from $\mathcal{R}^f_{\langle r,s\rangle}$ with $r\in 2\mathbb{N}$ (highlighted):
\begin{align}
\text{AMM}_{8,5}\ \&\ \text{DMM}_{8,5} \ : \qquad 
\begin{tikzpicture}[baseline=(current  bounding  box.center)]
  \kac{8}{5};
  \foreach\r in {1,3,5,7}{
  \draw[red, ultra thick] (\r-.4, .6) rectangle (\r+.4, 4.4);
  }
  \foreach\r in {2, 4, 6}{
  \fill[red!10!white] (\r-.4, .6) rectangle (\r+.4, 4.4);
  }
  \thiskac{1}{1}{0};
  \thiskac{2}{1}{-\frac{1}{32}};
  \thiskac{3}{1}{\frac14};
  \thiskac{4}{1}{\frac{27}{32}};
  \thiskac{5}{1}{\frac{7}{4}};
  \thiskac{6}{1}{\frac{95}{32}};
  \thiskac{7}{1}{\frac{9}{2}};
  \thiskac{1}{2}{\frac{7}{10}};
  \thiskac{2}{2}{\frac{27}{160}};
  \thiskac{3}{2}{-\frac{1}{20}};
  \thiskac{4}{2}{\frac{7}{160}};
  \thiskac{5}{2}{\frac{9}{20}};
  \thiskac{6}{2}{\frac{187}{160}};
  \thiskac{7}{2}{\frac{11}{5}};
 \end{tikzpicture}
\end{align}
The diagonal model AMM$_{8,5}$ has 14 primary fields $V^f_{\langle r,s\rangle}$ with $r=1,2,\dots,7$ and $s=1,2$. The non-diagonal model DMM$_{8,5}$ also has 14 primary fields: the 8 diagonal fields $V^f_{\langle 1,1\rangle},V^f_{\langle 3,1\rangle}, V^f_{\langle 5,1\rangle},V^f_{\langle 7,1\rangle},V^f_{\langle 1,2\rangle},V^f_{\langle 3,2\rangle}, V^f_{\langle 5,2\rangle},V^f_{\langle 7,2\rangle}$, and 6 non-diagonal fields:
\begin{align}
  \begin{array}{|c||c|c|c|c|c|c|}
  \hline 
  \text{Field} & V_{(0,\frac12)} & V_{(0,\frac32)} & V_{(2,-\frac32)} & V_{(2,-\frac12)} & V_{(2,\frac12)} & V_{(2,\frac32)} 
  \\
  \hline
  \text{Indices} & ( 4,2)( 4,2) & ( 4,1)( 4,1) &  ( 6,1)( 2,1) & ( 6,2)( 2,2)  & ( 2,2)( 6,2)  &  ( 2,1)( 6,1)
  \\
  \hline 
 \end{array}
\end{align}

\subsubsection{Generalized D-series minimal models}\label{sec:gdmm}

Let us consider a limit of D-series minimal models where $\frac{q}{p}\to \beta^2>0$. There is a natural way of taking the limit of the non-diagonal sector, which is to assume that non-diagonal fields $V_{(r,s)}$ have fixed indices:
\begin{align}
 \lim_{p,q\to\infty} \left. \mathcal{S}^{\text{DMM}_{p,q}} \right|_{\text{non-diagonal}} = 
 \frac12\left\{V_{(r,s)} \right\}_{ (r,s)\in 2\mathbb{Z}\times (\mathbb{Z}+\frac12)}\ .
\end{align}
For the left-moving Virasoro algebra, the DMM field $V_{(r,s)}$ is doubly degenerate with vanishing singular vectors at levels $(\frac{p}{2}+r)(\frac{q}{2}+s)$ and $(\frac{p}{2}-r)(\frac{q}{2}-s)$. Both levels tend to infinity, so the field becomes non-degenerate in the limit.

The diagonal sector of $\text{DMM}_{p,q}$ only differs from the spectrum $\mathcal{S}^{\text{AMM}_{p,q}}$ \eqref{samm} by the restriction $r\in 2\mathbb{N}+1$.
As we saw in Section \ref{sec:rwt}, there are several possible limits of A-series minimal models, including Liouville theory and generalized minimal models, and this also applies to our diagonal sector $\left. \mathcal{S}^{\text{DMM}_{p,q}} \right|_{\text{diagonal}} $. However, we will now lift this ambiguity by deriving the diagonal sector from the non-diagonal sector, using the fusion rule \eqref{vnvn}. At first sight, this fusion rule has the limit  
\begin{align}
 \lim_{p,q\to\infty} V_{(r_1,s_1)} V_{(r_2,s_2)}  \overset{?}{\sim} \sum_{r\overset{2}{=}|r_1-r_2|+1}^{\infty} \sum_{s\overset{2}{=}|s_1-s_2|+1}^{\infty} V^d_{\langle r,s\rangle}\ .
\end{align}
This is a sum over infinitely many degenerate fields, whose momenta $P_{(r,s)}$ are dense in the real line if $\beta^2\notin \mathbb{Q}$. The sum is therefore divergent. To obtain a finite limit, we need to compute
\begin{align}
 \lim_{p,q\to\infty} \frac{1}{q} V_{(r_1,s_1)} V_{(r_2,s_2)} \sim \frac12 \int_\mathbb{R} dP \ V_P\ .
\end{align}
This shows that in the limit CFT, the OPE of 2 non-diagonal fields is actually 
\begin{align}
 V_{(r_1,s_1)}V_{(r_2,s_2)} \sim \frac12 \int_{\mathbb{R}+i\epsilon} dP\ V_P\ ,
 \label{vrsvrs}
\end{align}
where the imaginary shift $i\epsilon$ is here for avoiding poles of OPE coefficients, just like in $c\leq 1$ Liouville theory \eqref{ireg}. This leads to the spectrum
\begin{align}
\boxed{\mathcal{S}^{\text{GDMM}_{\beta^2}} = \frac12 \left\{V_P\right\}_{P\in \mathbb{R}} \bigcup  \frac12\left\{V_{(r,s)} \right\}_{ (r,s)\in 2\mathbb{Z}\times (\mathbb{Z}+\frac12)}  }\ ,
\end{align}
and we call the corresponding CFT a \myindex{generalized D-series minimal model}\index{minimal model!generalized D-series---}\index{GDMM}. Due to the lack of $r\leftrightarrow s$ symmetry in the non-diagonal sector, this theory depends on $\beta^2$ rather than the central charge, i.e. it is not invariant under $\beta \to \beta^{-1}$. 
The OPEs of this CFT are determined by the conservation of diagonality: in addition to the OPEs of 2 non-diagonal fields \eqref{vrsvrs} and the OPE of 2 diagonal fields \eqref{idpv}, we have the mixed OPE 
\begin{align}
 V_{(r_1,s_1)}V_P \sim \frac12 \sum_{r\in 2\mathbb{Z}}\sum_{s\in \mathbb{Z}+\frac12} V_{(r,s)} \ . 
\end{align}
Like Liouville theory, generalized D-series minimal models cannot be analytically continued beyond $\beta^2\in \mathbb{R}_{>0}$. However, we can define these models by their spectra in the range $\Re\beta^2>0$ \eqref{rbp}, assuming that they exist. There is indeed evidence that they exist \cite{mr17}, although they have not been fully solved, and the OPEs \eqref{vrsvrs} may include discrete terms in addition to the integral \cite{rib19}.

\subsection{Loop CFTs}

In statistical physics, \textbf{loop models}\index{loop model} are a class of 2d lattice models whose states can be described in terms of configurations of non-intersecting loops. (See \cite{jac09} for a review.) Starting from the Ising model, many statistical models can indeed be reformulated in terms of loops, or directly constructed from loops.

In the critical limit, loop models become conformally invariant, and give rise to conformal field theories called \textbf{critical loop models}\index{loop model!critical---}. Since the definition of loop models is broad and vague, and taking the critical limit is nontrivial, the notion of a critical loop models does not mean much in CFT. We will shortly introduce the notion of a loop CFT, which captures a few crucial features of the known critical loop models. The price to pay for this more precise definition is that a loop CFT may not be the critical limit of any particular loop model.

There is a lot of choice about which observables to include in a loop model. In the case of the Ising model, if we consider only local observables, we obtain the minimal model $\text{AMM}_{4,3}$ in the critical limit. If we also consider non-local observables such as cluster connectivities, we obtain a much richer CFT.

For applications to statistical physics, what matters is less to define a consistent CFT than to compute correlation functions, which are then interpreted in terms of probabilities. We will sketch a few consistent CFTs in Section \ref{sec:models}, but our main focus will be on correlation functions, which will be the subject of Section \ref{sec:bcloop} and Section \ref{sec:cloop}.

\subsubsection{Definition and basic properties}

Let a \textbf{loop CFT}\index{loop CFT} be a 2d CFT with no extended chiral symmetry, such that:
\begin{enumerate}
 \item The degenerate field $V^d_{\langle 1,2\rangle}$ exists.
 \item The spectrum contains infinitely many non-diagonal fields.
 \item All OPE spectra are discrete.
\end{enumerate}
The crucial assumption is the existence of $V^d_{\langle 1,2\rangle}$; the other assumptions are meant to exclude other known CFTs. In particular, assumptions 2 excludes minimal models, assumption 3 excludes generalized D-series minimal models, and both assumptions 2 and 3 exclude Liouville theory.

It follows that the central charge obeys the constraint $\Re c<13$ \eqref{rbp}. And the existence of $V^d_{\langle 1,2\rangle}$ but not $V^d_{\langle 2,1\rangle}$ means that loop CFTs really depend on $\beta^2$ \eqref{cb}, not on $c$. We expect that this dependence is holomorphic over the domain $\Re c<13$: thanks to the assumption that OPE spectra are discrete, correlation functions do not have the non-analyticities on the line $c\leq 1$ that we saw in the context of Liouville theory in Section \ref{sec:liou}.

Moreover, there are known constraints on fusion rules:
\begin{itemize}
\item OPEs that involve the degenerate field $V^d_{\langle 1,2\rangle}$ are given by Eqs. \eqref{vpope} and \eqref{vdvrs},
 \item the first Kac index is conserved modulo integers \eqref{sriz}.
\end{itemize}

\subsubsection{Extended spectrum}\label{sec:es}

Let the extended spectrum $\widehat{S}^\text{loop}$ be made of all primary fields with half-integer spins that are compatible with the degenerate field $V^d_{\langle 1,2\rangle}$:
\begin{align}
 \boxed{\widehat{\mathcal{S}}^\text{loop} = \left\{V^d_{\langle 1,s\rangle}\right\}_{s\in\mathbb{N}^*}  \bigcup \left\{V_{(r,s)}\right\}_{\substack{r\in \frac12\mathbb{N}^*\\ s\in\frac{1}{2r}\mathbb{Z}}}\bigcup \left\{ V_P\right\}_{P\in\mathbb{C}} } \ .
 \label{sloop}
\end{align}
We include all degenerate fields that are generated from $V^d_{\langle 1,2\rangle}$ by fusion, and all possible diagonal fields. When it comes to non-diagonal fields, the first index $r$ must be half-integer according to Eq. \eqref{sriz}. The reflection relation $V_{(r,s)}=V_{(-r,-s)}$ \eqref{refl} allows us to assume $r\geq 0$, and the sector $r=0$ is redundant with the diagonal sector, since 
\begin{align}
 \boxed{V_{(0,s)} = V_{\frac12\beta^{-1}s}  \quad \text{i.e.} \quad V_P = V_{(0,2\beta P)}} \ .
 \label{sP}
\end{align}
On the other hand, $V_{(r,0)}$ is a genuinely non-diagonal field, whose fusion with $V^d_{\langle 1,2\rangle}$ differs from that of the diagonal field with the same dimensions $V_{\frac12\beta r}$. 
Non-diagonal fields with integer spins include:
\begin{subequations}
\label{vex}
\begin{align}
 & V_{(\frac12, 0)}, V_{(\frac12,\pm 2)}, V_{(\frac12, \pm 4)}, \dots
 \\
 & V_{(1,0)}, V_{(1,\pm 1)}, V_{(1,\pm 2)}, \dots 
 \\
 & V_{(\frac32, 0)}, V_{(\frac32, \pm\frac23)}, V_{(\frac32,\pm \frac43)} , V_{(\frac32,\pm 2)}, \dots
 \\
 & V_{(2, 0)}, V_{(2,\pm\frac12)}, V_{(2,\pm 1)}, V_{(2,\pm\frac32)}, \dots 
\end{align}
\end{subequations}
Our simple extended spectrum hides important subtleties: 
\begin{itemize}
 \item We characterize fields by their conformal dimensions, but there could be several independent fields with the same dimensions, in other words there could be nontrivial field multiplicities. This is known to occur in the loop CFTs of Section \ref{sec:models}, where the multiplicities may be described in terms of representations of global symmetry groups.
 \item Knowing which primary fields belong to $\widehat{S}^\text{loop}$ is not enough for determining how the conformal algebra acts on $\widehat{S}^\text{loop}$. For $(r,s)\in \mathbb{N}^*\times \mathbb{Z}^*$, the primary field $V_{(r,s)}$ has a singular vector, and the Verma module $\mathcal{V}_{\Delta_{(r,s)}}\otimes\bar{\mathcal{V}}_{\Delta_{(r,-s)}}$ is not the only representation that includes $V_{(r,s)}$. The correct representation can be determined by starting from $V_{(r,0)}$ and repeatedly taking OPEs with $V^d_{\langle 1,2\rangle}$. This leads to a logarithmic representation $\mathcal{W}_{(r,|s|)}$ that contains both $V_{(r,s)}$ and $V_{(r,-s)}$, and is not left-right factorized \cite{nr20}.
 To some extent, the structure of that representation can alternatively be derived from lattice loop models \cite{glhjs20}.
 By convention we label the resulting family $\left(\mathcal{W}_{(r,s)}\right)_{\substack{r\in \mathbb{N}^*\\ s\in\mathbb{N}}}$ with natural integers $s$, where we define
 $\mathcal{W}_{(r,0)}= \mathcal{V}_{\Delta_{(r,0)}}\otimes\bar{\mathcal{V}}_{\Delta_{(r,0)}}$.
In Section \ref{sec:log} we will deduce the corresponding logarithmic conformal blocks from the analytic properties of correlation functions.
\end{itemize}

\subsubsection{$O(n)$ CFT, $PSU(n)$ CFT, Potts CFT}\label{sec:models}

In any loop CFT, the spectrum must be a subset of the extended spectrum $\widehat{S}^\text{loop}$. In particular, three known loop CFTs give rise to integer-spin subsets:
\begin{align}
\mathcal{S}^{O(n)} &= \left\{V^d_{\langle 1,s\rangle}\right\}_{s\in 2\mathbb{N}+1} \bigcup \left\{V_{(r,s)}\right\}_{\substack{r\in \frac12\mathbb{N}^*\\ s\in\frac{1}{r}\mathbb{Z}}}  \ ,
\label{son}
 \\
 \mathcal{S}^{PSU(n)} &= \left\{V^d_{\langle 1,s\rangle}\right\}_{s\in\mathbb{N}^*} \bigcup \left\{V_{(r,s)}\right\}_{\substack{r\in \mathbb{N}^*\\ s\in\frac{1}{r}\mathbb{Z}}}  \ ,
 \label{sun}
 \\
 \mathcal{S}^\text{Potts} &= \left\{V^d_{\langle 1,s\rangle}\right\}_{s\in\mathbb{N}^*} \bigcup \left\{V_{(r,s)}\right\}_{\substack{r\in \mathbb{N}+2\\ s\in\frac{1}{r}\mathbb{Z}}} \bigcup  \left\{ V_{P_{(0,s)}}\right\}_{s\in \mathbb{N}+\frac12}\ .
 \label{spotts}
\end{align}
The spectra of the $O(n)$ and $PSU(n)$ models are natural constructions: we start with a degenerate field $V^d\in\{V^d_{\langle 1,3\rangle},V^d_{\langle 1,2\rangle}\}$, and add non-diagonal fields $V_{(r,s)}\in\widehat{S}^\text{loop}$ such that $V_{(r,s)}$ and the fields that appear in $V^dV_{(r,s)}$ have integer spins.
On the other hand, the Potts spectrum has no simple explanation in our formalism. We can only remark that it is obtained from the $PSU(n)$ spectrum by removing the non-diagonal fields $\left\{V_{(1,s)}\right\}_{s\in\mathbb{Z}}$ and adding the diagonal fields $\left\{ V_{P_{(0,s)}}\right\}_{s\in \mathbb{N}+\frac12}$, see Section \ref{sec:coup} for a justification from statistical physics.  

Each one of these CFTs has:
\begin{itemize}
 \item A modular-invariant torus partition function. Having an integer-spin spectrum is a necessary condition for modular invariance.
 \item A group of global symmetries that commutes with conformal symmetry: the orthogonal group $O(n)$, the unitary group $PSU(n)$, or the symmetric group $S_Q$ in the case of the Potts CFT. The parameters of these groups are related to the central charge:
 \begin{align}
  \boxed{n = -2\cos\left(\pi \beta^2\right)} \quad , \quad \boxed{Q = 4\cos^2\left(\pi \beta^2\right)}\ .
  \label{nQ}
 \end{align}
 The group structure actually exists only if $n,Q\in\mathbb{N}^*$. However, what matters for the CFT is the category of finite-dimensional representations, which can be defined for any $n,Q\in\mathbb{C}$ \cite{gnjrs21, br19}.
\end{itemize}
Degenerate fields are invariant under global symmetries, while other fields can transform in nontrivial representations \cite{jrs22, rjrs24}.
For example, in the $O(n)$ CFT, the fields $V_{(\frac12,0)},V_{(1,0)}$ and $V_{(1,1)}$ transform under $O(n)$ as a vector, a symmetric 2-tensor and an antisymmetric 2-tensor respectively. There are 2 fields of the type $V_{(\frac32,0)}$, which transform as symmetric and antisymmetric 3-tensors. Let us diplay the behaviour of these fields and a few more, while writing finite-dimensional irreducible representations of $O(n)$ as Young diagrams:
\begin{subequations}
\label{von}
\begin{align}
 V_{(\frac12,0)} &: [1]\ , \label{lhz}
 \\
 V_{(1,0)} &: [2]\ ,
 \\
 V_{(1,1)} &: [11]\ , \label{loo}
 \\
 V_{(\frac32,0)}&: [3]+[111]\ ,
\label{l320}
\\
V_{(\frac32,\frac23)} &: [21]\ ,
\label{l3223}
\\
V_{(2,0)}&: [4]+[22]+[211]+[2]+[]\ ,
\label{l20}
\\
V_{(2,\frac12)}& : [31]+[211]+[11]\ ,
\\
V_{(2,1)} &: [31]+[22]+[1111]+[2]\ ,
\label{l21}
\\
V_{(\frac52,0)} &: [5]+[32]+2[311]+[221]+[11111]+[3]+2[21]+[111]+[1]\ .
\label{l52}
\end{align}
\end{subequations}
In particular, there are 2 fields of type $V_{(\frac52,0)}$ that transform in the irreducible representation $[311]$, and also 2 fields that transform in $[21]$. This suggests that field multiplicities become larger as the first index of $V_{(r,s)}$ increases.

The situation is similar in the $PSU(n)$ and Potts CFTs: let us only mention that the diagonal non-degenerate fields $V_{P_{(0,s)}}$ of the Potts CFT transform in the standard  representation of $S_Q$, whose dimension is $Q-1$. 
For $Q=2$, the Potts CFT describes observables of the critical Ising model, and includes the minimal model AMM$_{4,3}$. For $Q=3$, the Potts CFT describes observables of the critical 3-state Potts model, and includes the minimal model DMM$_{6,5}$. The diagonal fields $V_{P_{(0,\frac12)}}$ appears twice in the spectrum \eqref{dmm65} of DMM$_{6,5}$, under the names $V_{(0,\frac12)}$ and $V^f_{\langle 3,2\rangle}$, because the standard representation of $S_3$ has dimension $2$. In brief,
\begin{align}
 \mathcal{S}^{\text{AMM}_{4,3}}\subset \mathcal{S}^\text{Potts}_{Q=2} \quad , \quad \mathcal{S}^{\text{DMM}_{6,5}}\subset \mathcal{S}^\text{Potts}_{Q=3}\ .
\end{align}

\subsection{Completing the picture}

\subsubsection{Summary of spectra}

For each CFT, we indicate the values of $\beta^2$, of the Kac indices of degenerate fields, of the indices of non-diagonal fields, and of the momenta of diagonal fields:
\begin{align}
 \begin{array}{|c|c|c|c|c|}
  \hline 
  \text{CFT} & \beta^2  & V^{d/f}_{\langle r,s\rangle} & V_{(r,s)} & V_P
  \\
  \hline \hline 
  \text{GMM} & \mathbb{C}\backslash \mathbb{Q}  & \mathbb{N}^*\times \mathbb{N}^* &  - & -
  \\
  \hline 
  \text{Liouville} & \mathbb{C}^*  & - & - &  \mathbb{R}
  \\
  \hline 
  \widehat{\text{Liouville}} & \mathbb{C}^*  & \mathbb{N}^*\times \mathbb{N}^* & - & \mathbb{C}
  \\
  \hline
  \text{AMM} & \frac{q}{p} > 0  & 
  \begingroup\renewcommand*{\arraystretch}{1}
  \begin{array}{c} \mathbb{N}\times \mathbb{N}\\  (0,p)\times (0,q)\end{array}\endgroup & - &  -
  \\
  \hline 
  \text{RWT} & \frac{q}{p} > 0  & - & - &  \mathbb{R}
  \\
  \hline \hline 
  \text{DMM} & \begingroup\renewcommand*{\arraystretch}{1}
  \begin{array}{c}\frac{q}{p} > 0 \\ p \in 2\mathbb{N} \end{array}\endgroup
  & \begingroup\renewcommand*{\arraystretch}{1}
  \begin{array}{c}
  (2\mathbb{N}+1)\times \mathbb{N} \\ (0,p)\times (0,q)\end{array}\endgroup
  &  \begingroup\renewcommand*{\arraystretch}{1}
  \begin{array}{c} 2\mathbb{Z}\times (\mathbb{Z}+\frac12) \\ (-\frac{p}{2},\frac{p}{2})\times (-\frac{q}{2},\frac{q}{2}) \end{array}\endgroup & - 
  \\
  \hline 
  \text{GDMM} & \left\{\Re\beta^2>0\right\} & - & 2\mathbb{Z}\times (\mathbb{Z}+\frac12) & \mathbb{R} 
  \\
  \hline\hline 
  \widehat{\text{Loop}} & \left\{\Re\beta^2>0\right\} & \{1\}\times \mathbb{N}^* & \frac12\mathbb{N}^* \times \frac{1}{2r}\mathbb{Z}  & \mathbb{C} 
  \\
  \hline 
  O(n) & \left\{\Re\beta^2>0\right\} & \{1\}\times(2\mathbb{N}+1) &  \frac12\mathbb{N}^* \times \frac{1}{r}\mathbb{Z} & - 
  \\
  \hline 
  PSU(n) & \left\{\Re\beta^2>0\right\} & \{1\}\times \mathbb{N}^* &  \mathbb{N}^* \times \frac{1}{r}\mathbb{Z} & - 
  \\
  \hline 
  \text{Potts} & \left\{\Re\beta^2>0\right\} & \{1\}\times \mathbb{N}^* &  (\mathbb{N}+2) \times \frac{1}{r}\mathbb{Z} & 
  \{P_{(0,s)}\}_{s\in\mathbb{N}+\frac12}
  \\
  \hline 
 \end{array}
 \label{sostab}
\end{align}
In the case of loop CFTs, this picture is most probably not complete. In addition to the extended spectrum of $\widehat{\text{Loop}}$, it is possible to defined a backbone field: a diagonal primary field whose conformal dimension does not have a simple expression in terms of Kac indices \cite{nqsz23}. In fact, from its dimension, it looks like this field may not be compatible with the existence of any degenerate field. Moreover, when studying correlation functions, we will see that the degenerate field $V^d_{\langle 2,1\rangle}$, which is absent from our extended spectrum, seems to play a role in the CFT, see Section \ref{sec:bcloop}.

\subsubsection{Limits of diagonal CFTs}\label{sec:lod}

Taking limits played an important role in deriving some of the CFTs of Section \ref{sec:diag}. 
Let us summarize the limits that relate Liouville theory, (generalized) minimal models, and Runkel--Watts-type theories: 
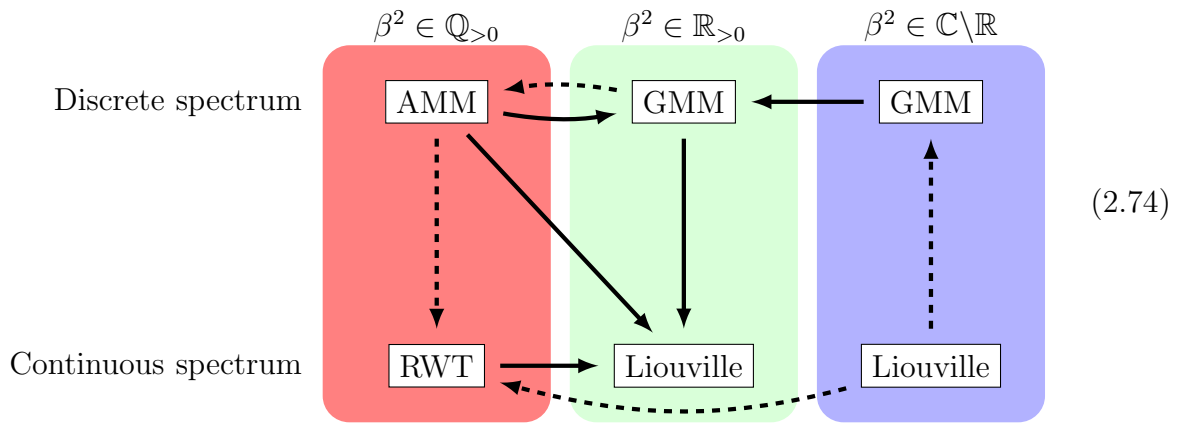
\begin{align}
\begin{tikzpicture}[scale = .25, baseline=(current  bounding  box.center)]
\fill[red!50, rounded corners = 10] (-6, -10) rectangle (6, 10);
\fill[green!15, rounded corners = 10] (7, -10) rectangle (19, 10);
\fill[blue!30, rounded corners = 10] (20, -10) rectangle (32, 10);
\node at (0, 11) {$\beta^2\in\mathbb{Q}_{>0}$};
\node at (13, 11) {$\beta^2\in\mathbb{R}_{>0}$};
\node at (26, 11) {$\beta^2\in \mathbb{C}\backslash \mathbb{R}$};
\node[left] at (-6.5, 7) {Discrete spectrum};
\node[left] at (-6.5, -7) {Continuous spectrum};
\draw (0, 7) node[draw, fill = white] (mm) {AMM};
\draw (13, 7) node[draw, fill = white] (gmm1) {GMM};
\draw (26, 7) node[draw, fill = white] (gmm2) {GMM};
\draw (0, -7) node[draw, fill = white] (rwt) {RWT};
\draw (13, -7) node[draw, fill = white] (clo) {Liouville};
\draw (26, -7) node[draw, fill = white] (liou) {Liouville};
\draw [ultra thick, shorten <= 2mm, shorten >= 2mm, -latex, 
       out = -10, in = -170] (mm) to (gmm1);
\draw [ultra thick, shorten <= 2mm, shorten >= 2mm, -latex, dashed,
       out = 170, in = 10] (gmm1) to (mm);
\draw [ultra thick, shorten <= 2mm, shorten >= 2mm, -latex] (gmm2) to (gmm1);
\draw [ultra thick, shorten <= 2mm, shorten >= 2mm, -latex] (mm) to (clo);
\draw [ultra thick, shorten <= 2mm, shorten >= 2mm, -latex, dashed] (mm) to (rwt);
\draw [ultra thick, shorten <= 2mm, shorten >= 2mm, -latex] (rwt) to (clo);
\draw [ultra thick, shorten <= 2mm, shorten >= 2mm, -latex] (gmm1) to (clo);
\draw [ultra thick, shorten <= 2mm, shorten >= 2mm, -latex,
       out = -165, in = -15, dashed] (liou) to (rwt);
\draw [ultra thick, shorten <= 2mm, shorten >= 2mm, -latex, dashed] (liou) to (gmm2);
\end{tikzpicture}
\label{lims}
\end{align}
Somewhat subjectively, we have distinguished limits that are straightforward (full arrows) from limits that are subtle (dashed arrows). The subtle limits are:
\begin{itemize}
 \item $\boxed{\text{AMM}\to \text{RWT}}$ (Section \ref{sec:rwt}): This involves fine-tuning the parameters $p,q$ of the minimal models, as well as the parameters $r,s$ of the primary fields.
  \item $\boxed{\text{Liouville}\to \text{RWT}}$ (Section \ref{sec:rwt}): This requires a detailed analysis of structure constants. 
 \item $\boxed{\text{GMM}\to \text{AMM}}$ (\cite{rib18}): This requires a detailed analysis of correlation functions, and only some of the GMM correlation functions tend to AMM correlation functions. 
 \item $\boxed{\text{Liouville}\to \text{GMM}}$ (Section \ref{sec:dtdf}): This requires a detailed analysis of correlation functions.
\end{itemize}

\subsubsection{Other exactly solvable CFTs}

Let us mention a few other solvable CFTs:
\begin{itemize}
 \item \myindex{E-series minimal models}\index{minimal model!E-series---} exist for $\beta^2=\frac{q}{p}$ with $p=12, 18, 30$ \cite{fms97}. They have finitely many primary fields, which belong to fully degenerate representations from the Kac table. 
They have 2 independent degenerate fields $\left\{V^f_{\langle r,1\rangle},V^f_{\langle 1,2\rangle}\right\}$ with $r=4,5,7$ respectively.
The 3-point structure constants are known explicitly for $p=12$, but not for $p=18,30$ \cite{nr25}.

A-series, D-series and E-series minimal models form the full set of rational CFTs that exist on any Riemann surface, also called \myindex{Virasoro minimal models}\index{minimal model!Virasoro---}\index{Virasoro!---minimal model}. The names of the 3 series come from the A-D-E classification of modular invariant torus partition functions, which provides a derivation of the spectra \cite{cz09}. In Section \ref{sec:sesc} we rederived the A-series and D-series spectra from the study of fusion rules, but we do not know how to do it for the E-series.

\item \myindex{Logarithmic minimal models}\index{minimal model!logarithmic---}\index{logarithmic!---minimal model} may be defined for $\beta^2\in \mathbb{Q}_{>0}$ by the assumption that their primary fields belong to the extended Kac table, i.e. they have integer Kac indices. (However, this definition is not universally accepted, and the same name is sometimes applied to other vaguely defined and poorly understood CFTs \cite{prz06}.) The spectra of logarithmic minimal models include logarithmic representations. Their spectra and fusion rules are not fully known. It is in fact not clear how many such CFTs exist at any given central charge. Some of them are probably limits of generalized minimal models, see Section \ref{sec:gmm}.

\item At $c=1$, the \myindex{compactified free boson}\index{compactified free boson} and its $\mathbb{Z}_2$ orbifold the \myindex{Ashkin--Teller model}\index{Ashkin--Teller model} are better understood in terms of an extended symmetry algebra (an abelian affine Lie algebra), although they can also be solved in terms of Virasoro symmetry only \cite{nr21}. They have a degenerate field $V^f_{\langle 1, 3\rangle}=V^f_{\langle 3,1\rangle}$, which however behaves as the identity field under fusion with non-diagonal fields. As a result, the constraints \eqref{sriz} are not obeyed. These models have a continuous parameter: the compactification radius $R\in\mathbb{C}^*$. For a special value of $R$, the compactified free boson is a $c\to 1$ limit of the $O(n)$ CFT, and the Ashkin--Teller model is a $c\to 1$ limit of the Potts CFT. Compactified free bosons also exist for $c\in\mathbb{C}$, but then $R$ is quantized \cite{rib14}. 
\end{itemize}

\section{Analytic bootstrap}\label{sec:ab}

In Section \ref{sec:csope}, we have worked out how local conformal symmetry constrains operator product expansions. This was done by using the associativity of operator products of the type $TVV$, where $T$ is the energy-momentum tensor. Similarly, we will now study how degenerate fields constrain structure constants, by using the associativity of 
operator products of the type $V^d_{\langle 2,1\rangle}VV$ or $V^d_{\langle 1,2\rangle}VV$.  
This will determine how structure constants behave under momentum shifts of the type $P\to P+ \beta$ or $P\to P+ \beta^{-1}$. 

\subsection{Degenerate 4-point functions}

Associativity of OPEs is equivalent to crossing symmetry of 4-point functions, so let us consider a 4-point function that involves a degenerate field. Since fusion products of degenerate representations \eqref{rrsvp} are finite, our 4-point function is a combination of finitely many conformal blocks in any one of the 3 channels $s,t,u$. This makes crossing symmetry equations particularly simple, and allows them to be solved analytically. 
And since any degenerate field $V^d_{\langle r,s\rangle}$ can be obtained from the 2 basic degenerate fields $V^d_{\langle 2,1\rangle}$ and $V^d_{\langle 1,2\rangle}$ by fusion, the crossing symmetry equations for $\left<V^d_{\langle r,s\rangle}V_1V_2V_3\right>$ follow from the 2 cases $(r,s)=(2,1)$ and $(r,s)=(1,2)$, which are related to one another by $\beta\to \beta^{-1}$.  

\subsubsection{BPZ equations}

As we have seen in Section \ref{sec:cor}, for any creation operator $L\in\mathcal{L}$, local conformal symmetry determines descendant correlation functions $\left<LV_0\cdots \right>$ in terms of the corresponding primary correlation functions $\left<V_0\cdots \right>$.
In the case of 4-point functions, we have 
\begin{align}
 \Big<LV_0(z)V_1(0)V_2(\infty)V_3(1)\Big> = P_L(z,\partial_z) \Big<V_0(z)V_1(0)V_2(\infty)V_3(1)\Big>\ ,
\end{align}
where $P_L(z,\partial_z)$ is a differential polynomial whose degree in $\partial_z$ is the level $|L|$ of the descendant field. If now our field is degenerate $V_0= V^d_{\langle r,s\rangle}$, then the singular vector equation $L_{\langle r,s\rangle}V^d_{\langle r,s\rangle}=0$ \eqref{lvdz} leads to a differential equation of order $rs$,
\begin{align}
 P_{L_{\langle r,s\rangle}}(z,\partial_z) \Big<V^d_{\langle r,s\rangle}(z)V_1(0)V_2(\infty)V_3(1)\Big> = 0\ .
\end{align}
This is called a Belavin--Polyakov--Zamolodchikov equation or \myindex{BPZ equation}\index{BPZ equation}.  
Let us derive the BPZ equation for a 4-point function  of the degenerate field $V_{\langle 2,1\rangle}$, together with 3 other primary fields:
\begin{align}
 G(z)=\Big<V_{\langle 2,1\rangle}^d(z)V_{1}(0)V_{2}(\infty)V_{3}(1)\Big>\ .
 \label{goz}
\end{align}
The relevant creation operator is $L_{\langle 2,1\rangle} = L_{-1}^2 -\beta^2L_{-2}$ \eqref{ars}, and the insertion of $L_{-2}$ is determined by the local Ward identity \eqref{lwi}. However, this identity involves derivatives in the positions of $V_{i}$, which are now fixed to $0,\infty,1$. We can eliminate these derivatives using global Ward identities. A technical trick to do this is to rewrite the local Ward identity as $G_\infty(z) = G_{z}(z) + G_0(z) + G_1(z)$, where we define
\begin{align}
 G_{z_0}(z) = \frac{1}{2\pi i} \oint_{z_0} \frac{y(y-1)}{y-z} \Big< T(y)V_{\langle 2,1\rangle}(z)V_{1}(0)V_{2}(\infty)V_{3}(1)\Big> \ .
\end{align}
We then compute $G_{z_0}(z)$ using the OPE $T(y)V_\Delta(z_0)$ \eqref{tvd}. The case $T(y)V_\Delta(\infty) = \frac{\Delta}{y^2} V_\Delta(\infty) + O(\frac{1}{y})$ is obtained from the case $z_0\in\mathbb{C}$ using the global conformal transformation $z\to \frac{1}{z}$ and the definition \eqref{vdi} of $V_\Delta(\infty)$. We find 
\begin{subequations}
\begin{align}
 G_\infty(z) &= \Delta_2 G(z) \quad , \quad G_0(z) = \frac{\Delta_1}{z}G(z)\quad , \quad G_1(z) = \frac{\Delta_3}{1-z}G(z)\ ,
 \\
 G_z(z) &= \left(\Delta_{(2,1)} + (2z-1)\partial_z\right) G(z) + z(z-1)\Big<L_{-2}V_{\langle 2,1\rangle}(z)V_{1}(0)V_{2}(\infty)V_{3}(1)\Big>
 \ .
\end{align}
\end{subequations}
Combining our rewritten local Ward identity with $L_{\langle 2,1\rangle}V^d_{\langle 2,1\rangle}=0$, and using $L_{-1}^2 V^d_{\langle 2,1\rangle}(z) = \partial_z^2 V^d_{\langle 2,1\rangle}(z)$ \eqref{pvlv}, we obtain the BPZ equation
\begin{align}
 \left(\frac{z(z-1)}{\beta^2}\frac{\partial^2}{\partial z^2} + (2z-1)\frac{\partial}{\partial z} + \frac{\Delta_1}{z} + \Delta_{(2,1)}-\Delta_2  +\frac{\Delta_3}{1-z} \right) G(z) = 0 \ .
\end{align}

\subsubsection{Hypergeometric conformal blocks}

According to the degenerate fusion rule \eqref{rvvp}, the decomposition of $G(z)$ into $s$-channel left-moving conformal blocks is a combination of 2 blocks with momenta $P_1\pm \frac{\beta}{2}$. The situation is similar in the $t$- and $u$-channels, and we write the resulting blocks as $\mathcal{F}^{(x)}_\pm(z)$ with $x\in\{s,t,u\}$:
\begin{align}
\mathcal{F}^{(s)}_\pm = 
 \begin{tikzpicture}[baseline=(current  bounding  box.center), very thick, scale = .35]
\draw (-1,2) node [left] {$1$} -- (0,0) -- node [above] {$P_1\pm\frac{\beta}{2}$} (4,0) -- (5,2) node [right] {$2$};
\draw (-1,-2) node [left] {$\langle 2,1\rangle\!$} -- (0,0);
\draw (4,0) -- (5,-2) node [right] {$3$};
\end{tikzpicture}
\quad 
\mathcal{F}^{(t)}_\pm = 
\begin{tikzpicture}[baseline=(current  bounding  box.center), very thick, scale = .35]
 \draw (-2,3) node [left] {$1$} -- (0,2) -- node [left] {$P_3\pm\frac{\beta}{2}$} (0,-2) -- (-2, -3) node [left] {$\langle 2,1\rangle\!$};
\draw (0,2) -- (2,3) node [right] {$2$};
\draw (0,-2) -- (2, -3) node [right] {$3$};
\end{tikzpicture}
\quad 
\mathcal{F}^{(u)}_\pm = 
\begin{tikzpicture}[baseline=(current  bounding  box.center), very thick, scale = .35]
\draw (-2,1) node [left] {$1$} -- (0,0) -- node [left] {$P_2\pm \frac{\beta}{2}$} (0,-4) ;
\draw (0,0)-- (2,-5) node [right] {$3$};
\draw (-2,-5) node [left] {$\langle 2,1\rangle\!$} -- (0,-4);
\draw (0,-4) -- (2,1) node [right] {$2$};
\end{tikzpicture} 
\end{align}
The conformal blocks provide 3 bases $\left\{\mathcal{F}^{(x)}_\pm(z)\right\}_\pm$ of solutions of the BPZ equation. Each solution is characterized by its asymptotic behaviour \eqref{stu} in a limit $z\to z_0$ with $z_0\in\{0,\infty,1\}$. 
For $z_0=0,1$ this behaviour is of the type $z^{\delta_1}$ and $(1-z)^{\delta_3}$, with exponents that we compute using Eq. \eqref{dp}:
\begin{align}
 \delta_i=\Delta\left(P_i+\tfrac{\beta}{2}\right) - \Delta(P_i) -\Delta_{(2,1)} = \beta P_i + \tfrac{1-\beta^2}{2}\ .
\end{align}
Introducing $\varphi(z) = z^{-\delta_1}(1-z)^{-\delta_3}$, we deduce that the differential equation for $\varphi(z) G(z)$ has holomorphic solutions near $z=0$ and near $z=1$, namely $\varphi(z)\mathcal{F}^{(s)}_+(z)$ and $\varphi(z)\mathcal{F}^{(t)}_+(z)$ respectively. By a tedious but straightforward calculation, we find that the equation for $\varphi(z) G(z)$ is a hypergeometric differential equation
\begin{align}
 \left(z(1-z)\frac{\partial^2}{\partial z^2} + \left[C-(A+B+1)z\right]\frac{\partial}{\partial z} -AB\right) \varphi(z) G(z) = 0 \ ,
\end{align}
with the parameters 
\begin{align}
 \{A,B\} = \left\{\tfrac12+\beta(P_1\pm P_2+P_3)\right\}_\pm \quad , \quad  C=1+2\beta P_1\ .
\end{align}
This allows us to write our conformal blocks in terms of the \myindex{hypergeometric function}\index{hypergeometric function}  ${}_2F_1(A,B,C,z)=\sum_{n=0}^\infty \left(\prod_{i=0}^{n-1}\frac{(A+i)(B+i)}{C+i}\right)\frac{z^n}{n!}$,
\begin{subequations}
\begin{align}
 \mathcal{F}_+^{(s)}(z) &= z^{\beta P_1 +\frac{1-\beta^2}{2}}(1-z)^{\beta P_3 + \frac{1-\beta^2}{2}} {}_2F_1\left(\tfrac12+ \beta P_{123},\tfrac12+\beta P_{13}^2,1+2\beta P_1,z\right) \ , 
 \\
 \mathcal{F}_+^{(t)}(z) &= z^{\beta P_1 +\frac{1-\beta^2}{2}}(1-z)^{\beta P_3 + \frac{1-\beta^2}{2}} {}_2F_1\left(\tfrac12+ \beta P_{123},\tfrac12+\beta P_{13}^2,1+2\beta P_3,1-z\right) \ , 
 \\
 \mathcal{F}_+^{(u)}(z) &= z^{-\beta P_{23} -\frac{\beta^2}{2}}(z-1)^{\beta P_3 + \frac{1-\beta^2}{2}} {}_2F_1\left(\tfrac12+ \beta P_{123},\tfrac12+\beta P_{23}^1,1+2\beta P_2,\tfrac{1}{z}\right) \ , 
\end{align}
\end{subequations}
where we used the notations $P_{123}=P_1+P_2+P_3$ and $P_{12}^3 = P_1+P_2-P_3$. The block $\mathcal{F}^{(x)}_-$ is obtained from $\mathcal{F}^{(x)}_+$ by $P_i\to -P_i$ where $i=1,3,2$ for $x=s,t,u$. 

The change of basis that relates the $s$- and $t$-channel conformal blocks is of the type
\begin{align}
 \mathcal{F}^{(s)}_{\epsilon_1} = \sum_{\epsilon_3=\pm} F_{\epsilon_1,\epsilon_3} \mathcal{F}^{(t)}_{\epsilon_3}\ ,
 \label{fsfft}
\end{align}
where the size 2 matrix $\left(F_{\epsilon_1,\epsilon_3}\right)_{\epsilon_1,\epsilon_3=\pm}$ is a \myindex{degenerate fusing matrix}\index{degenerate!---fusing matrix}. This matrix has the coefficients
\begin{align}
 \boxed{F_{\epsilon_1,\epsilon_3} = \frac{\Gamma(1+2\beta\epsilon_1P_1)\Gamma(-2\beta \epsilon_3P_3)}{\prod_\pm \Gamma(\frac12 +\beta \epsilon_1P_1 \pm \beta P_2 -\beta \epsilon_3P_3)}}\ ,
 \label{fee}
\end{align}
and the determinant 
\begin{align}
 \det F = -\frac{P_1}{P_3}\ .
 \label{detf}
\end{align}
The degenerate fusing matrix is built from the \myindex{Gamma function}\index{Gamma function}\index{function!Gamma---} $\Gamma(x)=\int_0^\infty t^{x-1}e^{-t}dt$: a meromorphic function of $x\in\mathbb{C}$ with simple poles for $x\in -\mathbb{N}$ and no zeros, which obeys $\Gamma(x+1)=x\Gamma(x)$ and $\Gamma(x)\Gamma(1-x)=\frac{\pi}{\sin(\pi x)}$.

\subsubsection{Crossing symmetry}

We will now assemble hypergeometric conformal blocks into crossing-symmetric 4-point functions. We assume that fusion with $V^d_{\langle 2,1\rangle}$ acts diagonally on $V_1,V_2,V_3$, i.e. the left and right momenta $P_i,\bar P_i$ are shifted by the same amount $\frac{\beta}{2}$ or $-\frac{\beta}{2}$. (The case where they are shifted by opposite amounts is equivalent, and can be obtained by $\bar P_i\to -\bar P_i$.) This assumption is fulfilled by diagonal fields $V_P$, but also by non-diagonal fields $V_{(r,s)}$ with momenta of the type $(P,\bar P)= (P_{(r,s)},P_{(r,-s)})$.
Then the degenerate 4-point function \eqref{goz} decomposes into degenerate $x$-channel conformal blocks as 
\begin{align}
 \Big<V_{\langle 2,1\rangle}^dV_{1}V_{2}V_{3}\Big> 
 = \sum_\pm d^{(x)}_\pm \left|\mathcal{F}^{(x)}_\pm \right|^2  
 \ ,
\end{align}
where we use the modulus squared notation \eqref{fdz2}.
Crossing symmetry is the equality of the 3 decompositions $x=s,t,u$, and leads to relations between the degenerate 4-point structure constants $d^{(x)}_\pm$. In particular, applying the change of basis \eqref{fsfft} to the $s$-channel decomposition, and comparing with the $t$-channel decomposition, we obtain 
\begin{align}
 \forall \epsilon,\bar\epsilon \in \{+,-\}\ , \qquad \sum_\pm d_\pm^{(s)} F_{\pm,\epsilon}\bar F_{\pm, \bar\epsilon} = \delta_{\epsilon,\bar\epsilon} d_\epsilon^{(t)}\ . 
 \label{feeb}
\end{align}
The 2 cases $\epsilon = -\bar\epsilon$ give us 2 linear relations between $d^{(s)}_+$ and $d^{(s)}_-$, whose compatibility amounts to 
\begin{align}
 \frac{F_{++}F_{--}}{F_{+-}F_{-+}} = \frac{\bar F_{++}\bar F_{--}}{\bar F_{+-}\bar F_{-+}}\ .
 \label{ffff}
\end{align}
With our degenerate fusing matrix \eqref{fee}, we find 
\begin{align}
 \frac{F_{++}F_{--}}{F_{+-}F_{-+}} = \prod_\pm \frac{\cos \pi \beta(P_1\pm P_2-P_3)}{\cos \pi \beta(P_1\pm P_2+P_3)}\ . 
\end{align}
Thanks to the conservation of the second Kac index $s=\beta(\bar P-P)$ modulo integers \eqref{ssiz}, this expression satisfies the compatibility condition \eqref{ffff}. Then the crossing symmetry relations \eqref{feeb} determine all the structure constants $d^{(s)}_\pm,d^{(t)}_\pm$ in terms of one of them,
\begin{align}
 \frac{d^{(s)}_-}{d^{(s)}_+} = -\frac{F_{++}\bar F_{+-}}{F_{-+}\bar F_{--}}\quad , \quad 
 \frac{d^{(t)}_+}{d^{(s)}_+} = \frac{F_{++}}{\bar F_{--}} \det \bar F \quad , \quad 
 \frac{d^{(t)}_-}{d^{(s)}_+} = -\frac{\bar F_{+-}}{F_{-+}} \det F\ .
\end{align}
Explicitly, these relations read 
\begin{subequations}
\begin{align}
  \frac{d^{(s)}_-}{d^{(s)}_+} = -\frac{\Gamma(1+2\beta P_1)}{\Gamma(1-2\beta P_1)}\frac{\Gamma(1+2\beta\bar P_1)}{\Gamma(1-2\beta\bar P_1)} (-)^{2s_2}
 \frac{\prod_{\pm,\pm}\Gamma\left(\tfrac12 -\beta\bar P_1\pm\beta \bar P_2\pm\beta \bar P_3)\right)}{\prod_{\pm,\pm}
 \Gamma\left(\tfrac12 +\beta P_1\pm \beta P_2\pm\beta P_3)\right)}\  ,
 \label{dmdp}
\end{align}
\begin{align}
 \frac{d^{(t)}_+}{d^{(s)}_+} = \frac{\Gamma\left(1+2\beta P_1\right)}{\Gamma\left(-2\beta\bar P_1\right)} 
 \frac{\Gamma\left(-2\beta P_3\right)}{\Gamma\left(1+2\beta\bar P_3\right)} 
  \frac{\prod_\pm\Gamma\left(\frac12 -\beta\bar P_1 \pm \beta\bar P_2 +\beta\bar P_3\right)}{\prod_\pm\Gamma\left(\frac12+\beta P_1 \pm \beta P_2 -\beta P_3\right)}\ ,
 \label{dpdp}
\end{align}
\begin{align}
 \frac{d^{(t)}_-}{d^{(s)}_+} = \frac{\Gamma\left(1+2\beta \bar P_1\right)}{\Gamma\left(-2\beta P_1\right)} 
 \frac{\Gamma\left(2\beta \bar P_3\right)}{\Gamma\left(1-2\beta P_3\right)} 
  \frac{\prod_\pm\Gamma\left(\frac12 -\beta P_1 \pm \beta P_2 -\beta P_3\right)}{\prod_\pm\Gamma\left(\frac12+\beta \bar P_1 \pm \beta \bar P_2 +\beta \bar P_3\right)}\ .
 \label{dtdp}
\end{align}
\end{subequations}
We will also need Eq. \eqref{dmdp} in the case of $\left<V_{\langle 2,1\rangle}^d V_1V_{\langle 2,1\rangle}^d V_1\right>$, where 2 of the fields are degenerate and the remaining 2 are equal:
\begin{align}
\frac{d^{(s)}_-}{d^{(s)}_+} =
-\frac{\Gamma\left(1+2\beta P_1\right)}{\Gamma\left(1-2\beta P_1\right)} \frac{\Gamma\left(1+2\beta \bar P_1\right)}{\Gamma\left(1-2\beta \bar P_1\right)} 
 \frac{\Gamma\left(\beta^2-2\beta P_1\right)\Gamma\left(1-\beta^2-2\beta P_1\right)}{\Gamma\left(\beta^2+2\beta \bar P_1\right)\Gamma\left(1-\beta^2+2\beta \bar P_1\right)}\ .
 \label{ddb}
\end{align}
Therefore, crossing symmetry allows us to determine degenerate 4-point functions up to an overall constant factor, by imposing relations on 4-point structure constants. Next we will deduce shift equations for the 2-point and 3-point structure constants.

\subsection{Shift equations for structure constants}

We will now derive shift equations for 3-point structure constants $C_{ijk}$. 
When it comes to 2-point structure constants, shift equations and their solutions can be deduced using Eq. \eqref{bicii}. 

\subsubsection{Degenerate OPE structure constants}

Let us introduce \myindex{degenerate OPE structure constants}\index{structure constant!degenerate OPE---} $c_i^\pm$ such that 
\begin{align}
V_{\langle 2,1\rangle}^d V_i \sim c^+_i V_{i^+} + c^-_i V_{i^-}\ ,
\end{align}
where we use the notation $i^\pm$ for the labels of the primary fields that appear in the degenerate OPE. 
The reflection relation \eqref{refl} determines $c_i^-$ from $c_i^+$:
\begin{align}
 c_i^- = c^+_{-i}\ ,
 \label{cim}
\end{align}
where we use the notation $-i$ for the label of a reflected field. 
With these notations, the degenerate 4-point structure constants are:
\begin{align}
 \begin{tikzpicture}[baseline=(base), very thick, scale = .4]
 \coordinate (base) at (0, -5);
\draw (-1,2) node [left] {$1$} -- (0,0) -- node [above] {$1^\pm$} (4,0) -- (5,2) node [right] {$2$};
\draw (-1,-2) node [left] {$\langle 2,1\rangle$} -- (0,0);
\draw (4,0) -- (5,-2) node [right] {$3$};
\node at (1.5, -5) {$d^{(s)}_\pm = \colorboxed{red}{c^\pm_1}\, \colorboxed{red}{C_{1^\pm23}}  $};
\draw[dashed, ->, red] (1.3,-4) to [out=90, in=-70] (.2, -.3);
\draw[dashed, ->, red] (3.5,-4.1) to [out=90, in=-110] (3.8, -.3);
\end{tikzpicture} 
\quad
\begin{tikzpicture}[baseline=(base), very thick, scale = .4]
 \coordinate (base) at (0, -5);
\draw (-1,2) node [left] {$1$} -- (0,0) -- node [above] {$1^\pm$} (4,0) -- (5,2) node [right] {$\langle 2,1\rangle$};
\draw (-1,-2) node [left] {$\langle 2,1\rangle$} -- (0,0);
\draw (4,0) -- (5,-2) node [right] {$1$};
\node at (1.5, -5) {$d^{(s)}_\pm = \colorboxed{red}{c^\pm_1}\, \colorboxed{red}{e^{\mp i\pi s_1}c^\mp_{1^\pm}}\, \colorboxed{red}{B_{1}} $};
\draw[dashed, ->, red] (-.2,-4) to [out=70, in=-70] (.2, -.3);
\draw[dashed, ->, red] (6,-4) to [out=150, in=-130] (5, -2.2);
\draw[dashed, ->, red] (3,-3.8) to [out=90, in=-110] (3.8, -.3);
\end{tikzpicture} 
\quad
\begin{tikzpicture}[baseline=(base), very thick, scale = .4]
\coordinate (base) at (0, -4);
\draw (-2,5) node [left] {$1$} -- (0,4) -- node [left] {$3^\pm$} (0,0) -- (-2,-1) node [left] {$\langle 2,1\rangle$};
\draw (2, 5) node [right] {$2$} -- (0,4);
\draw (0,0) -- (2,-1) node [right] {$3$};
\node at (0, -4) {$d^{(t)}_\pm = \colorboxed{red}{e^{\pm i\pi s_3}c_3^\pm}\, \colorboxed{red}{C_{123^\pm}}  $};
\draw[dashed, ->, red] (0,-2.9) to [out=90, in=-70] (0, -.4);
\draw[dashed, ->, red] (3,-3) to [out=50, in=-30] (.4, 3.8);
\end{tikzpicture} 
\label{cs}
\end{align}
Notice that the OPEs $V^d_{\langle 2,1\rangle}V_i$ and $V_iV^d_{\langle 2,1\rangle}$ are related by a permutation, under which the OPE structure constant picks up a factor $e^{i\pi(S_i-S_{i^\pm})} = e^{i\pi s_i}$, using Eq. \eqref{sdiff1}. 
This explains the factors $e^{\mp i\pi s_1}$ and $e^{\pm i\pi s_3}$, where $s_i\in\frac12\mathbb{Z}$.
In the case of the 4-point function $\left<V_{\langle 2,1\rangle}^d V_iV_{\langle 2,1\rangle}^d V_i\right>$, we therefore have
\begin{align}
 \frac{d^{(s)}_-}{d^{(s)}_+}= (-)^{2s_i}\frac{c_i^-}{c_{i^+}^-} \frac{c_{i^-}^+}{c_i^+}\ .
 \label{ddcccc}
\end{align}
Since its left-hand side is explicitly known by Eq. \eqref{ddb}, this equation is a constraint on the degenerate OPE structure constants.
It can be checked that the following expressions solves this constraint, and also obey the reflection relation \eqref{cim}: 
\begin{align}
 c^+_i = \beta^{2\beta(P_i+\bar P_i)} \frac{\Gamma(-2\beta P_i)}{\Gamma(1+2\beta \bar P_i)} \quad , \quad c^-_i = \beta^{-2\beta(P_i+\bar P_i)} \frac{\Gamma(2\beta P_i)}{\Gamma(1-2\beta \bar P_i)}\ ,
 \label{cpcm}
\end{align}
This solution is far from unique, because in the bootstrap approach we are always free to perform \myindex{field renormalizations}\index{field!---renormalization} using arbitrary $z$-independent coefficients $\lambda_i$,
\begin{align}
 V_i(z) \to \lambda_i V_i(z)\quad \implies \quad c_i^\pm \to \lambda_{\langle 2,1\rangle}\frac{\lambda_i}{\lambda_{i^\pm}}c_i^\pm \ , \  B_i \to \lambda_i^2B_i \ , \ C_{ijk}\to \lambda_i\lambda_j\lambda_k C_{ijk}\ .
 \label{vlv}
\end{align}
Instead of choosing a solution for $c_i^\pm$, and therefore a particular field normalization, 
we could work with renormalization-invariant quantities, which cannot involve $c_i^\pm$. For example, the $s$-channel 4-point structure constant $D^{(s)}_k = \frac{C_{12k}C_{k34}}{B_k}$ is invariant under renormalization of the channel field $V_k$. And $D^{(s)}_{k^+}$ can be written as a combination of degenerate 4-point structure constants of the type $d^{(s)}_\pm$ \eqref{cs}, where $c_i^\pm$ factors cancel. (To see this, use the identity $c_i^- B_{i^-} = (-)^{s_i} c^+_{i^-} B_i$, which follows from Eqs. \eqref{b21} and \eqref{ccb}).

In a CFT with 2 degenerate fields, the associativity of OPEs of the type 
$V^d_{\langle 2,1\rangle}V^d_{\langle 1,2\rangle}V_i$ leads to relations between 
the degenerate OPE coefficients for the 2 degenerate fields. (See \cite[Exercise 3.2]{rib14} for the case of Liouville theory.) These relations are equivalent to the compatibility of the shift equations from $V^d_{\langle 2,1\rangle}$ and $V^d_{\langle 1,2\rangle}$ for $B_i,C_{ijk}$. Here, we will directly show their compatibility by finding solutions.

\subsubsection{Explicit shift equations}

Combining the crossing symmetry equation \eqref{dmdp}, and \eqref{ddb} with the determination \eqref{cs} of degenerate 4pt structure constants and the choice \eqref{cpcm} of degenerate OPE structure constants, we obtain
\begin{align}
 \boxed{\frac{C_{1^-23}}{C_{1^+23}} = 
 (-)^{2s_2}\beta^{4\beta^2r_1} \frac{\prod_{\pm,\pm}\Gamma\left(\tfrac12 -\beta\bar P_1\pm\beta \bar P_2\pm\beta \bar P_3\right)}{\prod_{\pm,\pm}
 \Gamma\left(\tfrac12 +\beta P_1\pm \beta P_2\pm\beta P_3\right)}} \ ,
 \label{sh-mp}
\end{align}
This is called a \myindex{shift equation}\index{shift equation}, because it determines how structure constants behave when the momenta $P_1,\bar P_1$ of a field are shifted by $\beta$, under the condition $s_i\in \frac12 \mathbb{Z}$ \eqref{ssiz}. If now $s_i\in\mathbb{Z}$, then not only $S_{i^-}-S_{i^+}\in\mathbb{Z}$ but also $S_{i}-S_{i^+}\in\mathbb{Z}$. The fields $V_i$ and $V_{i^+}$ can both have integer spins, and we may want to relate their structure constants. From Eq. \eqref{dpdp}, we deduce an equation for shifts of momenta by $\frac{\beta}{2}$,
\begin{align}
 \boxed{\frac{C_{123^+}}{C_{1^+23}} = 
 (-)^{s_3} \beta^{2\beta^2(r_1-r_3)}
 \frac{ \prod_\pm\Gamma\left(\frac12 -\beta\bar P_1 \pm \beta\bar P_2 +\beta\bar P_3\right)}{ \prod_\pm\Gamma\left(\frac12+\beta P_1 \pm \beta P_2 -\beta P_3\right)}}\ .
 \label{sh-pp}
\end{align}
Our 2 shift equations were associated to the degenerate field $V^d_{\langle 2,1\rangle}$. To obtain the shift equations associated to $V^d_{\langle 1,2\rangle}$ instead, we may perform the substitutions $\beta\to \beta^{-1}$ and $r\leftrightarrow s$, which imply $(P,\bar P)\to (-P,\bar P)$. We obtain 
\begin{align}
 \frac{C_{1^-23}}{C_{1^+23}} = 
 (-)^{2r_2}\beta^{-4\beta^{-2}s_1} \frac{\prod_{\pm,\pm}\Gamma\left(\tfrac12 -\beta^{-1}\bar P_1\pm\beta^{-1} \bar P_2\pm\beta^{-1} \bar P_3\right)}{\prod_{\pm,\pm}
 \Gamma\left(\tfrac12 -\beta^{-1} P_1\pm \beta^{-1} P_2\pm\beta^{-1} P_3\right)} \ ,
 \label{sh-mp2}
\end{align}
\begin{align}
\frac{C_{123^+}}{C_{1^+23}} = 
 (-)^{r_3} \beta^{-2\beta^{-2}(s_1-s_3)}
 \frac{ \prod_\pm\Gamma\left(\frac12 -\beta^{-1}\bar P_1 \pm \beta^{-1}\bar P_2 +\beta^{-1}\bar P_3\right)}{ \prod_\pm\Gamma\left(\frac12-\beta^{-1} P_1 \pm \beta^{-1} P_2 +\beta^{-1} P_3\right)}\ .
 \label{sh-pp2}
\end{align}

\subsection{Solutions of shift equations}\label{sec:essc}

Since shift equations produce Gamma functions, their solutions should be written in terms of Barnes' \myindex{double Gamma function}\index{double Gamma function}\index{function!double Gamma---} $\Gamma_\beta$, which obeys
\begin{align}
\frac{\Gamma_\beta(x+\beta)}{\Gamma_\beta(x)} = \sqrt{2\pi}\frac{\beta^{\beta x-\frac12}}{\Gamma(\beta x)}
\quad , \quad 
\frac{\Gamma_\beta(x+\beta^{-1})}{\Gamma_\beta(x)} = \sqrt{2\pi}\frac{\beta^{\frac12-\beta^{-1}x}}{\Gamma(\beta^{-1}x)} \ .
\label{gshift}
\end{align}
The double Gamma function also obeys $\Gamma_\beta(x)= \Gamma_{\beta^{-1}}(x)$. It is a meromorphic function of $x\in\mathbb{C}$, with simple poles for $x\in -\beta\mathbb{N}-\beta^{-1}\mathbb{N}$ and no zeros. It is also a meromorphic function of $\beta\in \mathbb{C}\backslash i\mathbb{R}$, undefined for $\beta\in i\mathbb{R}$. To see that these values of $\beta$ are problematic, notice that $\text{closure}\left(-\beta\mathbb{N}-\beta^{-1}\mathbb{N}\right)\underset{\beta^2\in \mathbb{R}_{<0}-\mathbb{Q}} = i\mathbb{R}$, i.e. the simple poles would become dense in the imaginary axis.

Using the double Gamma function, we will write solutions of the shift equations for 3-point structure constants. We will deduce solutions for 2-point structure constants using $B_k\propto C_{Ikk}$ \eqref{bicii}, where the shift equations allow a $k$-independent factor.

A convenient identity that follows from Eq. \eqref{gshift} is:
\begin{align}
 \frac{\prod_\pm \Gamma_\beta\left(x \pm (y-\frac{\beta}{2})\right)}{\prod_\pm \Gamma_\beta\left(x \pm (y+\frac{\beta}{2})\right)} = \beta^{-2\beta y}\frac{\Gamma\left(\beta x+\beta y - \frac12 \beta^2\right)}{\Gamma\left(\beta x-\beta y -\frac12 \beta^2\right)}\ .
\end{align}

\subsubsection{Diagonal CFTs with 2 degenerate fields}\label{sec:dtdf}

Let us start with Liouville theory for $c\in (-\infty ,1]$ i.e. $\beta^2\in\mathbb{R}_{>0}$. 
We assume that the 3-point structure constant $C_{P_1,P_2,P_3}$ is a meromorphic function of the momenta, and a continuous function of $\beta$.
The shift equations determine $C_{P_1,P_2,P_3}$ modulo a factor that is invariant under $P_i\to P_i+\beta$ and $P_i\to P_i+\beta^{-1}$. 
If $\beta^2\notin \mathbb{Q}$, this factor must be constant on the complex $P_i$-plane. By continuity in $\beta$, the structure constant is unique on $c\in (-\infty,1]$ modulo an unimportant $P_i$-independent factor. The unique solution turns out to be a holomorphic function of $P_i\in\mathbb{C}$ and $\beta\in \mathbb{C}\backslash i\mathbb{R}$. We then deduce $B_{P_1}\propto C_{P_{(1,1)},P_1,P_1}$,
\begin{align}
 \boxed{B_{P_1} = \prod_{\pm,\pm}\Gamma_\beta^{-1}\left(\beta^{\pm 1}\pm 2P_1\right)} 
  \ \  , \ \
  \boxed{C_{P_1,P_2,P_3} =\prod_{\pm,\pm,\pm} \Gamma_\beta^{-1}\left(\tfrac{\beta+\beta^{-1}}{2} \pm P_1\pm P_2\pm P_3\right)}
 \ . 
 \label{bc}
\end{align}
The expression for $C_{P_1,P_2,P_3}$ obeys the shift equation \eqref{sh-mp} with $s_i=0$ and $P_i=\bar P_i$.
It is invariant under $P_i\to -P_i$, i.e. it obeys the reflection relation \eqref{refl}.
And it is invariant under $\beta \to \beta^{-1}$, so it also obeys the shift equation \eqref{sh-mp2}. 
Remarkably, the 2-point and 3-point functions can be analytically continued beyond the half-line $c\in (-\infty ,1]$, while $N$-point functions with $N\geq 4$ cannot, as follows from our discussion of the OPE in Section \ref{sec:liou}. 

The expressions \eqref{bc} for 2-point and 3-point structure constants are also valid in generalized minimal models, and in A-series minimal models. In such models, solutions of shift equations are unique, up to a field-independent factor. In the case of $C_{P_{(r_1,s_1)},P_{(r_2,s_2)},P_{(r_3,s_3)}}$, our shift equations \eqref{sh-mp} and \eqref{sh-pp}, together with the invariance under permutations of the 3 fields, determine all integer shifts that preserve $r_1+r_2+r_3\bmod 2$. But the fusion rules \eqref{rrsr} imply that the 3-point structure constant exists only if $r_1+r_2+r_3\equiv 1\bmod 2$, and can therefore be deduced from $C_{P_{(1,1)},P_{(1,1)},P_{(1,1)}}$ using shift equations. In other words, the ratio $\frac{C_{P_{(r_1,s_1)},P_{(r_2,s_2)},P_{(r_3,s_3)}}}{C_{P_{(1,1)},P_{(1,1)},P_{(1,1)}}}$ can be rewritten in terms of Gamma functions. (See \cite[Section 3.2.1]{rib14} for explicit expressions.) In particular, this shows that the 3-point structure constants are well-defined for any $\beta\in \mathbb{C}^*$, including in the regime $\beta\in i\mathbb{R}$ where $\Gamma_\beta$ is not defined.

In (generalized) minimal models, the structure constants are always non-vanishing, even though the function $\Gamma_\beta^{-1}$ has zeros. For the 3-point structure constant, this is again because of the fusion rules. 
The factor $\Gamma_\beta^{-1}\left(\frac{1+\sum_i\epsilon_ir_i}{2}\beta + \frac{1-\sum_i \epsilon_is_i}{2}\beta^{-1}\right)$ (with $\epsilon_i\in \{+,-\}$) of $C_{P_{(r_1,s_1)},P_{(r_2,s_2)},P_{(r_3,s_3)}}$ would indeed vanish if the 2 integers $\frac{1+\sum_i\epsilon_ir_i}{2}$ and $\frac{1-\sum_i \epsilon_is_i}{2}$ were both negative, but the fusion rules imply that the 3 integers $\sum_i\epsilon_i,\sum_i\epsilon_ir_i,\sum_i\epsilon_is_i$ all have the same (nonzero) sign. Conversely, if the fusion rules are violated, then $C_{P_{(r_1,s_1)},P_{(r_2,s_2)},P_{(r_3,s_3)}}$ vanishes in some cases, but not in some others. This has been a source of puzzlement \cite{zam05}. In the bootstrap approach, there is no puzzle, because structure constants make sense only provided fusion rules are obeyed. 

We have chosen normalizations such that the structure constants \eqref{bc} have simple expressions, and are holomorphic functions of $\beta,P_i$. In (generalized) minimal models, it is common to choose normalizations such that $V^d_{\langle 1,1\rangle}$ is the identity field, and such that 2-point functions are trivial. Calling $\widetilde{B}_{\langle r_1,s_1\rangle},\widetilde{C}_{\langle r_1,s_1\rangle\langle r_2,s_2\rangle \langle r_3,s_3\rangle}$ the corresponding structure constants, these conditions amount to 
\begin{align}
 \widetilde{B}_{\langle r_1,s_1\rangle} = \widetilde{C}_{\langle 1,1\rangle\langle r_2,s_2\rangle \langle r_2,s_2\rangle} = 1\ .
\end{align}
These conditions are fulfilled by 
\begin{align}
 \widetilde{C}_{\langle r_1,s_1\rangle\langle r_2,s_2\rangle \langle r_3,s_3\rangle} 
 = \frac{C_{P_{(r_1,s_1)},P_{(r_2,s_2)},P_{(r_3,s_3)}}}{C_{P_{(1,1)},P_{(1,1)},P_{(1,1)}}}
 \frac{B_{P_{(1,1)}}^\frac32}
 {\sqrt{B_{P_{(r_1,s_1)}}B_{P_{(r_2,s_2)}} B_{P_{(r_3,s_3)}}}}\ ,
 \label{wtc}
\end{align}
where the undetermined sign of $\sqrt{B_{P_{(r,s)}}}$ comes from the field renormalization $V^d_{\langle r,s\rangle}\to - V^d_{\langle r,s\rangle}$.
For example, in the Ising minimal model at $\beta^2=\frac34$, the nontrivial structure constant is $\widetilde{C}_{\langle 2,1\rangle\langle 2,1\rangle\langle 1,2\rangle}=\frac12$ (or $-\frac12$ if we renormalize $V^d_{\langle 1,2\rangle}$). In this case, the double Gamma functions not only combine into Gamma functions, but further simplify into a rational number.

Finally, let us consider Liouville theory with $c\in \mathbb{C}\backslash (-\infty,1]$. This includes the half-line $c\in [25,\infty)$ i.e. $\beta\in i\mathbb{R}^*$, where the function $\Gamma_\beta$ is not defined.
Instead, we should use the function $\hat{\Gamma}_\beta(x) = \frac{1}{\Gamma_{i\beta}(i\beta -ix)}$, which obeys shift equations that are very similar to the equations \eqref{gshift} for $\Gamma_\beta(x)$: 
\begin{align}
 \frac{\hat\Gamma_\beta(x+\beta)}{\hat\Gamma_\beta(x)} = \sqrt{2\pi}\frac{(i\beta)^{\beta x-\frac12}}{\Gamma(\beta x)}
\quad , \quad 
\frac{\hat\Gamma_\beta(x+\beta^{-1})}{\hat\Gamma_\beta(x)} = \sqrt{2\pi}(i\beta)^{\frac12-\beta^{-1}x}\Gamma(1-\beta^{-1}x) \ .
\label{hgshift}
\end{align}
Since we now have powers of $i\beta$ instead of $\beta$, we need to slightly modify the degenerate OPE structure constants \eqref{cpcm}, for example $c_i^+ = \beta^{4\beta P_i}\frac{\Gamma(-2\beta P_i)}{\Gamma(1+2\beta P_i)} \to (i\beta)^{4\beta P_i}\frac{\Gamma(-2\beta P_i)}{\Gamma(1+2\beta P_i)}$. The shift equation for structure constants \eqref{sh-mp} is similarly modified, i.e. $\beta^y \to (i\beta)^y$. Its unique solution $\hat B_{P_1},\hat C_{P_1,P_2,P_3}$ is obtained from $B_{P_1},C_{P_1,P_2,P_3}$ \eqref{bc} by $\Gamma_\beta \to \hat\Gamma_\beta$. Explicitly,
\begin{align}
 \boxed{\hat B_{P_1} = \prod_\pm \Gamma_b\left(\pm 2iP_1\right)\Gamma_b\left(Q\pm 2iP_1\right)} \quad , \quad 
 \boxed{\hat C_{P_1,P_2,P_3} = \prod_{\pm,\pm,\pm} \Gamma_b\left(\tfrac{Q}{2} \pm iP_1\pm iP_2\pm iP_3\right)}\ ,
 \label{bci}
\end{align}
where we use the notations 
\begin{align}
 b = i\beta \quad , \quad Q = b + b^{-1} \ . 
 \label{bQ}
\end{align}
By a field renormalization \eqref{vlv}, the 2-point and 3-point structure constants become 
\begin{align}
 \hat B^\text{DOZZ}_{P_1} = \frac{\Gamma_b\left(- 2iP_1\right)\Gamma_b\left(Q+ 2iP_1\right)}{\Gamma_b\left( 2iP_1\right)\Gamma_b\left(Q- 2iP_1\right)} \quad , \quad 
 \hat C^\text{DOZZ}_{P_1,P_2,P_3} = \frac{\prod_{\pm,\pm,\pm} \Gamma_b\left(\tfrac{Q}{2} \pm iP_1\pm iP_2\pm iP_3\right)}{\prod_{k=1}^3 \Gamma_b(2iP_k)\Gamma_b(Q-2iP_k)}\ ,
 \label{dozz}
\end{align}
which are the expressions found by Dorn--Otto and Zamolodchikov--Zamolodchikov when first solving Liouville theory: in particular, their formula for the 3-point structure constant is called the \myindex{DOZZ formula}\index{DOZZ formula}. In this normalization, the 2-point structure constant coincides with the reflection coefficient, i.e. it obeys the relations 
\begin{align}
 \hat B^\text{DOZZ}_{P_1}\hat B^\text{DOZZ}_{-P_1}=1 \quad , \quad  \hat C^\text{DOZZ}_{P_1,P_2,P_3} = \hat B^\text{DOZZ}_{P_1}\hat C^\text{DOZZ}_{-P_1,P_2,P_3} \quad , \quad V_{P_1} = \hat B^\text{DOZZ}_{P_1} V_{-P_1}\ . 
\end{align}
A remarkable property of Liouville theory in the DOZZ normalization is that degenerate fields are limits of non-degenerate fields,
\begin{align}
 \lim_{P\to P_{(r,s)}} V_P = V^d_{\langle r,s\rangle} \ .
 \label{limvp}
\end{align}
This relation is natural in the path integral construction of Liouville theory \cite{zz95}, but not necessarily in the bootstrap approach: it is not true in Liouville theory with $c\in (-\infty, 1]$, and for $c\in\mathbb{C}\backslash (-\infty,1]$ it requires a particular field normalization. It is thanks to this relation that generalized minimal models can be obtained as limits of Liouville theory, as announced in Section \ref{sec:lod}. This relation follows from the analytic properties of the structure constants, as we will now show in the case $(r,s)=(1,1)$. (See \cite[Section 3.1.5]{rib14} for the general case $r,s\in\mathbb{N}^*$.) Since $2iP_{(1,1)}=Q$ and $\Gamma_b(x)$ has a pole at $x=0$, we have $\hat C^\text{DOZZ}_{P_{(1,1)},P_2,P_3} = 0$ and it may seem that $\lim_{P_1\to P_{(1,1)}} V_{P_1} = 0$. However, in the OPE
\begin{align}
V_{P_1}V_{P_2} \sim \frac12 \int_{\mathbb{R}}dP\ \frac{\hat C^\text{DOZZ}_{P_1,P_2,P}}{\hat B^\text{DOZZ}_{P}}V_P\ , 
\end{align}
the limit does not commute with the integral,
because of the poles of $\hat C^\text{DOZZ}_{P_1,P_2,P}$ as a function of $P$. In particular, each factor $\Gamma_b(\frac{Q}{2}-iP_1\pm (iP_2-iP))$ has a pole at $P= P_2\pm (P_{(1,1)}-P_1)$. In the limit $P_1\to P_{(1,1)}$, these 2 poles coincide at $P=P_2$, with 1 pole coming from above the integration line and 1 from below. As a result, the limit of the integral is not zero:
\begin{align}
 \lim_{P_1\to P_{(1,1)}}V_{P_1}V_{P_2} \sim \lim_{P_1\to P_{(1,1)}}  \underset{P=P_1+P_2-P_{(1,1)}}{\operatorname{Res}}\frac{\hat C^\text{DOZZ}_{P_1,P_2,P}}{\hat B^\text{DOZZ}_{P}}V_P=V_{P_2}\ ,
\end{align}
which implies Eq. \eqref{limvp} in the case $(r,s)=(1,1)$. 

\subsubsection{Non-diagonal CFTs with 2 degenerate fields}

In (generalized) D-series minimal models, non-diagonal fields $V_{(r,s)}$ have indices $r\in 2\mathbb{Z}$ and $s\in\mathbb{Z}+\frac12$.
Due to the conservation of diagonality \eqref{ddd}, we have either $0$ or $2$ non-diagonal fields in a non-vanishing 3-point function. Let us begin with the latter case. We introduce the ansatz 
\begin{align}
 \boxed{C_{P_1,(r_2,s_2),(r_3,s_3)} = (-)^{\frac{r_3}{2}}\prod_{\pm,\pm}
 \Gamma_\beta^{-1}\left(\tfrac{\beta+\beta^{-1}}{2}+P_1 \pm P_2\pm P_3\right) 
 \Gamma_\beta^{-1}\left(\tfrac{\beta+\beta^{-1}}{2}-P_1 \pm \bar P_2\pm \bar P_3\right)}\ ,
 \label{cdnn}
\end{align}
whose sign prefactor ensures the correct behaviour under permutations \eqref{b21}. Moreover, we have the identity
\begin{align}
 C_{P_1,(r_2,s_2),(r_3,s_3)}=C_{-P_1,(r_2,s_2),(r_3,s_3)} \ ,
 \label{cpcmp}
\end{align}
which follows from the cases $(r,s)=(r_2\pm r_3,s_2\pm s_3)$ of the identity 
\begin{align}
 r,s\in\mathbb{Z} \implies \frac{\prod_\pm S_\beta\left(\frac{\beta+\beta^{-1}}{2} +P_1 \pm P_{(r,s)}\right)}{\prod_\pm S_\beta\left(\frac{\beta+\beta^{-1}}{2} +P_1 \pm P_{(r,-s)}\right)} =(-)^{rs}\ ,
 \label{dsr}
\end{align}
where we introduced the \myindex{double Sine function}\index{double Sine function}\index{function!double Sine---}
\begin{align}
 S_\beta(x) = \frac{\Gamma_\beta(x)}{\Gamma_\beta(\beta + \beta^{-1}-x)}\quad \implies \quad \frac{S_\beta(x+\beta)}{S_\beta(x)} = 2\sin(\pi\beta x)\ .
 \label{sb}
\end{align}
Having worked out the basic properties of our ansatz \eqref{cdnn}, let us check that it obeys the shift equations. The spectrum \eqref{sdmm} of D-series minimal models is invariant under $s\to s+1$, both in the diagonal and non-diagonal sectors. Therefore, when it comes to the shifts from $V^d_{\langle 1,2\rangle}$, we have to check the 2 shift equations \eqref{sh-mp2} and \eqref{sh-pp2}. This is straightforward, if we remember $r_2,r_3\in 2\mathbb{Z}$, leading to trivial sign factors $(-)^{r_i}=1$. We will now deal with the subtler case of the shifts from $V^d_{\langle 2,1\rangle}$. The spectrum of D-series minimal models is invariant under $r\to r+2$ but not $r\to r+1$, so we only have to deal with the shift equation \eqref{sh-mp}. If we shift a non-diagonal field, our ansatz \eqref{cdnn} behaves as
\begin{align}
 \frac{C_{(r_1-1,s_1),(r_2,s_2),P_3}}{C_{(r_1+1,s_1),(r_2,s_2),P_3}} = - \beta^{4\beta^2r_1}\frac{\prod_{\pm,\pm} \Gamma(\frac12-\beta \bar P_1\pm \beta\bar P_2\pm \beta P_3)}{\prod_{\pm,\pm} \Gamma(\frac12+\beta P_1\pm \beta P_2\pm \beta P_3)}\ , 
\end{align}
where the overall sign comes from $\prod_\pm \frac{\cos \pi\beta(P_1\pm P_2-P_3)}{\cos\pi\beta(\bar P_1\pm \bar P_2-P_3)} = \prod_\pm (-1)^{s_1\pm s_2} = -1$. This agrees with Eq. \eqref{sh-mp} with $\bar P_3=P_3$ and $(-)^{2s_2}=-1$. If we now shift the diagonal field, our ansatz leads to 
\begin{align}
 \frac{C_{P_1-\frac{\beta}{2},(r_2,s_2),(r_3,s_3)}}{C_{P_1+\frac{\beta}{2},(r_2,s_2),(r_3,s_3)}} =\beta^{8\beta P_1}\frac{\prod_{\pm,\pm} \Gamma(\frac12-\beta P_1\pm \beta\bar P_2\pm \beta \bar P_3)}{\prod_{\pm,\pm} \Gamma(\frac12+\beta P_1\pm \beta P_2\pm \beta P_3)}\ .
\end{align}
Now this ratio differs from Eq. \eqref{sh-mp} with $\bar P_1=P_1$ by an overall sign. 
In the case of D-series minimal models, 
let us cancel this sign.
Diagonal fields are degenerate in this case, so they can be labelled by Kac indices, and subjected to the field renormalization
$V^d_{\langle r,s\rangle}\to (-)^{\frac{r+1}{2}} V^d_{\langle r,s\rangle}$. 
This renormalization flips the sign of Eq. \eqref{sh-mp}, while leaving the other shift equations unchanged. As a result, the ansatz \eqref{cdnn} is now a solution of the shift equations. On the other hand, $C_{P_{(r_1,s_1)},P_{(r_2,s_2)},P_{(r_3,s_3)}}$ is no longer a solution
of the shift equations for the diagonal 3-point structure constant: the solution has an extra sign prefactor,
\begin{align}
 \boxed{C^\text{DMM}_{\langle r_1,s_1\rangle\langle r_2,s_2\rangle\langle r_3,s_3\rangle} = (-)^\frac{r_1+r_2+r_3+1}{2} C_{P_{(r_1,s_1)},P_{(r_2,s_2)},P_{(r_3,s_3)}}}\ , 
 \label{cdmm}
\end{align}
where $C_{P_1,P_2,P_3}$ is given in Eq. \eqref{bc}.

Now that we know the 3-point structure constants for D-series minimal models, let us deduce generalized D-series minimal models. As argued in Section \ref{sec:gdmm}, this amounts to taking the limit $P_{(r_1,s_1)}\to P_1\in\mathbb{R}$ in the diagonal sector. This is easily done in the case of $C_{P_1,(r_2,s_2),(r_3,s_3)}$, which is a meromorphic function of $P_1$, and is therefore still valid in the limit. When it comes to the diagonal 3-point structure constant \eqref{cdmm}, the sign prefactor does not have a smooth limit, and the structure constant becomes a distribution in the limit \cite{rib19},
\begin{align}
 \boxed{C^\text{GDMM}_{P_1,P_2,P_3} = \beta C_{P_1,P_2,P_3}\sum_{n\in\mathbb{Z}}(-)^n \frac{\prod_{i=1}^3 \cos(2\pi n\beta P_i)}{\cos(\pi n\beta^2)}}\ .
 \label{cgdmm}
\end{align}
The sum over $n$ is divergent, but it converges when inserted in the $s$-channel decomposition of a correlation function with non-diagonal fields, for example $\left<V_{P_1}V_{P_2}V_{(r_3,s_3)}V_{(s_4,s_4)}\right>$. 

Without the field renormalization $V^d_{\langle r,s\rangle}\to (-)^{\frac{r+1}{2}} V^d_{\langle r,s\rangle}$, would we have obtained a smooth diagonal 3-point structure constant $C_{P_1,P_2,P_3}$, and a distribution in the non-diagonal sector? Of course not: renormalizations cannot make such a difference. To see this, we may study the 4-point function $\left<V_{P_1}V_{P_2}V_{(r_3,s_3)}V_{(r_4,s_4)}\right>$ in the $s$-channel: whatever the normalization, the 4-point structure constant has a factor $C^\text{GDMM}_{P_1,P_2,P_s}$ in the limit. Our field renormalization only makes this clearer at the level of 3-point structure constants. 

Sign subtleties do not affect 2-point structure constants. For diagonal fields, the same expression $B_{P_1}$ \eqref{bc} as in $c\leq 1$ Liouville theory is still valid. For non-diagonal fields, we apply the relation $B_{(r,s)}\propto C_{P_{(1,1)},(r,s),(r,s)}$ to our ansatz \eqref{cdnn}, and find 
\begin{align}
 \boxed{B_{(r,s)} = (-)^\frac{r}{2} \prod_\pm \Gamma_\beta^{-1}\left(\beta\pm 2P\right)\Gamma_\beta^{-1}\left(\beta^{-1}\pm 2\bar P\right)}\ ,
 \label{nbrs}
\end{align}
with $\{P,\bar P\}=\{P_{(r,s)},P_{(r,-s)}\}$.

\subsubsection{Loop CFTs}\label{sec:bcloop}

From the structure of the loop CFTs' extended spectrum $\widehat{S}^\text{loop}$ \eqref{sloop}, it is clear that shift equations cannot completely determine the structure constants:
\begin{itemize}
 \item In the case of non-diagonal fields $V_{(r,s)}$, the dependence on $r$ is not constrained, because the spectrum $\widehat{S}^\text{loop}$ does not include the degenerate field $V_{\langle 2,1\rangle}^d$. We only have $V^d_{\langle 1,2\rangle}$, whose shift equations determine how structure constants behave under $s\to s+2$, whereas $s$ takes fractional values.
 \item In the case of diagonal fields $V_P$, the shift equations from $V^d_{\langle 1,2\rangle}$ determine how structure constants behave under $P\to P+\beta^{-1}$. 
\end{itemize}
The formula for the 3-point structure constants of loop CFTs is one of many solutions of the shift equations. It is of the type $C=\omega \check C$, where $\omega\in \{-1,1\}$ is a sign factor, and $\check C$ is
a product of double Gamma functions of combinations of Kac indices, involving the absolute value of $\sum_i\epsilon_i r_i$ \cite{nrj23, jnrr25}:
\begin{align}
\boxed{\check{C}_{(r_1,s_1)(r_2,s_2)(r_3,s_3)} =\prod_{\epsilon_1,\epsilon_2,\epsilon_3=\pm} \Gamma_\beta^{-1} \left(\tfrac{\beta+\beta^{-1}}{2} + \tfrac{\beta}{2}\left|\textstyle{\sum_i} \epsilon_ir_i\right| + \tfrac{\beta^{-1}}{2}\textstyle{\sum_i} \epsilon_is_i\right)}\ .
 \label{cref}
\end{align}
This expression is invariant under permutations, and obeys the shift equations only modulo signs. A sign factor $\omega$ is therefore needed to enforce the permutation relation \eqref{b21} and the shift equations. This sign factor does not have a simple explicit expression: rather, it is characterized by its invariance under cyclic permutations, and by its behaviour under shifts \cite{nrj23}:
\begin{align}
 \frac{\omega_{(r_1,s_1+1)(r_2,s_2)(r_3,s_3)}}{\omega_{(r_1,s_1-1)(r_2,s_2)(r_3,s_3)}} &\ =\  (-)^{2r_3}(-)^{\max(2r_1, 2r_2, 2r_3,r_1+r_2+r_3)} \ ,
 \label{tpt}
 \\
 \frac{\omega_{(r_1,s_1+1)(r_2,s_2)(r_3,s_3+1)}}{\omega_{(r_1,s_1)(r_2,s_2)(r_3,s_3)}} &\underset{r_i\in\mathbb{N}^*}{=}
 (-)^{\max(r_3, r_2-r_1)}\ .
\label{tppt}
\end{align}
The expression $C=\omega \check{C}$ for the 3-point structure constants is also valid if some of the fields are diagonal, via the identification $V_P = V_{(0,2\beta P)}$ \eqref{sP}. In particular, the relation \eqref{bicii} between 2-point and 3-point structure constants leads to
\begin{align}
 \boxed{B_{(r,s)} = (-)^{rs}\prod_\pm \Gamma_\beta^{2}\left(\beta^{\pm 1}\right)\cdot \check{C}_{(0,2\beta P_{(1,1)})(r,s)(r,s)} =  (-)^{rs}\prod_{\pm,\pm}\Gamma_\beta^{-1}\left(\beta^{\pm 1}+\beta r \pm \beta^{-1}s\right)}  \ ,
  \label{bref}
\end{align}
where we include the $r,s$-independent prefactor $\prod_\pm \Gamma_\beta^{2}\left(\beta^{\pm 1}\right)$ so that $B_{(0,2\beta P)}=B_P$ (for the $B_P$ of Eq. \eqref{bc}), and the sign factor has the explicit expression $\omega=(-)^{rs}$ in this case.

In contrast to CFTs with 2 degenerate fields, loop CFTs are not solved by the determination of 2-point and 3-point structure constants. This is because we do not know the structure or even the existence of OPEs, which would be needed to reduce $N$-point functions to 2-point and 3-point functions. In particular, 4-point structure constants are not combinations of 2-point and 3-point structure constants as in Eq. \eqref{stu}. However, they differ from the combinations by relatively simple factors, as we will see in Section \ref{sec:plw}.

The appearance of double Gamma functions in loop CFTs is remarkable, as this special function is the solution of 2 independent shift equations, while loop CFTs only have one degenerate field $V^d_{\langle 1,2\rangle}$. In particular, in the case of 3 diagonal fields, loop CFTs have the same 3-point structure constants $C_{P_1,P_2,P_3}$ \eqref{bc} as Liouville theory with $c\leq 1$, since
\begin{align}
\omega_{(0,2\beta P_1)(0,2\beta P_2)(0,2\beta P_3)} = 1 \quad , \quad \check{C}_{(0,2\beta P_1)(0,2\beta P_2)(0,2\beta P_3)} = C_{P_1,P_2,P_3}\ .
\end{align}
This coincidence of structure constants  has been a source of puzzlement \cite{ijs15, rib22}. Ultimately, the coincidence must be due to $C_{P_1,P_2,P_3}$ being the unique solution of the shift equations from $V^d_{\langle 2,1\rangle}$ and $V^d_{\langle 1,2\rangle}$. But in loop CFTs, why would $C_{P_1,P_2,P_3}$ obey the shift equation from $V^d_{\langle 2,1\rangle}$? This degenerate field does not belong to the spectrum, and we cannot easily add it because its OPE with non-diagonal fields $V_{(r,s)}$ would generate fields with non-integer spins.

\subsubsection{Summary table}\label{sec:table}

Let us summarize the solutions of shift equations for 2-point and 3-point structure constants. For each solution we indicate to which models it applies. Lighter green means an alternative field normalization. In the models' names, ``(G)MM'' means A-series minimal models and generalized minimal models, and ``Liou'' means Liouville theory. 
\begin{center}
\begin{tabular}{|l|c|c|c|c|c|c|c|c|}
\hline 
 Constants & Eqs. & (G)MM & Liou$_{c\leq 1}$ & RWT & Liou & DMM & GDMM & Loop 
 \\
 \hline
  $B_{P_1}$ & \eqref{bc} & \cellcolor{green!80!black} & \cellcolor{green!80!black} & \cellcolor{green!80!black} &  & \cellcolor{green!80!black} & \cellcolor{green!80!black} & \cellcolor{green!80!black}
  \\
  \hline 
  $C_{P_1,P_2,P_3}$& \eqref{bc} & \cellcolor{green!80!black} & \cellcolor{green!80!black} & \cellcolor{green!80!black} &  &  &  & \cellcolor{green!80!black}
  \\
  \hline 
  $\widetilde{C}_{\langle r_1,s_1\rangle\langle r_2,s_2\rangle\langle r_3,s_3\rangle}$ & \eqref{wtc} & \cellcolor{green!60} & & & & & &
  \\
  \hline 
  $\hat{B}_{P_1},\hat{C}_{P_1,P_2,P_3}$ &  \eqref{bci} & & & & \cellcolor{green!80!black} & & &
  \\
  \hline 
  $\hat{B}^\text{DOZZ}_{P_1},\hat{C}^\text{DOZZ}_{P_1,P_2,P_3}$ & \eqref{dozz} & & & & \cellcolor{green!60} & & &
  \\
  \hline 
  $B_{(r,s)}$ & \eqref{nbrs} & & & & & \cellcolor{green!80!black} & \cellcolor{green!80!black} &
  \\
  \hline 
  $C_{P_1,(r_2,s_2),(r_3,s_3)}$ & \eqref{cdnn} & & & & & \cellcolor{green!80!black} & \cellcolor{green!80!black} &
  \\
  \hline 
  $C^\text{DMM}_{\langle r_1,s_1\rangle\langle r_2,s_2\rangle\langle r_3,s_3\rangle}$ & \eqref{cdmm} &
  & & & & \cellcolor{green!80!black} & &
  \\
  \hline 
  $C^\text{GDMM}_{P_1,P_2,P_3}$ & \eqref{cgdmm} &
  & & & & & \cellcolor{green!80!black} &
  \\
  \hline 
  $B_{(r,s)}$ & \eqref{bref} & & & & & & & \cellcolor{green!80!black}
  \\
  \hline 
  $\check{C}_{(r_1,s_1)(r_2,s_2)(r_3,s_3)}$ & \eqref{cref} & & & & & & & \cellcolor{green!80!black}
  \\
  \hline 
\end{tabular}
\end{center}

\section{Numerical bootstrap}

In the analytic bootstrap approach of Section \ref{sec:ab}, we have used the crossing symmetry of 4-point functions that include degenerate fields, and deduced constraints on 2-point and 3-point structure constants. For 4-point functions without degenerate fields, crossing symmetry equations involve infinite sums and are hard to deal with analytically. We will thus need to solve them numerically. 

\subsection{Approaches to crossing symmetry}

\subsubsection{Achievements and limitations of analytic approaches}\label{sec:alaa}

It is not easy to directly evaluate the crossing symmetry equations \eqref{seteu}: conformal blocks are complicated, and their sums are even more complicated. Another approach is to construct 4-point functions in channel-independent ways, and then show that they can be decomposed into conformal blocks. This has led to analytic proofs of crossing symmetry in CFTs that can be constructed by perturbing a free bosonic CFT:
\begin{itemize}
 \item In A-series minimal models, 4-point functions can be constructed as Coulomb gas integrals \cite{df84}.
 \item In Liouville theory with $c\geq 25$, 4-point functions can be constructed from chiral vertex operators \cite{tes03b}.
 \item Again in Liouville theory with $c\geq 25$, 4-point functions can be constructed as expectation values in a probabilistic Gaussian free field theory \cite{ckrv05}. 
\end{itemize}
From these cases, all diagonal CFTs with 2 degenerate fields can be obtained by analytic continuation and taking limits \eqref{lims}, although mathematical rigor is lost in the process. In particular, the probabilistic construction only makes sense for $c\geq 25$.

Constructing Liouville theory with $c\leq 1$ would be particularly challenging, because in that case $\lim_{P\to P_{(1,1)}}V_P(z) = V_{P_{(1,1)}}(z)$ is a $z$-dependent, non-degenerate field. (Compare with \eqref{limvp} for $c\in\mathbb{C}\backslash(-\infty, 1)$.) But in the construction of $V_P$ as $e^{\alpha\varphi}$, our limit corresponds to $\alpha\to 0$ and leads to the identity field. And the limit can hardly fail to commute with the functional integral over $\varphi(z)$, which must yield correlation functions that are holomorphic in $\alpha$.

In this text we will focus on a numerical approach to crossing symmetry. This approach is applicable to any CFT whose spectrum is exactly known, and therefore to all the exactly solvable CFTs that we have considered. 
A similar approach is also applicable to integrable conformal gauge theories such as $\mathcal{N}=4$ super-Yang--Mills theory in 4 dimensions; these gauge theories however have spectra that are denser than in our 2-dimensional CFTs, requiring more sophisticated tools \cite{cgjp22}.

\subsubsection{Various flavours of numerics}\label{sec:vfn}

We consider a 4-point function $\left<V_1V_2V_3V_4\right>$, where the 4 primary fields are characterized by their conformal dimensions.
We assume that we exactly know the spectrum in each channel. If we do not know the structure constants, we view crossing symmetry as a system of linear equations \eqref{seteu} for the 4-point structure constants, normalized by the condition $D_{k_0}^{(s)}=1$ for some $k_0$. 

In the following diagram, we sketch what we can learn from numerically solving crossing symmetry, and give examples from exactly solvable CFTs. Potts connectivities are 4-point functions of the type $\Big<V_{P_{(0,\frac12)}}^4\Big>$, which have been an important testing ground for numerical bootstrap techniques. 
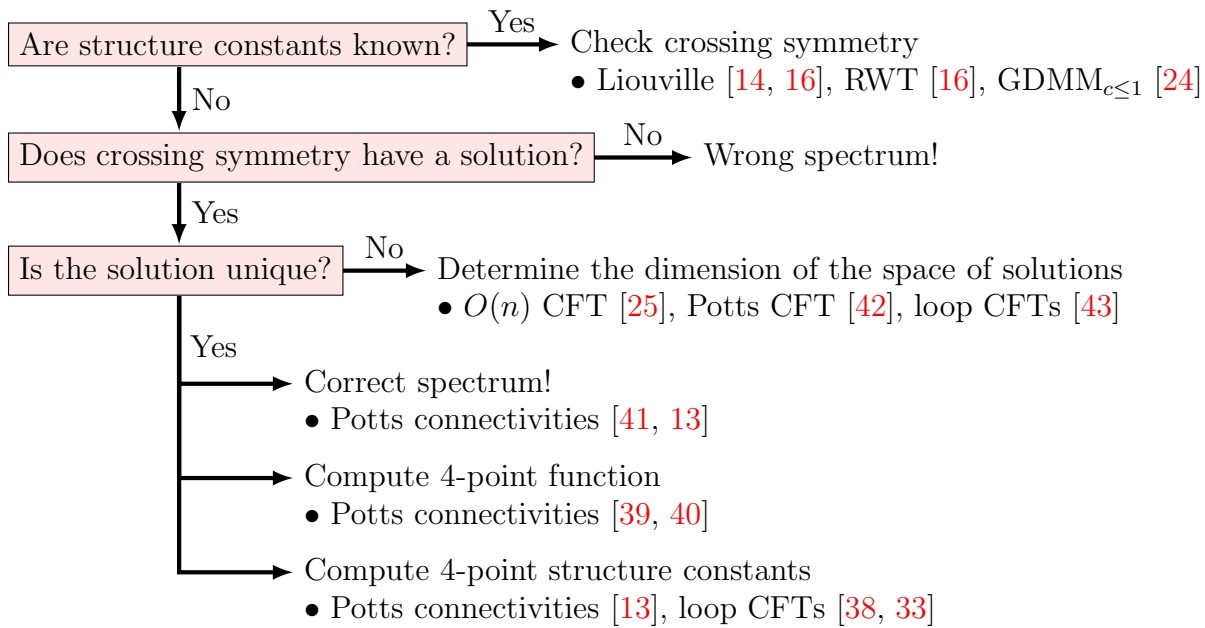
\begin{center}
 \begin{tikzpicture}[baseline=(base), scale = .5]
\node [right] at (7.5, -15) {$\bullet$ Potts connectivities \cite{nr20}, loop CFTs \cite{rib22, nrj23}};
 \node [right] at (7.5, -14) {Compute 4-point structure constants};
\draw [ultra thick, -latex] (4.5, -6) -- (4.5, -14) -- (7.5, -14);
 \node [right] at (7.5, -12.5) {$\bullet$ Potts connectivities \cite{prs16, prs19}};
 \node [right] at (7.5, -11.5) {Compute 4-point function};
 \draw [ultra thick, -latex] (4.5, -6) -- (4.5, -11.5) -- (7.5, -11.5);
\node [right] at (4.5, -8) {Yes};
  \node [right] at (7.5, -10) {$\bullet$ Potts connectivities \cite{hjs20, nr20}};
 \node [right] at (7.5, -9) {Correct spectrum!};
 \draw [ultra thick, -latex] (4.5, -6) -- (4.5, -9) -- (7.5, -9);
 \node [right] at (11, -7) {$\bullet$ $O(n)$ CFT \cite{gnjrs21}, Potts CFT \cite{niv22}, loop CFTs \cite{gjnrs23}};
 \node [right] at (11, -6) {Determine the dimension of the space of solutions};
 \draw [ultra thick, -latex] (8.8, -6) -- node[above]{No} (11, -6);
\draw (0, -6) node[right, draw, fill = red!10] {Is the solution unique?};
 \draw [ultra thick, -latex] (4.5, -3.6) -- node[right]{Yes} (4.5, -5.3);
 \draw (18, -3) node[right] {Wrong spectrum!};
\draw [ultra thick, -latex] (15.4, -3) -- node[above]{No} (18, -3);
\draw [ultra thick, -latex] (4.5, -.6) -- node[right]{No} (4.5, -2.3);
\draw (0, -3) node[right, draw, fill = red!10] {Does crossing symmetry have a solution?};
\draw (14.5, 0) node[right] {Check crossing symmetry};
\draw (14.5, -1) node[right] {$\bullet$ Liouville \cite{zz95, rs15}, RWT \cite{rs15}, GDMM${}_{c\leq 1}$ \cite{rib19}};
\draw [ultra thick, -latex] (12, 0) -- node[above]{Yes} (14.5, 0);
\draw (0, 0) node[right, draw, fill = red!10] {Are structure constants known?};
 \end{tikzpicture}
\end{center}

\subsubsection{Factorization of 4-point structure constants}\label{sec:factor}

In order to solve a CFT, we need to determine 2-point and 3-point structure constants. Solving crossing symmetry equations \eqref{seteu} gives us access to coefficients of conformal blocks. In the presence of nontrivial field multiplicities, these coefficients need not coincide with the factorized 4-point structure constants $D^{(s)}_k = C^k_{12}C_{k34}$ \eqref{stu}: rather, they are linear combinations thereof,
\begin{align}
 D^{(s)}_{\Delta,\bar\Delta} = \sum_{k\in\mathcal{S}^{(s)}\left|\substack{\Delta_k=\Delta\\ \bar{\Delta}_k=\bar\Delta}\right.} C^k_{12}C_{k34} \ .
 \label{dcc}
\end{align}
This is a sum of as many terms as the multiplicity of the primary field $V_{\Delta,\bar\Delta}$ in the $s$-channel spectrum $\mathcal{S}^{(s)}$.

This raises the problem of the \myindex{factorization of structure constants}\index{factorization of structure constants}, which means determining $C_{ijk}$ from $D^{(x)}_{\Delta,\bar\Delta}$, or just constraining $D^{(x)}_{\Delta,\bar\Delta}$ by requiring that it can be rewritten in terms of some $C_{ijk}$. If we know nothing about field multiplicities, factorization is empty constraint. (Any function of 2 variables has a decomposition of the type $f(x,y)=\sum_{i=1}^m g_i(x)h_i(y)$, if we allow the number of terms $m\in\mathbb{N}\cup \infty$ to be large enough.)
In principle, knowing the spectrum means knowing field multiplicities. In practice however, in loop CFTs, we know the spectrum of conformal dimensions without knowing the multiplicities.

If the multiplicity $m(V_{\Delta,\bar\Delta})$ of $V_{\Delta,\bar\Delta}$ in the spectrum $\mathcal{S}$ is finite, it can in principle be determined from the values of $D^{(s)}_{\Delta,\bar\Delta}$ for all possible 4-point functions $\left<V_{k_1}V_{k_2}V_{k_3}V_{k_4}\right>$:
\begin{align}
 m\left(V_{\Delta,\bar\Delta}\right) = \text{rank}\left(\left.D^{(s)}_{\Delta,\bar\Delta}\right|_{\left<V_{k_1}V_{k_2}V_{k_3}V_{k_4}\right>}\right)_{(k_1,k_2),(k_3,k_4)\in\mathcal{S}^2\times \mathcal{S}^2}\ .
\end{align}
In principle it is impossible to compute the rank of an infinite matrix, but 
in practice it is enough to study a few 4-point functions for determining a given multiplicity. This has been done in a few examples of diagonal 4-point functions in loop CFTs \cite{rib22}.

\subsection{Interchiral symmetry} \label{sec:icb}

As we have seen in Section \ref{sec:ab}, degenerate fields play a crucial role by constraining structure constants via shift equations. This suggests that we could extend the conformal algebra into an  \myindex{interchiral algebra}\index{interchiral!---algebra}, which would also include degenerate fields \cite{grs12}. In the same way as the conformal algebra is generated by the energy-momentum tensors $T,\bar T$, leading to conformal Ward identities, an interchiral algebra would be generated by $T,\bar T,V^d_{\langle 1,2\rangle}$ or $T,\bar T,V^d_{\langle 2,1\rangle},V^d_{\langle 1,2\rangle}$, and we would consider shift equations as interchiral Ward identities.

The extra generators do not commute with the conformal algebra (i.e. the OPE $TV^d_{\langle 1,2\rangle}$ is nontrivial), so interchiral symmetry is not a global symmetry: rather, it is an extension of spacetime symmetry, just like supersymmetry. This extension is natural in lattice loop models, where it is easy to construct a Temperley--Lieb algebra whose indecomposable representations tend to indecomposable representations of the interchiral algebra in the critical limit. On the other hand, there are no known lattice objects that tend to indecomposable representations of the conformal algebra \cite{gjs20}.

Using interchiral symmetry rather than conformal symmetry, the spectra \eqref{sostab} can be rewritten in terms of a smaller number of larger representations. In particular, each spectrum of an A-series or D-series minimal model is a combination of 4 (or fewer) interchiral representations. It would be interesting to reformulate the definition and classification of minimal models in terms of the representation theory of an interchiral algebra.

Here we will focus on the extended spectrum $\widehat{\mathcal{S}}^\text{loop}$ \eqref{sloop}, which involves the following representations of the interchiral algebra generated by $T,\bar T,V^d_{\langle 1,2\rangle}$:
\begin{subequations}
\label{icreps}
 \begin{alignat}{2}
 \widetilde{\mathcal{R}}^d_{\langle 1,s\rangle} &= \bigoplus_{k\in \mathbb{N}} \mathcal{R}^d_{\langle 1,s+2k \rangle} \otimes \overline{\mathcal{R}}^d_{\langle 1, s+2k\rangle} 
 && \ , \ s\in \{1, 2\}\ , 
 \label{wrd}
 \\
  \widetilde{\mathcal{V}}_{(r,s)} &= \bigoplus_{k\in\mathbb{Z}} \mathcal{V}_{P_{(r,s+2k)}}\otimes \overline{\mathcal{V}}_{P_{(-r,s+2k)}} && \ , \  r\in \tfrac12\mathbb{N}^*,\ s\in \tfrac{1}{2r}\mathbb{Z}\cap (-1, 1],\ (r,s)\notin \mathbb{N}^2\ , 
 \label{wvrs}
 \\
 \widetilde{\mathcal{W}}_{(r,s)} &= \bigoplus_{k\in \mathbb{N}} \mathcal{W}_{(r, s+2k)} && \ , \ r\in \mathbb{N}^*,\ s\in \{0,1\}\ ,
 \label{wlog}
 \\
   \widetilde{\mathcal{V}}_P &= \bigoplus_{k\in\mathbb{Z}} \mathcal{V}_{P+k\beta^{-1}}\otimes \overline{\mathcal{V}}_{P+k\beta^{-1}} && \ , \ P\in \mathbb{C}\ ,
 \label{wvp}
 \end{alignat}
\end{subequations}
where $\mathcal{R}_{\langle r,s\rangle}$ is a degenerate representation of the Virasoro algebra, $\mathcal{V}_P$ a Verma module, and $\mathcal{W}_{(r,s)}$ a logarithmic representation as described in Section \ref{sec:es}. In fact, we would like to define an interchiral algebra by the conditions that 
\begin{itemize}
 \item the representations \eqref{icreps} are irreducible,
 \item under reasonable assumptions, there are no other irreducible representations. 
\end{itemize}
We will sketch a tentative construction of an interchiral algebra, but we will stop short of developing the technical machinery of interchiral symmetry. The concept can nevertheless be useful: for example, we may ask which boundary conditions break or preserve interchiral symmetry. Most importantly for us, we will compute interchiral blocks, which are what we need for the numerical bootstrap.

\subsubsection{Towards an interchiral algebra}

By definition, the interchiral algebra is supposed to shift momenta of primary fields. We therefore need to replace the Virasoro generator $L_0$, whose eigenvalues are conformal dimensions, with a generator whose eigenvalues are momenta. This can be done by rewriting the Virasoro algebra in terms of an \myindex{abelian affine Lie algebra $\hat{\mathfrak{u}}_1$}\index{abelian affine Lie algebra $\hat{\mathfrak{u}}_1$}. The algebra $\hat{\mathfrak{u}}_1$ is the chiral symmetry algebra of free bosonic CFTs, which have a $\mathfrak{u}_1$ symmetry, and where the momentum is conserved.
However, this algebra can also be used for studying non-free CFTs, giving rise to Coulomb gas techniques. (See \cite[Section 4.1]{rib14} for a quick review.)

The algebra $\hat{\mathfrak{u}}_1$ has generators $(J_n)_{n\in\mathbb{Z}}$ and relations 
\begin{align}
 [J_m,J_n] = -\frac12 n\delta_{m+n,0}\ . 
\end{align}
For any $c\in\mathbb{C}$, a Virasoro algebra $\mathfrak{V}$ with central charge $c$ can be constructed as a subalgebra of the universal enveloping algebra of $\hat{\mathfrak{u}}_1$:
\begin{subequations}
\begin{align}
 L_n &= \sum_{m\in{\mathbb{Z}}} J_{n-m}J_m + \left(\beta-\beta^{-1}\right)(n+1)J_n\ , \qquad (n\neq 0)\ ,
\label{lnj}
\\
L_0 &=2\sum_{m=1}^\infty J_{-m}J_m +J_0^2+\left(\beta-\beta^{-1}\right)J_0 \ ,
\label{lzj}
\end{align}
\end{subequations}
where $\beta$ was defined from $c$ in Eq. \eqref{cb}.
(The central term in $[L_n,L_{-n}]$ is formally divergent, and becomes finite if the generators $J_m$ act on a state $V$ such that $\exists m_0, J_{m>m_0}V=0$.)
Let us define an \textbf{affine-primary state}\index{affine-primary state} $V_P$ of momentum $P$ by
\begin{align}
 J_0 V_P  = \left(P-\tfrac12\left(\beta-\beta^{-1}\right) \right) V_P \quad , \quad J_{n>0} V_P = 0\ .
\end{align}
Then $V_P$ is also a primary state, whose conformal dimension is computed from the last 2 terms of Eq. \eqref{lzj}, reproducing Eq. \eqref{dp}.

Now, let us extend the algebra $\hat{\mathfrak{u}}_1\times \overline{\hat{\mathfrak{u}}_1}$ by adding generators $D^\pm$ such that 
\begin{align}
 [J_0,D^\pm] = [\bar J_0,D^\pm] = \pm \beta^{-1}D^\pm \quad , \quad [J_{n\neq 0},D^\pm]=[\bar J_{n\neq 0},D^\pm]=0 \ .
\end{align}
We can deduce the commutation relations of $D^\pm$ with Virasoro generators,
\begin{align}
 [L_n,D^+] =2\beta^{-1}D^+J_n +\delta_{n,0}D^+ &\quad , \quad [L_n,D^-] =2\beta^{-1}J_nD^- +\delta_{n,0}D^- \ ,
 \\
 [\bar L_n,D^+] =2\beta^{-1}D^+\bar J_n +\delta_{n,0}D^+ &\quad , \quad [\bar L_n,D^-] =2\beta^{-1}\bar J_nD^- +\delta_{n,0}D^- \ .
\end{align}
Due to the generators $D^\pm$, the algebra is no longer chirally factorized into a left-moving and a right-moving factor: this is why it is called interchiral. We could have preserved chiral factorization by defining independent left-moving and right-moving generators $D^\pm,\bar D^\pm$, but such generators would produce states with spins that are not half-integer. 

By construction, the generators $D^\pm$ act on affine-primary states as $D^\pm V_P \propto V_{P\pm \beta^{-1}}$. Therefore, the representation $\widetilde{\mathcal{V}}_P$ \eqref{wvp} is now irreducible as required. However, our algebra is still way too large, and has way too many representations. To correct this, we may impose additional constraints. A natural constraint is
\begin{align}
 [D^+, D^-] = 2\beta^{-1}(J_0+\bar J_0) \ , 
\end{align}
so that $\left(\beta(J_0+\bar J_0),\beta D^+,\beta D^-\right)$ generate the Lie algebra $\mathfrak{sl}_2$. This constraint is not yet enough, because continuous representations of $\mathfrak{sl}_2$ come with 2 parameters, whereas our representations $\widetilde{\mathcal{V}}_P$ depend on only 1 parameter. As an additional constraint, we may set the quadratic Casimir to zero,
\begin{align}
 2(J_0+\bar J_0)^2 + D^+D^- + D^-D^+ = 0 \ .
\end{align}
This value of the quadratic Casimir is natural from the point of view of the interchiral identity representation $\widetilde{\mathcal{R}}^d_{\langle 1,1\rangle}$ \eqref{wrd}, whose interchiral primary field $V^d_{\langle 1,1\rangle}$ is an eigenvector of $J_0+\bar J_0$ with eigenvalue $0$. For this value of the Casimir, such a vector is annihilated by $D^-$, and is the lowest-weight state of a discrete representation of $\mathfrak{sl}_2$. As a result, $\widetilde{\mathcal{R}}^d_{\langle 1,1\rangle}$ is a semi-infinite combination of Virasoro representations, as opposed to the more generic infinite combination $\widetilde{\mathcal{V}}_P$. 

Our tentative construction of an interchiral algebra is surely not the final word on the subject. Remaining issues include:
\begin{itemize}
\item Explain algebraically what is special with having $D^\pm$ shift momenta by $\beta^{-1}$, as opposed to other numbers.
 \item Explain why $\widetilde{\mathcal{R}}^d_{\langle 1,2\rangle}$ is a semi-infinite combination.
 \item Predict the structure of the logarithmic representations $\mathcal{W}_{(r,s)}$ in Eq. \eqref{wlog}, which are supposed to be deduced from the Verma modules $\mathcal{W}_{(r,0)}$ by interchiral symmetry. We will see in Section \ref{sec:log} how this works at the level of conformal blocks, but an algebraic derivation would be welcome. 
 \item Define the fusion of interchiral representations, from which shift equations for 3-point structure constants should follow.
\end{itemize}

\subsubsection{Interchiral blocks}\label{sec:ib}

Just like a Virasoro block is the sum of contributions of states in a representation of the Virasoro algebra, an \myindex{interchiral block}\index{block!interchiral---}\index{interchiral!---block} is associated to a representation of the interchiral algebra. For example, in the case of the representation $\widetilde{\mathcal{V}}_P$ \eqref{wvp}, the interchiral block $\widetilde{\mathcal{F}}_P$ is a linear combination of the conformal blocks $\left|\mathcal{F}_{P+k\beta^{-1}}\right|^2$, where the coefficients are dictated by shift equations.

In any 4-point function of an A-series or D-series minimal model, all $s$-channel 4-point structure constants are related by shift equations, and the 4-point function involves a single $s$-channel interchiral block. In minimal models, interchiral blocks only differ from correlation functions by overall constants. On the other hand, when solving crossing symmetry equations in loop CFTs, using interchiral blocks reduces the number of unknown structure constants. We will focus on computing the interchiral blocks that are relevant to loop CFTs, i.e. the interchiral blocks that package the shift equations from the degenerate field $V^d_{\langle 1,2\rangle}$.  

Let us compute $s$-channel interchiral blocks for the 4-point function $\left<\prod_{i=1}^4 V_{(r_i,s_i)}\right>$. In terms of Virasoro blocks, the interchiral block for the $s$-channel interchiral representation $\widetilde{\mathcal{V}}_{(r,s)}$ \eqref{wvrs} reads 
\begin{align}
 \boxed{\widetilde{\mathcal{G}}_{(r,s)} = \sum_{k\in \mathbb{Z}} \frac{D^{(s)}_{(r,s+2k)}}{D^{(s)}_{(r,s)}} \mathcal{F}_{P_{(r,s+2k)}} \bar{\mathcal{F}}_{P_{(-r,s+2k)}}}\ . 
 \label{gtrs}
\end{align}
To make this explicit, it remains to compute the ratios of 4-point structure constants. This is in principle a straightforward consequence of the shift equation for the 3-point structure constant \eqref{sh-mp2}. However, for certain values of the parameters, this shift equation 
is finite after cancelling poles from the numerator and denominator. Such cancellations can be avoided by writing the shift equations differently. In fact, in shifts of the 3-point structure constant $\check C$ \eqref{cref}, no cancelling poles occur. The price to pay is that $\check C$ only solves shift equations up to signs. Taking these signs into account, the shift equation \eqref{sh-mp2} may be rewritten as
\begin{subequations}
 \label{cbshift}
 \begin{multline}
 \frac{C_{(r_1,s_1-1)(r_2,s_2)(r_3,s_3)}}{C_{(r_1,s_1+1)(r_2,s_2)(r_3,s_3)}} = (-)^{2r_3}(-)^{\max(2r_1,2r_2,2r_3,r_1+r_2+r_3)} \beta^{-4\beta^{-2}s_1} 
 \\ \times 
 \prod_{\epsilon_2,\epsilon_3=\pm} 
 \frac{\Gamma\left(\frac12 + \frac12 |\epsilon_2r_2+\epsilon_3 r_3-r_1| + \frac{\beta^{-2}}{2}(\epsilon_2s_2+\epsilon_3s_3-s_1)\right)}{\Gamma\left(\frac12 + \frac12 |\epsilon_2r_2+\epsilon_3 r_3+r_1| + \frac{\beta^{-2}}{2}(\epsilon_2s_2+\epsilon_3s_3+s_1)\right)}\ ,
 \label{cshift}
\end{multline}
where $2r_i, r_1+r_2+r_3\in\mathbb{Z}$ from Eq. \eqref{sriz}. The shift equation for the 2-point structure constant follows from the 3-point case via Eq. \eqref{bicii},
\begin{align}
 \frac{B_{(r,s-1)}}{B_{(r,s+1)}} = (-)^{2r} \beta^{-8\beta^{-2}s} \prod_{a\in \{0,1,\beta^{-2},1-\beta^{-2}\}} \frac{\Gamma(a+r-\beta^{-2}s)}{\Gamma(a+r+\beta^{-2}s)}\ . 
 \label{bshift}
\end{align}
\end{subequations}
From these shift equations, we can assemble shift equations for 
\begin{align}
 D_{(r,s)}^{(s)} = \frac{C_{(r_1,s_1)(r_2,s_2)(r,s)}C_{(r,s)(r_3,s_3)(r_4,s_4)}}{B_{(r,s)}}\ , 
\end{align}
and therefore the coefficients of the interchiral block $ \widetilde{\mathcal{F}}_{(r,s)}$ \eqref{gtrs}. 

The shifts of 2-point and 3-point structure constants are combinations of Gamma functions, and $\Gamma(x)$ has poles for $x\in -\mathbb{N}$.
If all involved fields are non-diagonal, i.e. of the type $V_{(r,s)}$ with $r\in\frac12\mathbb{N}^*$ and $s\in\frac{1}{2r}\mathbb{Z}$, the arguments of the Gamma functions are never in $-\mathbb{N}$ (assuming $\beta^2\notin \mathbb{Q}$), and the shifts are finite. If some fields are diagonal, i.e. of the type $V_{(0,2\beta P)}$ with $P\in \mathbb{C}$, then we can hit the poles of Gamma functions. We will consider such cases as accidental and not discuss them further, except when the $s$-channel field is a degenerate field $V^d_{\langle 1,s\rangle}\sim V_{(0,-\beta^2+s)}$ with $s\in\mathbb{N}^*$. 
In this case, the 3-point ratio \eqref{cshift} for $\left<V^d_{\langle 1,s\rangle}V_{(r_1,s_1)}V_{(r_2,s_2)}\right>$ vanishes if $r_1=r_2$ and $|s_1-s_2|=s$. This means that the interchiral block does not contain a term for $V^d_{\langle 1,|s_1-s_2|-1\rangle}$, consistently with the fusion rule \eqref{vdvrs}. Our 4-point function can have $s$-channel degenerate fields if and only if 
\begin{align}
 r_1=r_2 \quad , \quad r_3=r_4 \quad , \quad s_1-s_2,s_3-s_4\in \mathbb{Z} \quad , \quad s_3-s_4\equiv s_3-s_4\bmod 2\ , 
\end{align}
in which case the degenerate interchiral block is 
\begin{align}
 \boxed{\widetilde{\mathcal{G}}^d = \sum_{k\in \mathbb{N}} \frac{D^{(s)}_{(1,s_0+2k)}}{D^{(s)}_{(1,s_0)}}\left| \mathcal{F}_{P_{(1,s_0+2k)}} \right|^2\ ,  \quad \text{with} \quad s_0 =  \max(|s_1-s_2|,|s_3-s_4|)+1}\ . 
\end{align}

\subsection{Computation of Virasoro blocks}

In order to solve crossing symmetry equations numerically, we need to compute Virasoro blocks. It is possible to compute Virasoro blocks from their definition \eqref{sdec}, but this is awfully inefficient. We will now review the more efficient recursive representation due to Alexey Zamolodchikov \cite{zam87b}, in the case of 4-point $s$-channel blocks on the sphere. The $t$-channel and $u$-channel blocks can then be deduced using Eq. \eqref{stotu}. For $N$-point blocks on arbitrary Riemann surfaces, see \cite{ccy17}.

The idea of Zamolodchikov's recursion is to characterize Virasoro blocks by their analytic properties as functions of the channel dimension, in particular their poles, residues, and behaviour at $\infty$. 

\subsubsection{Poles and residues}

As a function of the channel dimension $\Delta$, the $s$-channel Virasoro block $\mathcal{F}_\Delta(z)$ has a simple pole at any $\Delta=\Delta_{(r,s)}$ with $r,s\in\mathbb{N}^*$. 
The pole is due to the existence of the singular vector $L_{\langle r,s\rangle}V_{\Delta_{(r,s)}}$, which is a primary state of dimension $\Delta_{(r,-s)}$. The block's residue at this pole is the sum of the contributions of the singular vector and its descendant states, therefore the residue is itself a Virasoro block,
\begin{align}
 \boxed{\underset{\Delta=\Delta_{(r,s)}}{\operatorname{Res}} \mathcal{F}_\Delta(z) = R_{r,s}\mathcal{F}_{\Delta_{(r,-s)}}(z)}\ .
 \label{resf}
\end{align}
The coefficient $R_{r,s}$ is called a \myindex{Virasoro block residue}\index{block!---residue}. 
According to Eq. \eqref{fnd}, the block's coefficients can be written in terms of the inverse Shapovalov form, which has a pole with residue \eqref{smo}, and descendant 3-point functions \eqref{gl}, which are regular at $\Delta=\Delta_{(r,s)}$. This leads to the expression \begin{align}
\boxed{R_{r,s} = \frac{c^{r,s}(P_1,P_2)c^{r,s}(P_3,P_4)}{b^{r,s}}}\ , 
 \label{rrs}
\end{align}
where the factors are 
\begin{align}
 c^{r,s}(P_1,P_2) = g^{L_{\langle r,s\rangle}}_{\Delta_{(r,s)},\Delta_1,\Delta_2} \quad , \quad b^{r,s} =
 S'_{L_{(r,s)},L_{(r,s)}}(\Delta_{(r,s)})\ .
\end{align}
Let us determine $b^{r,s}$ and $c^{r,s}$ explicitly.
Using the expression \eqref{sll} of the Shapovalov form as a 2-point function, we write
\begin{align}
 S_{L_{(r,s)},L_{(r,s)}}(\Delta) = \frac{ \left<L_{(r,s)}V_P(0)L_{(r,s)}V_P(\infty)\right>}{\left<V_P(0)L_{(r,s)}V_P(\infty)\right>} \Big<V_P(0)L_{(r,s)}V_P(\infty)\Big> \ . 
\end{align}
The first factor has a finite value at $P=P_{(r,s)}$, which coincides with $c^{r,s}(P_{(r,-s)},P_{(1,1)})$, because $P_{(1,1)}$ is the momentum of the identity field. The second factor has a simple zero, and behaves like $(-)^{rs} c^{r,s}(P, P_{(1,1)})$, where the sign prefactor comes from exchanging the positions of the 2 fields. As a result, we find an expression for $b^{r,s}$ in terms of $c^{r,s}$:
\begin{align}
 b^{r,s} = (-)^{rs} c^{r,s}\left(P_{(r,-s)},P_{(1,1)}\right)  \frac{1}{2P_{(r,s)}}\left.\frac{\partial}{\partial P} c^{r,s}\left(P,P_{(1,1)}\right)\right|_{P=P_{(r,s)}}\ . 
 \label{bcdc}
\end{align}
To determine $c^{r,s}$, let us deduce how it behaves under $r\to r+1$ from the change of basis \eqref{fsfft} between $s$-channel and $t$-channel degenerate 4-point conformal blocks. The idea is to use this relation for 4-point functions of the types $\left<V^d_{\langle 2,1\rangle}V_{P_{(r,-s)}}V_{P_1}V_{P_2}\right>$ and $\left<V^d_{\langle 2,1\rangle}V_{P_{(r,s)}}V_{P_1}V_{P_2}\right>$. More precisely, let us compare the coefficients $F_{++}$ and $\widetilde{F}_{++}$ that relate the following conformal blocks:
\begin{subequations}
\begin{align}
  \begin{tikzpicture}[baseline=(current  bounding  box.center), very thick, scale = .4]
\draw (-1,2) node [left] {$P_{(r,-s)}$} -- (0,0) -- node [above] {$P_{(r+1,-s)}$} (4,0) -- (5,2) node [right] {$P_1$};
\draw (-1,-2) node [left] {$\langle 2,1\rangle\!$} -- (0,0);
\draw (4,0) -- (5,-2) node [right] {$P_2$};
\end{tikzpicture}
\quad \overset{F_{++}}{\longrightarrow} \quad 
\begin{tikzpicture}[baseline=(current  bounding  box.center), very thick, scale = .4]
 \draw (-2,3) node [left] {$P_{(r,-s)}$} -- (0,2) -- node [left] {$P_2+\frac{\beta}{2}$} (0,-2) -- (-2, -3) node [left] {$\langle 2,1\rangle\!$};
\draw (0,2) -- (2,3) node [right] {$P_1$};
\draw (0,-2) -- (2, -3) node [right] {$P_2$};
\end{tikzpicture}
\end{align}
\begin{align}
\begin{tikzpicture}[baseline=(current  bounding  box.center), very thick, scale = .4]
\draw (-1,2) node [left] {$P_{(r,s)}$} -- (0,0) -- node [above] {$P_{(r+1,s)}$} (4,0) -- (5,2) node [right] {$P_1$};
\draw (-1,-2) node [left] {$\langle 2,1\rangle\!$} -- (0,0);
\draw (4,0) -- (5,-2) node [right] {$P_2$};
\end{tikzpicture}
\quad \overset{\widetilde{F}_{++}}{\longrightarrow} \quad 
\begin{tikzpicture}[baseline=(current  bounding  box.center), very thick, scale = .4]
 \draw (-2,3) node [left] {$P_{(r,s)}$} -- (0,2) -- node [left] {$P_2+\frac{\beta}{2}$} (0,-2) -- (-2, -3) node [left] {$\langle 2,1\rangle\!$};
\draw (0,2) -- (2,3) node [right] {$P_1$};
\draw (0,-2) -- (2, -3) node [right] {$P_2$};
\end{tikzpicture}
\end{align}
\end{subequations}
The relation $V_{P_{(r,-s)}}\propto L_{\langle r,s\rangle}V_{P_{(r,s)}}$ leads to a relation between $F_{++}$ and $\widetilde{F}_{++}$. In fact, it may seem that these 2 coefficients coincide, since the relation \eqref{fsfft} is the same for a descendant block as for the corresponding primary block. However, the coefficient $F_{++}$ is only valid for blocks that are normalized such that their asymptotics are as in Eq. \eqref{stu}. In order to correctly normalize blocks that involve descendant fields $L_{\langle r,s\rangle}V_{P_{(r,s)}}$ and/or $L_{\langle r+1,s\rangle}V_{P_{(r+1,s)}}$, we must multiply them with factors of the type $c^{r,s}$. As a result, we have 
\begin{align}
 \frac{c^{r+1,s}(P_1,P_2)}{c^{r,s}(P_1,P_2+\frac{\beta}{2})} \propto \frac{\widetilde{F}_{++}}{F_{++}} \propto \beta^{-2s}\frac{\prod_{\pm} \Gamma(\frac12+\beta P_{(r,-s)} \pm \beta P_1 -\beta P_2)}{\prod_{\pm} \Gamma(\frac12+\beta P_{(r,s)} \pm \beta P_1 -\beta P_2)} \ , 
 \label{cros}
\end{align}
where we use the expression \eqref{fee} for $F_{++}$ and $\widetilde{F}_{++}$. By $\propto$ we mean that equalities are modulo $P_1,P_2$-independent factors. Unlike the residue $R_{r,s}$, such factors are not invariant under renormalizations $L_{\langle r,s\rangle} \to \lambda_{r,s}(\beta)L_{\langle r,s\rangle}$. We have used this freedom, and chosen factors that will make $c^{r,s}$ simple. 

To solve the shift equation \eqref{cros}, and the analogous equation for $\frac{c^{r,s+1}}{c ^{r,s}}$, we initialize the recursion by determining $c^{1,1}(P_1,P_2) =\prod_\pm (P_1\pm P_2) $ from Eq. \eqref{fl1}. We then find 
\begin{align}
 \boxed{c^{r,s}(P_1,P_2) = \prod_{j\overset{2}{=}1-r}^{r-1} \prod_{k\overset{2}{=}1-s}^{s-1} \prod_\pm \left(P_2\pm P_1+ P_{(j,k)}\right) = \frac{\prod_{\pm,\pm}\Gamma_\beta\left(\frac{\beta+\beta^{-1}}{2} +P_2\pm P_1 \pm P_{(r,s)}\right)}
 {\prod_{\pm,\pm}\Gamma_\beta\left(\frac{\beta+\beta^{-1}}{2} +P_2\pm P_1 \pm P_{(r,-s)}\right)}}\ , 
 \label{crs}
\end{align}
where the second expression uses the double Gamma function \eqref{gshift}. As a result of Eq. \eqref{dsr}, we have $c^{r,s}(P_2,P_1)=(-)^{rs} c^{r,s}(P_1,P_2)$ and therefore $c^{r,s}(P_1,-P_2)=c^{r,s}(P_1,P_2)$. From Eq. \eqref{bcdc}, we then deduce 
\begin{align}
 \boxed{b^{r,s} = \frac{-(-)^{rs}}{2P_{(r,s)}P_{(0,0)}} \prod_{j=1-r}^r\prod_{k=1-s}^s 2P_{(j,k)} = 
 \frac{-\prod_\pm \Gamma_\beta\left(\beta \pm 2P_{(r,s)}\right)}{P_{(r,s)}\Gamma_\beta\left(\beta+2P_{(r,-s)}\right)\operatorname{Res}_{\beta-2P_{(r,-s)}}\Gamma_\beta}}\ .
 \label{brs}
\end{align}
While the product formulas are adequate for numerically computing the Virasoro block residues $R_{r,s}$, the expressions in terms of the double Gamma function are convenient for elucidating their relations with structure constants of exactly solvable CFTs, such as Eq. \eqref{dpld}.

Let us explicitly write these quantities for $rs\leq 4$. Since $c^{r,s},b^{r,s}$ are invariant under $\left\{\begin{smallmatrix} r\leftrightarrow s \ \ \ \\ \beta \to \beta^{-1}\end{smallmatrix}\right.$, we may assume $r\geq s$. With the help of the identity 
\begin{align}
 \prod_{\pm,\pm}\left(P_2\pm P_1\pm \ell\beta\right) = \left(\Delta_1-\Delta_2\right)^2 -2\ell^2\beta^2\left(\Delta_1+\Delta_2-1\right) +\ell^2(\ell^2-1)\beta^4-\ell^2\ ,
\end{align}
where conformal dimensions and momenta are related by Eq. \eqref{dp}, we find
\begin{subequations}
\begin{align}
 c^{1,1}(P_1,P_2) &= \Delta_1-\Delta_2\ ,
 \\
 c^{2,1}(P_1,P_2) &= \left(\Delta_1-\Delta_2\right)^2 -\tfrac12\beta^2\left(\Delta_1+\Delta_2-1\right) -\tfrac{3}{16}\beta^4-\tfrac14\ ,
 \\
 c^{3,1}(P_1,P_2) &= \left[\Delta_1-\Delta_2\right]\left[\left(\Delta_1-\Delta_2\right)^2 -2\beta^2\left(\Delta_1+\Delta_2-1\right) -1 \right]\ , 
 \\
 c^{4,1}(P_1,P_2) &= c^{2,1}(P_1,P_2)\times \left[\left(\Delta_1-\Delta_2\right)^2 -\tfrac92\beta^2\left(\Delta_1+\Delta_2-1\right) +\tfrac{45}{16}\beta^4-\tfrac94\right]\ ,
\end{align}
\begin{multline}
 c^{2,2}(P_1,P_2) = \left(\Delta_1-\Delta_2\right)^4 +\tfrac{c-13}{6}\left(\Delta_1-\Delta_2\right)^2\left(\Delta_1+\Delta_2-1\right)
 \\
 +\tfrac{(c-1)(c-25)}{144}\left(\Delta_1+\Delta_2-1\right)^2
 -\left(1+\tfrac{(c-1)(c-25)}{96}\right)\left(\Delta_1-\Delta_2\right)^2
 \\
 -\tfrac{(c-1)(c-13)(c-25)}{1152}\left(\Delta_1+\Delta_2-1\right) + \tfrac{(c-1)(c-9)(c-17)(c-25)}{36864}\ ,
\end{multline}
\end{subequations}
where the central charge $c$ is given by Eq. \eqref{cb}.
And we have 
\begin{subequations}
 \begin{align}
  b^{1,1} &= -2 \ ,
  \\
  b^{2,1} &= -4\left(1-\beta^4\right)\ ,
  \\
  b^{3,1} &= -24\left(1-\beta^4\right)\left(1-4\beta^4\right)\ ,
  \\
  b^{4,1} &= -288\left(1-\beta^4\right)\left(1-4\beta^4\right)\left(1-9\beta^4\right)\ ,
  \\
  b^{2,2} &= 8\left(\beta^2-\beta^{-2}\right)^2\left(1-4\beta^4\right)\left(1-4\beta^{-4}\right) = -\frac{2}{81}(c+2)(c-1)(c-25)(c-28)\ .
 \end{align}
\end{subequations}

\subsubsection{The nome}

In order to write Zamolodchikov's recursion for Virasoro blocks, and their large $\Delta$ asymptotics, it is convenient to replace the cross-ratio $z$ with the \myindex{nome}\index{nome} $q$, a function of $z$ built from the hypergeometric function: 
\begin{align}
 \boxed{q=e^{i\pi \tau}} \quad \text{with} \quad \boxed{\tau = i \frac{{}_2F_1(\frac12,\frac12,1,1-z)}{{}_2F_1(\frac12,\frac12,1,z)}}\ .
\end{align}
The inverse relation is 
\begin{align}
 z = \frac{\theta_2^4(q)}{\theta_3^4(q)} \quad , \quad 1-z = \frac{\theta_4^4(q)}{\theta_3^4(q)}\quad ,\quad 
 {}_2F_1\left(\tfrac12,\tfrac12,1,z\right) =\theta_3^2(q)\ , 
\end{align}
where $\theta_k(q)$ are 
\myindex{Jacobi theta functions}\index{Jacobi theta function}\index{function!Jacobi theta---}
\begin{align}
 \theta_2(q) = \sum_{n=-\infty}^\infty q^{(n+\frac12)^2} \quad , \quad \theta_3(q) = \sum_{n=-\infty}^\infty q^{n^2}\quad , \quad \theta_4(q) = \sum_{n=-\infty}^\infty (-1)^n q^{n^2} \ ,
\end{align}
which obey the identities 
\begin{align}
 \sum_{k=2,3,4}(-1)^k \theta_k^4(q) = 0 \quad , \quad \prod_{k=2,3,4} \theta_k(q) = 2q^\frac14\prod_{m=1}^\infty \left(1-q^{2m}\right)^3\ . 
 \label{thetaids}
\end{align}
This implies in particular the relation 
\begin{align}
 q\left(\tfrac{z}{z-1}\right) = -q(z)\ , 
\end{align}
as well as the asymptotic behaviour 
\begin{align}
 z(q) \underset{q\to 0}{=} 16q + 128q^2 +O\left(q^3\right) \quad ,\quad q(z)\underset{z\to 0} = \frac{z}{16} + \frac{z^2}{32} + O\left(z^3\right)\ . 
\end{align}
Let us describe how the map $z\mapsto q$ acts on a few features of the complex $z$-plane. The features in question are related to the $S_3$ subgroup of global conformal transformations that permutes $z=0,1,\infty$, called the crossing symmetry group in \cite{lsswy15}. The elements of this group are the following 6 global conformal maps:
\begin{align}
 \renewcommand{\arraystretch}{1.5}
 \begin{array}{|r||c|c|c|c|c|c|}
  \hline 
  \text{Map} & z & \frac{z}{z-1} & \frac{1}{z} & 1-z & \frac{1}{1-z} & 1-\frac{1}{z} 
  \\
  \hline 
  \text{Order} & 1  & 2 & 2 & 2 & 3 & 3
  \\
  \hline 
  \text{Fixed points} & \overline{\mathbb{C}} & 0, 2 & 1 , -1 & \infty, \frac12 & e^{\pm i\frac{\pi}{3}} & e^{\pm i\frac{\pi}{3}}
  \\
  \hline 
 \end{array}
 \label{S3}
\end{align}
We will now plot these fixed points in the complex $z$-plane, and plot their images in $q(\mathbb{C})$. We also plot the circles $|z|=1,|1-z|=1$ and the line $\left|\frac{z}{z-1}\right|=1$, and their images, which split $\mathbb{C}$ and $q(\mathbb{C})$ into 6 fundamental domains of $S_3$:
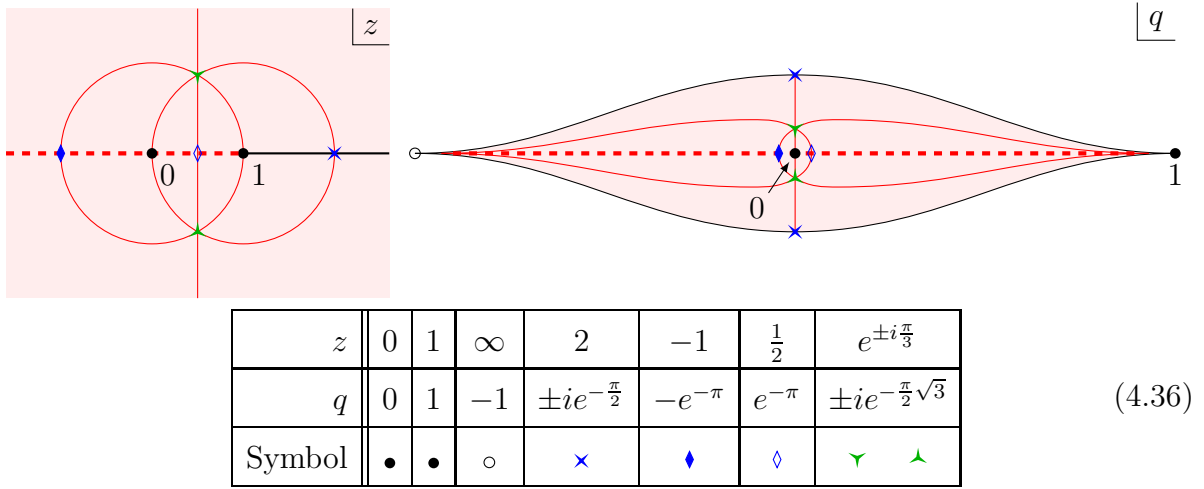
\begin{multline}
 \begin{tikzpicture}[baseline=(base), scale = 1.2]
  \coordinate (base) at (0, 0);
  \fill[red!7] (-1.6, -1.6) rectangle (2.6, 1.6);
  \draw (2.6, 1.2) -- (2.2, 1.2) -- (2.2, 1.6);
  \node at (2.4, 1.4) {$z$};¨
  \draw[red] (0, 0) circle [radius = 1];
  \draw[red] (1, 0) circle [radius = 1];
  \draw[red] (.5, -1.6) -- (.5, 1.6);
  \draw[thick] (1, 0) node[fill, circle, minimum size = 1.4mm, inner sep = 0]{} -- (2.6, 0);
  \draw[ultra thick, dashed, red] (-1.6, 0) -- (1, 0);
  \node[below] at (1, 0){$1$};
  \node[below] at (0, 0){$0$};
  \draw(0, 0) node[fill, circle, minimum size = 1.4mm, inner sep = 0]{};
  \node at (.5, 0) [blue, draw, kite, kite vertex angles = 50, scale = .25] {};
  \node at (-1, 0) [blue, fill, draw, kite, kite vertex angles = 50, scale = .25] {};
  \node at (2, 0) [blue, fill, star, star points = 4, star point ratio = .2, scale = .7] {};
  \node at (.5, .866) [green!70!black, fill, star, star points = 3, star point ratio = .2, scale = .7]{};
  \node at (.5, -.866) [rotate = 180, green!70!black, fill, star, star points = 3, star point ratio = .2, scale = .7]{};
 \end{tikzpicture}
 \ \ 
 \begin{tikzpicture}[baseline=(base), scale = 5]
 \coordinate (base) at (0, 0);
 \draw (1, .3) -- (.9, .3) -- (.9, .4);
 \node at (.95, .35) {$q$};
  \filldraw[fill = red!7] (-1, 0) to [out = 0, in = 180] (0, .208) to [out = 0, in = 180] (1, 0) to [out = 180, in =0] (0, -.208) to [out = 180, in =0] (-1, 0);
  \draw[red] (0, -.208) -- (0, .208);
  \draw[red] (-.93, 0) to [in = 180, out = 0] (-.115, .0893) to [out = 0, in = 150] (0, .066) to [out = -30, in = 90] (.043, 0) to [out = -90, in = 30] (0, -.066) to [out = -150, in = 0] (-.115, -.0893) to [out = 180, in = 0] (-.93, 0);
  \draw[red] (.93, 0) to [in = 0, out = 180] (.115, .0893) to [out = 180, in = 30] (0, .066) to [out = -150, in = 90] (-.043, 0) to [out = -90, in = 150] (0, -.066) to [out = -30, in = 180] (.115, -.0893) to [out = 0, in = 180] (.93, 0);
  \draw[ultra thick, dashed, red] (-.9, 0) -- (.9, 0);
  \draw(0, 0) node[fill, circle, minimum size = 1.4mm, inner sep = 0]{};
  \draw(1, 0) node[fill, circle, minimum size = 1.4mm, inner sep = 0]{};
  \node[below] at (1, 0) {$1$};
  \node at (-.1, -.145) {$0$};
  \draw[-latex] (-.07, -.105) -- (-.015, -.022);
  \node at (-1, 0) [draw, circle, minimum size = 1.4mm, inner sep = 0]{};
  \node at (0, .208) [blue, fill, star, star points = 4, star point ratio = .2, scale = .7] {};
  \node at (0, -.208) [blue, fill, star, star points = 4, star point ratio = .2, scale = .7] {};
  \node at (0, .066) [green!70!black, fill, star, star points = 3, star point ratio = .2, scale = .7]{};
  \node at (0, -.066) [rotate = 180, green!70!black, fill, star, star points = 3, star point ratio = .2, scale = .7]{};
   \node at (.043, 0) [blue, draw, kite, kite vertex angles = 50, scale = .25] {};
  \node at (-.043, 0) [blue, fill, draw, kite, kite vertex angles = 50, scale = .25] {};
 \end{tikzpicture}
\\
\renewcommand{\arraystretch}{1.5}
 \begin{array}{|r||c|c|c|c|c|c|c|}
  \hline 
  z & 0 &  1 & \infty & 2 & -1 & \frac12 & e^{\pm i\frac{\pi}{3}}
  \\
  \hline 
  q & 0 & 1 & -1 & \pm ie^{-\frac{\pi}{2}} & -e^{-\pi} & e^{-\pi} & \pm ie^{-\frac{\pi}{2}\sqrt{3}}
  \\
  \hline 
  \text{Symbol} & 
  \begin{tikzpicture}
   \draw(0, 0) node[fill, circle, minimum size = 1.4mm, inner sep = 0]{};
  \end{tikzpicture}
 & 
 \begin{tikzpicture}
   \draw(0, 0) node[fill, circle, minimum size = 1.4mm, inner sep = 0]{};
  \end{tikzpicture}
 & 
 \begin{tikzpicture}
   \node at (0, 0) [draw, circle, minimum size = 1.4mm, inner sep = 0]{};
 \end{tikzpicture}
 & 
 \begin{tikzpicture}
   \node at (0, 0) [blue, fill, star, star points = 4, star point ratio = .2, scale = .7] {};
 \end{tikzpicture}
 & 
 \begin{tikzpicture}
  \node at (0, 0) [blue, fill, draw, kite, kite vertex angles = 50, scale = .25] {};
 \end{tikzpicture}
 & 
 \begin{tikzpicture}
  \node at (0, 0) [blue, draw, kite, kite vertex angles = 50, scale = .25] {};
 \end{tikzpicture}
 &  
 \begin{tikzpicture}
  \node at (0, 0) [green!70!black, fill, star, star points = 3, star point ratio = .2, scale = .7]{};
  \node at (.8, 0) [rotate = 180, green!70!black, fill, star, star points = 3, star point ratio = .2, scale = .7]{};
 \end{tikzpicture}
 \\
 \hline 
 \end{array}
 \qquad\qquad
 \label{zqval}
 \end{multline}
In the $z$-plane, the branch cut of the hypergeometric function is taken to be $(1,\infty)$, whose image is the boundary of $q(\mathbb{C})$. The whole complex $z$-plane is mapped into a domain that is much smaller than the unit disc $|q|<1$: as a result, conformal blocks tend to converge faster as power series in $q$ than in $z$.
 
\subsubsection{Zamolodchikov's recursion}\label{sec:zrr}

As a function of the nome, the Virasoro block 
may be rewritten as a prefactor, times a power series $H_\Delta(q)$, which has a trivial asymptotic behaviour:
\begin{align}
 \boxed{\mathcal{F}_\Delta(q) = (16q)^{\Delta-\frac{c-1}{24}} \left[\prod_{k=2,3,4} \left(\theta_k^{-4}(q)\right)^{\Delta_k-\frac{c-1}{24}+(-1)^k \Delta_1} \right] H_\Delta(q)}\ ,
 \label{fdq}
\end{align}
\begin{align}
 \text{where} \quad \boxed{H_\Delta(q) = 1 + \sum_{N=1}^\infty h_N(\Delta)(16q)^N}  \quad , \quad \boxed{\lim_{\Delta\to\infty} H_\Delta(q) = 1}\ .
 \label{hdq}
\end{align}
These formulas mean that 
$\log \mathcal{F}_\Delta(q) \underset{\Delta\to\infty}{=} O(\Delta)$, and that the $O(\Delta)$ and $O(1)$ terms have simple expressions in terms of the nome. 
In Section \ref{sec:bld}, we will sketch why this is the case. For the moment, let us complete the determination of the Virasoro block.
From the poles and residues \eqref{resf} of the Virasoro block $\mathcal{F}_\Delta(z)$, we deduce the poles and residues of the power series $H_\Delta(q)$ \eqref{fdq},
\begin{align}
 \underset{\Delta=\Delta_{(r,s)}}{\operatorname{Res}} H_\Delta(q) = (16q)^{rs} R_{r,s}H_{\Delta_{(r,-s)}}(q)\ .
\end{align}
Since we moreover know its large $\Delta$ asymptotic behaviour \eqref{hdq}, we can rewrite $H_\Delta(q)$ as a sum over its poles,
\begin{align}
 \boxed{H_\Delta(q) = 1 + \sum_{r,s=1}^\infty \frac{(16q)^{rs} R_{r,s}}{\Delta-\Delta_{(r,s)}} H_{\Delta_{(r,-s)}}(q)}\ . 
 \label{hrec}
\end{align}
Together with Eq. \eqref{fdq}, this is called \myindex{Zamolodchikov's recursion}\index{Zamolodchikov's recursion} for Virasoro blocks. (There is another recursion for $\mathcal{F}_\Delta(z)$, obtained by summing over its poles as function of $c$ instead of $\Delta$: to distinguish the 2 recursions, we may call them the $\Delta$-recursion and the $c$-recursion.)

The recursive reprentation completely determines the series $H_\Delta(q)$, order by order in powers of $16q$. Let us write the first 4 coefficients, using $\Delta_{(1,1)}=0$ and Eq. \eqref{drms} for simplifying some expressions: 
\begin{subequations}
\begin{align}
 h_1(\Delta) &= \frac{R_{1,1}}{\Delta}\ , 
 \\
 h_2(\Delta) &= \frac{R_{1,1}^2}{\Delta} + \frac{R_{2,1}}{\Delta-\Delta_{(2,1)}}+\frac{R_{1,2}}{\Delta-\Delta_{(1,2)}} \ , 
 \end{align}
 \begin{multline}
 h_3(\Delta) = \frac{R_{1,1}^3}{\Delta} + \frac{\left(3\Delta-\Delta_{(2,1)}\Delta_{(2,-1)}\right)R_{1,1}R_{2,1}}{\Delta(\Delta-\Delta_{(2,1)})(1-\Delta_{(2,1)})\Delta_{(2,-1)}}  
 \\
 + \frac{\left(3\Delta-\Delta_{(1,2)}\Delta_{(1,-2)}\right)R_{1,1}R_{1,2}}{\Delta(\Delta-\Delta_{(1,2)})(1-\Delta_{(1,2)})\Delta_{(1,-2)}} 
 + \frac{R_{3,1}}{\Delta-\Delta_{(3,1)}} +\frac{R_{1,3}}{\Delta-\Delta_{(1,3)}} 
 \ ,
\end{multline}
\begin{multline}
 h_4(\Delta) = \frac{R_{1,1}^4}{\Delta} + \frac{\left(4\Delta - \Delta_{(2,1)}(\Delta_{(2,-1)}+1)\right)R_{1,1}^2 R_{2,1}}{\Delta(\Delta-\Delta_{(2,1)})(1-\Delta_{(2,1)})\Delta_{(2,-1)}} + \frac{\left(4\Delta-\Delta_{(1,2)}(\Delta_{(1,-2)}+1)\right)R_{1,1}^2R_{1,2}}{\Delta(\Delta-\Delta_{(1,2)})(1-\Delta_{(1,2)})\Delta_{(1,-2)}} 
 \\
 + \frac{R_{2,1}^2}{2(\Delta-\Delta_{(2,1)})} + \frac{\left(4\Delta-(\Delta_{(2,1)}-\Delta_{(1,2)})^2 -2(\Delta_{(2,1)}+\Delta_{(1,2)})\right)R_{2,1}R_{1,2}}{(\Delta-\Delta_{(2,1)})(\Delta-\Delta_{(1,2)})(\Delta_{(2,-1)}-\Delta_{(1,2)})(\Delta_{(1,-2)}-\Delta_{(2,1)})} 
 \\
 + \frac{R_{1,2}^2}{2(\Delta-\Delta_{(1,2)})} + \frac{\left(4\Delta-\Delta_{(3,1)}\Delta_{(3,-1)}\right)R_{1,1}R_{3,1}}{\Delta(\Delta-\Delta_{(3,1)})(1-\Delta_{(3,1)})\Delta_{(3,-1)}}  
 + \frac{\left(4\Delta-\Delta_{(1,3)}\Delta_{(1,-3)}\right)R_{1,1}R_{1,3}}{\Delta(\Delta-\Delta_{(1,3)})(1-\Delta_{(1,3)})\Delta_{(1,-3)}} 
\\
+  \frac{R_{4,1}}{\Delta-\Delta_{(4,1)}}+\frac{R_{2,2}}{\Delta-\Delta_{(2,2)}} +\frac{R_{1,4}}{\Delta-\Delta_{(1,4)}} \ .
\end{multline}
\end{subequations}
In Zamolodchikov's recursion, some properties of the blocks are not manifest. In particular, $h_N(\Delta)$ is a meromorphic function of the central charge, whose poles depend on $\Delta$. However, the recursion writes $h_N(\Delta)$ as a sum of terms that are not always invariant under $\beta\to \beta^{-1}$, and therefore not meromorphic in $c$. Moreover, due to the denominator of $R_{r,s}$ \eqref{rrs}, and also due to factors of the type $\frac{1}{\Delta_{(r,-s)}-\Delta_{(r',s')}}$, these terms have $\Delta$-independent poles at rational values of $\beta^2$. As a result, the recursion is only valid if $\beta^2\notin\mathbb{Q}$. Finding a recursion for $\beta^2\in \mathbb{Q}$ is an open problem, whose simplest incarnation is in the case of 1-point Virasoro blocks on the torus \cite{sto22}. 

\subsubsection{Behaviour for large channel dimensions} \label{sec:bld}

The asymptotic behaviour \eqref{fdq}, \eqref{hdq} of Virasoro blocks in the limit $\Delta\to\infty$ is the most important piece of the derivation of Zamolodchikov's recursion. As far as we know, there is no complete derivation of this behaviour in the literature: in particular, Zamolodchikov's original argument \cite{zam87b} is elliptic. However, there are ideas that could well lead to a derivation: the algebraic idea of \myindex{exponentiation}\index{exponentiation of Virasoro blocks}, which states that $\log\mathcal{F}_\Delta$ grows only linearly in $\Delta$, and the conformal map to the pillow geometry, which explains the relevance of the nome. 

In the expansion \eqref{fsl} of the Virasoro block as powers of $z$, the second-order coefficient \eqref{f2d} behaves as $f_2(\Delta)= O(\Delta^2)$, and higher-order coefficients involve higher powers of $\Delta$. 
The linear growth of $\log\mathcal{F}_\Delta$ is therefore a nontrivial phenomenon, resulting from cancellations of faster-growing contributions to $\mathcal{F}_\Delta$. Virasoro blocks exponentiate also in limits where $c\to\infty$ while $\Delta,\Delta_i\to \infty$, starting from the \myindex{heavy limit}\index{heavy limit} where $\frac{\Delta}{c},\frac{\Delta_i}{c}$ are kept fixed \cite{al24}.

Exponentiation is not easy to deduce from the computation of blocks as sums over Virasoro descendants, which leads to expressions for their coefficients in terms of 2 vectors and 1 matrix, see Eq. \eqref{f2d} for the case of the $O(z^2)$ coefficient. It is better to sum over another basis of the Verma module $\mathcal{V}_\Delta$, namely the \myindex{oscillator basis}\index{oscillator basis} of descendants for the abelian affine Lie algebra $\hat{\mathfrak{u}}_1$, which is related to the Virasoro algebra by Eq. \eqref{lnj}. Since $\hat{\mathfrak{u}}_1$ is abelian, the corresponding matrix is diagonal, and easy to compute. This has led to a proof of exponentiation in the heavy limit \cite{bdk19}, which can be generalized to other limits, including the limit $\Delta\to \infty$ we are interested in. 

Once we know that Virasoro blocks exponentiate, it remains to compute the terms in $O(\Delta)$ and $O(1)$ of $\log \mathcal{F}_\Delta$. According to Eq. \eqref{fdq}, these terms are (relatively) simple functions of the nome. We will now review a geometrical interpretation of these functions \cite{msz15}, which suggests how they could be derived. The idea is to flatten the 4-punctured sphere into a \myindex{pillow}\index{pillow}, i.e. a double cover of a parallelogram of size $\pi\times \pi \tau$ or equivalently a $\mathbb{Z}_2$ quotient of a torus of size $2\pi\times 2\pi\tau$, such that the 4 punctures are mapped to the pillow's corners:
\begin{align}
 \begin{tikzpicture}[scale = .5, baseline=(base)]
 \coordinate (base) at (0, 0);
  \draw [thick, red] plot [domain = 0:-180, smooth] ({4*cos(\x)}, {sin(\x)});
  \draw [thick, red, dashed] plot [domain = 0:180, smooth] ({4*cos(\x)}, {sin(\x)});
  \draw (0, 0) circle [radius = 4];
  \draw (-1.8,-2) node [fill, circle, minimum size = 1.6mm, inner sep = 0]{} node[below left] {$0$};
  \draw (2.8, -1.8) node [fill, circle, minimum size = 1.6mm, inner sep = 0]{} node[below left] {$z$};
  \draw (-1.5, 2.5) node [fill, circle, minimum size = 1.6mm, inner sep = 0]{} node[below left] {$\infty$};
  \draw (2.4, 1.9) node [fill, circle, minimum size = 1.6mm, inner sep = 0]{} node[below left] {$1$};
  \draw (0, -5) node {$x$-sphere $\overline{\mathbb{C}}$};
 \end{tikzpicture}
 \hspace{2cm}
 \begin{tikzpicture}[scale = .5, baseline=(base)]
 \coordinate (base) at (0, 0);
  \draw [thick, red] (-3, 0) to [out = -5, in = -175] (4, 0);
   \draw [thick, red, dashed] (-3, 0) to [out = 5, in = 175] (4, 0);
  \draw (-4, -3) node [below left] {$0$} node [fill, circle, minimum size = 1.6mm, inner sep = 0]{} -- (-2,3) node [above left] {$\pi\tau$} node [fill, circle, minimum size = 1.6mm, inner sep = 0]{} -- (5, 3) node [above right] {$\pi(\tau +1)$} node [fill, circle, minimum size = 1.6mm, inner sep = 0]{} -- (3, -3) node [below right] {$\pi$} node [fill, circle, minimum size = 1.6mm, inner sep = 0]{} -- cycle; 
  \draw (.5, -5) node {$u$-pillow $\left. \frac{\mathbb{C}}{2\pi\mathbb{Z}+2\pi\tau\mathbb{Z}}\right/\left(u\to -u\right)$};
 \end{tikzpicture}
 \label{fig:pillow}
\end{align}
The relation between the coordinate $x$ on the sphere and the coordinate $u$ on the pillow is 
\begin{align}
 u(x) = \frac{1}{{}_2F_1\left(\tfrac12,\tfrac12,1,z\right)} \int_0^x \frac{dx'}{\sqrt{x'(x'-1)(x'-z)}} \ ,
\end{align}
where the $z$-dependent prefactor is such that $u(z)=\pi$, and the square root's sign ambiguity leads to the identification $u\sim -u$. To compute $s$-channel blocks on the pillow, we imagine that time runs upwards, so that any constant time slice has length $2\pi$ (as drawn on Figure \eqref{fig:pillow}). Such a slice scans the whole pillow if time runs for a duration $\pi\tau$. This leads to the factor $e^{i\pi\tau\Delta}=q^\Delta$ in the conformal block, where the $s$-channel dimension $\Delta$ is an eigenvalue of the $s$-channel Hamiltonian. 

The conformal transformation from the sphere to the pillow modifies conformal blocks, which get multiplied by $c,\Delta_i$-dependent factors. These factors must be regularized, because the transformation is singular, and the 4 fields sit at the singularities \cite[Appendix D]{msz15}. In $\mathcal{F}_\Delta(q)$ \eqref{fdq}, these factors appear as exponentials of $\Delta_i$ with $i=1,2,3,4$, coming from the transformation of the 4 fields, and exponentials of $c$, coming from the Weyl anomaly. The regularized pillow Virasoro blocks are therefore
\begin{align}
 \mathcal{F}_\Delta^\text{pillow}(q) = (16q)^{\Delta-\frac{c}{24}} \prod_{m=1}^\infty \left(1-q^{2m}\right)^{-\frac12} H_\Delta(q)\ , 
 \label{fpillow} 
\end{align}
where we used Eq. \eqref{thetaids} for simplifying the $c,\Delta_i$-independent factors. 

Therefore, the derivation of the $\Delta\to\infty$ asymptotics of Virasoro blocks is surely easier on the pillow than on the sphere. The infinite product in Eq. \eqref{fpillow} might naturally emerge from a calculation using the oscillator basis. More fundamentally, it would be interesting to understand why blocks simplify in the pillow geometry, and whether this is the unique geometry with this property. 

\subsection{Numerically solving crossing symmetry}

Assuming we know the spectrum $\mathcal{S}$, the 
crossing symmetry equations \eqref{seteu} form a system of linear equations for the 4-point structure constants $D^{(x)}_k$. There are infinitely many unknowns if the spectrum is infinite. There are also infinitely many equations: 2 equations ($s=t$ and $t=u$) for each value of the cross-ratio $z\in \overline{\mathbb{C}}\backslash\{0,1,\infty\}$. The \myindex{number of solutions}\index{number of solutions} of crossing symmetry is
\begin{align}
 \mathcal{N} = \dim\left\{\left(D^{(x)}_k\right)_{k,x}\middle|\text{crossing symmetry is obeyed}\right\}\in \mathbb{N}\cup \{\infty\}\ . 
 \label{nsol}
\end{align}
For numerical calculations on a computer, we need a system of finitely many equations with finitely many unknowns. Symmetries, including interchiral symmetry, can be used for reducing the number of unknowns, but do not make this number finite. We therefore need to truncate the system, by ignoring fields whose total conformal dimension exceeds a given number $\Lambda$. From the $\Lambda$-dependence of the solutions of the truncated system, we can see whether they converge towards a solution of the original system. And we can also infer the dimension of the space of solutions. 

\subsubsection{Reducing the number of unknowns}

Some 4-point functions $\left<\prod_{i=1}^4 V_{\Delta_i,\bar\Delta_i}\right>$ have discrete $\mathbb{Z}_2$ symmetries that reduce the number of unknowns. We will sketch the ideas here, and refer the reader to \cite[Section 2.4]{nrj23} for technical details:
\begin{itemize}
 \item The \myindex{parity transformation}\index{parity transformation} $z\to \bar z$ exchanges the left- and right-moving Virasoro algebras, and therefore acts on non-diagonal fields as $V_{\Delta,\bar\Delta}\to V_{\bar\Delta,\Delta}$. If our 4-point function is parity-invariant i.e. if $\Delta_i=\bar\Delta_i$, then the space of solutions of crossing symmetry equations decomposes into a parity-even subspace with $D^{(x)}_{\Delta,\bar\Delta}=D^{(x)}_{\bar\Delta,\Delta}$, and a parity-odd subspace with $D^{(x)}_{\Delta,\bar\Delta}=-D^{(x)}_{\bar\Delta,\Delta}$. 
 \item If $(\Delta_1,\bar\Delta_1)=(\Delta_2,\bar\Delta_2)$ in our 4-point function, then 
the permutation $z_1\leftrightarrow z_2$ is a symmetry of the crossing symmetry equations. Again, this allows us to decompose the space of solutions into permutation-even and permutation-odd subspaces. Of course, if the fields $V_{\Delta_1,\bar\Delta_1}$ and $V_{\Delta_2,\bar\Delta_2}$ are actually identical, then only permutation-even solutions are relevant. However, if we allow nontrivial field multiplicities, then these 2 fields may differ while having the same conformal dimensions, and permutation-odd solutions may also be relevant.
Our permutation acts on 4-point structure constants as 
\begin{align}
\left(D^{(s)}_{\Delta,\bar\Delta},D^{(t)}_{\Delta,\bar\Delta},D^{(u)}_{\Delta,\bar\Delta}\right) \mapsto \left((-)^{S}D^{(s)}_{\Delta,\bar\Delta},(-)^{S+S_1+S_3}D^{(u)}_{\Delta,\bar\Delta},(-)^{S+S_1+S_3}D^{(t)}_{\Delta,\bar\Delta}\right)\ ,
\end{align}
so that permutation-even solutions only involve $s$-channel fields with even conformal spins $S\in 2\mathbb{Z}$. 
\end{itemize}

Let us now assume that the degenerate field $V^d_{\langle 1,2\rangle}$ exists. 
As we saw in Section \ref{sec:icb}, this determines how 4-point structure constants $D^{(x)}_{(r,s)}$ behave under $s\to s+2$, and we can rewrite the crossing symmetry equations \eqref{seteu} as 
\begin{align}
 \forall z\in \overline{\mathbb{C}}\backslash\{0,1,\infty\} \ , \quad \sum_{k\in\widetilde{\mathcal{S}}^{(s)}} D_k^{(s)} \widetilde{\mathcal{G}}^{(s)}_{\Delta_k,\bar\Delta_k}(z) = \sum_{k\in\widetilde{\mathcal{S}}^{(t)}} D_k^{(t)} \widetilde{\mathcal{G}}^{(t)}_{\Delta_k,\bar\Delta_k}(z) = \sum_{k\in\widetilde{\mathcal{S}}^{(u)}} D_k^{(u)} \widetilde{\mathcal{G}}^{(u)}_{\Delta_k,\bar\Delta_k}(z)\ ,
\end{align}
where $\widetilde{\mathcal{G}}^{(x)}_{\Delta,\bar\Delta}$ is now an interchiral block \eqref{gtrs}, and $\widetilde{\mathcal{S}}^{(x)}=\left.\mathcal{S}^{(x)}\right|_{-1<s\leq 1}$ is a \myindex{reduced spectrum}\index{spectrum!reduced---}. In loop CFTs, the reduced spectrum is still infinite, but the list \eqref{vex} of non-diagonal fields $V_{(r,s)}$ with integer spins reduces to $2r$ fields for any given value of $r\in\frac12\mathbb{N}^*$:
\begin{subequations}
\label{vexr}
\begin{align}
 & V_{(\frac12, 0)},
 \\
 & V_{(1,0)}, V_{(1,1)}, 
 \\
 & V_{(\frac32, 0)}, V_{(\frac32, \pm\frac23)}, 
 \\
 & V_{(2, 0)}, V_{(2,\pm\frac12)}, V_{(2,1)}.
\end{align}
\end{subequations}
The 4-point function $\left<\prod_{i=1}^4 V_{(r_i,s_i)}\right>$ is parity-invariant if $r_is_i=0$. It is parity-invariant 
modulo interchiral symmetry if $s_i\in\mathbb{Z}$, and permutation-invariant modulo interchiral symmetry if $r_1=r_2$ and $s_1-s_2\in 2\mathbb{Z}$. A $\mathbb{Z}_2$ invariance modulo interchiral symmetry still induces a $\mathbb{Z}_2$ action on the space of solutions, and focussing on an eigenspace of that action still reduces the number of unknowns by roughly half. However, eigenvectors no longer have a simple behaviour under $z\to \bar z$ or $z_1\leftrightarrow z_2$.

\subsubsection{Truncating to a finite system}

Given a spectrum $\mathcal{S}$ and a cutoff $\Lambda \in \mathbb{R}_+$, let us define the \myindex{truncated spectrum}\index{spectrum!truncated---}
\begin{align}
 \mathcal{S}_\Lambda = \left\{ k\in \mathcal{S}\middle| \Re\left(\Delta_k+\bar\Delta_k\right)<\Lambda \right\} \ . 
 \label{sl}
\end{align}
We assume that truncated channel spectra $\mathcal{S}^{(x)}_\Lambda$ are finite, i.e. there are finitely many primary states in each channel. This is necessary for our numerical method to apply, but not for consistency: in Liouville theory, truncated spectra are continuous. Moreover, we need to assume that the terms that we truncate out tend to zero as $\Lambda\to +\infty$. In Liouville theory, it can be shown that $D^{(x)}_\Delta \left|\mathcal{F}^{(x)}_\Delta(q)\right|^2 \underset{\Delta\to+\infty}{\propto} |q|^{2\Delta}$ with $|q|<1$ \cite{rs15}, as a consequence of the behaviour of Virasoro blocks \eqref{fdq} and of structure constants. In loop CFTs, a similar behaviour is observed numerically.

The truncation scheme remains the same if we use interchiral blocks rather than conformal blocks: in each interchiral block, we keep only the conformal blocks whose total conformal dimension obeys the cutoff \eqref{sl}. 

We then have to approximate conformal blocks so that they can be computed in finite time. To do this, we use Zamolodchikov's recursion for Virasoro blocks, and truncate the resulting infinite series. In order for the block-truncating errors to be comparable to the spectrum-truncating errors, we should use the same cutoff $\Lambda$, now applied to descendant states rather than primary states. This means that in the series $H_\Delta(q)$ \eqref{hdq}, we should keep the term of order $q^N$ provided
\begin{align}
 \Re\left(\Delta +\bar\Delta\right) + N + \bar N <\Lambda\ . 
\end{align}
However, it is impractical to perform correlated truncations of the left-moving and right-moving blocks, with truncation parameters that moreover depend on the channel dimensions $\Delta,\bar\Delta$. In practice, the lowest dimension $\min_{k\in\mathcal{S}}\Re\left(\Delta_k+\bar\Delta_k\right)$ is typically not far from zero, and we simply truncate all blocks to order $\Lambda$, i.e. we keep terms such that 
\begin{align}
 N,\bar N < \Lambda \ . 
\end{align}

Finally, now that we have a finite number of unknowns, we can afford to use only finitely many equations. One possibility is to expand the crossing symmetry equations near a point $z_0$, and to keep finitely many terms in the expansion. In order to solve the bootstrap equations that involve only the $s$-channel and $t$-channel decompositions, it is natural to use $z_0=\frac12$, as this is the finite fixed point of $z\mapsto 1-z$. In order to solve our 3-channel bootstrap equations, $z_0 = e^{i\frac{\pi}{3}}$ would be more natural, see Table \eqref{S3}. Another possibility is to refrain from expanding around a point, and to use a number of different values of $z$ instead. These values should be chosen reasonably far from one another, from the singularities $z=0,1,\infty$, and also from the real line $z\in\mathbb{R}$, where the conformal blocks $\mathcal{G}^{(x)}_{\Delta,\bar\Delta}$ and $\mathcal{G}^{(x)}_{\bar\Delta,\Delta}$ coincide. Under these conditions, we can draw these values randomly, resulting in a finite set $Z_\Lambda$.

After these truncations and approximations, the crossing symmetry equations \eqref{seteu} reduce to a finite system of the type 
\begin{align}
 \forall z\in Z_\Lambda\ , \quad 
 \sum_{k\in\mathcal{S}^{(s)}_\Lambda} D_k^{(s)} \mathcal{G}^{(s)}_{\Lambda|\Delta_k,\bar\Delta_k}(z) 
 = \sum_{k\in\mathcal{S}^{(t)}_\Lambda} D_k^{(t)} \mathcal{G}^{(t)}_{\Lambda|\Delta_k,\bar\Delta_k}(z) 
 = \sum_{k\in\mathcal{S}^{(u)}_\Lambda} D_k^{(u)} \mathcal{G}^{(u)}_{\Lambda|\Delta_k,\bar\Delta_k}(z)\ .
 \label{fziz}
\end{align}
The number of unknowns is $\sum_{x\in\{s,t,u\}}\left|\mathcal{S}^{(x)}_\Lambda\right|$, and the number of equations is $2|Z_\Lambda|+1$, if we add a normalization condition such as $D_{k_0}^{(s)}=1$. We choose $Z_\Lambda$ such that these two numbers coincide. Then there is a unique solution $\left(D_k^{(x)|\Lambda,Z_\Lambda}\right)_{k,x}$.

\subsubsection{Studying one solution}\label{sec:sos}

In the limit $\Lambda\to \infty$, the solutions $\left(D_k^{(x)|\Lambda,Z_\Lambda}\right)_{k,x}$ of the truncated system converge towards an exact solution $\left(D_k^{(x)}\right)_{k,x}$ of crossing symmetry provided 
\begin{align}
 \forall k,x,Z_\Lambda\ , \quad \lim_{\Lambda\to \infty} D_k^{(x)|\Lambda,Z_\Lambda} = D_k^{(x)}\ .
\end{align}
However, it is computationally expensive to use large values of the cutoff $\Lambda$ in order to test convergence. In practice, we instead diagnose convergence by studying how strongly $D_k^{(x)|\Lambda,Z_\Lambda}$ depend on $Z_\Lambda$. To do this quantitatively, we introduce the \myindex{deviation}\index{deviation} of a structure constant for 2 choices $Z_\Lambda^{(1)},Z_\Lambda^{(2)}$ of sets of points:
\begin{align}
 \epsilon_k^{(x)|\Lambda} = \left| 1- \frac{D_k^{(x)|\Lambda,Z^{(1)}_\Lambda}}{D_k^{(x)|\Lambda,Z^{(2)}_\Lambda}}\right|\ .
\end{align}
While the deviation depends on $Z_\Lambda^{(1)},Z_\Lambda^{(2)}$, we are only interested in its order of magnitude, which typically does not, so we omit this dependence in the notation $\epsilon_k^{(x)|\Lambda}$. In order to minimize the computational cost, we can choose two sets $Z_\Lambda^{(1)},Z_\Lambda^{(2)}$ that differ only by a few points.

When computing deviations, we encounter the following situations, depending on the number $\mathcal{N}$ \eqref{nsol} of solutions of crossing symmetry:
\begin{align}
\renewcommand{\arraystretch}{1.5}
 \begin{tabular}{|l|l|}
 \hline 
  Situation & Deviation
  \\
  \hline \hline 
  $\mathcal{N}\neq 1$ & $\log \epsilon_k^{(x)|\Lambda} \sim -2$
  \\
  \hline 
  $\mathcal{N}=1$, $D_k^{(x)}\neq 0$ & $\log \epsilon_k^{(x)|\Lambda} \sim \Re(\Delta_k+\bar\Delta_k) - \Lambda$
  \\
  \hline 
  $\mathcal{N}=1$, $D_k^{(x)}=0$ & $\log \epsilon_k^{(x)|\Lambda} \sim 0$
  \\
  \hline 
 \end{tabular}
 \label{les}
\end{align}
If there is no solution or if there are several solutions, deviations remain quite large, irrespective of $\Lambda$. A unique solution leads to small deviations, but only for structure constants that are nonzero.

For example, consider the 4-point function $\left<V_{(\frac32,\frac23)}V_{(\frac12,0)}V_{(1,0)}V_{P_{(0,\frac{4+i}{10})}}\right>$ in a loop CFT at $\beta = \frac{5}{4+i}$. Taking interchiral symmetry into account, we write the spectra
\begin{align}
 \widetilde{S}^{(s)} = \left\{V_{P_{(0,\frac{1+i}{10})}}\right\}\cup \left\{V_{(r,s)}\right\}_{\substack{r\in\mathbb{N}^*\\ s\in \frac{1}{r}\mathbb{Z} \\ -1<s\leq 1 }} \quad ,\quad \widetilde{S}^{(t)}=\widetilde{S}^{(u)} = \left\{V_{(r,s)}\right\}_{\substack{r\in \frac52+\mathbb{N}\\ s\in \frac{1}{r}\mathbb{Z} \\ -1<s\leq 1 }}\ ,
\end{align}
and impose the normalization condition $D^{(s)}_{P_{(0,\frac{1+i}{10})}}=1$. The resulting crossing symmetry equations have a unique solution: we will justify this in Section \ref{sec:cf}, where that solution is characterized by the combinatorial map 
$
\begin{tikzpicture}[baseline=(base), scale = .25]
   \uvertex 
   \draw (0, 0) to [out = 70, in = 0] (0, 3.5) to [out = 180, in = 110] (0, 0);
   \draw (0, 0) -- (0, 3);
   \draw (3, 3) to [out = -70, in = 0] (3, -.5) to [out = 180, in = -110] (3, 3);
  \end{tikzpicture}
$
and the signature $\sigma = (0,\frac52,\frac52)$. We will now display numerical data, obtained with the following numerical parameters:
\begin{align}
\renewcommand{\arraystretch}{1.3}
 \begin{array}{|r|c|c|}
  \hline 
  \Lambda & 30 & 70 
  \\
  \hline 
  \#\text{digits} & 24 & 56 
  \\
  \hline
  \text{running time} & \sim 200s & \sim 4200s
  \\
  \hline 
 \end{array}
\end{align}
Times are indicated for Python code \cite{rn23} running on an old desktop computer. For each value of $\Lambda$, we display the real parts of $s$- and $t$-channel structure constants together with their deviations, for all $r\leq \frac72$ as well as $(r,s)\in\left\{(4,0),(4,\frac14),(\frac92,0),(\frac92,\frac29)\right\}$. To save space, we show only about $15$ of the $24$ or $56$ digits:
\begin{align}
\renewcommand{\arraystretch}{1.1}
 \begin{array}{|l||r|r||r|r|}
  \hline 
  (r, s) & \epsilon_{(r,s)}^{(s)|30}  & \Re D_{(r,s)}^{(s)|30} & \epsilon_{(r,s)}^{(s)|70}& \Re D_{(r,s)}^{(s)|70}
  \\
  \hline 
(1,0) &    10^{-18} &  \underline{0.82242208337639668}719 & 10^{-45} &  0.82242208337639669048 \\
(1,1) &    10^{-17} &  -\underline{1.45489355316066448}03 &  10^{-45} &  -1.4548935531606644841 \\
(2,0) &    10^{-13} &  -\underline{0.0005823102081532}698 &  10^{-40} &  -0.0005823102081533416 \\
(2,\frac12) &    1.4 &  -8.9601788211935\times 10^{-17} &   1.8 & -8.924630755397\times 10^{-44} \\
(2,1) &    10^{-13} &  \underline{0.000592788853931}72 & 10^{-40} &  0.00059278885393164\\
(2,-\frac12) &    1.3 &  4.7020237099170\times 10^{-17} &  1.7 & -2.961377839207\times 10^{-44} \\
(3,0) &    10^{-6} &  -\underline{5.6619}023163079\times 10^{-11} &  10^{-32} &  -5.6619788484916 \times 10^{-11}\\
(3,\frac13) &    1.5 &  6.8733214151865\times 10^{-15} &  2.4  & 1.3028663659915\times 10^{-41} \\
(3,\frac23) &    1.2 &  4.2685780225150\times 10^{-15} &  2.4  & -1.45320554160523\times 10^{-43} \\
(3,1) &    10^{-6} &  \underline{1.47299}16053293\times 10^{-9} &  10^{-33} &  1.4730032519748\times 10^{-9} \\
(3,-\frac13) &    1 &  -1.9732803811763\times 10^{-15} &  1  & -2.1375702240518\times 10^{-41} \\
(3,-\frac23) &    1 &  -8.3839892985966\times 10^{-15} & 0.9 &  -6.7477400595850\times 10^{-42} \\
(4,0) &    0.6 &  4.9465341395038\times 10^{-10} &  10^{-19} &  -2.5225903605711\times 10^{-17} \\
(4,\frac14) &    0.6 &  -3.7885783283454\times 10^{-9} &  1.4 &  -2.95230287596625\times 10^{-35} \\
\hline \hline 
 (r, s) & \epsilon_{(r,s)}^{(t)|30}  & \Re D_{(r,s)}^{(t)|30} & \epsilon_{(r,s)}^{(t)|70}  & \Re D_{(r,s)}^{(t)|70} 
\\
\hline 
(\frac52,0) &    10^{-14} &  -\underline{0.0028701010781127}2060 &  10^{-43} &  -0.00287010107811285690 \\
(\frac52,\frac25) &    10^{-14} &  -\underline{0.0040384970890702}9441 &  10^{-43} &  -0.00403849708907038200 \\
(\frac52,\frac45) &    10^{-14} &  \underline{0.001531792632911}45239 &  10^{-42} &  0.00153179263291161481 \\
(\frac52,-\frac25) &    10^{-14} &  \underline{0.003524120330565}7769 &  10^{-43} &  0.0035241203305659808 \\
(\frac52,-\frac45) &    10^{-14} & \underline{0.0069570147283432}9452 &  10^{-43} & 0.006957014728343194327917 \\
(\frac72,0) &    10^{-8} &  \underline{0.000004444583}14475906 &  10^{-35} &  0.00000444458367504768 \\
(\frac72,\frac27) &    10^{-8} &  \underline{0.000004121521}383854464 &   10^{-35} &  0.000004121520804451911 \\
(\frac72,\frac47) &    10^{-8} &  \underline{0.000003513910}538266622 &  10^{-35} &  0.000003513910885535083 \\
(\frac72,\frac67) &    10^{-8} &  \underline{0.0000028787942}66748157 &  10^{-35} &  0.000002878794192189494 \\
(\frac72,-\frac27) &    10^{-8} &  \underline{0.000004029073}578395895 &   10^{-35} &  0.000004029073277638004 \\
(\frac72,-\frac47) &    10^{-8} &  \underline{0.0000029308939}45610566 &   10^{-35} &  0.000002930894076496366 \\
(\frac72,-\frac67) &    10^{-8} &  \underline{0.0000017184668}36222695 &  10^{-35} &  0.000001718466769985650\\
(\frac92,0) &    0.5 &  2.2616940515940\times 10^{-8} &  10^{-23} &  3.230122853228 \times 10^{-12}\\
(\frac92,\frac29) &    0.5  & -3.5998726869265\times 10^{-8} &  10^{-23} &  3.298879973039\times 10^{-12} \\
\hline 
\end{array}
\end{align}
Let us interpret these results:
\begin{itemize}
 \item
 Whenever a deviation is small, which in the case $\Lambda=30$ means $\leq 10^{-6}$, we expect that we obtain an approximation of the corresponding structure constant. For example, $\epsilon_{(\frac72,\frac27)}^{(t)|30}\sim 10^{-8}$ suggests that $D_{(\frac72,-\frac27)}^{(t)|30}$ gives about $8$ correct digits of the exact solution $D_{(\frac72,-\frac27)}^{(t)}$. This is confirmed by comparing with $D_{(\frac72,-\frac27)}^{(t)|70}$, which has the same $7$ leading nonzero digits. In all $\Lambda = 30$ results, we have underlined the digits that agree with $\Lambda = 70$ results: the number of correct digits is always accurately estimated by the deviation.
 \item
 When a deviation is $O(1)$, we have either a vanishing structure constant, or a non-vanishing structure constant that can only be estimated with a larger cutoff.
 For example, the value $\epsilon^{(s)|30}_{(3,\frac13)}=1.5$ indicates that $D^{(s)}_{(3,\frac13)}=0$: this can be seen by comparing with the result $\epsilon^{(s)|30}_{(3,0)}=10^{-6}$
for a field with a similar total conformal dimension, which should have a similar deviation if $D^{(s)}_{(3,\frac13)}\neq 0$,  according to Table \eqref{les}. This is confirmed by the $\Lambda = 70$ results $\epsilon_{(3,\frac13)}^{(s)|70} = O(1)$ and $D_{(3,\frac13)}^{(s)|70} \ll D_{(3,\frac13)}^{(s)|30}$.
 Similarly, the $\Lambda = 30$ results are enough for concluding $D^{(s)}_{(2,\pm\frac12)} =D^{(s)}_{(3,\pm \frac13)} = D^{(s)}_{(3,\pm \frac23)}=0$.
 \item
 In the cases $\epsilon^{(s)|30}_{(4,\star )},\epsilon^{(t)|30}_{(\frac92,\star)} = O(1)$, the $\Lambda =30$ results are not enough for knowing whether the corresponding structure constants vanish. The $\Lambda =70$ results show that $D^{(s)}_{(4,s)} = 0 \iff s \notin \mathbb{Z}$ and $D^{(t)}_{(\frac92,\star)} \neq 0$. Based on all these results, it is natural to conjecture $\forall r,s,\ D^{(s)}_{(r,s)} = 0 \iff s \notin \mathbb{Z}$.
\end{itemize}

\subsubsection{Counting solutions}

We would like to determine the dimension $\mathcal{N}$ \eqref{nsol} of the space of solutions, and to find a basis of solutions. 

If we view the system of crossing symmetry equations as an infinite matrix $\mathcal{C}$, then $\mathcal{N}$ is the number of singular values of $\mathcal{C}$ that vanish. Can we deduce $\mathcal{N}$ from the matrix $\mathcal{C}_\Lambda$ of a truncated system? The truncation has 2 effects on the set of singular values: making the set finite, and making the vanishing elements nonzero but tiny. Therefore, if $\mathcal{N}<\infty$, then $\mathcal{N}$ is the number of \textit{tiny} singular values of $\mathcal{C}_\Lambda$ for $\Lambda$ \textit{large enough}. But large matrices tend to have many small singular values, which tend to zero as $\Lambda\to \infty$. The tiny singular values tend to zero faster than the others, but we do not have a quantitative criterion for characterizing them. (See \cite{gjnrs23} for examples.) 

In practice, instead of studying singular values, we count solutions by a rather pedestrian method, which does not require high values of $\Lambda$. To begin with, let us determine whether $\mathcal{N}>0$. For a solution $\left(D_k^{(x)|\Lambda,Z_\Lambda}\right)_{k,x}$ of the truncated system \eqref{fziz}, let $Z^{(x)}_\Lambda(z)$ be the corresponding $x$-channel 4-point function. For $z_0\notin Z_\Lambda$, we compute the violation of crossing symmetry 
\begin{align}
 \delta_\Lambda(z_0) = \max_{x\in\{t,u\}} \left|Z^{(x)}_\Lambda(z_0)-Z^{(s)}_\Lambda(z_0)\right|\ . 
\end{align}
We expect this violation to be small if $\mathcal{N}>0$, and larger if $\mathcal{N}=0$. To state this  quantitatively, we introduce the quantity 
\begin{align}
 \eta_\Lambda = \max_{k\in \mathcal{S}_\Lambda^{(s)}} \left|\epsilon_k^{(s)|\Lambda} D_k^{(s)|\Lambda,Z_\Lambda}\mathcal{G}^{(s)}_{\Lambda|\Delta_k\bar\Delta_k}(z_0)\right| \ .
\end{align}
Then we expect
\begin{itemize}
 \item $\mathcal{N}=0\implies \delta_\Lambda(z_0) = O(\eta_\Lambda)$: the large deviation $\epsilon_k^{(s)|\Lambda}$ measures the failure of crossing symmetry.
 \item $\mathcal{N}>1\implies \delta_\Lambda(z_0) \ll \eta_\Lambda$: the large deviation reflects the existence of several solutions, and the 3 channels should agree up to a small discrepancy that is due to our truncation of the system. 
\end{itemize}
Once we know that $\mathcal{N}>1$, the idea is to add $\mathcal{N}$ equations to the crossing symmetry equations. Then the solution becomes unique, leading to small deviations as in Section \ref{sec:sos}. Without loss of generality, we may use equations of the type $D^{(x)}_k=0$, in addition to the normalization condition $D^{(s)}_k=1$. This is equivalent to removing primary fields from the spectrum, until we have small deviations. When this is achieved, we have $\mathcal{N}\leq \mathcal{R}+1$, where $\mathcal{R}$ is the number of removed fields. We have an inequality rather than an equality, because it may happen that some of our additional equations are in fact obeyed by all solutions, and are therefore redundant with crossing symmetry equations. To eliminate this possibility, we should test whether removing all $\mathcal{R}$ primary fields is necessary for having small deviations. If the set of removed fields is minimal, then $\mathcal{N}= \mathcal{R}+1$.

The method of removing primary fields is computationally cheap, because it only involves deleting columns of the matrix $\mathcal{C}_\Lambda$, without recomputing $\mathcal{C}_\Lambda$, or changing the cutoff $\Lambda$. This method is however pedestrian, because it involves inspecting deviations for various sets of removed fields, and determining a minimal set by trial and error. It would be nice to find a simple way of determining $\mathcal{N}$ from $\mathcal{C}_\Lambda$.

The method of removing primary fields can also work if $\mathcal{N}=\infty$. For example, in the case of 4-point connectivities in the Potts CFT, using only the first equation in \eqref{seteu} (i.e. leaving the $u$-channel unconstrained) leads to an infinite-dimensional space of solutions, instead of the finite-dimensional spaces that otherwise occur in loop CFTs. Nevertheless, it is possible to determine a basis of solutions, where each element corresponds to a set of removed fields that is infinite before truncation \cite{nr20}.

\section{Correlation functions in loop CFTs} \label{sec:cloop}

\subsection{Statistical sums over loop ensembles}

Let us sketch the construction of correlation functions as statistical sums over loop ensembles. We will focus on features that survive in the critical limit, and are therefore relevant to loop CFTs. In particular, we will use combinatorial maps for describing combinatorial features of loop configurations. In loop CFTs, it turns out that combinatorial maps parametrize solutions of crossing symmetry.

\subsubsection{Sums over configurations of non-intersecting loops}

Loop models have correlation functions of the type
\begin{align}
 Z(\mathcal{E}) = \sum_{E\in\mathcal{E}} W(E)\ , 
\end{align}
where $E$ is a set of non-intersecting closed loops on the Riemann sphere $\overline{\mathbb{C}}$, belonging to an ensemble $\mathcal{E}$ of such sets, and $W$ is a weight function.

If we discretize $\overline{\mathbb{C}}$ and replace it with a finite lattice, the ensemble $\mathcal{E}$ becomes finite, and each configuration $E$ is made of finitely many loops: $|\mathcal{E}|, |E|<\infty$. Then the sum $Z(\mathcal{E})$ is manifestly finite, and can be numerically evaluated on a computer. The price to pay for the lattice discretization is that conformal symmetry is broken, and reemerges only in a large lattice size limit --- provided $Z(\mathcal{E})$ does have a limit. 

An alternative is the mathematical notion of a conformal loop ensemble, which is an infinite ensemble $\mathcal{E}$ of configurations of infinitely many loops: $|\mathcal{E}|=|E|=\infty$. This ensemble is conformally invariant by construction. 

We will not need a precise definition of a loop ensemble: we only assume $|E|<\infty$ for technical simplicity. We assume that the energy $\log W(E)$ is extensive: then the weight of a configuration is a product of weights of individual loops $w(\ell)$. On the sphere, all loops are topologically the same, and must have the same weight by conformal symmetry. Therefore,
\begin{align}
 W(E) = \prod_{\ell \in E } w(\ell) = n^{|E|}\ ,
 \label{woe}
\end{align}
where $n$ is the \myindex{contractible loop weight}\index{loop weight!contractible---}.

\subsubsection{Legless punctures}

Our correlation function $Z(\mathcal{E})$ is the analog of a 0-point function in CFT. In order to define the analog of an $N$-point function,
we introduce $N$ \myindex{legless punctures}\index{legless puncture} on our sphere with positions $z_1,\dots, z_N$, which will modify the weights of loops.
Without violating conformal symmetry, we can let the weight of a loop depend on how the loop splits the set $\{z_1,\dots, z_N\}$ into 2 subsets, i.e. on the loop's combinatorial properties. It cannot however depend on the loop's topology in $\overline{\mathbb{C}}\backslash \{z_1,\dots, z_N\}$, because we want $Z(\mathcal{E})$ to be a single-valued function of $z_1,\dots, z_N$. For example, in the case $N=4$, the following 2 loops are topologically different but combinatorially equivalent, and must therefore have the same weight: 
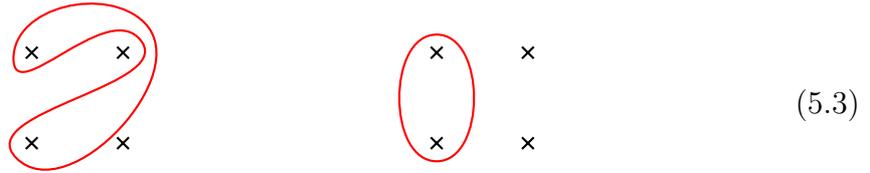
\begin{align}
\begin{tikzpicture}[baseline=(base), scale = .4]
\coordinate (base) at (0, 1);
 \draw (0, 0) node[cross]{};
  \draw (3, 0) node[cross]{};
  \draw (0, 3) node[cross]{};
  \draw (3, 3) node[cross]{};
 \draw[red, thick] (-.5, -.5) to [out = -45, in = -45] (3.7, 4) to [out = 135, in = 90] (-.6, 2.8) to [out = -90, in = 135] (3.5, 3.5) to [out = -45, in = 135] (-.5, -.5);
\end{tikzpicture}
\qquad \qquad \qquad 
\begin{tikzpicture}[baseline=(base), scale = .4]
\coordinate (base) at (0, 1);
 \draw (0, 0) node[cross]{};
  \draw (3, 0) node[cross]{};
  \draw (0, 3) node[cross]{};
  \draw (3, 3) node[cross]{};
 \draw[red, thick] (0, -.6) to [out = 0, in = 0] (0, 3.6) to [out = 180, in = 180] (0, -.6);
\end{tikzpicture}
\label{tvsc}
\end{align}
This leads to 8 combinatorially inequivalent classes of loops, with 8 different weights: contractible loops with weight $n$, $i$-loops around the puncture at $z_i$ with weight $w_i$, and $s$-loops, $t$-loops, $u$-loops that split the 4 punctures in pairs:
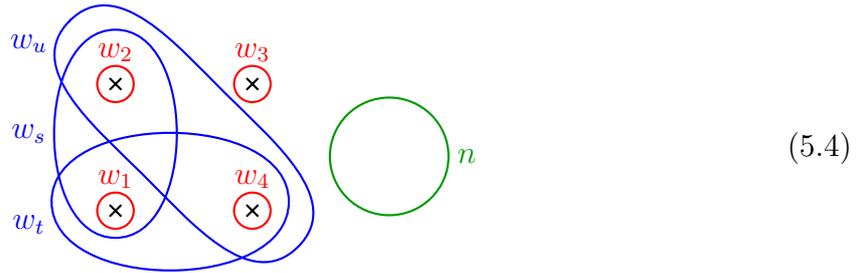
\begin{align}
 \begin{tikzpicture}[baseline=(base), scale = .6]
\coordinate (base) at (0, 1);
 \draw (0, -.2) node[cross]{};
  \draw (3, -.2) node[cross]{};
  \draw (0, 2.6) node[cross]{};
  \draw (3, 2.6) node[cross]{};
  \draw[red, thick] (0, -.2) circle [radius = .4]; 
   \draw[red, thick] (3, -.2) circle [radius = .4]; 
    \draw[red, thick] (0, 2.6) circle [radius = .4]; 
   \draw[red, thick] (3, 2.6) circle [radius = .4]; 
   \draw[blue, thick] (0, -.8) to [out = 0, in = 0] (0, 3.8) to [out = 180, in = 180] (0, -.8);
   \draw[blue, thick] (-1.4, 0) to [out = 90, in = 90] (3.8, 0) to [out = -90, in = -90] cycle;
   \draw[blue, thick] (-1, 4) to [out = 45, in = 135] (2.4, 2.4) to [out = -45, in = 45] (4, -1) to
   [out = -135, in = -45] (.6, .6) to [out = 135, in = -135] cycle;
   \draw[thick, green!80!black] (6, 1) circle [radius = 1.3];
   \node[blue] at (-1.9, 1.5) {$w_s$};
   \node[blue] at (-1.9, -.5) {$w_t$};
   \node[blue] at (-1.9, 3.5) {$w_u$};
   \node[green!80!black] at (7.7, 1) {$n$};
   \node[red] at (0, .5) {$w_1$};
   \node[red] at (0, 3.3) {$w_2$};
    \node[red] at (3, .5) {$w_4$};
   \node[red] at (3, 3.3) {$w_3$};
  \end{tikzpicture}
\end{align}
Since loops do not intersect, a configuration $E$ that contains $s$-loops has no $t$-loops or $u$-loops. 

\subsubsection{Legged punctures and combinatorial maps}

We also introduce \myindex{legged punctures}\index{legged puncture} (sometimes called watermelon operators), where open loops can end. Such a puncture has a valency $2r\in\mathbb{N}^*$, with $r=0$ corresponding to the previously introduced legless punctures. In the presence of $N$ punctures, the number of open loops is
\begin{align}
 \sum_{i=1}^N r_i \in \mathbb{N} \ . 
 \label{srin}
\end{align}
Conformal symmetry allows us to specify how punctures are connected by open loops: this information is called a \myindex{combinatorial map}\index{combinatorial map} \cite{gjnrs23}. A combinatorial map is a graph whose vertices correspond to punctures, and whose edges correspond to open loops. For example, there exist 4 combinatorial maps with vertices of valencies $3,2,1,0$:
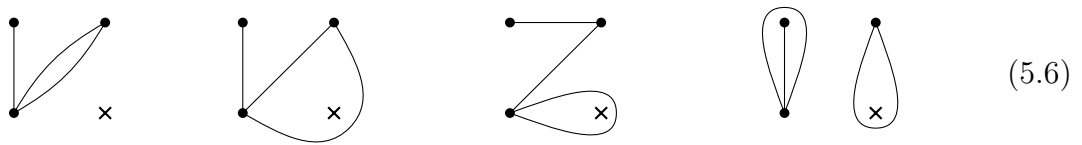
\begin{align}
  \begin{tikzpicture}[baseline=(base), scale = .4]
   \uvertex 
   \draw (0, 0) -- (0, 3);
   \draw (0, 0) to [out = 60, in = -150] (3, 3) to [out = -120, in = 30] (0, 0);
  \end{tikzpicture}
\qquad \qquad 
\begin{tikzpicture}[baseline=(base), scale = .4]
   \uvertex 
   \draw (0, 3) -- (0, 0) -- (3, 3);
   \draw (0, 0) to [out = -30, in = -135] (3.5, -0.5) to [out = 45, in = -60] (3, 3);
  \end{tikzpicture}
  \qquad  \qquad 
\begin{tikzpicture}[baseline=(base), scale = .4]
   \uvertex 
   \draw (0, 3) -- (3, 3) -- (0, 0);
   \draw (0, 0) to [out = 20, in = 90] (3.5, 0) to [out = -90, in = -20] (0, 0);
  \end{tikzpicture}
  \qquad  \qquad 
\begin{tikzpicture}[baseline=(base), scale = .4]
   \uvertex 
   \draw (0, 0) to [out = 70, in = 0] (0, 3.5) to [out = 180, in = 110] (0, 0);
   \draw (0, 0) -- (0, 3);
   \draw (3, 3) to [out = -70, in = 0] (3, -.5) to [out = 180, in = -110] (3, 3);
  \end{tikzpicture}
  \label{4maps}
\end{align}
Notice that we forbid edges that can be pulled inside vertices, as the corresponding open loops are indistinguishable from contractible loops. This is the reason why we do not count the following map:
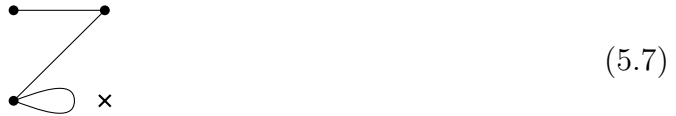
\begin{align}
 \begin{tikzpicture}[baseline=(base), scale = .4]
   \uvertex 
   \draw (0, 3) -- (3, 3) -- (0, 0);
   \draw (0, 0) to [out = 20, in = 90] (2, 0) to [out = -90, in = -20] (0, 0);
  \end{tikzpicture}
\end{align}
To a combinatorial map $M$, we associate the ensemble $\mathcal{E}_M$ of configurations of non-intersecting open and closed loops, such that the open loops connect the punctures according to $M$. The map $M$ dictates which types of closed loops can exist: for example, the last map in Eq. \eqref{4maps} allows closed loops with weights $n,w_4,w_s$. Conformal symmetry allows the weight of a loop configuration to depend on the angles of incidence of open loops at punctures. 
More precisely, at any puncture we have a weight factor that depends on an angular momentum $s$,
\begin{align}
 \boxed{w_{(r,s)}(E) = \exp \left\{\tfrac{i}{2} s\textstyle{\sum}_{k=1}^{2r}\theta_{k}\right\}}\ .
 \label{wrs}
\end{align}
where the angles are measured with respect to a reference direction, and a reference open loop whose angle is $\theta_1$:
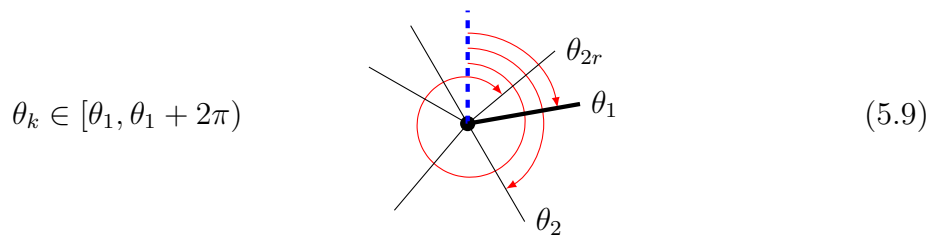
\begin{align}
 \theta_{k}\in [\theta_{1}, \theta_{1}+2\pi)
\qquad \qquad 
 \begin{tikzpicture}[baseline=(base), scale = .5]
  \coordinate (base) at (0, 0);
  \draw [domain=0:410,variable=\t,smooth,samples=35,red,-latex]
        plot ({90-\t}: {1.6- \t/1000});
  \draw[-latex, red] (0, 2.4) arc (90:10:2.4);
  \draw[-latex, red] (0, 2) arc (90:-60:2);
  \draw (0, 0) node[fill, circle, minimum size = 2mm, inner sep = 0]{};
  \foreach\t in {10, 40, 120, 150, 230, 300}{
  \draw (0, 0) -- ++(\t:3);
  }
  \draw[ultra thick, blue, dashed] (0, 0) -- ++(90:3);
  \draw[ultra thick] (0, 0) -- ++(10:3);
  \node[right] at ++(10:3) {$\theta_{1}$};
  \node[right] at ++(300:3) {$\theta_{2}$};
  \node[right] at ++(40:3) {$\theta_{2r}$};
 \end{tikzpicture}
 \label{titt}
\end{align}
The reference direction would define each angle modulo $2\pi$. Thanks to the reference open loop and the cyclic ordering of open loops around the puncture, our angles are actually defined modulo a collective shift $(\theta_k) \to (\theta_k+2\pi)$. This shift leaves our weight factor invariant provided 
\begin{align}
 rs\in \mathbb{Z}\ . 
\end{align}
Under a change of reference direction or reference open loop, the weight of a configuration $E$ changes by an $E$-independent factor.

We therefore define correlation functions that depend on the punctures $z_1,\dots,z_N$ with their parameters $(r_1,s_1),\dots, (r_N,s_N)$, on the map $M$ and on the weights of closed loops $\{w_k\}$:
\begin{align}
 \boxed{Z\Big(\{z_i\},\{(r_i,s_i)\},\{w_k\}\Big|\mathcal{E}_M\Big) =\sum_{E\in\mathcal{E}_M} \prod_{i=1}^N w_{(r_i,s_i)}(E) \prod_{\substack{\ell\in E\\ \ell \text{ closed}}} w(\ell)} \ .
 \label{zbig}
\end{align}

\subsubsection{Cases of the $O(n)$, $PSU(n)$ and Potts models}\label{sec:coup}

In the $O(n)$ model, all possible legged punctures are allowed. But there are no legless punctures: all closed loops have the same weight $n$, whether or not they are contractible. The loop model is obtained by reformulating a model of spins on a lattice, where each spin is an $n$-dimensional vector that transforms in the fundamental representation of the group $O(n)$. 

In the $PSU(n)$ model, we add the restriction that all loops (closed and open) can be oriented such that any 2 neighbouring loops have opposite orientations. Equivalently, the Riemann sphere should be bicolorable, with each loop separating 2 regions of different colors. This allows only punctures with even valencies $2r\in 2\mathbb{N}^*$:
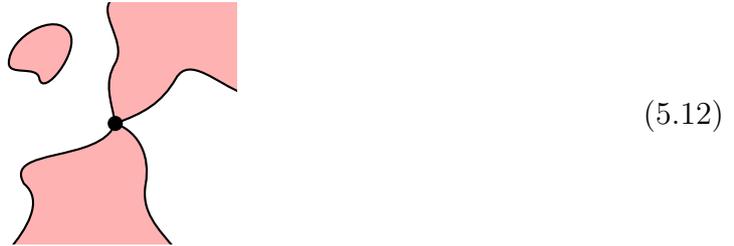
\begin{align}
 \begin{tikzpicture}[baseline=(current  bounding  box.center), scale = .4]
 \clip (-4, -4) rectangle (4, 4);
  \filldraw[thick, fill = red!30] (0, 0) to [out = 100, in = -120] (0, 2) to [out = 60, in = -150] (0, 4.2) -- (4.2, 4.2) -- (4.2, 1) to [out = 160, in = 60] (2, 1.5) to [out = -120, in = 20] (0, 0);
  \filldraw[thick, fill = red!30] (0, 0) to [out = -20, in = 80] (1, -2) to [out = -100, in = 130] (2, -4.2) -- (-3.5, -4.2) to [out = 50, in = -40] (-3, -2) to [out = 120, in = -110] (0, 0);
  \filldraw[thick, fill = red!30] (-3.5, 2) to [out = 90, in = 120] (-1.5, 3) to [out = -60, in = -80] (-2.5, 1.5) to [out = 100, in = -90] (-3.5, 2);
   \draw (0, 0) node[fill, circle, minimum size = 2mm, inner sep = 0]{};
 \end{tikzpicture}
\end{align}
Moreover, we introduce legless punctures that detect the color of the region they belong to. Since that color does not change if we add 2 closed loops around the puncture, the puncture's weight must obey $w^2=n^2$, therefore $w=-n$. 

The Potts model is the same as the $PSU(n)$ model, except that the fundamental geometrical objects are clusters instead of loops, with loops forming the boundaries of clusters. In general, loops and clusters are equivalent: in the above example, a puncture of valency 4 is a point where 2 large clusters meet, where a cluster is large if it cannot be pulled inside the puncture. In the following example, an open loop can be pulled inside the puncture, and we have a squeezed cluster rather than the meeting of 2 different clusters:
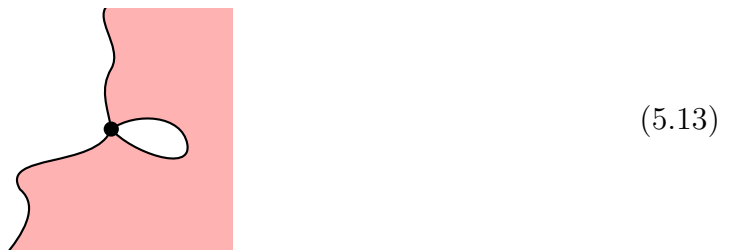
\begin{align}
 \begin{tikzpicture}[baseline=(current  bounding  box.center), scale = .4]
  \clip (-4, -4) rectangle (4, 4);
  \filldraw[thick, fill = red!30] (0, 0) to [out = 100, in = -120] (0, 2) to [out = 60, in = -150] (0, 4.2) -- (4.2, 4.2) -- (4.2, -4.2) -- (-3.5, -4.2) to [out = 50, in = -40] (-3, -2) to [out = 120, in = -110] (0, 0) to [out = -50, in = -80] (2.5, -.5) to [out = 100, in = 35] (0, 0);
   \draw (0, 0) node[fill, circle, minimum size = 2mm, inner sep = 0]{};
 \end{tikzpicture}
\end{align}
However, there is a subtle difference between loops and clusters in the case of a puncture of valency 2. In the following 2 examples, which are equivalent in the loop model, the puncture belongs to a large cluster in the first case but not in the second case:
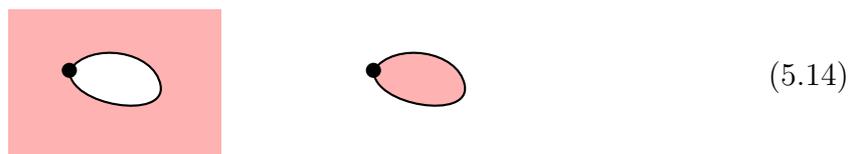
\begin{align}
 \begin{tikzpicture}[baseline=(current  bounding  box.center), scale = .4]
  \fill[red!30] (-2, -3) rectangle (5, 2);
  \filldraw[thick, fill = white] (0, 0) to [out = -90, in = -80] (3, -.5) to [out = 100, in = 55] (0, 0);
   \draw (0, 0) node[fill, circle, minimum size = 2mm, inner sep = 0]{};
   \filldraw[thick, fill = red!30] (10, 0) to [out = -90, in = -80] (13, -.5) to [out = 100, in = 55] (10, 0);
   \draw (10, 0) node[fill, circle, minimum size = 2mm, inner sep = 0]{};
 \end{tikzpicture}
\end{align}
The correct condition for a puncture to belong to 1 large cluster is in fact that no closed loop goes around that puncture. 
Therefore, the Potts model has punctures with valencies $2r\in 2(\mathbb{N}+2)$, and legless punctures of weight $w=0$.

\subsection{Relation with loop CFTs}

There is much evidence that loop CFTs describe the critical limit of loop models, including direct numerical checks. But it is not easy to prove this statement, or even to find which CFT is the limit of which loop model. In fact, it is already tricky to derive the relation between the central charge $c$ and the contractible loop weight $n$. See for example \cite[Appendix A]{grz18} for a discussion of Kondev's original argument, and for an attempt to patch its perceived flaws with the help of the torus partition function.

We will now assume that loop CFTs are critical limits of loop models, and give a simple derivation of the relation between their parameters. We will start with the relation between non-diagonal primary fields $V_{(r,s)}$ and legged punctures. This is relatively straightforward, because both objects have discrete parameters. We will then deduce the relation between diagonal fields $V_P$ and legless punctures, and finally the relation between the central charge and the contractible loop weight.

\subsubsection{Relations between punctures and primary fields}

Our starting assumption is that in the critical limit, the punctures tend to fields of a loop CFT.
Since the correlation functions \eqref{zbig} do not depend on choices of local coordinates, the resulting fields must be primary. We first focus on legged punctures. We have used the same notation $r,s$ for these punctures' parameters as for the non-diagonal fields $V_{(r,s)}$. Let us justify this notation by showing that $r,s$ take the same values in the loop model as in the CFT:
\begin{itemize}
 \item In the CFT, as a result of the existence of the degenerate field $V^d_{\langle 1,2\rangle}$, the first Kac index $r$ is half-integer, and conserved modulo $\mathbb{Z}$, see Eq. \eqref{sriz}. For a legged puncture, $2r$ is the valency, and it obeys the same constraints, see Eq. \eqref{srin}.
 \item In the CFT, $rs$ is the conformal spin, and it is integer for bosonic fields. Under a rotation around a legged puncture, which we interpret as $\theta_k\mapsto \theta_k + \theta$, the weight factor \eqref{wrs} picks up a factor $e^{irs\theta}$: according to Eq. \eqref{veis}, this means that our legged puncture has the conformal spin $rs$.
\end{itemize}
Therefore, a legged puncture can be identified with a non-diagonal primary field $V_{(r,s)}$.

Next, let us consider a legless puncture of weight $w$. We again assume that it tends to a primary field. Since $w$ is a continuous parameter, this must be a diagonal field. To determine its momentum, let us use the idea \eqref{sP} that a diagonal field is a non-diagonal field with $r=0$. The analogous idea for punctures is that the weight $w$ of a closed loop around a legless puncture is related to the weight factor \eqref{wrs} for a rotation by $\theta = 2\pi$ around a legged puncture. In fact, since the closed loop is not oriented, we have to add the rotations by $2\pi$ and $-2\pi$. This gives us the \myindex{loop weight}\index{loop weight} $w=2\cos(\pi s)$, equivalently
\begin{align}
 \boxed{w(P) = 2\cos(2\pi\beta P)}\ .
 \label{wP}
\end{align}
Finally, the identity field $V^d_{\langle 1,1\rangle}$ should not change the weight of loops.
Therefore, the contractible loop weight is
\begin{align}
 \boxed{n= w\left(P_{(1,1)}\right) = -2\cos(\pi \beta^2)} \ .
 \label{ncb}
\end{align}
This is the same $n$ as in the $O(n)$ CFT \eqref{nQ}, because in the lattice $O(n)$ model, the contractible loop weight is the dimension of the vector representation.

We do not discuss the loop model interpretation of the nontrivial degenerate fields $V^d_{\langle 1,s\geq 2\rangle}$, because we will not study their correlation functions. These fields cannot simply correspond to legless punctures, if only because $w(P_{(1,3)})=w(P_{(1,1)})=n$.

\subsubsection{Correlation functions}\label{sec:cf}

According to our identification of punctures with primary fields, the loop model $N$-point function \eqref{zbig} corresponds to a loop CFT $N$-point function of the type $\left<\prod_{i=1}^N V_{(r_i,s_i)}\right>$. In the bootstrap approach, $N$-point functions are defined as single-valued solutions of crossing symmetry equations. To write these equations, we need to know the spectrum in any channel. We introduce the following notations for spectra:
\begin{align}
 \mathcal{S}_{\sigma} \underset{\sigma\in \frac12\mathbb{N}^*}{=} \left\{V_{(r,s)}\right\}_{\substack{r\in \sigma+\mathbb{N} \\ s\in\frac{1}{r}\mathbb{Z}}} \quad , \quad \mathcal{S}_0(P) = \left\{V_{P+k\beta^{-1}}\right\}_{k\in\mathbb{Z}} \bigcup \left\{V_{(r,s)}\right\}_{\substack{r\in\mathbb{N}^* \\ s\in\frac{1}{r}\mathbb{Z}}}\ .
 \label{sxs}
\end{align}
Let $\mathcal{Z}$ be the space of solutions of crossing symmetry for $\left<\prod_{i=1}^N V_{(r_i,s_i)}\right>$, where in each channel the spectrum is either $\mathcal{S}_0(P)$ or $\mathcal{S}_\frac12$, as determined by the conservation of $r\bmod 1$. This spectrum includes all non-diagonal fields that are allowed by the conservation of $r\bmod 1$, plus a family of diagonal fields whenever $r$ is integer. 

Given a 4-point combinatorial map $M$, let us define the $x$-channel \myindex{signature}\index{combinatorial map!signature of a---} $\sigma^{(x)}\in\frac12\mathbb{N}$ as half the minimum number of lines in $M$ that are crossed by an $x$-loop. For example, the map 
$
\begin{tikzpicture}[baseline=(base), scale = .25]
 \vertices
  \draw (0, 0) -- (0, 3);
  \draw (3, 0) -- (3, 3);
  \draw (0, 0) to [out = 70, in = 0] (0, 3.5) to [out = 180, in = 110] (0, 0);
 \end{tikzpicture}
 $
has signature $(\sigma^{(s)},\sigma^{(t)},\sigma^{(u)})=(0,2,2)$. Then let us define the spectrum
\begin{align}
\mathcal{S}_M = (\mathcal{S}^{(s)}_M, \mathcal{S}^{(t)}_M, \mathcal{S}^{(u)}_M)= (\mathcal{S}_{\sigma^{(s)}},\mathcal{S}_{\sigma^{(t)}},\mathcal{S}_{\sigma^{(u)}})\ .
\end{align}
In the case $\sigma^{(x)}=0$, there can exist $x$-loops with weight $w_x=2\cos(2\pi\beta P_x)$, and we take the $x$-channel spectrum to be $\mathcal{S}_0(P_x)$. 

We conjecture that each combinatorial map $M$ with vertices of valencies $(2r_1,\dots, 2r_N)$ corresponds to a solution $Z_M = \lim_\text{critical} Z(\mathcal{E}_M)$ of the crossing symmetry equations with spectra $\mathcal{S}_M$, and that these solutions form a basis of the space of solutions \cite{gjnrs23}:
\begin{align}
 \mathcal{Z} = \operatorname{Span}\left\{ Z_M\right\}_{M}\  .
\end{align}
For example, the existence of the 4 combinatorial maps
\eqref{4maps} implies that for any allowed values of the parameters $s_1,s_2,s_3,P_4$, the space of 4-point functions of the type $\left<V_{(\frac32,s_1)}V_{(1,s_2)}V_{(\frac12,s_3)}V_{P_4}\right>$ has dimension 4. 

A subtelty is that crossing symmetry equations for $Z_M$ may in fact have several linearly independent solutions. To see this, notice that signatures lead to a natural partial ordering on combinatorial maps, with $M\geq M' \iff \forall x, \sigma^{(x)}(M)\geq \sigma^{(x)}(M')$. If $M\geq M'$, then $\forall x, \ \mathcal{S}_M^{(x)} \subset \mathcal{S}_{M'}^{(x)}$, and both $Z_M,Z_{M'}$ solve the crossing symmetry equations with spectrum $\mathcal{S}_{M'}$. This happens in particular if 2 maps have the same signature, for example 
$
\begin{tikzpicture}[baseline=(base), scale = .25]
 \vertices
  \draw (0, 0) -- (0, 3);
  \draw (0, 0) -- (3, 3);
  \draw (0, 0) -- (3, 0);
 \end{tikzpicture}
$ and $
 \begin{tikzpicture}[baseline=(base), scale = .25]
 \vertices
  \draw (0, 0) -- (3, 0);
  \draw (0, 0) -- (0, 3);
  \draw (0, 0) to [out = -30, in = -135] (3.5, -.5) to [out = 45, in = -80] (3, 3);
  \end{tikzpicture}
$. On the other hand, if $M$ is strictly maximal for our ordering, i.e. if $\nexists M'\neq M, M'\geq M$, we expect that the corresponding crossing symmetry equations have a 1d space of solutions, proportional to $Z_M$.

Can we interpret the elements of our space $\mathcal{Z}$ as $4$-point functions $\left<\prod_{i=1}^4 V_{(r_i,s_i)}\right>$? A potential puzzle is that the 4-point functions should only depend on the 4 fields $V_{(r_i,s_i)}$, whereas our solutions can depend on channel weights, for example on $w_s$ if $\sigma^{(s)}=0$. The proposed explanation is that our $4$-point functions involve not only 4 primary fields, but also a combinatorial defect with parameter $w_s$ \cite{rib22}. In the presence of this defect, $s$-channel diagonal fields $V_{P_s}$ such that $w(P_s)=w_s$ appear in the $s$-channel spectrum, and the structure constants of non-diagonal fields may depend on $w_s$. By combinatorial defect we mean a defect that only depends on which punctures are inside or outside the defect. This implies that the defect is topological, but not all topological defects are combinatorial, see Figure \eqref{tvsc}. In the $O(n)$ CFT on the torus, adding a combinatorial defect does not spoil modular invariance, but the partition function becomes a function of the defect's weight.

In particular CFTs, there may exist more or fewer solutions than combinatorial maps. 
For example, the map 
$
\begin{tikzpicture}[baseline=(base), scale = .25]
 \vertices
  \draw (0, 0) -- (0, 3);
  \draw (3, 0) -- (3, 3);
  \draw (0, 0) to [out = 70, in = 0] (0, 3.5) to [out = 180, in = 110] (0, 0);
 \end{tikzpicture}
 $
corresponds to a solution with 1 $s$-channel diagonal field.
 In the $O(n)$ model, all diagonal fields are degenerate, and their fusion rules would forbid this solution, since $V^d_{\langle 1,s\rangle}\notin V_{(\frac32,s_1)}V_{(\frac12,s_2)}$ \cite{gnjrs21}. In the Potts model on the other hand, the spectrum includes degenerate fields of weights $n,-n$ and diagonal fields of weight $0$. In some 4-point functions, diagonal fields with different weights can propagate in the same channel, leading to more solutions than maps \cite{niv22}. 
 
\subsubsection{Critical limits}

Starting from a statistical model on a finite lattice, the large lattice limit may give rise to a conformal field theory, provided the lattice couplings are set to certain critical values. The same loop CFT can arise as the critical limit of statistical models with various lattice geometries, various rules for drawing loops on the lattice, and various parameters. These non-universal features are forgotten in the critical limit, and all these models belong to the same universality class. (See the book \cite{car96} for an introduction to these concepts.)

The existence of relevant fields in the CFT influences the critical limit. A primary field is called \myindex{relevant}\index{field!relevant---} if $\Re(\Delta+\bar\Delta)<2$, \myindex{marginal}\index{field!marginal---} if $\Re(\Delta+\bar\Delta)=2$, or \myindex{irrelevant}\index{field!irrelevant---} if $\Re(\Delta+\bar\Delta)>2$. 
Any given lattice model has a finite set of couplings, which correspond to relevant fields of the CFT, and have to be tuned to their critical values for the critical limit to exist. If there exist relevant fields that do not correspond to any coupling, the critical limit ceases to exist, unless it is protected by some symmetry. Since the 
dimensions of primary fields depend on the parameter $n$ or $Q$, there exist \myindex{critical regions}\index{critical region} in the complex $n$-plane or $Q$-plane where a given model has a critical limit, and the boundary of these regions are \myindex{marginality lines}\index{marginality line}, where a field becomes marginal. 

Let us plot the marginality lines of a few primary fields, which will give us an idea on how critical regions look like. For simplicity, we focus on degenerate diagonal fields $V^d_{\langle 1,s\rangle}$. In the $O(n)$, $PSU(n)$ and Potts models, these fields are invariant under the global symmetry, which therefore does not protect the critical limit from the corresponding couplings. The marginality line of $V^d_{\langle r,s\rangle}$ is
\begin{align}
 M_{(r,s)} = \Big\{\Re\Delta_{(r,s)} = 1 \Big\} = \Big\{\Re \left(P_{(r+1,s-1)}P_{(r-1,s+1)}\right) = 0\Big\} \ . 
\end{align}
In the complex $\beta^2$-half-plane, the line $M_{(1,s)}=\left\{\Re\beta^{-2}=\frac{2}{s-1}\right\}$ is a circle with a center at $\beta^2=\frac{s-1}{4}$, which goes through the origin. The field $V^d_{(1,s)}$ is irrelevant inside the circle, and relevant outside. Here we plot the circles for $s=2,3,\dots, 9$ in the right half-plane \eqref{rbp}, with dashed lines for the even values, which do not appear in the $O(n)$ CFT:
\begin{align}
 \begin{tikzpicture}[baseline=(current  bounding  box.center), scale = 3.5]
 \draw [-latex] (-.3, 0) -- (2.6, 0) node[below]{$\beta^2$};
 \node at (-.5, -.5) {$O(n)$ model:};
 \node[draw, fill = black!5] at (.5, -.5) {dense};
 \node[draw] at (1.5, -.5) {dilute};
 \node[draw, fill = green!5] at (2.5, -.5) {multicritical};
 \clip (-.3, -.3) -- (-.3, 1.3) -- (2.4, 1.3) -- (2.4, -.3) -- cycle;
 \fill [color = green!10] (0, -1.3) -- (2.4, -1.3) -- (2.4, 1.3) -- (0, 1.3) -- cycle;
  \fill [color = white] (1, 0) circle [radius = 1];
 \fill [color = black!10] (.5, 0) circle [radius = .5];
 \fill [color = black!30] (0, -1.3) -- (-.3, -1.3) -- (-.3, 1.3) -- (0, 1.3) -- cycle;
 \draw[ultra thick] (0, -1.3) -- (0, 1.3);
 \draw[thick] (.5, -1.3) -- (.5, 1.3);
 \draw[thick] (1, -1.3) -- (1, 1.3);
 \draw[thick] (1.5, -1.3) -- (1.5, 1.3);
 \draw[thick] (2, -1.3) -- (2, 1.3);
 \draw[ultra thick, green!70!black] (1, -1) -- (1,1.5);
  \draw [-latex] (-.3, 0) -- (0, 0) node [below left] {$0$} -- (1, 0) node [below left] {$1$} -- (2, 0) node[below left] {$2$} -- (2.6, 0);
   \draw [ultra thick, red] (0, 0) -- (1, 0); 
  \foreach \s in {2,4,6,8}{
  \draw [ultra thick, dashed, blue] ({(\s-1)/4}, 0) circle [radius = (\s-1)/4];
  \draw [ultra thick, blue] ({\s/4}, 0) circle [radius = \s/4];
  }
 \node at (1, 0) [fill, circle, scale = .5] {};
 \node at (0, 0) [fill, circle, scale = .5] {};
 \end{tikzpicture} 
 \label{phases}
\end{align}
When it comes to the complex $n$-plane, we first notice that the relation \eqref{ncb} maps any strip $k<\Re \beta^2\leq k+1$ to the full complex $n$-plane. 
This is one reason why the same CFT can describe different phases of a statistical model. For example, the $O(n)$ model's dense phase is usually defined for $\beta^2\in (0,1)$, and the dilute phase for $\beta^2\in (1,2)$, with $\beta^2>2$ corresponding to multicritical phases. The $O(n)$ model has 1 lattice coupling $K$, contributing a factor $K^{\#\{\text{occupied edges}\}}$ to the weight function \eqref{woe}, and corresponding to the field $V^d_{\langle 1,3\rangle}$. In the dense phase, that field is irrelevant, and the critical limit is obtained for a range of values of the coupling. In the dilute phase it is relevant, and the critical limit is obtained for the critical value of the coupling, which is $K_c = \frac{1}{\sqrt{2+\sqrt{2-n}}}$ for the honeycomb lattice. In multicritical phases, criticality is achieved by adding extra couplings to the model, and fine-tuning their values.

In the diagram \eqref{phases} we have extended these phases to the complex $\beta^2$-plane, by taking marginality lines as phase boundaries. Let us now plot the marginality lines in the $n$-plane that corresponds to $0<\Re \beta^2<1$:
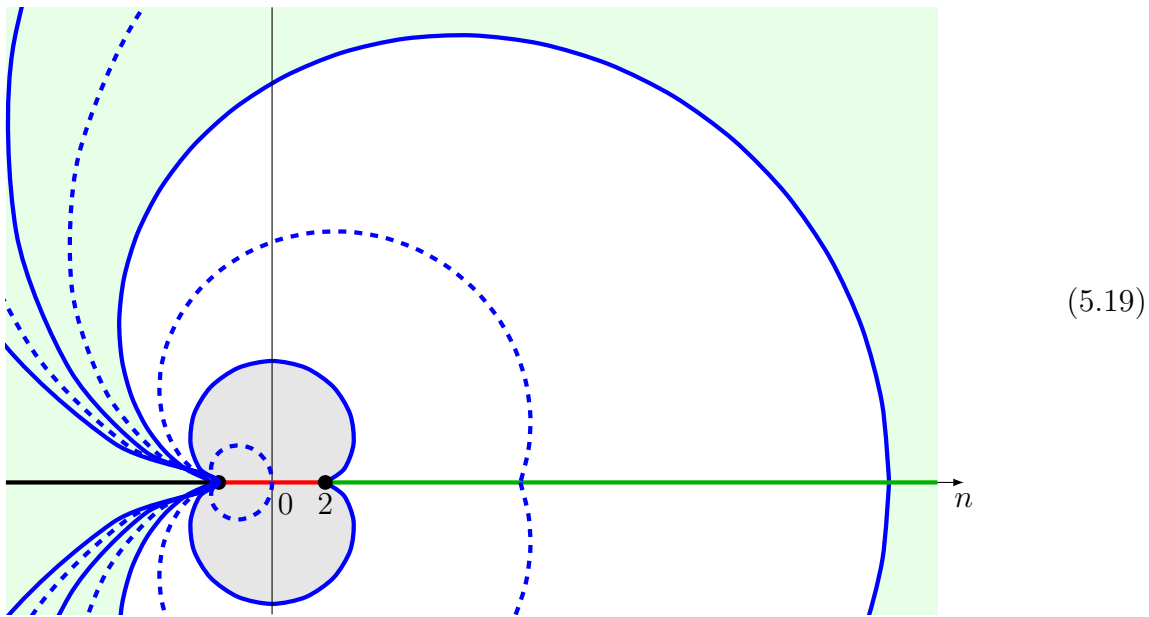
\newcommand{\margs}[2]{
\draw[domain = {-2/(#1-1)}:{2/(#1-1)}, smooth, variable = \t, #2]
plot({-2*cos(pi/2*(#1-1)*\t r)*cosh(pi/2*(#1-1)*sqrt(abs(\t)-\t*\t)}, {2*sin(pi/2*(#1-1)*\t r)*sinh(pi/2*(#1-1)*sqrt(abs(\t)-\t*\t)});
}
\begin{align}
 \begin{tikzpicture}[baseline=(current  bounding  box.center), scale = .35]
\clip (-10, 18) -- (27, 18) -- (27, -5) -- (-10, -5) -- cycle;
\fill [color = green!10] (-10, 18) -- (25, 18) -- (25, -5) -- (-10, -5) -- cycle;
\margs{5}{ultra thick, blue, samples = 50, fill = white}
\margs{3}{ultra thick, blue, fill = black!10}
\draw[-latex] (-10, 0) -- (26, 0) node[below]{$n$};
\draw (0, -5) -- (0, 18);
\draw[ultra thick, green!70!black] (2, 0) -- (25, 0);
\draw [ultra thick] (-10, 0) -- (-2, 0);
\draw [ultra thick, red] (-2, 0) -- (2, 0); 
\node[below right] at (-.2, 0) {$0$};
\node[below] at (2, 0) {$2$};
\node at (2, 0) [fill, circle, scale = .5] {};
\node at (-2, 0) [fill, circle, scale = .5] {};
 \draw[domain = 0:360, smooth, variable = \t, samples = 30, ultra thick, dashed, blue]
   plot({-2*sin(pi*(1+cos(\t))/4 r)*cosh(pi*sin(\t)/4)}, {2*cos(pi*(1+cos(\t))/4 r)*sinh(pi*sin(\t)/4)});
 \margs{4}{ultra thick, dashed, blue, samples = 50}
\margs{6}{ultra thick, dashed, blue}
\margs{7}{ultra thick, blue}
\margs{8}{ultra thick, dashed, blue}
\margs{9}{ultra thick, blue}
\end{tikzpicture}
\end{align}
Numerical investigations of the zeros of the partition function suggest that the gray region within the criticality line $M_{(1,3)}$ is indeed the critical region of certain lattice loop models \cite{bjjz22}. 
In fact, in the critical limit, zeros of the partition function condense not only on criticality lines, but also inside critical regions, on lines where the ground state changes. The ground state, i.e. the state with the lowest dimension (in real part), can be $V^d_{\langle 1,1\rangle}, V_{(1, 0)}, V_{(2,0)}$ or $V_{P_{(0,\frac12)}}$, depending on $\beta^2$ and on the model. On a line where the 2 lowest states have the same dimension, the free energy is singular. 

A puzzle has arisen in the $O(n)$ CFT, when it was found that there is 1 field $V_{(2,0)}$ that is invariant under $O(n)$ \cite{gz20}. From the point of view of the renormalization group flow triggered by the field $V^d_{\langle 1,3\rangle}$, this field is \textit{dangerously irrelevant}, i.e. it becomes relevant along the flow, and could prevent the flow from reaching the dense phase from the dilute phase. But numerical studies show that the flow does reach the dense phase. The puzzle is resolved by noticing that $V^d_{\langle 1,3\rangle}$ is degenerate, and cannot produce the non-diagonal field $V_{(2,0)}$ by repeated fusion, even if $O(n)$ symmetry would allow it. (See \cite{js23} for a formulation of this argument in the language of topological defects.) Therefore, it is global symmetry itself that is \textit{dangerously irrelevant} to the understanding of this flow.

If we now consider correlation functions, the existence of a critical limit might depend on the case, and in particular on the weights $w$ of legless punctures. And which diagonal fields $V_P$ do we obtain in the limit? Since the relation \eqref{wP} between $w$ and $P$ is not bijective, the answer is not obvious. (A simple possibility would be $0\leq \Re(\beta P)\leq \frac12$.)

\subsection{Analytic properties of structure constants}

Whenever crossing symmetry equations have a unique solution, numerical bootstrap calculations can give us numerical data for the corresponding 4-point structure constants. From these data, we would like to infer exact expressions. To do this, we will constrain structure constants by making assumptions on their analytic properties. These assumptions will determine each structure constant up to a polynomial in loop weights.

We consider a 4-point function $\left<\prod_{i=1}^4 V_{(r_i,s_i)}\right>$, associated to a combinatorial map $M$. If the $s$-channel signature is $\sigma=0$, the $s$-channel spectrum $\mathcal{S}_0(P)$ \eqref{sxs} depends on a channel momentum $P$. Then the decomposition of our 4-point function into $s$-channel conformal blocks reads
\begin{align}
 Z(P) =  \sum_{k\in\mathbb{Z}} D_{P+k\beta^{-1}} \left|\mathcal{F}_{P+k\beta^{-1}}\right|^2 +\sum_{r=1}^\infty \sum_{s\in\frac{1}{r}\mathbb{Z}} D_{(r,s)}(P) \mathcal{G}_{(r,s)}\ .
 \label{zop}
\end{align}
Besides the Virasoro blocks $\mathcal{F}_P$, we introduce conformal blocks $\mathcal{G}_{(r,s)}$ that gather the contributions of non-diagonal primary fields $V_{(r,s)}$ and their descendants, with
\begin{align}
 (r,s)\notin\mathbb{N}^*\times \mathbb{Z}^*\quad \implies\quad \mathcal{G}_{(r,s)} = \mathcal{F}_{P_{(r,s)}}\bar{\mathcal{F}}_{P_{(r,-s)}}\ .
\end{align}
For $(r,s)\in\mathbb{N}^*\times \mathbb{Z}^*$, the primary field $V_{(r,s)}$ has a singular vector. So there exist several representations of the conformal algebra to which $V_{(r,s)}$ could belong, giving rise to different conformal blocks.

\subsubsection{Holomorphicity in the channel momentum}\label{sec:hcm}

Analytic bootstrap equations from the existence of $V^d_{\langle 1,2\rangle}$ imply $Z(P)=Z(P+\beta^{-1})$, so $Z(P)$ is in fact a function of the channel weight $w(P)$ \eqref{wP}, but we find it more convenient to write it as a function of $P$. In particular, we have $D_{(r,s)}(P)=D_{(r,s)}(P+\beta^{-1})$. Moreover, the analytic bootstrap equations 
determine how $D_P$ behaves under $P\to P+\beta^{-1}$. Let us assume that diagonal 4-point structure constants factorize into 2-point and 3-point structure constants
\begin{align}
 D_P = \frac{\check{C}_{(0,2\beta P)(r_1,s_1)(r_2,s_2)}\check{C}_{(0,2\beta P)(r_3,s_3)(r_4,s_4)}}{B_{P}}\ ,
 \label{dccb}
\end{align}
where $B$ is given in Eq. \eqref{bc} and $\check{C}$ in Eq. \eqref{cref}.
(We use $\check{C}$ rather than the full 3-point structure constant $C=\omega \check{C}$, as the sign factor $\omega$ is invariant under $P\to P+\beta^{-1}$.)
For any given 4-point function $Z(P)$, this assumption only amounts to an overall choice of normalization.

Then we further assume that $Z(P)$ is a holomorphic function of $P\in\mathbb{C}$. From the point of view of loop models, this is a natural assumption, because $Z(P)$ is the critical limit of polynomials in loop weights. From the point of view of the CFT, this is a nontrivial assumption, because the diagonal terms have poles.
To begin with, according to Eq. \eqref{resf}, the Virasoro block $\mathcal{F}_P$ has a simple pole at $P=P_{(r,s)}$ with $r,s\in \mathbb{N}^*$, so that 
\begin{align}
 \mathcal{F}_P = \frac{R^{\#}_{r,s}}{P-P_{(r,s)}} \mathcal{F}_{P_{(r,-s)}} + \mathcal{F}^\text{reg}_{P_{(r,s)}} + O(P-P_{(r,s)}) \qquad \text{with} \qquad R^{\#}_{r,s} = \frac{R_{r,s}}{2P_{(r,s)}}\ .
 \label{freg}
\end{align}
Therefore, $\left|\mathcal{F}_P\right|^2$ has a double pole, with a residue proportional to $\left|\mathcal{F}_{P_{(r,-s)}}\right|^2$. Now, $\lim_{P\to P_{(r,s)}} Z(P)$ involves another contribution proportional to $\left|\mathcal{F}_{P_{(r,-s)}}\right|^2$, from the term $k=s$ in Eq. \eqref{zop}. And the structure constant $D_{P+s\beta^{-1}}$ does have a double pole at $P=P_{(r,s)}$, because $B_P$ has a double zero at $P=P_{(r,-s)}$. The cancellation of the double poles of $Z(P)$ requires
\begin{align}
 R^{\#}_{r,s}\bar{R}^{\#}_{r,s}D_{P_{(r,s)}} + \lim_{P\to P_{(r,-s)}}\left(P-P_{(r,-s)}\right)^2 D_P = 0 \ .
 \label{dpld}
\end{align}
It can be checked that this equation is satisfied by the Virasoro block residues $R_{r,s}$ \eqref{rrs}. This equation is relevant not only to loop CFTs, but also to (generalized) minimal models \cite{rib18}.

Simple poles of $Z(P)$ must cancel too, i.e. $\operatorname{Res}_{P=P_{(r,s)}}Z(P)=0$ for any $(r,s)\in \mathbb{N}^*\times\mathbb{Z}$. At $P=P_{(r,0)}$, $B_P$ has a simple zero, and $D_P\left|\mathcal{F}_P\right|^2$ has a simple pole with a residue proportional to $\left|\mathcal{F}_{P_{(r,0)}}\right|^2$. This is  cancelled by the term $D_{(r,0)}(P) \mathcal{G}_{(r,0)}$, provided $D_{(r,0)}(P)$ has a simple pole, and 
\begin{align}
  \underset{P=P_{(r,0)}}{\operatorname{Res}} D_{(r,0)}(P) + \underset{P=P_{(r,0)}}{\operatorname{Res}}D_P = 0 \ .
\end{align}
Finally, let us consider the simple pole of $\left|\mathcal{F}_{P}\right|^2$ at $P=P_{(r,s)}$ with $r,s\in \mathbb{N}^*$. The residue at this pole includes the term 
$\mathcal{F}_{P_{(r,s)}}^\text{reg}\bar{\mathcal{F}}_{P_{(r,-s)}}$, which must also appear in $\mathcal{G}_{(r,s)}$ as the contribution of $V_{(r,s)}$ and some of its descendants. This leads to the relation
\begin{align}
 \underset{P=P_{(r,s)}}{\operatorname{Res}} D_{(r,s)}(P) + \bar R^{\#}_{r,s} D_{P_{(r,s)}} = 0\ .
 \label{rdp}
\end{align}
In fact, the equation $\operatorname{Res}_{P=P_{(r,s)}}Z(P)=0$ with $r,s\in \mathbb{N}^*$ leads to the determination of the conformal block $\mathcal{G}_{(r,s)}$, which turns out to be logarithmic. 

\subsubsection{Logarithmic conformal blocks}\label{sec:log}

For $r,s\in\mathbb{N}^*$, the residue $\operatorname{Res}_{P=P_{(r,s)}}D_P\left|\mathcal{F}_{P}\right|^2$ 
involves a number of terms, which cannot be canonically separated into the blocks 
$\mathcal{G}_{(r,s)}$ and $\mathcal{G}_{(r,-s)}$. Rather, this residue is a logarithmic conformal block, which sums up the contributions of all the states in a logarithmic representation that includes both primary fields $V_{(r,s)}$ and $V_{(r,-s)}$, see \cite{nr20} for more detail on that representation. By convention we call $\mathcal{G}_{(r,s)}$ the logarithmic conformal block, and set $\mathcal{G}_{(r,-s)}=0$, so that
\begin{align}
 \mathcal{G}_{(r,s)} = \frac{1}{\bar R^{\#}_{r,s} D_{P_{(r,s)}}} \left( \operatorname{Res}_{P=P_{(r,s)}} + \operatorname{Res}_{P=P_{(r,-s)}}\right)D_P\left|\mathcal{F}_{P}\right|^2 \ . 
\end{align}
More explicitly, using Eq. \eqref{dpld}, this can be expressed as 
\begin{multline}
  \mathcal{G}_{(r,s)} = \left(\mathcal{F}^\text{reg}_{P_{(r,s)}} -R^{\#}_{r,s} \mathcal{F}'_{P_{(r,-s)}}\right) \bar{\mathcal{F}}_{P_{(r,-s)}} 
  + \frac{R^{\#}_{r,s}}{\bar{R}^{\#}_{r,s}} \mathcal{F}_{P_{(r,-s)}} \left(\bar{\mathcal{F}}^\text{reg}_{P_{(r,s)}} -\bar R^{\#}_{r,s} \bar{\mathcal{F}}'_{P_{(r,-s)}}\right)
  \\
  + R^{\#}_{r,s} \left( \frac{D'_{P_{(r,s)}}}{D_{P_{(r,s)}}} - \lim_{P\to P_{(r,-s)}}\left[\frac{2}{P-P_{(r,-s)}} +\frac{D'_P}{D_P}\right]\right)
  \left|\mathcal{F}_{P_{(r,-s)}}\right|^2 
  \ ,
\end{multline}
where the derivatives are with respect to $P$, and
$\mathcal{F}^\text{reg}_{P_{(r,s)}}$ was defined in Eq. \eqref{freg}.
The logarithmic nature of the block $\mathcal{G}_{(r,s)}$ comes from the asymptotic behaviour $\mathcal{F}_P(z)\underset{z\to 0}{\propto} z^{P^2} $ \eqref{stu} of Virasoro blocks. When taking derivatives in $P$, or regularizing near a pole, this power-like prefactor gives rise to terms in $\log z$. Logarithmic terms are however absent if $R^{\#}_{r,s}=\bar R^{\#}_{r,s}=0$, which occurs if fusion rules allow the degenerate field $V^d_{\langle r,s\rangle}$ to propagate in the $s$-channel of our 4-point function. In this case, the logarithmic block reduces to 
\begin{align}
 \mathcal{G}_{(r,s)} \underset{R^{\#}_{r,s}=\bar R^{\#}_{r,s}=0}{=} 
 \mathcal{F}_{P_{(r,s)}}  \bar{\mathcal{F}}_{P_{(r,-s)}} + 
 \frac{R^{\#}_{r,s}}{\bar{R}^{\#}_{r,s}} \mathcal{F}_{P_{(r,-s)}} \bar{\mathcal{F}}_{P_{(r,s)}} \ .
\end{align}
This is a combination of the blocks for the Verma modules of $V_{(r,s)}$ and $V_{(r,-s)}$. Their relative coefficient $\frac{R^{\#}_{r,s}}{\vphantom{\tilde{|}}\bar{R}^{\#}_{r,s}}$ can be determined either by taking a limit in $s_i$ with $r_i$ fixed \cite{gnjrs21}, or by using the shift equations \eqref{cbshift}, applied to $\frac{R^{\#}_{r,s}}{\vphantom{\tilde{|}}\bar{R}^{\#}_{r,s}} = \frac{D_{(r,-s)}}{D_{(r,s)}}$. 
Just like Virasoro blocks, logarithmic blocks can be combined into interchiral blocks of the type 
\begin{align}
 \widetilde{\mathcal{G}}_{(r,s_0)} = \sum_{k\in 2\mathbb{N}} \frac{D_{(r,s_0+2k)}}{D_{(r,s_0)}} \mathcal{G}_{(r,s_0+2k)}\ , 
\end{align}
where $s_0\in \{0,1\}$, and the ratios of 4-point structure constants are again determined by the shift equations \eqref{cbshift}. 

\subsubsection{Polynomials in loop weights}\label{sec:plw}

Given our 4-point function $\left<\prod_{i=1}^4 V_{(r_i,s_i)}\right>$ and combinatorial map $M$ with $s$-channel signature $\sigma$, let us study the $s$-channel 4-point structure constants $D_{(r,s)}$. A natural ansatz for $D_{(r,s)}$ is a combination of 2-point and 3-point structure constants $D_k \underset{\text{ansatz}}{=} \frac{C_{12k}C_{k34}}{B_k}$ from Eq. \eqref{stu}. It turns out that this ansatz is incorrect, but it obeys the same shift equations as $D_k$. This implies that $D_k$ is given by the ansatz, times a shift-invariant function.

Our 4-point function can depend on the momenta of diagonal fields $V_P$: either one of the 4 fields $V_{(r_i,s_i)}$ with $r_i=0$, or a channel field if $\sigma =0$. A function of the momentum $P$ that is invariant under shifts $P\to P+\beta^{-1}$ and reflections $P\to -P$ must be a function of the loop weight $w(P)$ \eqref{wP}. Therefore, in order to complete the ansatz, we must determine functions of loop weights.

We first focus on the weight $w=w(P)$ of a channel diagonal field. We have argued in Section \ref{sec:hcm} that $D_{(r,s)}$ is meromorphic in $P$, therefore $D_{(r,s)}$ is also meromorphic in $w=w(P)$. In fact $D_{(r,s)}$ is holomorphic unless $s\in \mathbb{Z}$, in which case $D_{(r,s)}$ has a simple pole at $w=w(P_{(r,s)})$. The residue can be computed using Eqs. \eqref{rdp} and \eqref{dccb}, leading to
\begin{align}
 D_{(r,s)} =  \frac{\check{C}_{(r,s)(r_1,s_1)(r_2,s_2)}\check{C}_{(r,s)(r_3,s_3)(r_4,s_4)}}{B_{(r,s)}\kappa^{r,s}} \left[d_{(r,s)}  - \delta_{\sigma,0}\delta_{s\in\mathbb{Z}}\frac{(-)^{(r+1)s}\rho^{r,s}_{(r_1,s_1)(r_2,s_2)}\rho^{r,s}_{(r_4,s_4)(r_3,s_3)}}{w-w(P_{(r,s)})}\right] \, .
 \label{sdrs}
\end{align}
In this expression, the second term accounts for the simple pole at $w=w(P_{(r,s)})$. We have written the corresponding residue in terms of our ansatz $\frac{\check{C}\check{C}}{B}$, times trigonometric factors $\frac{\rho\rho}{\kappa}$, where we define
\begin{align}
\rho^{r,s}_{(r_1,s_1)(r_2,s_2)}= (-)^{s\min(r,|r_1-r_2|)\delta_{r_1<r_2}} \prod_\pm \prod_{j\overset{1}{=}-\frac{r-1-|r_1\pm r_2|}{2}}^{\frac{r-1-|r_1\pm r_2|}{2}} 2\cos\pi\left(j\beta^2+\tfrac{s-s_1\mp s_2}{2}\right)\ ,
 \label{rhop}
\end{align}
and
\begin{align}
\kappa^{0,s} = \frac{1}{2\sin^2(\pi s)} \ , \quad \kappa^{\frac12,s}=1 \ , \quad
\kappa^{r,s} \underset{r\geq 1}{=} \frac{2^{2r-1}}{\sin\pi(\text{frac}(r)+s)}\prod_{j\overset{1}{=}1-r}^{r-1} \sin\pi(j\beta^2 +s) \ .
 \label{kappa}
\end{align}
Here $\kappa^{0,s}$ is chosen such that $\kappa^{0,2\beta P}B_{(0,2\beta P)}=B_P$. We can write $\kappa^{r,s}$ in a manifestly polynomial form:
\begin{align}
 \kappa_{r,s} \ &\underset{r\in \mathbb{N}^*}{=}\  2\prod_{j=1}^{r-1}\prod_{k=0}^{j-1} \left(n^2 -4\cos^2\pi \tfrac{k+s}{j}\right) \ ,
 \\
 \kappa_{r,s} \ &\underset{r\in \mathbb{N}+\frac32}{=}\ \frac{(-)^{r-\frac12}}{\cos\pi s} \prod_{j\overset{2}=1}^{2r-2} \prod_{k=0}^{j-1}\left(n+2\cos 2\pi \tfrac{k+s}{j}\right)\ .
\end{align}
After subtracting the pole term, our structure constant $D_{(r,s)}$ \eqref{sdrs} is proportional to a quantity that we called $d_{(r,s)}$, which should therefore be holomorphic in the channel weight $w$.
Numerical results suggest that $d_{(r,s)}$ is not only holomorphic in $w$, but in fact polynomial. Furthermore, the quantity $\rho^{r,s}$ is polynomial in the loop weights $w_i=\cos(\pi s_i)$ of the fields $V_{(r_i,s_i)}$ with $r_i=0$. And both $\rho^{r,s},\kappa^{r,s}$ are polynomial in the contractible loop weight $n$ \eqref{ncb}. This suggests that $d_{(r,s)}$ might be polynomial in $n,w_i$ as well. Numerical results are consistent with this suggestion, not only in the case $\sigma=0, s\in\mathbb{Z}$ where $D_{(r,s)}$ depends on a channel weight and has a pole, but in all cases.

We conjecture that $d_{(r,s)}$ is polynomial in all loop weights $n,w_i,w$.
Thanks to this conjecture, we can extract analytic expressions for $d_{(r,s)}$ from numerical results for $D_{(r,s)}$.

\subsection{Examples of 4-point structure constants}\label{sec:ex4pt}

\DeclareRobustCommand\picz{
$
\begin{tikzpicture}[baseline=(base), scale = .25]
 \qvertex;
 \end{tikzpicture}
 $
}
\DeclareRobustCommand\pice{
$
\begin{tikzpicture}[baseline=(base), scale = .25]
  \vertices
  \draw (0, 0) -- (0, 3);
  \draw (3, 0) -- (3, 3);
 \end{tikzpicture}
$}
\DeclareRobustCommand\picj{
$
\begin{tikzpicture}[baseline=(base), scale = .25]
  \vertices
  \draw (0, 0) -- (3, 3) to [out = -120, in = 30] (0, 0);
  \draw (0,0) -- (0, 3);
  \draw (0, 3) to [out = 30, in = 135] (3.5, 3.5) to [out = -45, in =60] (3, 0);
 \end{tikzpicture}
 $}

We will now write the first few 4-point structure constants for 4-point functions that correspond to the combinatorial maps \picz\ , \pice\ and \picj . 
More examples can be found in \cite{nrj23}. 

\subsubsection[Case of $\left<\prod_{i=1}^4 V_{P_i}\right>$]{Case of $\left<\prod_{i=1}^4 V_{P_i}\right>$ \picz}\label{sec:cod}

We have a unique solution of crossing symmetry for any 4 fields $V_{P_i}$ and channel fields $V_{P_s},V_{P_t},V_{P_u}$. Due to their invariance under shifts $d^{(x)}_{(r,s)}=d^{(x)}_{(r,s+2)}$ and parity $d^{(x)}_{(r,s)}=d^{(x)}_{(r,-s)}$, we need only write structure constants for $0\leq s\leq 1$:
\begin{subequations}
\begin{align}
& d_\text{diag}^{(s,t,u)} = 1\ ,
\label{ddiag}
\\
& d_{(1, 0)}^{(s)} = w_t+w_u \quad , \quad d_{(1,0)}^{(t)}= w_s+w_u \quad ,\quad d_{(1,0)}^{(u)} = w_s+w_t\ ,
 \\
& d_{(1, 1)}^{(s)} = w_u-w_t \quad , \quad d_{(1,1)}^{(t)}= w_u-w_s \quad ,\quad d_{(1,1)}^{(u)} = w_s-w_t\ .
\end{align}
\begin{multline}
 2d^{(s)}_{(2, 0)} = (n^2-4)\left[w_t^2 + w_u^2 +2w_s -4\right]
 -(n-2)(w_t+w_u)(w_1+w_2)(w_4+w_3)
 \\
 -(n+2)(w_t-w_u)(w_1-w_2)(w_4-w_3)\ ,
 \label{q2,0}
\end{multline}
\begin{multline}
 2d^{(s)}_{(2,\frac12)} = n^2(w_u^2 - w_t^2) +n(w_t-w_u)(w_1+w_2)(w_4+w_3) 
 \\
 +n(w_t+w_u)(w_1-w_2)(w_4-w_3)\ ,
 \label{q2,12}
\end{multline}
\begin{multline}
2d^{(s)}_{(2, 1)} = (n^2-4)\left[w_t^2 + w_u^2 -2w_s -4\right] 
-(n+2)(w_t+w_u)(w_1+w_2)(w_4+w_3)
\\
-(n-2)(w_t-w_u)(w_1-w_2)(w_4-w_3)\ ,
 \label{q2,1}
\end{multline}
\end{subequations}
Formulas for $d^{(s,t,u)}_{(3,*)}$ are known but rather complicated.

In the diagonal 4-point function $\left<\prod_{i=1}^4 V_{P_i}\right>$, it is particularly easy to study the factorization \eqref{dcc} of 4-point structure constants, thanks to the presence of the continuous variables $P_i$. In particular, we see that $d^{(s)}_{(2,*)}$ is a sum of 3 factorized terms. Due to pole terms, $D^{(s)}_{(2,0)}$ and $D^{(s)}_{(2,1)}$ are sums of 4 terms, while $D^{(s)}_{(2,\frac12)}$ is still a sum of 3 terms. And the presence of several terms is not due to combinatorial defects: in the case $w_s=w_t=w_u=n$ where the defects are absent, $D^{(s)}_{(2,*)}$ loses only 1 term.

\subsubsection[Case of $\left<V_{(\frac12, 0)}^4\right>$]{Case of $\left<V_{(\frac12, 0)}^4\right>$ \pice}

We now have a diagonal field $V_{P_s}$ in the $s$-channel, but no diagonal fields in the other channels. Structure constants depend on $n$ and on $w_s$. 
They are known up to $r=4$ in the $s$-channel and $r=3$ in the $t,u$-channels. By invariance under the permutation $V_1\leftrightarrow V_2$, we have $rs\equiv 1\bmod 2\implies D^{(s)}_{(r,s)} = 0$:
\begin{subequations}
\label{011}
\begin{align}
 d^{(s)}_\text{diag} &= 1 
 \ , \\
 d^{(s)}_{(1,0)} &=  0 
 \ , \\
 d^{(s)}_{(2, 0)} &=n^2 -4
\ ,\\
 d^{(s)}_{(2, 1)} &= -(n^2 - 4)
  \ ,\\
3d^{(s)}_{(3, 0)} &= -8n^2(n-2)^2(n+2)
\ ,\\
3d^{(s)}_{(3,  \frac23)} &= 4(n^2 - 1)(n^2 - 3)(n - 2)
\ , \\
2d^{(s)}_{(4, 0)} &=  n^2(n - 2)^3(n+1)^2(n+ 2)
\nonumber
\\&\hspace{3cm}
\times
\left[
w_s(n+2)^2(n-1)^2
+2n^4 -6n^2 -8n + 16
\right]
\ , \\
2d^{(s)}_{(4, \frac12)}
&=
-n^3(n^2-2)(n^2-3)
\nonumber
\\&\hspace{1.5cm}
\times
\left[w_sn(n^2-2)(n^2-3)-4n^4+4n^3+8n^2+4n-16\right]
\, , \\
2d^{(s)}_{(4,1)} &=
n^2(n^2-4)^3(n-1)^2\left[w_s (n+1)^2 -2n^2-8n-10\right]
\ .
\end{align}
\aline
\begin{align}
 d^{(t)}_{(1,0)} &= 1 
 \ , \\
 d^{(t)}_{(1,1)} &= -1 
 \ , \\
 2d^{(t)}_{(2, 0)} &= (n - 2)\left[w_s(n + 2) - 8\right]
  \ ,\\
2d^{(t)}_{(2, \frac12)} &= -n(w_sn-4)
  \ ,\\
2d^{(t)}_{(2, 1)} &=  w_s(n^2-4)
\ ,\\
3d^{(t)}_{(3, 0)} &=  n^2(n-2)^2
\left[
w^2_s(n+2)^2 - 4w_s(n+2)+ n^2 +8
\right]
\ ,\\
3d^{(t)}_{(3,  \frac13)} &= 
-(n-1)^2(n + 1)
\nonumber
\\
&\hspace{-0.5cm}
\times
\left[
(n^2 - 3)(n + 1)w^2_s
-2(n+1)(2n - 3)w_s
-2(n-2)(n^2+4n+1)
\right]
\ , \\
3d^{(t)}_{(3,  \frac23)} &= 
(n^2 - 3)(n^2 - 1)
\left[w^2_s(n^2-1)-2w_s(2n-1)-2(n+1)(n-2)
\right]
\ , \\
3d^{(t)}_{(3, 1)} &= 
-(n^2 - 4)^2
\left[
w^2_sn^2 -4w_sn + (n+2)^2
\right]
\ .
\end{align}
\end{subequations}

\subsubsection[Case of $\left<V_{(\frac32,0)}V_{(1,1)}V_{(1,0)}V_{(\frac12,0)}\right>$]{Case of $\left<V_{(\frac32,0)}V_{(1,1)}V_{(1,0)}V_{(\frac12,0)}\right>$ \picj}

In this case there are no channel diagonal fields, and the structure constants only depend on $n$. To write the structure constants with $r\leq 3$, we use the golden ratio $\varphi = \frac{1+\sqrt{5}}{2} = 2\cos\left(\frac{\pi}{5}\right)$:
\begin{subequations}
\label{3221}
\begin{align}
d^{(s)}_{(\frac12,0)}
&= 0\ ,
\\
3d^{(s)}_{(\frac32, 0)}
&= n + 2
\ , \\
3d^{(s)}_{(\frac32, \frac23)}
&= -2(n -1 )
\ , \\
5d^{(s)}_{(\frac52, \frac25)}
&= 4\cos\left(\tfrac{\pi}{10}\right)\left[-n^4+4n^2-2 + \varphi(n+1)(n^2-3)\right]
\ ,\\
5
d^{(s)}_{(\frac52, \frac45)}
&= 4\cos\left(\tfrac{3\pi}{10}\right)
\left[n^4 - 4n^2 +2 +\varphi^{-1}(n + 1)(n^2 - 3)\right]
\ .
\end{align}
\aline
\begin{align}
d_{(1,0)}^{(t)} &= d^{(t)}_{(1,1)} = d^{(t)}_{(2,0)} = d^{(t)}_{(3,0)}=0 \ ,
\\
2d_{(2, \frac12)}^{(t)}
& = \sqrt{2}n^2
\ ,\\
d_{(2, 1)}^{(t)}
&=n^2 - 4
\ ,\\
3d_{(3, \frac13)}^{(t)}
&=
(n -4)(n-1)(n+1)^2(n^2 - 3)
\ , \\
3d_{(3, \frac23)}^{(t)}
&=
\sqrt{3}(n^2 - 1)^2(n + 2)(n -3)
\ , \\
3d_{(3, 1)}^{(t)}
&=
2n^2(n^2 - 4)^2
\ .
\end{align}
\aline
\begin{align}
d^{(u)}_{(\frac12, 0)}
&=1
\ , \\
d^{(u)}_{(\frac32,0)} &= d^{(u)}_{(\frac52,0)} = 0\ ,
\\
3d^{(u)}_{(\frac32, \frac23)}
&= -2\sqrt{3}
\ , \\
5d^{(u)}_{(\frac52, \frac25)}
&= 4\cos\left(\tfrac{\pi}{10}\right)\left[n^2-2+\varphi\right]
\ ,\\
5
d^{(s)}_{(\frac52, \frac45)}
&= 4\cos\left(\tfrac{3\pi}{10}\right)
\left[-n^2+2+\varphi^{-1}\right]
\ .
\end{align}
\end{subequations}
The coefficients of these polynomials are trigonometric numbers, related to the values of the second Kac index.

\subsection{Outlook}

\subsubsection{Towards solving loop CFTs}

In the conformal bootstrap approach, solving loop CFTs on the sphere means being able to compute $N$-point functions for any $N\in \mathbb{N}^*$. If OPEs exist, this problem reduces to knowing the 3-point structure constants. (See Section \ref{sec:cscb}.) However, 4-point structure constants \eqref{sdrs} are more complicated than products of 3-point structure constants. The factorization of 4-point structure constants can in principle be explored numerically, as we discussed in Section \ref{sec:factor}. But a case-by-case analysis of factorization would be tedious and not necessarily very illuminating, if we do not understand the underlying algebraic structure.

In fact, it is not even clear that OPEs exist in loop CFTs. The fact that 4-point functions are combinations of conformal blocks is a necessary condition for the existence of OPEs, but it is not sufficient. This condition may seem empty at first sight, because conformal blocks form a basis of functions of the cross-ratio. However, only blocks that correspond to channel fields in the loop CFTs' spectrum appear, and this makes the condition nontrivial. More fundamentally, we may try to deduce the existence of OPEs from the definition of statistical loop models. Technically, this might be done by constructing lattice loop models, and studying the algebraic properties of the resulting diagram algebras, see \cite{im25} for work in that direction.

Since correlation functions correspond to combinatorial maps, OPEs and factorization should be formulated in terms of combinatorial maps. For any $r_1,r_2,r_3\in \frac12\mathbb{N}$ such that $\sum r_i\in\mathbb{N}$ there exists exactly one 3-point combinatorial map on the sphere with vertices of valencies $2r_1,2r_2,2r_3$. But it is not clear how to reconstruct the finitely many 4-point maps for a given 4-point function $\left<\prod_{i=1}^4 V_{(r_i,s_i)}\right>$, from such 3-point maps.

If and when loop CFTs can be solved at generic central charge, there will remain the challenge of taking rational limits. Such limits are relevant to applications such as percolation ($\beta^2=\frac23$) or polymers ($\beta^2=\frac12$). These special cases are more difficult than the general case, because representations of the Virasoro algebra have more complicated structures, see for instance \cite{hs21}. Hints of these algebraic complications are visible in the fusion rules of degenerate representations, which are simpler for generic $\beta^2$ \eqref{rrrs} than for rational $\beta^2$ \eqref{rfrf}. In the case of a 4-point function, we know that conformal blocks diverge in the limit $\beta^2\to \frac{q}{p}\in\mathbb{Q}$, but we expect that the divergences cancel so that the 4-point function converges. When a linear combination of divergent blocks $\mathcal{G} = \sum_k D_k \mathcal{G}_k$ converges, the interpretation is that the corresponding modules of the conformal algebra combine into a bigger indecomposable module in the limit. Such phenomenons also occur in CFTs with 2 degenerate fields, in particular in the limit GMM\,$\to$\,AMM of Section \ref{sec:lod}.

\subsubsection{Global symmetries in $d=2$ and $d\geq 3$}

As we stated in Section \ref{sec:models}, the $O(n)$, $PSU(n)$ and Potts CFTs have global symmetries, described by the groups $O(n)$, $PSU(n)$ and $S_Q$ respectively (or their categories of representations if $n,Q$ are not integer). Usually, in physics, we would expect such unbroken symmetries to have a strong influence. Models with different symmetry groups would behave differently, and we would formulate our analysis in group-theoretic language: spectra as representations, correlation functions as invariant tensors.

However, the $O(n)$, $PSU(n)$ and Potts CFTs share many conformal dimensions and even some correlation functions.
This is because they are all formulated in terms of similar loop ensembles, see Section \ref{sec:coup}.
Moreover, if we decompose the spectrum of the $O(n)$ CFT into $O(n)$ representations, as we sketched in Eq. \eqref{von}, we find that a primary field $V_{(r,s)}$ with a given conformal dimension generally does not transform in an irreducible representation, but in a larger representation. This shows that the model's global symmetry is larger than $O(n)$. 

The origin of this larger symmetry is the requirement that loops do not intersect, as can be seen in the context of lattice models \cite{jrs22}. The lattice $O(n)$ model describes the dynamics of a periodic spin chain of length $L\in\mathbb{N}$, whose space of states is $\mathcal{H}_L=\left(\mathbb{R}^n\right)^{\otimes L}$. In the linear group $GL(\mathcal{H}_L)$, the commutant of $O(n)$ is the Brauer algebra $\mathcal{B}_L$, a diagram algebra where loops are allowed to intersect. Forbidding intersections, we would obtain a subalgebra of $\mathcal{B}_L$. The commutant of that subalgebra is larger than $O(n)$, and may be thought of as the global symmetry algebra of the lattice $O(n)$ model.

Invariant tensors of $O(n)$ can be represented in terms of loops that encode contractions of indices: for example, given 4 vectors $v_1,v_2,v_3,v_4\in \mathbb{R}^n$, we can build the 3 invariants 
\begin{align}
\begin{tikzpicture}[baseline=(base), scale = .4]
  \vertices
  \draw (0, 0) node [left]{$1$} -- (0, 3) node[left]{$2$};
  \draw (3, 0) node [right]{$4$} -- (3, 3) node[right]{$3$};
  \node at (1.5, -2) {$(v_1\cdot v_2)(v_3\cdot v_4)$};
 \end{tikzpicture}
 \hspace{2cm}
 \begin{tikzpicture}[baseline=(base), scale = .4]
  \vertices
 \draw (0, 3) -- (3, 3) ;
 \draw (0,0) -- (3,0);
 \node at (1.5, -2) {$(v_1\cdot v_4)(v_2\cdot v_3)$};
 \end{tikzpicture}
 \hspace{2cm}
 \begin{tikzpicture}[baseline=(base), scale = .4]
  \vertices
 \draw (0, 3) -- (3, 0) ;
  \draw (3, 3) -- (0, 0);
  \node at (1.5, -2) {$(v_1\cdot v_3)(v_2\cdot v_4)$};
 \end{tikzpicture}
\end{align}
where $(v_i\cdot v_j)$ is the scalar product of 2 elements of the vector representation $\mathbb{R}^n$ of $O(n)$, in other words the image of $v_i\otimes v_j$ under the equivariant map from $\mathbb{R}^n\otimes \mathbb{R}^n$ to the identity representation. In order to represent all invariant tensors, we must allow loops to intersect: the geometry of the plane where we draw the loops is not relevant to representation theory. Forbidding intersections amounts to eliminating the invariant tensors that violate the larger symmetry. 

When it comes to a 4-point function $\left<\prod_{i=1}^4 V_{(r_i,s_i)}\right>$, the number of 4-point invariant tensors is at least the number of solutions of crossing symmetry (= the number of combinatorial maps). If $\sum r_i = 2, 3$ these numbers coincide, while if $\sum r_i\geq 4$ the inequality becomes strict in most cases (for example, $6>5$ for $\big\langle V_{(\frac32,0)}^2V_{(\frac12,0)}^2\big\rangle$), and the number of solutions grows much slower with $r_i$ \cite{gjnrs23}. This slower growth is a manifestation of the larger symmetry at the level of 4-point functions.

In $d\geq 3$ dimensions, the $O(n)$ and $PSU(n)$ models can still be formulated in terms of loops, but
from a distance there is no way to tell whether 2 loops intersect or not. The case $d\geq 3$ is therefore similar to the case $d=2$ with intersecting loops, and we expect no larger symmetry. 
In the critical $O(n)$ model for $d\geq 3$ or $d=2$ with intersecting loops, we expect that a puncture with a given valency $2r$ gives rise to the same fields as in the $O(n)$ CFT, transforming in the same irreducible $O(n)$ representations. However, all their conformal dimensions are modified, due to the breaking of the larger global symmetry, the breaking of interchiral symmetry, and if $d\geq 3$ the breaking of local conformal symmetry down to global conformal symmetry. As a result, for any conformal dimension that appears in the spectrum, we expect 1 primary field that transforms in an irreducible representation of $O(n)$. If we write a 4-point function as a linear combination of 4-point invariant tensors, the coefficients are all solutions of crossing symmetry, and have no reason to be linearly dependent. Therefore, we expect as many solutions of crossing symmetry as 4-point invariant tensors. 

In the $d\geq 3$ Potts model, the boundaries of clusters are not loops, but objects of dimension $d-1\geq 2$. Therefore, we do not expect the critical Potts model to share nontrivial conformal dimensions with the critical $O(n)$ and $PSU(n)$ models.


\bibliographystyle{SciPost_bibstyle}
\bibliography{cft}

\pagebreak

\printindex

\end{document}